\documentclass{ucalgthes1}
\usepackage[letterpaper,top=1in, bottom= 1in, left= 1in, right= 1in]{geometry}
\usepackage{hyperref}
\usepackage{graphicx}
\usepackage{subfig}
\usepackage{verbatim} 
\usepackage{amsmath}
\usepackage{cite}
\usepackage{epigraph}
\newcommand{\noi}{\noindent}
\newcommand{\beq}{\begin{equation}}
\newcommand{\eeq}{\end{equation}}

\title{Dynamics of Plant Growth;\\ \bigskip A Theory Based on Riemannian Geometry }
\author{Julia Pulwicki}
\thesisyear{2015}
\thesis{thesis}    

\monthname{DECEMBER}
\dept{PHYSICS AND ASTRONOMY}
\degree{DOCTOR OF PHILOSOPHY}
\begin{document}
\makethesistitle
\pagenumbering{roman}     
\setcounter{page}{1}


\phantomsection
\altchapter{\bf{Abstract}}
In this work, a new model for macroscopic plant tissue growth based on dynamical Riemannian 
geometry is presented. 
We treat 1D and 2D tissues as continuous, deformable, growing geometries for sizes larger than 1mm.
The dynamics of the growing tissue are described by a set of coupled tensor equations in non-Euclidean (curved) 
space. These coupled equations represent a novel feedback mechanism between growth and curvature dynamics. 

For 1D growth, numerical simulations are compared to 
two measures of root growth. First, modular growth along the simulated root shows an elongation
zone common to many species of plant roots. Second, the relative elemental growth rate (REGR)
calculated \textit{in silico} exhibits temporal dynamics recently characterized in high-resolution root growth
studies but which thus far lack a biological hypothesis to explain them. Namely, the REGR can evolve from a single
peak localized near the root tip to a double-peak structure. In our model, this is a direct consequence of considering
growth as both a geometric reaction-diffusion process and expansion due to a distributed source of new materials.

In 2D, we study a circularly symmetric growing disk with emergent negative curvatures. These
results are compared against thin disk experiments, which are a proxy model for plant leaves. These 
results also apply to the curvature evolution and the inhomogeneous growth pattern of the \textit{Acetabularia} cap.  
Lastly, we extend the model to anisotropic disks and predict the growth dynamics for a 2D curved surface 
which develops an elongated shape with localized ruffling.

Our model also provides several measures of the dynamics of tissue growth. These include the time evolution
of the metric and velocity field, which are dynamical variables in the model, as well as expansion, shear and rotation 
which are deformation tensors characterizing the growth of the tissue. The latter are physically measurable quantities that
remain to be fully explored using modern tissue growth imagining techniques. 
\newpage
\phantomsection
\altchapter{\bf{Acknowledgments}}
Creating this work would not have been possible without the generous support from many sources.
Early discussions with Dr. Przemys\l{}aw Prusinkiewicz and Dr. Maya Paczuski as well as the students in their
groups (Dr. Adam Runions, Dr. Elliot Martin, Dr. Jacob Foster, Dr. Andrew Berdahl), 
laid out key ideas about geometry in plant development. 

Financial support came from the Natural Sciences and Engineering Research Council of Canada (NSERC) 
and from Alberta Innovates Technology Futures (AITF). 

The opportunity to learn more about condensed
matter physics and plant biology from experts in these fields came 
through two summer schools: the Boulder School for Condensed Matter and Materials Physics 2011 whose theme was
hydrodynamics in biological problems, and the Les Houches summer program in 2014 which focused on
integrating knowledge from a variety of biological specialties, from virus to ecosystem scales.

A huge thanks goes to my supervisor Dr. David Hobill, whose expertise 
in mathematics and physics made this framework tractable in the first place. 
Over the years, his guidance allowed for many exciting insights that all stemmed from the
unique idea of combining cosmology with plant biology. The experience of doing this kind of science has 
been immensely engaging and satisfying, and has spurred many new questions to pursue in the future.

Last, but certainly not least, this project would not have been possible without the support
of my family, friends and the Department of Physics and Astronomy at the University of Calgary. Each
of these communities played a part in helping me stay focused and balanced throughout the years of work it took to
complete this project.
\newpage
\epigraph{Discovery, like surprise, favours the well-prepared mind. Discovery is in its essence a matter of 
rearranging or transforming evidence in such a way that one is enabled to go beyond the evidence
so reassembled to new insights. It may well be that an additional fact or shred of evidence
makes this larger transformation possible. But it is often not even dependent on new information.}
{- J. Bruner}
\newpage
\begin{singlespace}
\newpage
\phantomsection
\tableofcontents
\pagestyle{plain}
\newpage
\phantomsection
\listoffigures
\pagestyle{plain}
\clearpage
\clearpage          
\end{singlespace}
\newpage
\phantomsection
\chapter*{\bf{List of Symbols, Abbreviations and Nomenclature}\hfill} \addcontentsline{toc}{chapter}{List of Symbols}
\listofsymbols
\pagestyle{plain}
\clearpage

\pagenumbering{arabic}

\chapter{Introduction}
\label{chapterIntro}

Biological systems operate across an astounding breadth of physical scales, from quantum and molecular processes at 
nanometer scales ~\cite{quantumPhotosynthesis} to architectures that grow to tens of meters in size. Over the past 60 years, 
the focus has been strongly biased toward molecular and cellular levels of understanding biology. However, important exceptions 
to this trend are the works of D'Arcy Thompson ~\cite{thompson1917} 
and Alan Turing ~\cite{turing1952} who, in different ways, examined mathematical patterns on a whole-system scale.

The goal of the work presented here is to establish a macroscopic theory of plant leaf growth. We start with two very 
simple observations about leaves and petals. The first is that many types of leaves and petals are not flat; by this
we mean these tissues have intrinsic curvature that cannot be laid flat without either tearing the tissue 
(as in the case of flattening an orange
peel), or overlapping of the tissue (as for a ruffled lettuce leaf). The second observation is that leaves and 
petals grow from millimeter to centimeter or meter scales apparently continuously, both in space and time. As will
be seen, these facts lead to remarkable implications for the mathematics and physics of the system. Furthermore, such
a model of macroscopic growth leads us to the hypothesis that feedback between growth and curvature is essential to the
development of plant structures with highly regular patterns of curvature.

\section{Biological motivation}
\label{sectionBioMotivation}

Several well studied, though relatively recent, biological arguments exist to motivate a macroscopic and 
continuum approach to plant leaf development. These are discussed below.

\subsection{Mechanotransduction in tissues}

Mounting evidence supports the idea that physical forces produce a measurable effect at all 
scales during tissue development. In experimental biology, this effect is known as 
\textit{mechanotransduction}, which can be summarized as the cascade of reactions a tissue has to a physical
force starting at the macroscopic scale down to the cellular, sub-cellular and genetic scales. 

A striking example of this is given in ~\cite{braam2005}, where the genetic response
of plants to touch has been reviewed. In one study, two groups of the same plant were grown. One group was
left to grow without any external mechanical stimuli. The other group was touched twice per day by the experimenter
who gently brushed the top of the plants with the palm of their hand. The resulting plants were markedly 
different; the unstimulated plants grew tall, while the touched plants were significantly shorter. Furthermore, the
stunted growth of the touched plants was correlated with the expression of an extended set of TOUCH genes, fully
2.5\% of the plant's genome.

At the sub-cellular level, responses to mechanical stress have also been seen in microtubule arrangement. 
Microtubules are micrometer long filaments inside the plant cell that, among other functions,
help the cell keep its shape. The studies reviewed in ~\cite{hamant2013microtubles} indicate that 
microtubule arrangement within plant cells is dynamically coordinated by the stresses experienced at the tissue scale.
The key observation in these studies is that when an external stress is put on a portion of plant tissue, 
the microtubules within each cell will realign to be 'parallel to the direction of maximal stress'.

Another candidate for a mechanosensing response can be seen in \textit{Drosophila} (fruit fly)
wing disks ~\cite{hufnagelWingDisk}. When the wing disk reaches maturity, cells will stop growing in unison 
throughout the disk. The lack of a time-dependent chemical signal (morphogen) and a nearly instantaneous response
of all the wing disk cells strongly suggests a mechanically mediated response to growth induced stress. 

Although the previous example involves animal cells, it is not unreasonable to think that a similar process could occur 
in plants.
Indeed, it has been observed in both plant and animal studies that organ size can be independent of the number
of cells in the organ ~\cite{sugimoto2003big} ~\cite{Day2000organSizeControl}. In other words, there is some 
tissue-scale mechanism than governs growth in addition to cellular level
processes. One possibility is that this regulation occurs as a result of mechanical stress. 

In all of the above examples, the most important feature of how forces affect development is that 
an organism's response to stress is direct and nearly instantaneous. That is to say, the response to 
a force is not mediated in the same way as other signaling pathways in a plant, namely the relatively slow processes of 
diffusion or active transport of a signaling molecule from one cell to another. 
Therefore, physics must play a direct and key role 
in the development of all kinds of tissues: embryos, organs, leaves, petals, roots. 
Indeed, biological systems live in a physical world, which means they must have mechanisms to respond
and adapt to the physical constraints and stimuli around them. The study of these pathways is an active field of 
research in both plant and animal systems ~\cite{hamant2013widespread}.

One example of this physics-based approach in biological modeling is the work of Green ~\cite{Green1996Phyllotactic-pa} 
~\cite{green1996transductions}. To understand the patterning of primordia 
on the meristem (phyllotaxis), Green analyses the mechanics of deformable, constrained disk and annulus geometries using a 
thin-disk approximation. The patterns of deformations, specifically peaks and troughs, allowed by these geometries are strikingly similar to
the ones seen on the meristem, suggesting that the mechanics and geometry of plant structures play a key
role in morphogenesis.

However, the exact nature of how mechanical signals propagate through tissues is not known. 
This is because plant tissues can be considered 'soft matter', in the same category as
gels and films whose internal structure accommodates dynamics somewhere in between rigid matter
and fluids. Indeed, it is the nonlinear viscoelasticity of soft matter that makes characterizing dynamics
from first principles so difficult. Often, phenomenological approaches combined with experimental
data are needed to understand these types of materials. The study of plant tissues from the perspective
of soft matter is only beginning to take shape within the scientific community. ~\cite{softMatterBook}
~\cite{instabilityOfGels}

\subsection{Plants as continuous structures}

In this section, we aim to motivate the study of plant leaves at scales larger than 1mm as 
continuous thin sheets using well known facts about plant leaves. More discussion on established work
using this paradigm will be found in Sections \ref{sectionSE} and \ref{sectionSE2}.

First, a simple comparison of the size of cells to the scale of curvature patterning suggests a continuum
approximation of the leaf surface is appropriate. In the model plant \textit{Arabidopsis}, leaf cells grow to about $1000 \mu$m$^2$ 
in size ~\cite{cellSizesDonnelly}. 
The resolution of the curvature patterns we seek to study (in particular, buckling and ruffling) 
is on the order of $1$mm$^2$, or $10^6 \mu$m$^2$~\cite{bucklingCascades}. 
Therefore, curvature patterns are collections of many cells,
which indicates that a continuum approximation is an appropriate
tool for analyzing the problem of macroscopic plant growth.

Another argument for a continuum approach is that the patterns of ruffling and buckling seen
at the millimeter and larger scale are continuous. The surfaces of even highly curved leaves are locally
smooth and connected. This suggests that leaves can be abstracted as  
differentiable manifolds, allowing for the use of powerful mathematical methods to provide a 
precise description of the system being studied. In particular, the tools from differential geometry
will allow the intrinsic curvature of the tissues to be well understood and incorporated into the growth
of the tissue over time.

Another useful fact in constructing our model is that macroscopic leaf growth is dominated by cell
expansion, not cell division ~\cite{rootsBook} ~\cite{exitFromProliferation}. Also of note is that cell expansion 
is not uniform across space or time. This process is not well understood and varies from species
to species. Furthermore, plant cells have fixed neighbours and cannot move relative to one another, so 
each cell is like a point on an expanding geometry, with distances between points growing 
and deforming over time. 

Lastly, cells in plant leaves are highly connected to each other via plasmodesmata. These 
'windows' between cells facilitate fast information exchange between cells, including the transfer
of large molecules such as proteins, RNA and DNA. Each cell has many such links to its 
neighbours ~\cite{taiz}. This high interconnectedness suggests that tissue and system scale communication
is efficient and evolutionarily favoured.  

Indeed, all of these arguments fall in the purview of an organismal theory of plants, which the botanist
Julius von Sachs summarizes as 'The plant forms cells, the cells do not form plants.' ~\cite{raven}

\subsection{Biological processes are feedback loops}

In recent years, the view that biological processes are linear, with a clear cause-and-effect behaviour,
has been replaced with a better understanding of the complexity of feedback mechanisms. These mechanisms operate at 
many different scales, and may interact across scales. 

Gene regulatory networks are widely known to be complicated feedback loops that regulate gene expression in cells. 
They are represented visually as a network of mRNA and protein nodes
with multiple links, but can also be described mathematically in terms of a set of coupled ordinary differential equations
~\cite{GRNmodeling}.

When coupled with chemical signaling across the organism, gene regulatory networks can be triggered 
in a way that supports pattern formation on the system scale ~\cite{monkGR}. Information about the state of 
a cell's neighbouring region can be transmitted by the diffusion of a chemical signal that can then 
activate a genetic pathway within the cell. The cell can pass on information about its new state using
another chemical signal, thereby affecting the development of its neighbours.

A specific example that illustrates this point is the self-organization of PIN transporters in plant 
meristems ~\cite{godin2008}.
In order to grow, a plant cell must be exposed to the growth hormone auxin. Auxin is transported
from cell to cell with the help of the transporter protein PIN. \textit{In silico} modelization of the 
meristem shows that if PIN localization within the cell responds to the flux of auxin through the cell, PIN 
transporters will dynamically align (polarize) to facilitate more auxin flow to an emerging organ on 
the meristem. This self-organization also leads to regular patterning of organ primordia on the meristem
(phyllotaxy). Hence, the flow of auxin changes the structure of each cell in the meristem to cause a 
positive feedback loop that supports large-scale tissue growth and patterning.

At the system level, a very important feedback loop was hypothesized by Turing in 1952 ~\cite{turing1952} and 
recently confirmed experimentally ~\cite{turingTest}. Turing's revolutionary idea was that large scale pattern formation in 
biological systems is a result of reaction-diffusion processes of chemicals called morphogens. Turing's original
paper was largely concerned with the dynamics of linear chemical reactions on a ring of discrete permeable cells. 
This model successfully predicted
the classes of pattern formation seen in the recent experimental study. 

The linear models are, however,
only an approximation of the much more complex nonlinear dynamics seen in experiment. Indeed, Turing 
acknowledged that nonlinear couplings would be more realistic for modeling biological systems, but these
would require numerical simulations that surpassed the computing power available at the time. Today, phase
diagrams produced from linear and nonlinear models show just how different the dynamics are between the
approximate model and a more realistic, nonlinear one ~\cite{turingTest}.

Lastly, of particular interest in the context of forces and feedback loops, is recent experimental evidence that 
mechanotransduction pathways are evolutionarily ancient and shared across what are now  
distantly related species.  The same mechanical mechanism for mesoderm 
specification has been shown to occur in both zebrafish and \textit{Drosophila}, which last shared a common
ancestor some 570 million years ago ~\cite{brunet2013}. What mechanotransduction pathways exist in plants remains an 
open question ~\cite{feelingGreen} ~\cite{PrusinkiewiczConstraints}.

\subsection{Patterning as a reaction-diffusion process}

Reaction-diffusion dynamics occur in many natural systems including those in chemistry and biology. The resulting phenomena 
are identified as having, among other behaviours, self-organizing spatial and temporal patterns emerging from 
a random initial state. Mathematically, the system's dynamics must have both a diffusive and reactive term:

\begin{equation}
\partial_t \mathbf{u} = \mathbf{D} \nabla^2 \mathbf{u} + \mathbf{R}(\mathbf{u}),
\label{eqnRD}
\end{equation}

\noi where $\mathbf{u}$ is a N-dimensional dynamical variable in space and time, $\mathbf{D}$ is a matrix of diffusion coefficients 
and $\mathbf{R}$ is a real vector functional
of $\mathbf{u}$ that may have many terms including nonlinearities and differential operators. We define Equation \ref{eqnRD} to be 
\textit{quasi-linear} if $D = D(\mathbf{u})$ and $\mathbf{R} = \mathbf{R} (\mathbf{u}, \nabla \mathbf{u})$.

Turing's work on morphogens falls into this class of equations, as do many studies of pattern formation
in biology ~\cite{RDinBio}. Models for coupling reaction-diffusion chemical dynamics with growth patterns on the
plant meristem and other tip growth structures have been studied by Harrison 
and Holloway ~\cite{harrison2012dynamic}. In the context of plant morphogenesis, the notion of reaction-diffusion dynamics is particularly
relevant because of the highly regular patterns of curvature seen in many leaves and petals. Indeed, the structure
of reaction-diffusion equations will guide the development of a set of equations  
describing plant growth (Chapter \ref{chapterEquations}).

\section{A note on modeling in biology}

Biological disciplines today are dominated by observation, with theory playing a rather secondary role.
In contrast, physics tends to place a more balanced emphasis on theory and experiment, so that the two drive 
each other to an ever more rigorous understanding of the world around us.

Often observations of biological systems are overwhelmingly rich and complicated, reflecting the inherent complexity
of life. The value of constructing a model for such a system is therefore to focus our attention on
the salient features of a certain problem. When we construct a framework that simplifies the system, 
we can then test the assumptions of that framework in the form of predictions.

Modeling a biological system can help guide experiments to be performed on a complicated system. Rather than
starting with all the details in the hopes of sifting out the important parts, we can hypothesize what those
important parts may be and then identify key processes occurring in vastly more complicated biological system.

A hugely successful example of this is Turing's model of morphogens ~\cite{turing1952}. By postulating the existence of chemical
messengers and how they interact, he constructed a theory that explained how large-scale patterns arise 
in biological systems. It was only decades later that his theory was fully verified by experiment ~\cite{turingTest}.

In the model of plant growth presented here, we make the simple observations that plant leaves grow and exhibit 
intrinsic curvature (recall how neither an orange peel nor a lettuce leaf can be laid flat without tearing or folding). 
To understand these shapes in a mathematically rigorous way requires certain tools from differential 
geometry. As it happens, these are the same tools used to understand the nature of curved spacetime 
first described by Einstein in his theory of general relativity. Because of this, there will be parallels between
the two systems - systems that are both geometric and dynamic. This was a huge shift in perspective 
for the science of cosmology a century ago, and hopefully dynamic geometry will shed some new light on plant growth as well.

\section{Established work}
\label{sectionEstablishedWork}

\subsection{Plant growth as a continuous process in space and time}
\label{sectionSE}

One of the earliest works on treating growing tissues as continuous media was done by
Silk and Erickson in 1979 ~\cite{silk1979}. In it they argue for the use of continuum mechanics in
studying plant growth, noting that certain structures such as the meristem are defined by 
their geometry and position on the plant, rather than a fixed set of cells. They also
use the fact that growth curves are time-continuous to motivate both temporal and 
spatial continuity at millimeter and larger length scales.

\subsection{Material transport causes leaf growth}
\label{sectionSE2}

A very simplified but sufficient view of how leaf development occurs at scales larger than 1mm is
that material from the stem is transported to the leaf, which enables cells in the leaf to expand. 
Material is also created locally in cells by transforming sunlight into carbohydrates (among other local
processes) and can be transported to neighbouring cells in a variety of ways.
In the continuum limit, this is enough to allow material transport in the leaf to be expressed as 
a conservation of mass equation with an associated source term.

Continuous material transport in the leaf also allows us to use the equations of viscous fluid flow to describe
the velocity field of a growing leaf ~\cite{silk1979}. Just as the boundary of the leaf expands with some velocity,
so too does every point inside the boundary. This is the velocity field of the leaf. The velocity field can
be thought of as having two contributing effects: one is the diffusion of material between cells through
a number of molecular transport pathways, and the other is the how material flow is altered by the expansion 
of the tissue itself.

\subsection{Stress, strain and growth are tensor quantities}
\label{sectionGrowthWithTensors}

\begin{figure}[ht]
    \centering
    \includegraphics[width=0.5\textwidth]{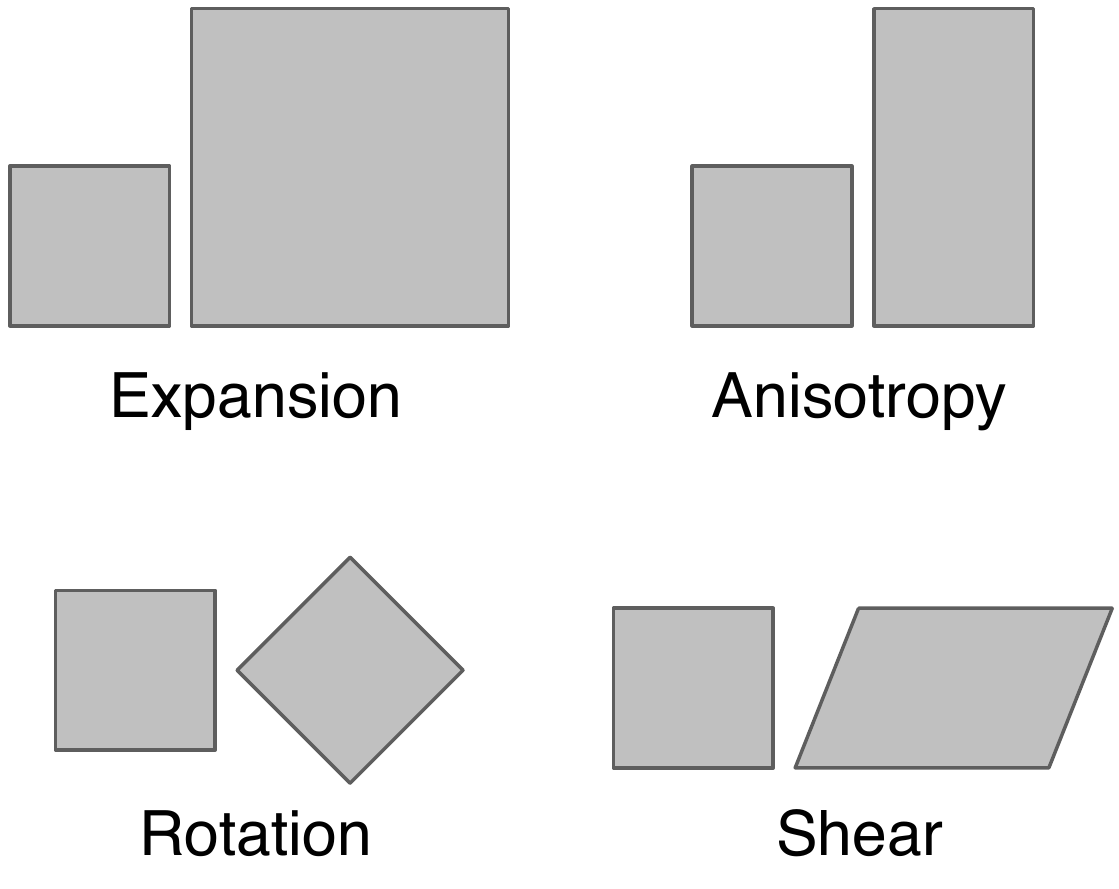}
    \caption{Some examples of rank-two tensor quantities.}\label{figureTensors}
\end{figure}

As early as 1984, Hejnowicz and Romberger discussed how tensors are the only mathematical objects
to fully describe the physical deformations experienced by a growing tissue: expansion,
rotation and shear ~\cite{hejnowicz1984} (see Figure \ref{figureTensors}). In addition, geometric 
effects such as anisotropy and inhomogeneity can also be considered as results of growth.

Indeed, tensors are used to describe the properties of both physical objects and geometry. Tensors can be
as simple as a single number, like the temperature at a specific point in a room. These are zero-rank tensors
which are also known as scalars.
A tensor with information about magnitude and a single direction is a first-rank tensor, or vector. Higher rank tensors can express 
physical quantities with multi-directional properties like stress, strain and growth or geometric quantities like curvature.
They are a very powerful way of encoding information about a set of points in a multidimensional space and are used 
in disciplines like fluid mechanics, elasticity theory and general relativity.

One example where tensors have been used to better understand a biological system is given in ~\cite{rootGTnakielski}.
Here, a time-independent growth tensor in curvilinear coordinates ~\cite{rootApexGTcoords} is used to model the 
tip of a radish root. This information is then used to generate the principle directions of growth 
at each cell, which dictate how the cells will divide and grow. While it was long postulated that 
cells divide along principle directions of growth, 
it was unclear how the cells coordinate these directions ~\cite{PDGinSAM}. By using a growth tensor, 
this model is able to 
build an algorithm that links a tissue-level property with cellular division and growth.

Another tensor based approach to biological growth is to consider elastic (reversible) deformations due to stress
as separate from the plastic (irreversible) deformations of growth itself ~\cite{stressDecomp}.
Both types of deformations are described using tensor quantities, but have very different
time scales over which they act, allowing a multiplicative decomposition of the two effects.
This approach has been successfully applied to simple growth modes such as constant isotropic growth,
but the computational complexity of this algorithm for more realistic scenarios has prevented these models 
from developing further ~\cite{SIAMreview}.

\subsection{Curved surfaces are a result of nonhomogeneous growth}

More recently, there is also strong empirical evidence showing that nonhomogeneous growth leads 
to curvature. 

In ~\cite{thinSheetShapes}, a thin disk of gel is treated with a heat-sensitive monomer whose concentration 
varies with radius. When heat is applied to the disk, the disk contracts proportionally to how much monomer
is present at a given radius. Thus, the disk can be manipulated to contract more along the edges than
at the center, causing a dome shape, or the contraction can occur more in the center than at the edges, 
causing ruffles along the edges. The amount of contraction is a physical representation of the geometry
of the disk, and thus the link between nonhomogeneous growth and intrinsic curvature can be studied.
These results will play a key role in interpreting the 2D geometries of the model developed here 
(Chapter \ref{chapter2DConstrainedSimulations}).

Other experiments on torn plastic sheets show similar results. When a plastic sheet is put under tension
and then torn in the direction perpendicular to the direction of tension, the result is ruffled edges
along the tear. If one then measures lengths along different directions on the sheet, it can be seen 
that the torn (ruffled) edge is longer than 
the flat edge ~\cite{leavesFlowersPlasticRuffling}. Moreover, the ruffles form a fractal cascade over several
orders of magnitude ~\cite{bucklingCascades} ~\cite{wrinklingByGeometry} which can be understood in terms
of minimizing the bending energy of the long edge under geometrical and elastic constraints ~\cite{AudolyBoudaoud}.
A similar analysis for a disk geometry where anisotropic growth results in curvature is discussed 
in ~\cite{anisotropicDiskGrowth}. Intuitively, these models show that buckling will always occur in systems 
where there is an excess of material along one edge that, for geometric reasons, cannot spread out to be flat.
For example, a lettuce leaf and a ballerina's skirt both have this property; if one tries to flatten a 
portion of the leaf or skirt, the rest becomes more buckled. The excess material along the outer edges 
cannot relax to a flat geometry, so it folds up in a way that distributes the curvature as evenly as possible
across the whole surface.
How the buckling manifests itself, whether through short, long or fractal buckling, depends on the material 
itself, in particular its elasticity and thickness. Given two sheets of equal thickness and equal excess
edge length, but different elasticity (for example, one made of 
tissue paper and the other of cardboard), the more elastic one will be easier to bend, and so produce 
more short-wavelength, low amplitude buckling than the less elastic one. Similarly, the thickness of the sheet will influence
how easy it is to bend the material; a thicker sheet is more difficult to bend, so it will tend to have
more long-wavelength buckling with higher amplitudes.

The idea behind these experiments is to correlate the final shape of a manipulated system (thin sheets) to
the shapes found in nature, in particular plant leaves and petals. Indeed, based on these experiments, it 
is obvious that intrinsic curvature is the result of nonhomogeneous growth processes. 

Nonhomogeneous growth has been studied from a genetic perspective ~\cite{geneticsOfGeometry}
~\cite{genesAndShape} ~\cite{genesAndCurvature}, and models for how genetically programmed growth rates affect 
tissue shape and curvature have been developed ~\cite{genesAndGeoModel} ~\cite{kuchen2012generation}. These authors postulate that 
growth processes and the stresses they produce could influence the genetic expression of the tissue 
through mechanotransduction, however no model has incorporated this feedback loop yet. Additionally,
these models of genetic control of curvature do not take into account how curvature effects react back
onto the growth mechanism.

\begin{figure}[h]
    \centering
    \includegraphics[width=0.9\textwidth]{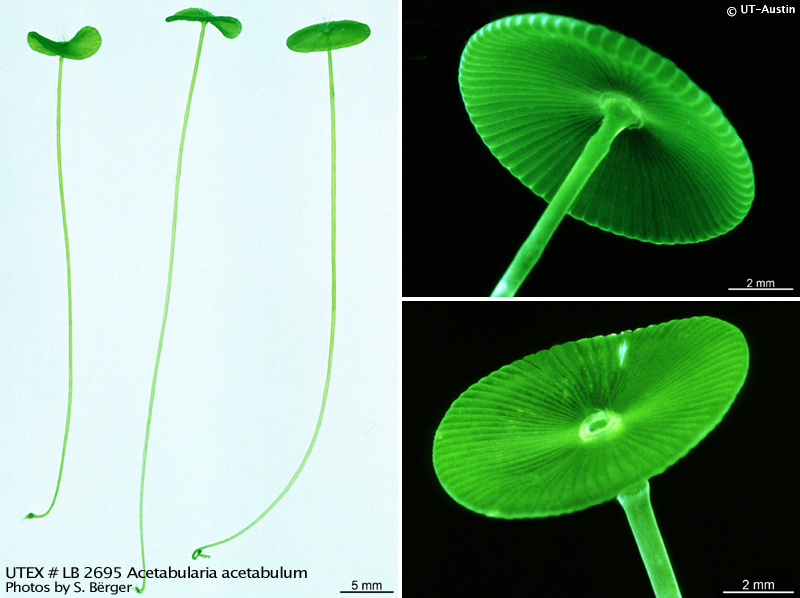}
    \caption{The algae \textit{Acetabularia} showing caps with different curvatures. With permission from UTEX Culture Collection of Algae, University of Texas, Austin.}\label{figureAcetaPhoto} 
\end{figure} 

A relevant model organism for studying the development of curvature is the single-celled algae \textit{Acetabularia}.
Despite being single-celled, it grows to centimeter length scales and has a cap that changes its curvature as it grows
(See Figure \ref{figureAcetaPhoto}). Tracking the growth of the cap has also revealed that the cap grows mostly in
an annular region around the main stalk, indicating highly nonhomogeneous growth processes at work  ~\cite{acetabularia}.
These characteristics will be revisited in Chapter \ref{chapter2DConstrainedSimulations} as we seek to understand the
dynamics of our simulated growth system.

\section{What is needed for a better model?}

The elements of continuity, material transport and nonhomogeneous growth are all important first steps in understanding
plant growth dynamics, but are incomplete without considering the physical effects of curved
geometries. Silk and Erickson consider only flat geometries for their continuum model, which
we show omits very important dynamics found only in systems with curvature.

The idea of using tensors to describe the geometry, growth, velocity field and scalar density of the plant leaf
is fully embraced in our model, and as will be seen in Chapter \ref{chapterEquations}, provides the basis
of a self-consistent physical model of plant growth. However, rather than being merely descriptive, tensor 
quantities in our model will be time dependent and evolve according to the structure of the 
dynamical equations of plant growth.

The thin-sheet experiments correlate the final shape of a plastic sheet with the shapes found
in plant leaves and petals, but do not encompass the growth process itself. 
Indeed, models developed to describe thin sheet experiments do not address the feedback that a changing
geometry imposes on the dynamics of the physical processes creating the curvature. In our model, this two-way 
feedback between growth and geometry is explicit and results in predictions on the large scale growth
dynamics of plant tissues.

\section{Chapter summary: \\ Geometry as the missing link}

\begin{figure}
    \centering
    \includegraphics[width=0.75\textwidth]{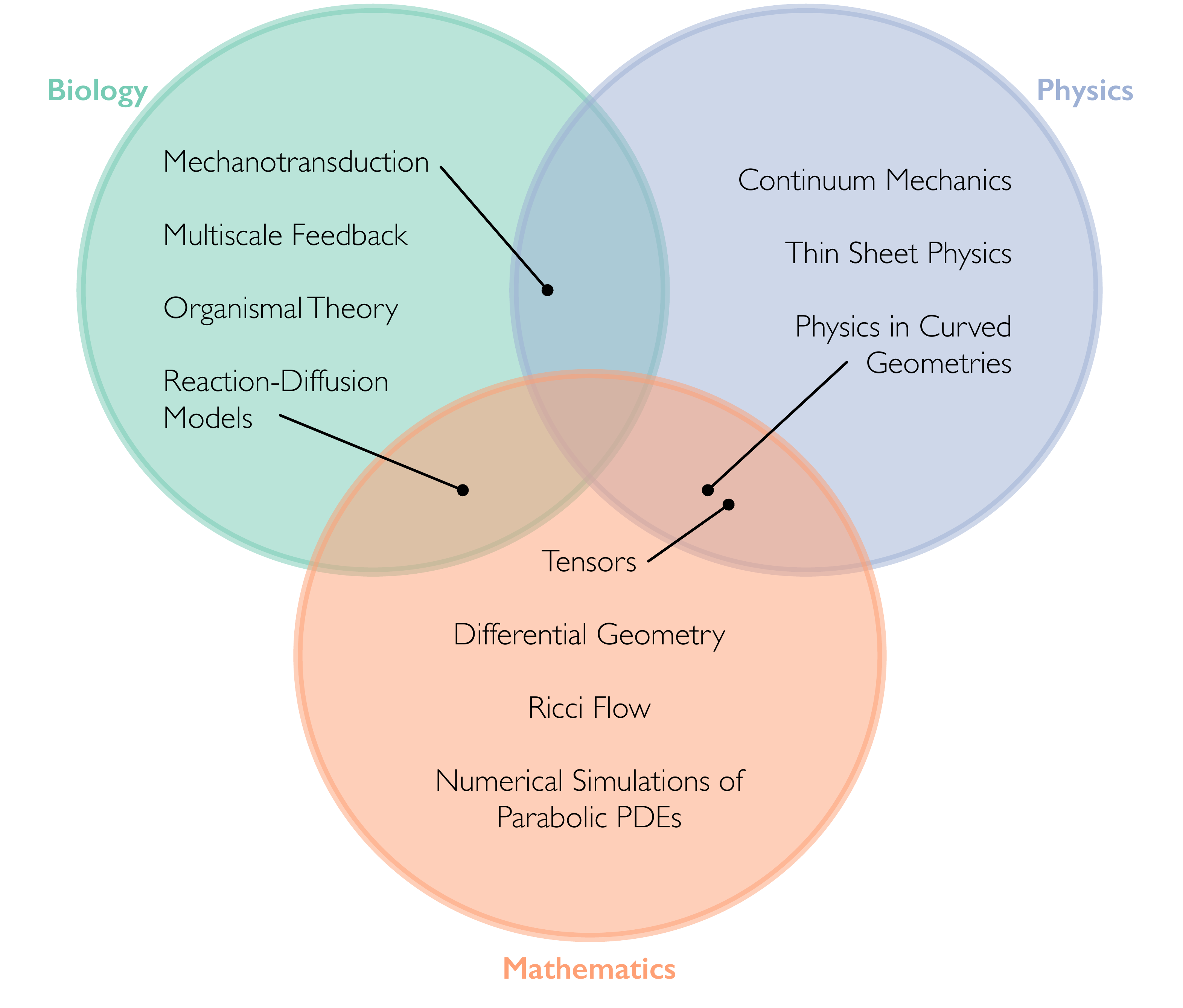}
    \caption{A conceptual representation of the ingredients for a physical model of plant leaf growth.}
    \label{figureVennTopics}
\end{figure}

As seen in this chapter, there is strong evidence from both biology and thin-sheet physics that 
tissue curvature in plant leaves and petals is a rich interplay of many factors, as depicted in Figure \ref{figureVennTopics}. 
To model this curved, continuous system requires specialized tools in mathematics and physics, which will be discussed 
in Chapter \ref{chapterEquations}. The numerical methods used to study the equations of plant growth will be 
discussed in Chapter \ref{chapterNumericalMethods}, and results for 1D and 2D models will be presented 
in Chapters \ref{chapter1DSimulations}, \ref{chapter2DConstrainedSimulations}, \ref{chapter2DFullSimulations}. 
Finally the implications of the numerical simulations and future directions of the work will be considered 
in Chapter \ref{chapterConclusions}.

\chapter{Developing a Physical Model}
\label{chapterEquations}

In this chapter, the physics and mathematics of our model are developed. 
Each assumption made about the physics of the plant leaf system provides a guide to the
appropriate mathematical tools needed to study it. Certain crucial facts about the system 
also aid in this process. The goal is to create a minimal set 
of assumptions which will then generate a realistic model of leaf growth dynamics.

\section{Tensors in physical systems}
\label{sectionTensors}

A fundamental property of tensors is that the quantities they relate through equations are 
coordinate independent, and they can be transformed between coordinate systems ~\cite{dinverno}. 
This is an important feature when constructing any physical theory. No matter what coordinates the 
system 'lives in', the relationships among the tensor quantities must remain the same. For the plant 
leaf model, this allows us the 
freedom to write down the tensor equations representing the physics of the system without worrying 
about how to represent the tensor components in the chosen space.

To illustrate the point, we can think of tracking the growth of the leaf by drawing small dots on 
the surface and measuring how each dot moves over time. The distance between dots tells us about
the geometry of the leaf. The velocity of each dot (both speed and direction) gives information 
about how the leaf is growing. 
Clearly, where we draw the dots does not impact the dynamics of the leaf. 
The dots are merely a coordinate system that helps us observe growth. There may be 
very convenient arrangements of dots that immediately bring to light key aspects of the physics 
occurring in the leaf. Other coordinates may obscure our understanding.

A corollary of the fact that tensors can be transformed between coordinate systems is that 
tensor equations must equate only tensors of equal \textit{valence}, and must have the same units. 
The valence $(p,q)$ denotes how many contravariant indices $p$ and covariant indices
$q$ a tensor has. For example, the tensor $A_j^i$ is a rank 2 tensor with valence $(1,1)$; $A_{ij}$
has valence $(0,2)$; $A^{ij}$ has valence $(2,0)$.

Transforming tensors between coordinate systems follows well defined rules. The simplest 
transformation is for scalar (rank 0) quantities, which stay the same in different coordinate systems:

\begin{equation}
\phi'(x'^1,x'^2,...,x'^n) = \phi(x^1,x^2,...,x^n).
\end{equation}

In an n-dimensional
differentiable manifold, we can define a contravariant vector 
quantity $A^i$ in the $x^i$ coordinates ($i = 1,2,...,n$). If we transform from the $x^i$ 
coordinates to another coordinate system, say $x'^i$, then the tensor $A^i$ also transforms
to $A'^i$:

\begin{equation}
A'^i = \sum_{j=1}^n \frac{\partial x'^i}{\partial x^j} A^j.
\end{equation}

The $\partial x'^i / \partial x^j$ term forms the $n \times n$ \textit{transformation matrix}, 
or \textit{Jacobian} that takes the $x^i$ coordinates to the $x'^i$ coordinates.

In practice, \textit{Einstein summation notation} is used to reduce the clutter in such equations. 
Any repeated index is understood to indicate a sum over $n$:

\begin{equation}
A'^i = \frac{\partial x'^i}{\partial x^j} A^j = \frac{\partial x'^i}{\partial x^b} A^b
\end{equation}

Higher rank tensors can be constructed out of covariant and contravariant 
tensors, and transformation rules can be written down for any tensor. Some relevant examples include:

Covariant tensor transformations:
\begin{equation}
A'_i = \frac{\partial x^j}{\partial x'^i} A_j.
\end{equation}

Second rank tensor transformations:
\begin{equation}
A'^{ij} = \frac{\partial x'^i}{\partial x^k} \frac{\partial x'^j}{\partial x^l} A^{kl}. 
\end{equation}

\begin{equation}
A'_{ij} = \frac{\partial x^k}{\partial x'^i} \frac{\partial x^l}{\partial x'^j} A_{kl}.
\end{equation}

Mixed rank tensor transformations:
\begin{equation}
A'^j_i = \frac{\partial x^k}{\partial x'^i} \frac{\partial x'^j}{\partial x^l} A_k^l.
\end{equation}

One mathematical aspect that becomes very important is how to do differential calculus
using tensor quantities. Let us first take a look at the partial derivative of a contravariant tensor in 
two different coordinate systems:

\begin{eqnarray}
\frac{\partial A'^i} {\partial x'^k} & = & \frac{\partial} {\partial x'^k} \left( \frac{\partial x'^i} {\partial x^j} A^j  \right) \nonumber \\
  & = & \frac{\partial x^l} {\partial x'^k} \frac{\partial} {\partial x^l} \left( \frac{\partial x'^i} {\partial x^j} A^j  \right) \nonumber \\
  & = & \frac{\partial x'^i} {\partial x^j} \frac{\partial x^l} {\partial x'^k} \frac{\partial A^j} {\partial x^l}
         + \frac{\partial^2 x'^i} {\partial x^j \partial x^l} \frac{\partial x^l} {\partial x'^k} A^j
\end{eqnarray}

The standard transformation rule that worked before now yields a non-tensorial quantity, namely the
second term on the right hand side of the equation. This presents a problem since partial
derivatives form the basis of more complex differential operators, such as the gradient, divergence, etc. If partial
derivatives of tensors cannot transform linearly between coordinate systems, then physics would act differently
in different coordinate systems.

To remedy the situation, we can introduce a set of terms that compensate for the non-tensorial behaviour
of partial derivatives in different coordinate systems called the \textit{connection coefficients},
which are denoted by $\Gamma_{jk}^i$. This allows for covariant derivatives to transform appropriately
between coordinate systems. The details of this are beyond the scope of this
short review, but can be found in ~\cite{dinverno}. We simply state that by using the connection coefficients, 
we can then define the \textit{covariant derivative}:

\begin{equation}
\nabla_k A^i = \partial_k A^i + \Gamma_{jk}^i A^j
\end{equation}

\noindent where $\partial_k A^i$ is short hand for $\partial A^i / \partial x^k$. 

As will be seen in the next section, the connection 
coefficients are directly linked to geometry of the space in which the tensor exists. 

Just as different tensors have different transformation rules, covariant derivatives act differently
on different tensors. 

For a scalar:
\begin{equation}
\nabla_k \phi = \partial_k \phi.
\end{equation}

For a covariant vector:
\begin{equation}
\nabla_k A_i = \partial_k A_i - \Gamma_{ik}^j A_j.
\end{equation}


\section{Physical tensors: expansion, rotation and shear}
\label{sectionMechTensors}

As a first example of how tensors can describe physical quantities of a system, we can look at three
tensors that are often used in continuum dynamics to describe transport phenomena: 
expansion, shear and rotation. Indeed, these formalize 
the notions introduced in Section \ref{sectionGrowthWithTensors} and will be useful later on 
when analyzing the stresses and deformations of simulated tissue growth. Also, tensor equations for these
quantities are needed so that dynamical laws governing the physics of deformable systems remain
coordinate independent.

A viscoelastic fluid flow can be described with a velocity 
field $v^a$ that varies over space and time. The same velocity field can be used to construct the expansion scalar
which is the covariant divergence of the velocity field:

\begin{equation}
\Theta = \nabla_a v^a,
\label{eqnExpansion}
\end{equation}

\noi the shear tensor:

\begin{equation}
\sigma_{ab} = \frac{1}{2} ( \nabla_a v_b + \nabla_b v_a), 
\label{eqnShear}
\end{equation}

\noi and the rotation tensor:

\begin{equation}
\omega_{ab} = \frac{1}{2} ( \nabla_a v_b - \nabla_b v_a).
\label{eqnRotation}
\end{equation}

\noi In each of these equations, $\nabla_a$ is the covariant derivative as presented in the previous section.

Intuitively, these deformations are summarized in Figure \ref{figureTensors}. The expansion scalar
measures the rate of relative increase in scale, the shear tensor measures an area preserving deformation
of shape change when forces act to skew the shape of the element, and the rotation tensor measures 
how much an element is rotated during deformation.

In viscoelastic systems, a continuous medium is described using fluid elements. As the medium is
deformed over time, shape of the fluid elements change as well, which 
can in turn be described using the velocity field of the medium. A full derivation
of these equations can be found in ~\cite{contMechBook}.

\section{Geometric tensors}
\label{sectionRiemannianGeo}

We now move on to the introduce the roles of tensors in curved spaces, and hence their roles in plant
tissues. Based on the 
discussion in Section \ref{sectionBioMotivation}, plant leaves can be viewed as continuous, curved and growing 2D surfaces.
An appropriate mathematical tool to describe such a system is Riemannian geometry. A thorough treatment of
this branch of mathematics is beyond the scope of this discussion, but the most salient points are 
presented here. Further details can be found in ~\cite{dinverno}.

The fundamental object in Riemannian geometry is the metric tensor. The metric tensor encodes all the geometric
information about a space. From it, basic quantities like the distance between two points or the curvature
at a given point can be calculated. We define the geometry of a leaf's surface using a 2D metric tensor.

In an $n$-dimensional geometry, we need $n$ coordinates to describe any given point. This can be 
written as $x^i = (x^1,x^2,...,x^n)$. 
If we wish to calculate the distance $s$ between two points in the geometry, we can do so by generalizing the 
Pythagorean theorem $ds^2 = dx^2 + dy^2$ by defining an infinitesimal 
distance $ds$ that is a function of the coordinate displacement $dx^i$ and the metric tensor
$g_{ik}$:

\begin{equation}
ds^2 = g_{ik} dx^i dx^k
\end{equation}

\noindent where $g_{ik}$ is an $n \times n$ tensor. Again, Einstein summation notation is
used to denote a sum over repeated indices. 

Similarly, the metric defines an inner product between vectors or components of higher rank tensors:

\beq
C = g_{ik} A^i B^k.
\eeq

It must be remembered that, in general, the coordinates themselves do not determine distances. 
For example, in a plane polar coordinate system, the angular coordinate $\theta$ is used to
distinguish between points on a circle of radius $R$. The distance between $\theta_2$ and $\theta_1$ along
the circle is given by $\Delta s = R \Delta \theta$ where $\Delta \theta = \theta_2 - \theta_1$. 
The factor multiplying the difference in the coordinates is represented by a component of the metric tensor.

Extending this example to a flat circular geometry, we choose $x^1 = r$ and $x^2 = \theta$. 
The distance between two infinitesimally separated points is then:

\begin{equation}
ds^2 = dr^2 + r^2 d \theta^2.
\end{equation}

The corresponding metric tensor for a flat 2D circular geometry in $(r,\theta)$ coordinates is given by
$g_{11} = g_{rr} = 1$, $g_{12} = g_{21} = g_{r \theta} = 0$, $g_{22} = g_{\theta \theta} = r^2$. 
This can also be written as a $2$-dimensional array:

\begin{equation}
 g_{ik} = 
\left(
\begin{array}{cc}
1 & 0 \\
0 & r^2
\end{array}
\right).
\end{equation}

For non-flat 2D geometries such as those describing plant leaves, the metric tensor is more general
and may involve arbitrary coordinates $(u,v)$:

\begin{equation}
 g_{ik} = 
\left(
\begin{array}{cc}
f(u,v) & h(u,v) \\
h(u,v) & g(u,v)
\end{array}
\right).
\end{equation}

In this case, the distance between two infinitesimally separated points is:

\begin{equation}
ds^2 = f \, du^2 + 2 h \, du \, dv + g \, dv^2.
\end{equation}

Another interesting issue regarding the representation of curved surfaces the use of intrinsic rather than extrinsic geometry. 
The leaves we see around us are all comfortably embedded in flat 3D space. However, under our 
assumptions of the leaf as a thin sheet, the dynamics of the system only require a 2D space. If 
we study the system in a 3D space external to the leaf, we have chosen an extrinsic (Eulerian) 
approach. If we choose a 2D space defined on the leaf itself, such as dots drawn on leaf, then 
we have chosen an intrinsic (Lagrangian) approach.

Two arguments exist to support the intrinsic approach. First, Silk and Erickson noted that patterns 
of curvature are stationary in intrinsic coordinates ~\cite{silk1979}, suggesting that the plant uses a 
local reference frame to guide its growth. 

The second argument follows from the fact that curvature leads to forces (more details on this in the next section)
and is an intrinsic property 
of a geometry. With clear evidence that plants respond to forces (Section \ref{sectionBioMotivation}), 
our hypothesis is that
plants must also be able to sense forces arising from curvature. These could conceivably be sensed locally by each 
cell, primarily because curvature contributes to expansion and shear in a 2D manifold ~\cite{Raychaudhuri}. 
Because these are properties intrinsic to the space itself (i.e. they are independent of coordinate choice), 
a 2D coordinate system is sufficient to describe the physics associated with a curved leaf geometry.

The role of extrinsic coordinates, however, cannot be altogether discounted. External forces, including 
bending, undoubtedly affect plant tissue. Furthermore, embedding a curved 2D geometry in a flat 3D space 
provides a sort of symmetry breaking for the curved system, causing buckling cascades that minimize the 
elastic and bending energies of the tissue ~\cite{AudolyBoudaoud} ~\cite{NEPmechanics}.

Since we are presently concerned with the effect of internal forces generated by nonhomogeneous growth of 
leaf tissue, we proceed with using a 2D metric and a 2D velocity field defined on the surface of the leaf,
with curvature defined intrinsically on the surface of the manifold.

\section{Curvature leads to forces}
\label{sectionCurvatureForces}

One of the great discoveries of 20th century science was that gravity is really an effect of curved 
spacetime. Indeed, general relativity uses the
same concepts of Riemannian geometry, tensors, continuity and differential geometry to formalize how curvature 
affects motion at the largest scales.

But first, let us return to Newton's Second Law:

\beq
F = m\mathbf{a} = m \frac{d\mathbf{v}}{dt} = m \frac{d^2 \mathbf{x}}{dt^2}.
\eeq

Again note that a force is related to a change in an object's velocity
vector. To show how curvature leads to forces, let us look at two illustrative examples.

\begin{figure}[ht]
    \centering
        \subfloat{
                \includegraphics[width=0.3\textwidth]{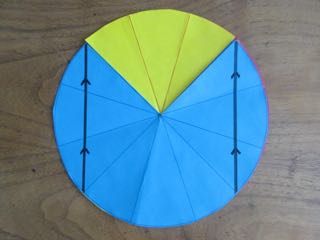}
        }        
        \subfloat{
                \includegraphics[width=0.3\textwidth]{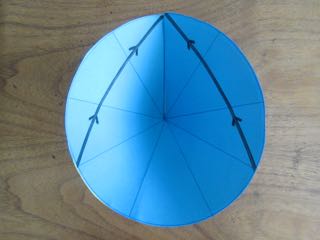}
        }
        \subfloat{
                \includegraphics[width=0.3\textwidth]{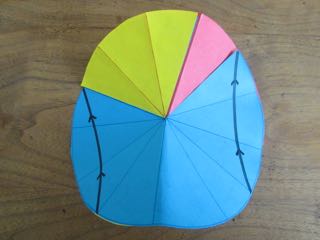}
        }
        \caption{From left to right, geometries with flat, positive and negative curvatures and 
        their corresponding 'straight' paths.}\label{figureGeodesics} 
\end{figure}

The first example is of two ants traveling on a surface. The ants walk in straight lines, which is locally
defined as taking steps of equal length on the right and left with each stride forward ~\cite{turtleGeo}. 
Figure \ref{figureGeodesics} shows three cases of such walks, also known as \textit{geodesics}.

What is happening? The three shapes represent three possible geometries, as defined by Gauss' 
\textit{Theorema Egregium}. The flat disk can be defined by the fact that it's circumference $C = 2 \pi r$. 
The middle shape is defined as a positively curved surface in which there is a deficit
in circumference so that $C<2 \pi r$ (the yellow area is removed from the flat disk). The last 
shape is a negatively curved surface in which there is an excess of circumference so that 
$C>2 \pi r$ (the red area is added to the flat disk). Because area measurements can be done without
leaving the surface of the manifold, the curvature calculated this way is an intrinsic property of the space.
There are also extrinsic measures of curvature, such as mean curvature, but these are not considered here.

In the flat geometry, two ants that start walking on parallel lines will remain the same 
distance from each other forever. On the positively curved geometry, the
ants appear to attract each other. On the negatively curved geometry, they appear to repel each other. In 
each case, the ants still walk in locally straight lines (the black line is never redrawn), 
but the velocity vector of the ants is clearly 
altered by geometry. Moreover, this change in velocity will be interpreted by the ants as a force, just
as we interpret curvature as the force of gravity. Again, these perceived forces are an intrinsic part 
of the curved manifold.

In relativity, this effect is formalized in the notion of geodesic deviation,
in which the separation of infinitesimally close geodesics is proportional to the curvature of the 
space itself. It is also related to the phenomenon of gravitational lensing, one of the first predictions
of general relativity to be observed experimentally. 

In biology, this effect could in principle impact any phenomenon that is described with a vector field.
In particular, in Section \ref{sectionVelFieldEqns}, we will consider how the flow of materials in a 
curved surface is altered through geometry by incorporating the geodesic flow of matter in a 
curved space into the velocity field of the growing leaf.

\begin{figure}[ht]
    \centering
        \subfloat{
                \includegraphics[width=0.3\textwidth]{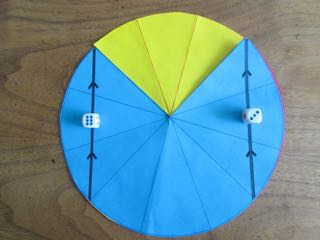}
        }        
        \subfloat{
                \includegraphics[width=0.3\textwidth]{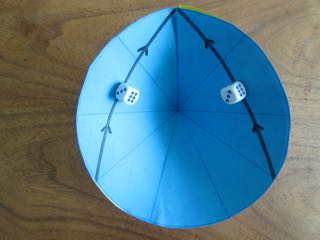}
        }
        \subfloat{
                \includegraphics[width=0.3\textwidth]{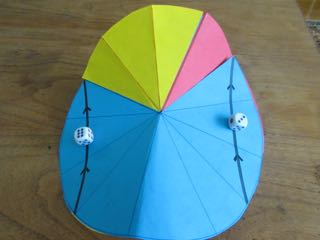}
        }
        \caption{From left to right, geometries with flat, positive and negative curvatures.}\label{figureNormalVectors} 
\end{figure} 

Let us look at a second example of how curvature can lead to force. Now assume the ants carry little cubes
that always have a face resting on the ground. Observe what happens to the cubes in Figure 
\ref{figureNormalVectors}.
In the flat geometry, we could swap the two cubes and they would still lie flat on the ground beneath
them. But in the positive and negative curvature spaces, this cannot be done without rotating the cubes 
to make them perpendicular to the ground again. Locally, the ants obey the rule of keeping the cubes
flat on the ground, but the intrinsic curvature of the space says that the orientation of the cubes  
changes from point to point.

This effect is formalized in differential geometry through connection coefficients $\Gamma_{jk}^i$. 
Connection coefficients quantify how much a vector or tensor changes when evaluated at different 
points on the manifold. In the previous example, the cubes represent vector or tensor quantities. 
Next, we show how connection coefficients are related to the metric of a curved space.

Given a curved space with metric $g_{ik}$, one can ask what is the natural motion of a particle in that
space. In other words, what are the minimal paths governing the motion of particles in a curved space?

Following Chapters 6.9 and 7.6 in ~\cite{dinverno}, this problem can be solved using a Lagrangian functional of the form 

\beq
L = \sqrt{g_{ik} \dot{x}^i \dot{x}^k}
\eeq

\noi and calculating the variation of

\beq
\int_{P1}^{P2} Ldu = \int_{P1}^{P2} ds = s
\eeq 

\noi where $u$ is a parameter along a smooth curve such that $\dot{x}^i = dx^i/du$, and $s$ is the interval 
between points $P1$ and $P2$ where, as before, $ds^2 = g_{ik} dx^i dx^k$.

Solving the Euler-Lagrange equations under the condition that $u=s$ gives us a second order differential 
equation for the extremal paths of particles traveling in a curved space, also called geodesics:

\beq
\ddot{x}^i + \Gamma_{jk}^i \dot{x}^j \dot{x}^k = 0.
\label{eqnGeodesics}
\eeq 

The connection coefficients are the result of partial derivatives of the metric that appear in the Euler-Lagrange equations:

\begin{equation}
\Gamma_{jk}^i = \frac{1}{2} g^{il} \left( \frac{\partial g_{jl}}{\partial x^k} + \frac{\partial g_{lk}}{\partial x^j}
				- \frac{\partial g_{jk}}{\partial x^l} \right). \label{eqnGammas}
\end{equation}

Hence, the connection coefficients are an intrinsic part of motion in a curved space. We will revisit this notion
when considering transport of material in a curved and growing tissue later in this chapter.

Another property of this particular form of the connection coefficients is that $\nabla_k g_{ik} = 0$. Connection coefficients
created from the Riemannian metric are often called Christoffel symbols or metric connections.

Recall from Section \ref{sectionTensors} that partial derivatives of tensors do not transform linearly
between coordinate systems. A corollary of this statement is that tensor quantities are affected by 
their location on a manifold, since moving to a different point on the manifold is mathematically similar  
to a coordinate transformation.
Now, instead of a coordinate transformation, we consider the spatial variation of a tensor $A^i$ 
in curved space. The connection coefficients act like the corrective term introduced before in 
the covariant derivative:

\begin{equation}
\nabla_k A^i = \partial_k A^i + \Gamma_{jk}^i A^j.
\end{equation}

Covariant derivatives introduce
curvature effects that are not possible in a flat system, but that are crucial to a correct mathematical
understanding of dynamics in a curved space. As will be seen, the covariant derivative will play an 
important role in modifying the continuity and transport equations.

Another powerful feature of Riemannian geometry is the ability to quantify curvature locally. In the 
research program started by Gauss and completed by his student Riemann in the mid-19th century, 
the search for methods to measure curvature without leaving the curved space itself were pursued 
in earnest. The result was a generalization of Euclidean geometry, where basic notions of distance, 
parallel lines and the circumference of a circle had to be redefined. 

Not too long after, at the beginning of the 20th century, Einstein showed that curved geometries
can produce the force of gravity. As we have seen, forces are described by the local effects they have 
on an object's momentum or velocity. Curvature can be described locally as a property of the metric 
tensor. One such measure of curvature is the \textit{Riemann tensor}:

\begin{equation}
R^a_{bcd} = \partial_c \Gamma_{bd}^a - \partial_d \Gamma_{bc}^a + \Gamma_{bd}^e \Gamma_{ec}^a - \Gamma_{bc}^e \Gamma_{ed}^a.
\label{eqnRiemannTensor}
\end{equation}

\noi A flat geometry is defined as one in which $R^a_{bcd}$ is identically zero. 

The Riemann tensor also acts as a measure of non-commutativity of the covariant derivative:

\beq
\nabla_a \nabla_b A_c - \nabla_b \nabla_a A_c = R_{abc}^d A_d.
\eeq

Note that if we substitute in the expression for the connection coefficients from Equation \ref{eqnGammas},
the Riemann tensor can be seen as a nonlinear function of the metric tensor, composed of second 
order derivatives and nonlinear first order derivatives of $g_{ik}$. 

Two other curvature tensors exist as well. The \textit{Ricci tensor}:

\begin{equation}
R_{ab} = R^c_{acb} = g^{cd}R_{acbd} = \partial_c \Gamma_{ab}^c - \partial_b \Gamma_{ac}^c + \Gamma_{ab}^d \Gamma_{dc}^c - \Gamma_{ac}^d \Gamma_{db}^c
\end{equation}

\noindent and the \textit{Ricci scalar}:
\begin{equation}
R = g^{ab}R_{ab}
\end{equation}

With a mathematical description of curvature in hand, we can now look at its effect on motion through
geodesic deviation. As mentioned earlier in the section, curvature directly impacts the geodesics of a
space. If we imagine a vector $\eta^i$ that quantifies the separation between two geodesics, then the
equation of geodesic deviation is:

\beq
\frac{D^2 \eta^i}{D \tau^2} = - R_{jkl}^i v^j v^k \eta^l
\eeq

\noi where $\tau$ is the proper time along the geodesics and $v^i = dx^i/d\tau$ ~\cite{dinverno}. The 
consequence of this equation is exactly what was seen in Figure \ref{figureGeodesics}, where a shortest
path between two points becomes warped by the geometry which the path must traverse.

In addition to quantifying the effect of curvature of a static geometry, dynamical systems can be constructed where the 
curvature changes over time. Indeed, general relativity is one of those systems. In it, matter and energy 
bend spacetime, while the geometry of spacetime dictates how matter and energy move. This is the essence of 
Einstein's field equations for general relativity:

\begin{equation}
R_{ab} - \frac{1}{2}g_{ab}R = \frac{8 \pi G}{c^4} T_{ab}
\end{equation}

\noindent where $T_{ab}$ is the energy-momentum tensor. This equation defines a tensor relationship 
for a 4D pseudo-Riemannian spacetime. The equations are hyperbolic in structure, which allows for wave dynamics and leads
to the prediction of gravitational waves.

More abstract \textit{curvature flows} can also be constructed. Another important though less well
known dynamical equation is the \textit{Ricci flow}:

\begin{equation}
\partial_t g_{ab} = - \kappa R_{ab}
\label{eqnRicciFlow}
\end{equation}

\noindent where $\kappa$ is a real-valued constant and the metric is purely Riemannian. For pure
Ricci flow as developed by Hamilton, $\kappa = 2$ in order to normalize the system such that a factor
of $1/2$ in the connection coefficients is removed.

The Ricci flow was developed by Hamilton in the 1980s in an attempt to solve the remaining parts 
of the Poincar\'{e} conjecture. Later, in the early 2000s, Perelman completed the work, thereby solving a 
100-year old problem in geometry and winning a Fields medal for the solution. What the equation does is 
dissipate the curvature of a metric in 
a diffusive-like way, allowing a highly convoluted geometry to be gradually deformed to a smoothed 
geometry, allowing direct analysis of the topology of a space. ~\cite{toppingRF}

A common feature of curvature flows is their quasi-linear partial differential structure. This 
makes solving them analytically very difficult even in the simplest scenarios, and often requires 
numerical techniques to tackle anything more than toy models.

What do curvature and Riemannian geometry mean in the biological context? As discussed previously, 
stress, strain and growth are all
tensor quantities. Therefore, these quantities will be affected by the local curvature of space. For 
example, a stress applied locally to a portion of negatively curved tissue will register differently than 
the same stress applied to a portion of positively curved tissue. Since cells react to forces through
mechanotransduction, it is reasonable to postulate that curvature will modulate a cell's response to any 
applied force. A similar argument can be made for the perception of strain and growth in curved spaces.
Indeed, such curvature effects were not considered
by Silk and Erickson, who proposed that the growth tensor is a gradient of the velocity field 
in a flat geometry. Petals and leaves are, however, intrinsically curved and their curvature changes
over time.

As we have seen, an important property of Riemannian geometry is the ability to describe calculus in 
curved spaces. This
may seem like an abstract mathematical point, but it has profound implications on physical structures 
like plant leaves as well as more abstract notions like spacetime. This is because calculus provides 
a local description of physical forces, so anything that may change how those local interactions behave,
such as a non-flat geometry, directly affects how the system experiences forces.

For our theory of leaf growth, the implication of considering curvature is that
in addition to tensors representing strain, growth, and material transport, we must also include 
terms that reflect how the system is affected by curvature. 

\section{The physics of material transport and growth models}

\subsection{The continuity equation}

As discussed in the previous chapter, Silk and Erickson argued that a continuum approach is appropriate when
studying large-scale plant growth ~\cite{silk1979}. Hence, a relevant equation to study is that of
mass continuity with a source term in 2D: 

\begin{equation}
\label{eqnContinuity}
\frac{\partial \rho}{\partial t} + \nabla \cdot (\rho \mathbf{v}) = S(\mathbf{x},t).
\end{equation}

In this equation, $\rho = \rho(\mathbf{x},t)$ is the mass density of the material, $t$ is time, 
$\mathbf{x}$ is the position vector, $S$ is the source function, and $\mathbf{v}$ is the velocity field of the material. 
The term $\rho \mathbf{v}$ is also known as the mass current $\mathbf{j}$. 

Equation \ref{eqnContinuity} is sufficient in flat geometries, but as seen in the previous sections, differential operators such
as the gradient $\nabla$ are influenced by geometry if the space is curved.  The curved-space
version of this equation can be written in tensor notation, and indeed all the terms including the gradient can be
generalized to their curvilinear equivalents ~\cite{RHB}:

\begin{equation}
\label{eqnContinuityWithTensors}
\frac{\partial \rho}{\partial t} + \nabla_i (\rho v^i) = S(x^i,t).
\end{equation}

The major difference between the two notations is that now $\nabla_i$ is a covariant derivative rather than a partial derivative.

Intuitively, the continuity equation states that the density of the material at a point $\mathbf{x}$ depends 
on the creation or destruction of material in a source or sink, and the flux of material into or out of a portion of space.
The significance of this equation in a growing medium will be revisited in Section \ref{sectionMassDensityEqns}.

\subsection{The transport equation}
\label{sectionNavierStokes}

Material transport in a viscoelastic system can also be understood from the perspective of Newton's Second Law, which gives 
a relation between forces and inertia: 

\begin{equation}
\mathbf{F} = m\mathbf{a} = m \frac{d\mathbf{v}}{dt} = m \frac{d^2 \mathbf{x}}{dt^2}.
\label{eqnN2nd}
\end{equation}

Motion, therefore, is a local phenomenon formalized using the methods of calculus. Here, infinitesimal time and spatial
changes govern how an object moves. A change in an object's velocity or momentum vector is proportional to a force.
This notion is just as applicable for macroscopic objects as it is for a fluid subdivided into infinitesimally 
small units of volume.

The ordinary derivative with respect to time in Equation \ref{eqnN2nd} can be rewritten in terms of partial derivatives using
the chain rule:

\begin{eqnarray}
\frac{d \mathbf{v}}{dt} & = & \frac{\partial \mathbf{v}}{\partial t} + \sum_{i=1}^n \frac{d x^i}{dt} \frac{\partial \mathbf{v}}{\partial x^i} \nonumber \\
  & = & \frac{\partial \mathbf{v}}{\partial t} + \mathbf{v} \cdot \nabla \mathbf{v}.
\end{eqnarray}

Hence, the time evolution of the velocity field can be understood in terms of the forces per unit mass $\mathbf{f} = \mathbf{F}/m$ and
the \textit{advection} $\mathbf{v} \cdot \nabla \mathbf{v}$:

\begin{equation}
\frac{\partial \mathbf{v}}{\partial t} = - \mathbf{v} \cdot \nabla \mathbf{v} + \mathbf{f}
\end{equation}

\noindent where the mass $m$ can be normalized to unity assuming a constant density tissue.

What can be said about the forces acting in a continuous system like plant tissue? 
They can be thought of as the external and internal
forces affecting the tissue. In this model, we assume that external forces such as gravity are negligible,
and direct mechanical contact forces are not considered. 
Internally, we assume that the pressure inside each cell is a constant in space and time (no pressure gradients) and that no
significant elastic restoring forces act at this scale. This allows for the construction of the simplest 
possible version of the transport equation, and allows the model to examine the effects of curvature as
separate from other possible physical effects. However, should terms representing pressure gradients and 
elasticity be needed in order to extend the applicability of the model in different regimes,
it would be possible to incorporate them into the transport equation.

We do note, however, that individual cells are known to undergo elastic and plastic 
deformations during growth, and measurements of externally induced tissue strain have been performed 
~\cite{hejnowicz1995tissuesStrains} ~\cite{leafBulkModulus}.
These elastic restoring forces generally do not affect tissues at the scales we consider here. For instance, 
cutting a leaf or petal tissue does not cause it to relax to a different shape, nor does stretching the tissue
result in deformations on the mm length scale. Indeed, the deformations are on the $\mu$m scale, and the relationship
between cellular level elasticity and tissue elasticity is still poorly understood ~\cite{leafBulkModulus}.
This is in contrast to highly elastic materials such as rubber, which can deform reversibly at the macroscopic
scales we consider. These assumptions also do not exclude the presence of stresses in the tissue due to growth, 
as discussed in Section \ref{sectionMechTensors}; as will be seen, expansion, shear and rotation tensors can be calculated
for the numerically simulated tissue growth models.

Yet another assumption is that the material inside the leaf is incompressible. Leaves are composed mostly of water,
while other structures like the mature cell wall are quite rigid ~\cite{taiz}. Growth of the cell walls does not 
occur by pure expansion of the cell wall material, but by the addition more material to the walls. 

Incompressibility is also associated with notions of elasticity and pressure gradients. Elasticity is defined as
$ \epsilon = V dP/dV = [1/V dV/dt]^{-1} (dP/dt)$. Having assumed constant pressure in the tissue previously, this
leads directly to zero elasticity of the material. Compressibility is quantified as the inverse of elasticity, so for
zero elasticity, we necessarily have an incompressible medium. Intuitively, one can think of a synthetic tissue made up 
of many balloons. The tissue is not constrained at the outer boundary. As the balloons expand, they can also add material to 
their walls (just like a cell), thereby maintaining constant pressure inside each balloon while at the same time 
accommodating ever more material (air) inside of them. 

The balloon analogy can be taken one step further when we notice that the balloons are also in local equilibrium 
with each other. Because of this, the elasticity of the individual balloons is not related to the elasticity
of the whole tissue in a simple way. If the balloons are connected in a such a way that they maintain 
the same neighbours over time (as is the case in plant tissue), externally stretching the tissue means 
stretching many connected balloons. Therefore, the individual elasticity of each cell contributes to the
elasticity of the tissue as a whole, but the effective elasticity of the tissue will be much lower.
This has in fact been measured to be the case in iris leaves, where the elasticity of the cell walls is
$29-100$Mpa while the elasticity of the leaf is several orders of magnitude higher, at 
$0.838-22.6$Gpa ~\cite{niklas1992plantMechBook}.

We do assume there is viscosity in the system, noting that as plant cells grow they
keep the same neighbours throughout late-stage development, which can be viewed as a sort of friction between
the elements in the tissue. Materials traveling through a plant tissue also experience viscosity by way of
diffusion and active transport processes. Transport happens across cell walls in a number of ways, including by way of carrier 
proteins, osmosis, or plasmodesmata. Transport of material also happens within the cell, again in 
a variety of chemical processes. We assume that the net macroscopic effect can be understood in terms of a diffusive process.

Viscosity is a standard term in the transport equation given by the Laplacian
of the velocity field $\nabla^2 \mathbf{v}$. Under the preceding assumptions, the velocity field evolves under 
advection and diffusion:

\begin{equation}
\label{eqnNavierStokes}
\frac{\partial \mathbf{v}}{\partial t} = - \mathbf{v} \cdot \nabla \mathbf{v} + D \nabla^2 \mathbf{v}
\end{equation}

\noindent where $D$ is the diffusion coefficient, which in the simplest scenario is a constant throughout 
the material. In more complex scenarios, the diffusion term can itself be a spatially varying tensor.

Indeed, Equation \ref{eqnNavierStokes} is a form of the Navier-Stokes equation for incompressible viscous flow in flat space. 
The curved-space version of this equation will be revisited in Section \ref{sectionVelFieldEqns}.

\section{Local interactions in open, driven systems}
\label{sectionKPZstructure}

In constructing this model of plant growth, two ideas have not come into play. Namely, the notions of 
closed and near-equilibrium systems. Having adopted the view that leaf growth is a result of material 
transport from the stem into the leaf that at large scales predominantly causes cell expansion, 
these two commonly used assumptions for many physical systems simply do 
not hold. The plant leaf is an open system because of the influx of new material. The system is far from 
equilibrium because it is driven by new material flowing into it. Moreover, cells manufacture materials
and consume energy locally in cellular processes, from photosynthesis which stores solar energy as sugar, 
to endoreduplication which produces multiple copies of a cell's DNA without cell division (mitosis).

For an open and driven system in which interactions are still local, we follow the general prescription 
given by Kardar that the dynamical equations are the 'fundamental object of interest' ~\cite{kardar}. Since 
the system still relies on local interactions, the general structure of its deterministic dynamical equations is: 

\begin{equation}
\label{eqnKardar}
\partial_t h(\mathbf{x},t) = v(h(\mathbf{x},t), \nabla h(\mathbf{x},t), ...)
\end{equation}

\noi where $h$ is a generalized field and $v$ is a velocity.

Such a structure can clearly lead to nonlinear interactions, where the field interacts with itself and its gradients
as determined by the structure of the velocity functional. 
Since we are dealing with a pattern forming 
biological system, this result is also consistent with reaction-diffusion dynamics in biology.

The idea of local interactions governing global behaviour is also consistent with the methods of differential
geometry. Local behaviour can be integrated over the entire structure to produce global topological characteristics.
For a plant leaf, each cell grows locally but its growth must be consistent with global mass continuity and 
velocity flow.  

\section{Leaf dynamics as nonlinear coupled tensor equations}

In considering what factors influence the time evolution of a leaf's geometry, two factors have already
come to light: growth and curvature. Combined with the notion that growth, curvature and the metric are all 
tensor quantities, we can begin to piece together the structure of our dynamical equations.

Namely, what is necessary to capture the feedback between geometry, growth and curvature is a set of 
nonlinear coupled tensor equations. 
In the following sections, the tensor formalism will bring together all the important 
biological and physical arguments put forward so far (forces, growth, continuity, curvature, geometry), 
steering us toward a specific mathematical structure.

A significant impact of adopting a formalism based on nonlinear coupled tensor equations is that,
in general, they do not have analytical solutions. There is no superposition principle for nonlinear 
systems that guarantees a complete set of basis functions from which a general solution can be constructed.
Instead, we must turn to numerical simulations, particularly in realistic situations.

This approach is also fraught with difficulty. Nonlinear systems can be inherently unstable, which
makes computing them not an easy task. 
Constructing a numerical simulation of our coupled tensor equations will require choosing a stable
finite differencing scheme. This is discussed further in Chapter \ref{chapterNumericalMethods}.

Based on the issues discussed above, we now present the specific equations that constitute the model.

\section{Defining the domain and boundaries of 1D and 2D tissue models}
\label{sectionPDEdomain}

\begin{figure}[h]
    \centering
    \includegraphics[width=0.75\textwidth]{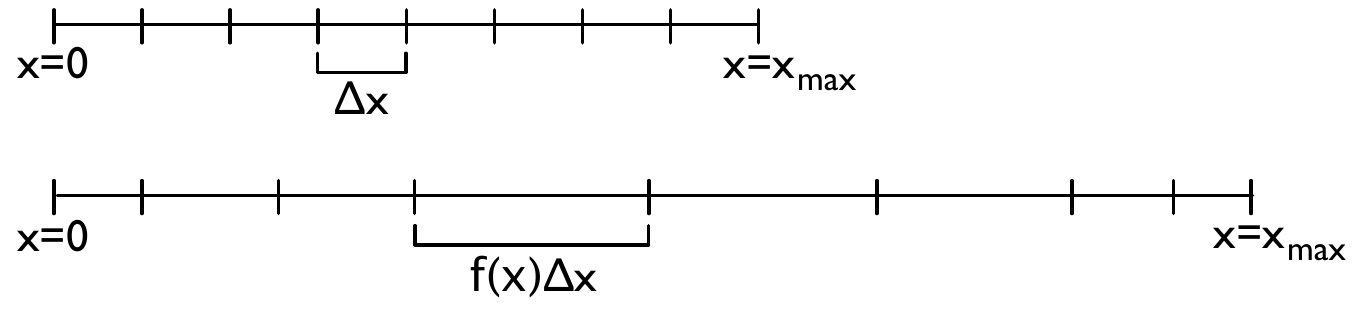}
    \caption{1D growth in Lagrangian coordinates. The effect of nonhomogeneous growth on initially equidistant coordinate positions is shown. The total length of the 1D manifold is the sum of all the distances between coordinate positions from the origin to the outer boundary.}
    \label{figure1Dgrowth}
\end{figure}

\begin{figure}
    \centering
    \includegraphics[width=0.9\textwidth]{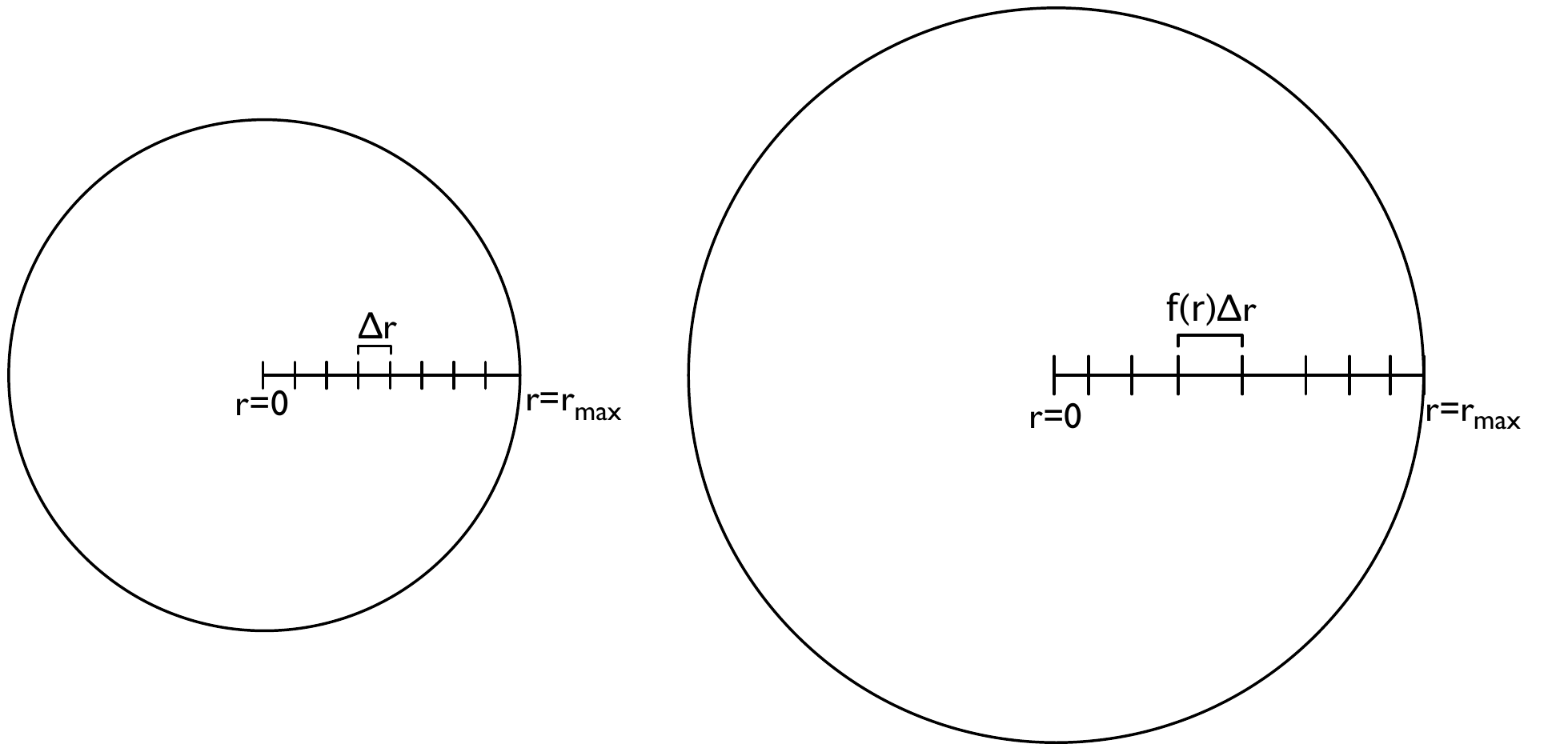}
    \caption{2D growth in a circular geometry using a Lagrangian formalism. The manifold can grow anywhere inside the boundary, including growth that leads to anisotropies like elongation.}
    \label{figure2Dgrowth}
\end{figure}

In modeling the growth of tissue, we choose a Lagrangian coordinate system that is fixed to the tissue
itself. As discussed previously
in this chapter, the Lagrangian formalism allows us to model the dynamics of growth in a way that is
intrinsic to the tissue itself, that is, without reference to external, fixed coordinates. 

This approach is analogous to the \textit{landmark method} used in experimental biology to measure 
growth of tissues in a nondestructive way. Initially, points (such as ink dots) are placed equidistantly on the tissue.
Over time, growth displaces the points, revealing where the tissue has grown. In general, 
biological growth is nonhomogeneous, which manifests itself as distances
between landmarks increasing nonhomogeneously as the system evolves. This method has been used for
nearly a century and continues to be utilized in more modern variations 
~\cite{avery1933structure} ~\cite{SilkSciAm} ~\cite{inkjetLeaves}.

Our mathematical model requires three dynamical variables: the mass density, a velocity field and a metric tensor.
These will be discussed in more detail in subsequent sections, but first we define the domain and 
boundary conditions underlying the model.

In our model, the tissue is defined on the domain $x=(0,x_{max})$ in 1D, and $r=(0,r_{max})$, 
$\theta = (0,2\pi)$ in 2D. Further discussion on why a circular domain is chosen for 2D models is 
discussed in Chapter \ref{chapter2DConstrainedSimulations}. The domain is subdivided into an integer 
number of fixed coordinate positions. As the tissue grows, the distances between coordinate positions 
also grow. This process is encoded in the metric tensor evolution discussed Section \ref{sectionMetricEvo}.
A similar approach is taken when studying cosmology and the expansion of the universe. In that system,
galaxies mark the fixed coordinates of the universe, and expansion is quantified by the rate at which galaxies move 
away from each other.

The outer boundary of the tissue stays fixed at $x_{max}$ in 1D and $r_{max}$ in 2D. Growth occurs
throughout the tissue inside of the boundary by increasing the distance between coordinate positions. The total
length or radius of the tissue is then the sum of all the distances between the origin and the outer boundary.

Time, represented by $t$, is not in principle bounded for these models. It has arbitrary units that can be
adapted to the magnitude of the diffusion coefficients. The model treats time as a continuous variable.

The boundary conditions for the model are based on both mathematical and physical considerations. As has been 
discussed, the equations forming the model will be differential equations involving second spatial derivatives
that represent diffusive processes. This necessarily requires the functions to be at least twice differentiable everywhere.
The dynamics also depend on reaction terms that are first spatial derivatives of the dynamical
variables. 

For the velocity field, we also assume the outer boundary of the tissue cannot allow material to flow out of it; 
the outer boundary therefore represents a condition on the diffusion term (second spatial derivative). 
One plausible condition for the outer boundary that satisfies all of the above arguments is a Neumann condition.
In 1D, for a velocity variable $v$, this would be $\partial_x v(t,x=x_{max})= \phi (t)$
where the function $\phi (t)$ is equivalent to $\partial_x v$ evaluated at $x=x_{max-1}$. In 2D,
the equivalent statement would be $\partial_r v(t,r=r_{max},\theta) = \phi (t,\theta)$ where
$\phi (t,\theta)$ is equivalent to $\partial_r v$ evaluated at $r=r_{max-1}$. These conditions 
quantify the notion that diffusion does not occur across the outer boundary while allowing the dynamical
variables to remain twice differentiable. 

For the inner boundary, a similar consideration of dynamics leads to the boundary conditions for the velocity at the origin.
Here, a Dirichlet condition is appropriate since we hold the origin fixed in place for both 1D and 2D models. Hence,
the velocity is null at the origin for all time.

The metric can in principle expand anywhere in the domain, including the origin, so a Neumann condition 
is more appropriate for this dynamical variable. One possible constraint on the metric components 
(for simplicity denoted by $f$) that also provides numerical stability in simulations discussed in future chapters is:  
$\partial_x f(t,x=0) = 0$, $\partial_x f(t,x=x_{max}) = 0$ (1D model);
$\partial_r f(t,r=0,\theta) = 0$, $\partial_r f(t,r=r_{max},\theta) = 0$ (2D model).

For 2D simulations, the angular coordinate $\theta$ has a periodic boundary condition, reflecting the 
symmetry of the imposed geometry.

\section{The mass density equation}
\label{sectionMassDensityEqns}

The first dynamical variable is the scalar mass density of the tissue $\rho( \textbf{x}, t)$. At each 
point $\textbf{x}$, the density is represented by a number (scalar) value. Also, the density $\rho$
may change over time $t$ according to the continuity equation.

\begin{equation}
\label{eqnDensityLong}
\frac{\partial \rho}{\partial t} + \nabla_i(\rho v^i) = S(\mathbf{x},t).
\end{equation}

The two terms are the same as in Equation \ref{eqnContinuity} for mass conservation, but generalized to
curved geometries because $\nabla_i$ is a covariant derivative rather than a flat-space partial derivative. We also note
that $S(\mathbf{x},t)$ is called the deposition rate in ~\cite{silk2014deposition}.

As a simplification, we assume the leaf has constant density in time and space. This is also consistent with
the assumption of incompressible fluid flow discussed in Section \ref{sectionNavierStokes} and ~\cite{NumericsBook}.
Indeed, plant tissues are made mostly of water and cell walls are rigid structures that grow by 
deposition of new cell wall material (rather than pure elastic stretching), further justifying this assumption
at large spatial scales.

Therefore we have:

\begin{eqnarray}
\label{eqnDensity}
\frac{\partial \rho}{\partial t} &=& 0 \nonumber \\
\nabla_i \rho & = & 0.
\end{eqnarray}

This means we assume the source term is in balance with the velocity field to create a constant density tissue. 
Biologically, this can be interpreted as the tissue drawing as many resources as are required to maintain its
growth pattern; more active cells will draw more resources than quiescent or mature cells.

It is important to note the implication of these assumptions, summarized in the form:

\begin{equation}
\label{eqnSourceFunction}
S(\mathbf{x},t) = \rho \nabla_i v^i = \rho \Theta.
\end{equation}

Recall from Section \ref{sectionMechTensors} that $\nabla_i v^i$ is the expansion scalar $\Theta$. Hence,
the influx of material from the distributed source directly influences the macroscopic, physical expansion of the tissue. The other
influence becomes the geometry of the manifold itself, since $\nabla_i$ is a covariant derivative calculated
based on the metric tensor of the curved 2D space in which the material is flowing.

\section{The velocity field equation}
\label{sectionVelFieldEqns}

The second dynamical variable in this model is the velocity field describing the transport of material throughout 
the leaf. Since we consider the system to be
open and driven, we must construct this equation using suitable local dynamical terms.

The most elegant way to understand how the geometry affects the velocity field is by considering 
the Lagrangian of a particle moving in curved space, as was done in Section \ref{sectionCurvatureForces}.
Expanding on this, we can extend Equation \ref{eqnGeodesics} to include external forces:

\beq
\ddot{x}^i + \Gamma_{jk}^i \dot{x}^j \dot{x}^k = f^i.
\eeq

Next, we replace $\dot{x}^j$ and $\dot{x}^k$ with the velocity vectors $v^j$ and $v^k$, and split the full 
time derivative $\ddot{x}^i$ into partial derivatives in space and time of the velocity vector $v^i$. 

\begin{equation}
\frac{\partial v^i}{\partial t} + v^k \nabla_k v^i + \Gamma_{jk}^i v^j v^k  = f^i. 
\end{equation}

The forces $f^i$ are the same internal and external forces discussed in Section \ref{sectionNavierStokes}.
If we again assume a viscosity term, an incompressible medium, and no elastic restorative forces, no velocity
dependent damping forces and no other external forces and pressure gradients across the tissue,
we obtain the following equation for the time evolution of the velocity field:

\begin{equation}
\frac{\partial v^i}{\partial t} = -\Gamma_{jk}^i v^j v^k - v^k \nabla_k v^i + c g^{jk} \nabla_j \nabla_k v^i. 
\label{eqnVelocityField}
\end{equation}

Comparing this equation to Equation \ref{eqnNavierStokes}, we see it is equivalent to the Navier-Stokes equation 
generalized to a curved geometry. Just as the flat space Navier-Stokes equation is a restatement of 
Newton's Second Law, all the terms in this equation can be understood in terms of forces acting
on the infinitesimal elements of viscoplastic tissue.

The first term contains the connection coefficient $\Gamma_{jk}^i$. 
When combined with the velocity field, $\Gamma_{jk}^i v^j v^k$ becomes the contribution that the curvature 
of the leaf makes to material transport. It can be thought of as a 'geometric force'. 

Another way to think of the $\Gamma_{jk}^i v^j v^k$ term is that it represents the shortest path a 
particle can take across a curved geometry. The direct impact for material transport is that, in a curved geometry,
the shortest path is no longer 'straight' in the conventional (flat space) sense. In other words,
the curvature of the leaf will influence how material flows across it. 

It is also important to note that even if the velocities themselves are small, the effect of the connection
coefficients can still produce a large change in a particle's motion. This is because the connection coefficients
are calculated based on spatial derivatives of the metric (Equation \ref{eqnGammas}). If the spatial derivatives 
of the metric are large, this can influence the transport of material even at small velocities. Indeed this 
will be the case in many of the simulations studied in later chapters.

The second term, representing advection, comes from the fact that both the outer boundary 
and every point inside the boundary experience a velocity as the leaf grows. This velocity field
is a result of material entering the leaf (i.e. the distributed source term from Equation \ref{eqnDensityLong}). As the
material flows through the tissue, it is also affected by the expanding geometry of the tissue.

The last term represents diffusive processes in the tissue. These arise from a host of cellular transport
processes, both passive and active, which then can be generalized to an effective, macroscopic material diffusion
throughout the tissue. Transport processes are also used to grow the tissue and generate curvature, which will be 
explored further in the next section. 

This modified form of material transport now includes nonlinear velocity terms. It can be thought of as a general 
reaction-diffusion equation for the velocity. Note too the presence of the covariant derivatives $\nabla_k$
which further couple the material transport to a curved geometry, allowing curvature to act as a force on the vector
field of the material velocity.

\section{The metric tensor equation}
\label{sectionMetricEvo}

The third and final dynamical variable is the metric tensor. The metric tensor characterizes the 
geometry of the leaf. Since we assume the leaf is a thin sheet, the metric is two dimensional.

\begin{equation}
\frac{\partial g_{ik}}{\partial t} = - \kappa R_{ik} + \kappa_1 [ \nabla_i v_k + \nabla_k v_i]
\label{eqnMetricTensor}
\end{equation}

\noi where $\kappa$ and $\kappa_1$ are coupling constants that at a fundamental level will depend on the system
under study. These may be determined by the environment and/or the genetics of the organism. 

Let us start with the last term, which is a symmetric growth tensor 

\beq
T_{ik} = \nabla_i v_k + \nabla_k v_i.
\label{eqnGT}
\eeq

The notion of a growth tensor of this form was originally 
introduced by Silk and Erickson ~\cite{silk1979}. Including this coupling provides feedback between the
velocity field and the metric tensor. Since the metric tensor and the Ricci tensor are symmetric 
($g_{ik} = g_{ki}$ and $R_{ik} = R_{ki}$), the growth tensor must also be constructed symmetrically.

Why is the growth tensor a gradient of the velocity field? If we think of two 
points on the leaf with the same velocity relative to each other (no gradient), they 
experience no change in the distance between them, so no growth has occurred. If, however,
the points move at different speeds or in different directions (non-zero gradient), 
then growth or contraction must be occurring between the two points. This also clearly indicates that
the growth tensor must affect the metric tensor since the metric tensor measures
distances between points.

The growth tensor is also closely related to the expansion, shear and rotation tensors (Equations 
\ref{eqnExpansion}, \ref{eqnShear} and \ref{eqnRotation}). This can be understood in the context of treating
growth as a deformation of the surface. It also indicates that growth causes strain in the tissue,
an important concept that will be revisited later on. In fact, the growth tensor $T_{ik}$ as defined in 
Equation \ref{eqnGT} is twice the strain rate tensor commonly used in continuum mechanics. 

The first term is the coupling of the metric to its own curvature. Including this term expresses
the notion that curvature directly influences the time evolution of the leaf's shape, as discussed
in Section \ref{sectionBioMotivation}. Alternatively, this equation can be interpreted as stating that 
growth leads to both metric expansion and curvature, an effect clearly described in thin sheet 
experiments (Section \ref{sectionEstablishedWork}). The negative sign is kept from Equation \ref{eqnRicciFlow}
and indicates that an increase in scale factor leads to negative curvature, which is consistent with experiments 
on torn plastic sheets ~\cite{leavesFlowersPlasticRuffling}.

The Ricci tensor components are also functions of first and second derivatives of the 
metric components. In this way,
the Ricci flow can be thought of as a reaction-diffusion term for the metric itself.

Readers familiar with differential geometry may wonder why the metric flow is coupled to the Ricci 
tensor rather than some other measure of the curvature like the Riemann tensor or the scalar 
curvature (lowered or raised as needed to become valence $(0,2)$ tensors). In fact, choosing any of these 
curvature couplings is equivalent in 2D. The Riemann tensor has only one unique component in 2D, 
so there is no information lost going from the Riemann tensor to the Ricci tensor $R_{ik}$ or scalar curvature $R$. 
Even the Gaussian curvature $K$ is directly proportional to the scalar curvature, $K = \frac{1}{2}R$.
Using the Ricci tensor is convenient because it has the same valence as the metric and growth tensors.

Quite importantly, Ricci flow allows the growing tissue to dissipate curvature because 
of its reaction-diffusion type behaviour. This is important in a growing tissue, since a 
region of very rapid growth causes strain on the 
neighbouring tissue as well as a localized area of high curvature. The ability to dissipate 
this area of high curvature prevents the tissue from tearing and excessive bending. 
If cells in the tissue were to
sense this area of high curvature and respond through a mechanotransduction pathway such 
that the curvature dissipates, then Ricci flow is merely another physical phenomenon of 
macroscopic leaf growth, much like the growth tensor.

Another interesting feature of coupling the time derivative of the metric to its own curvature is that even if the 
tissue is not growing (i.e. the growth tensor vanishes), the tissue still seeks the most uniform
shape possible through curvature diffusion. This is a general property of the Ricci flow equation.

Explicitly coupling curvature and growth to the geometry of the leaf can only be done
using rank two tensors of the same valence. Hence, our assumptions about the feedback between 
growth, curvature and geometry in plant leaves have led us directly to using a set of 
coupled nonlinear tensor equations. In particular, an expression emerges that allows
these three important elements to interact directly with each other, while at the same time excluding 
other couplings. For instance, $\partial_t g_{ik} = T_{ik}$ couples the growth tensor directly to the
metric, but neglects the reaction-diffusion terms of the Ricci tensor that allow areas of 
quick growth to dissipate their curvature. Another variation that is incompatible with the dynamical
nature of growth is $R_{ik} = T_{ik}$ which does not allow for the metric itself to grow over time,
only change shape. 

Indeed, the number of 2nd rank tensors that can be constructed out of dynamical vector quantities  
is quite limited. We can either use tensor products or derivatives, and the quantities thus created must
still be relevant to the physics of the problem. Given that we have assumed a scalar mass density that is 
constant in both space and time, derivatives and tensor products of this quantity will either be null or constant. 
The covariant derivatives of the velocity field have already been used to construct the symmetrized growth tensor.
A tensor product of the velocity field would yield terms related to an energy-momentum tensor that would have to 
obey a conservation law; as we have seen, the influx of material and energy into the leaf violates conservation laws,
so the inclusion of a velocity tensor product would not reflect the physics of the system. Lastly, the metric tensor
and its derivatives have already been included by way of the Ricci tensor. Hence, the equations already contain 
most, if not all the possible rank 2 tensors that can be constructed self-consistently from the dynamical variables
that make up the model.

A further refinement of which terms to include when constructing a dynamical model is related to the 
principle of \textit{minimal coupling} ~\cite{dinverno}. There are many tensors that could be made to fit the rank two, rank
one and rank zero equations, simply by raising and lowering indices. However, these do not necessarily represent 
any new physics acting on the system. Each term included in the equations introduced in this section is based on 
physical or biological arguments. The resulting mathematical structure is found to be consistent with
the dynamics of an open, driven system. Minimal coupling specifies that only the metric tensor can be coupled
to curvature; other tensors like our velocity field and mass density should have no explicit curvature
coupling terms of the type $\partial_t v^i = \kappa R^{ij}v_j$.

The possibility of other tensor couplings is, however, not strictly out of the question.
These could include physical forces that are not considered here, such as restoring forces of the type
$-k^{ij}x_j$ or retarding forces of the type $-B^{ij}v_j$ ~\cite{kinFlowsOnCurvedSurfaces}.
Should it be found that key dynamics are not present in the model as it is presented here, different
couplings can be explored that still satisfy the physical and biological assumptions of the model.

\section{Chapter summary: \\ The interplay of geometry, growth and curvature}
\label{sectionInterplay}

From these biological and physical considerations of mechanosensing, continuity, growth and curvature, 
we can now state the problems our theory seeks to explore. 

First, let us summarize the system of equations introduced in this chapter that are the foundation
of a tissue growth model based in dynamical Riemannian geometry:

\begin{eqnarray}
\frac{\partial \rho}{\partial t} & = & 0 \\[10pt] 
\frac{\partial v^i}{\partial t} & = & -\Gamma_{jk}^i v^j v^k - v^k \nabla_k v^i + c g^{jk} \nabla_j \nabla_k v^i \\[10pt]
\frac{\partial g_{ik}}{\partial t} & = & - \kappa R_{ik} + \kappa_1 [ \nabla_i v_k + \nabla_k v_i] 
\end{eqnarray}

This can be considered to be a 'minimal' model where the only forces considered are geometric and diffusive, 
and growth is driven by deposition of material from a distributed source.

\begin{figure}
    \centering
    \includegraphics[width=0.75\textwidth]{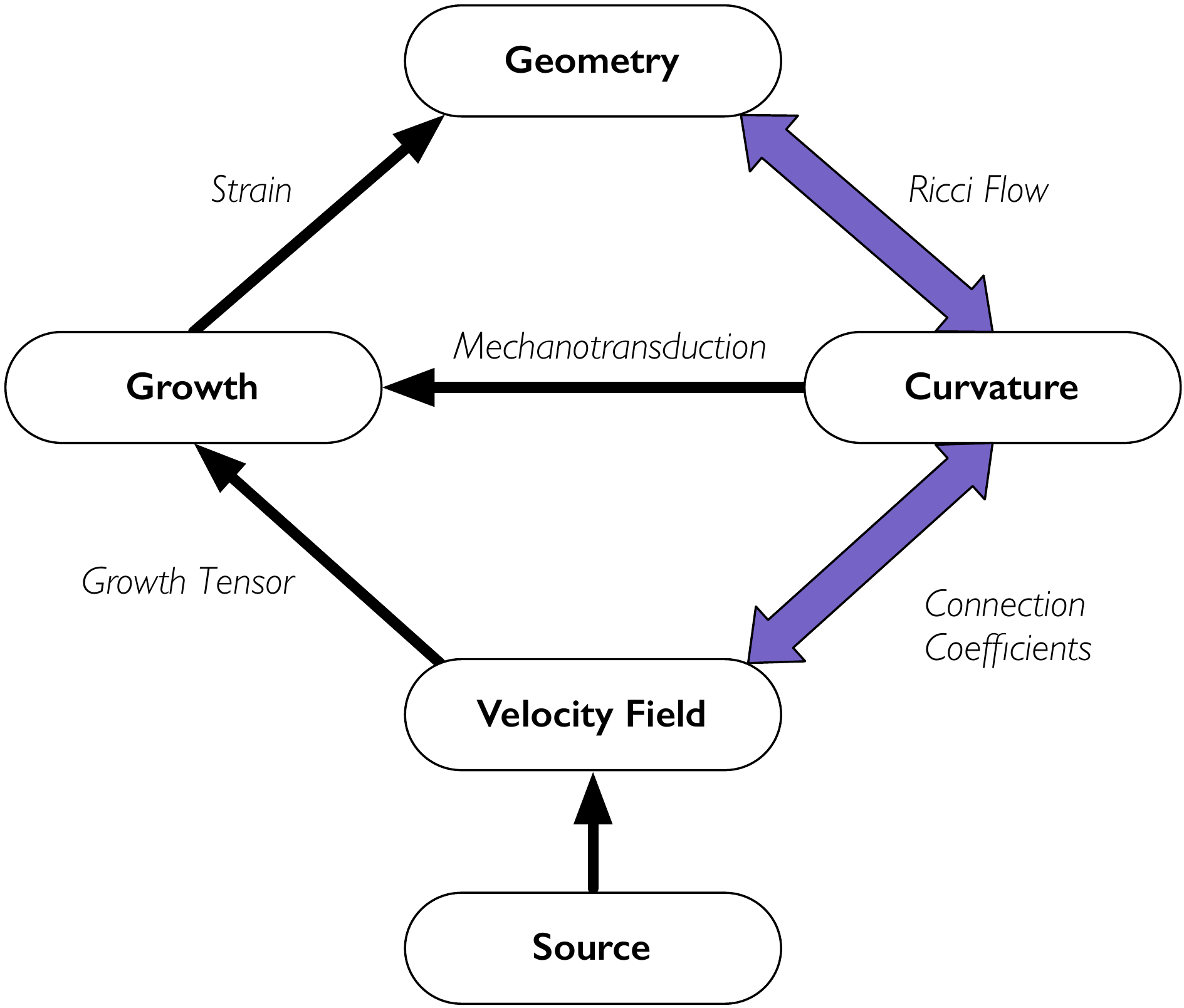}
    \caption{A diagram of how geometry, growth and curvature may be connected in plant growth. Purple arrows indicate novel concepts from modeling the leaf as a Riemannian geometry.} 
    \label{figurePDEdiagram}
\end{figure}

The central theme of our theory is the feedback between growth and curvature at macroscopic
scales. On the one hand, growth causes inhomogeneous and anisotropic changes in geometry that manifests itself as a 
deformation of the tissue; hence, growth leads to curvature. On the other hand, curvature generates
strain on the tissue, as real as any externally applied force. Through mechanotransduction,
these forces caused by curvature could be sensed by the tissue, creating a self-regulating system. This is
feedback is captured in Figure \ref{figurePDEdiagram}.

The details of mechanotransduction are not addressed by our theory as that is clearly the
domain of experimental biology. What is addressed is a new and rigorous mathematical model of how the 
feedback between growth and curvature can occur given the assumption that leaves act like
continuous, deformable, thin-sheet systems at some threshold scale.

What are the possible questions that can be asked with a mathematical model of plant growth?

For one, we would like to know if such a model results in shapes similar to those found in wild-type
and mutant plants. In other words, the shapes and structures that result from such a model form a 
landscape which can be explored by
varying the parameters and initial conditions in our set of coupled tensor equations. 
The extent to which a comparison might be done between model and real leaves 
will be discussed in Chapters \ref{chapter2DConstrainedSimulations} and \ref{chapterConclusions}.

We can also ask if our assumptions are sufficient to understand the feedback between geometry
and growth. If the solutions to our equations do not produce the shapes found in plant leaves,
then our basic understanding of the system is flawed.

Lastly, can our model predict some fundamental but as yet undiscovered part of plant growth? As mentioned
in Section \ref{sectionBioMotivation}, the dynamics of cell expansion as the leaf matures at scales 
larger than 1mm are not well
understood. Modeling this stage of leaf development mathematically may produce results that help guide experiments 
toward understanding the interplay of growth, geometry and mechanotransduction in both wild-type and mutant plants.

\chapter{Numerical Methods}
\label{chapterNumericalMethods}

In the previous chapter, three dynamical equations were developed to model the physics of a growing plant tissue:
the mass density equation, the velocity field equation and the metric tensor equation. Though initially developed
as coupled tensor equations, they can be rewritten component by component using continuous functions. In this way,
the coupled tensor equations become coupled partial differential equations (PDEs). More details on the functions 
themselves will be 
discussed in Chapters \ref{chapter1DSimulations}, \ref{chapter2DConstrainedSimulations} and \ref{chapter2DFullSimulations}.

The study of numerical methods to solve coupled nonlinear PDEs is a branch of mathematics and computation in
its own right. Here, we present an outline of the guiding principles and tests of the numerical methods used to study the 
dynamical equations in our model. 
All code was written in C using standard libraries and can be run on any computer with a UNIX platform. This was a 
programming choice that allows the code to be open-source and free of 'black box' solutions to sensitive 
numerical problems. Full source code is available from the author upon request.

\section{Finite element versus finite differencing methods}

In order to numerically model the dynamics of our system, several important factors must be considered.
On the one hand, we seek to accommodate the features of large scale leaf tissues we know to be true. These include 
continuous patterns at mm and larger scales, time continuous growth, nonlinear viscoelastic behaviour of both 
cells and the tissue continuum, and a nonhomogeneously expanding geometry with large changes in scale and curvature
(typically 1000-fold areal expansion from the end of cell proliferation to maturity ~\cite{cellSizesDonnelly}).
On the other hand, we must choose appropriate numerical methods that will be optimal for modeling this system. 
A priori, there is no obvious choice for how to best simulate this set of coupled partial differential equations.

Numerical simulations of PDEs in physical systems are generally handled either by finite element methods 
or finite differencing methods ~\cite{NumericsBook}.
Other more specialized techniques exist as well, but we restrict our attention to these two broad categories as they 
are widely adopted in scientific and engineering applications. 

In finite element methods, 
the physical space being modeled is subdivided into a grid of small areas or volumes. This approach  
mimics the plant system at the cellular level, where a continuous tissue is made up of cells that 
do not move relative to each other and have rigid walls.

However, several issues arise with adopting a finite element approach. Foremost is the question of 
growth. For a growing
manifold, the geometry of the system changes with every time step and is coupled to the dynamics within
the geometry. Hence, the elements of the grid itself would have to change at every time step in nontrivial
ways such as location-dependent rescaling, the addition of more elements throughout the grid and the boundary, 
anisotropic growth of each element, and so on. 

The use of a finite element method also requires that some quantity be conserved in the system being studied.
In viscoelastic theories, this is mass and momentum. For problems in fluid mechanics, this makes
intuitive sense both globally and locally. Globally, matter and energy cannot be created or destroyed. Locally,
the influx of matter and energy across one boundary of an element must balance the efflux across the other boundaries.
In a growing cell, however, this is not necessarily the case. An influx of material can cause growth rather than 
balanced efflux. It is not obvious what the conserved quantity should be for a tissue driven to grow by
external source terms, and whether such a term can be expressed in integral form as required for a 
finite element method.

Finally, the case for finite element methods is further weakened by the inherent assumption of an Eulerian
coordinate system. This is in fundamental conflict with our previous assumption of a Lagrangian coordinate
system for the plant leaf (Section \ref{sectionRiemannianGeo}). While not irreconcilable, a finite element
method would require nontrivial calculations to translate between the two coordinate frameworks. More importantly,
key features of plant growth that depend on Lagrangian coordinates, such as the stationary behaviour of the
elongation zone in roots and the curvature of a cotyledon ~\cite{SilkSciAm}, would be more difficult 
to compare if calculated in an Eulerian coordinate system.

In the continuum limit, certain factors come into play that suggest a finite differencing approach is more appropriate.
The equations are written in differential form, which is easily representable in a variety of finite differencing
schemes. In this way, a finite differencing scheme is the most direct way of numerically interpreting the 
mathematics of the problem. The requirement for a conserved, integrable quantity is also lifted.

Furthermore, the central physical aspects of tissue growth  
can all be formalized with continuous functions. Modeling the dynamics of these continuous 
functions in a curved, dynamical, growing geometry is similar to the problems encountered in numerical general 
relativity where matter and geometry interact via nonlinear feedback dynamics. There, progress has been 
made using finite differencing methods so it is not unreasonable to try an approach based on similar numerical methods.

Lastly, a finite differencing scheme can more easily accommodate calculations in Lagrangian coordinates.

\section{The Dufort-Frankel method}
\label{sectionDFmethod}

\begin{figure}[h]
    \centering
    \includegraphics[width=0.5\textwidth]{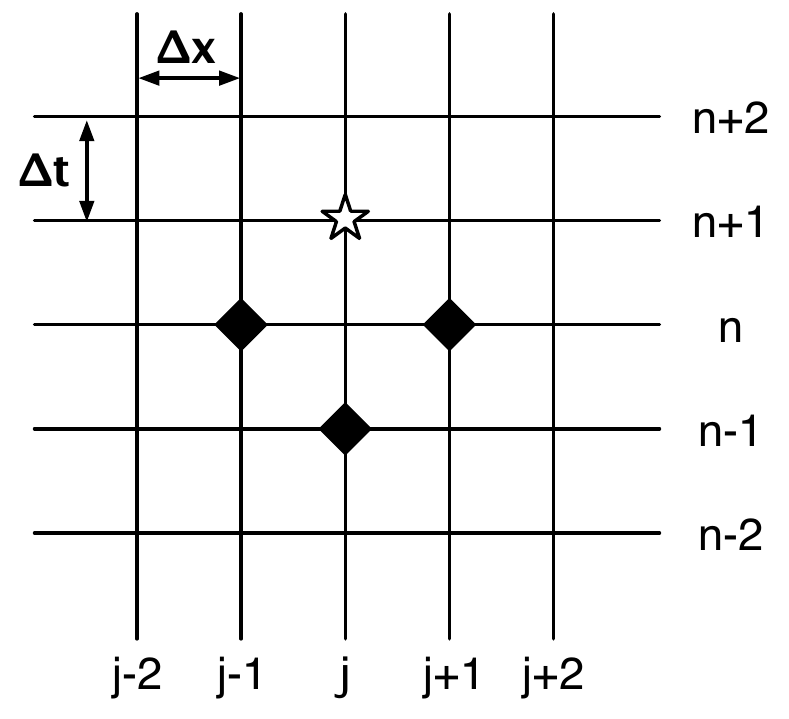}
    \caption{A representation of the Dufort-Frankel discretization method. The functional value at $t=(n+1)\Delta t$ and $x=j\Delta x$ is calculated from functional values at previous time steps.} \label{figDF}
\end{figure}

In light of the above arguments, we adopt the DuFort-Frankel finite difference method for the numerical
analysis where the equations have terms that are diffusive in one spatial dimension. That is to say,
this method is appropriate for 1D and constrained 2D geometries, but is not generally used where mixed
second derivatives appear. 

The Dufort-Frankel method is a semi-implicit differencing method that is particularly accurate for diffusive,
heat-like systems ~\cite{NumericsBook}. As we will see in the next chapters, the Ricci flow term is functionally similar to
a diffusive term when written out for each metric component. Hence the Dufort-Frankel method is chosen to 
accurately represent the dynamics of Ricci flow in the metric. The diffusion in the velocity field 
will also be treated using this method.

In general, finite differencing methods discretize the parameter space of continuous functions. In the 
formulas considered here, space and time become discrete parameters for the metric and velocity field components.

To represent this discretization, let us consider the continuous function $u$ with a spatial variable $x$ and time variable $t$.
We let the spatial and time variables be discretized in increments of size $\Delta x$ and $\Delta t$, and enumerate them using indices $j$ and $n$ 
such that

\begin{equation}
u(x,t) \rightarrow u(j \Delta x, n \Delta t) \rightarrow u_j^n. 
\end{equation}

For spatial derivatives, we use a central differencing:

\begin{equation}
\frac{\partial u}{\partial x} |_j^n \rightarrow \frac{ u_{j+1}^{n} -  u_{j-1}^{n}}{2 \Delta x}. \label{eqnCentralDiff}
\end{equation}

Let us look at how to treat the standard diffusion equation

\begin{equation}
\frac{\partial u}{\partial t} = D \frac{\partial^2 u}{\partial x^2} \label{eqnDiffusion}
\end{equation}

\noi where $D$ is the diffusion coefficient.

Using Equation \ref{eqnCentralDiff}, the standard finite difference scheme for a second-order spatial derivative yields: 

\begin{equation}
\frac{\partial^2 u}{\partial x^2} |_j^n \sim D \frac{u_{j+1}^n - 2u_j^{n} + u_{j-1}^n}{(\Delta x)^2}. \label{eqnEuler2nd}
\end{equation}

It is important to note that this standard differencing scheme uses terms evaluated all at the same time step $t=n\Delta t$.
Because of this, when expression \ref{eqnEuler2nd} is coupled to a time derivative as in the diffusion equation \ref{eqnDiffusion}, 
the method becomes unstable for $\Delta t/ (\Delta x)^2 \geq 1/2$ ~\cite{NumericsBook}. To avoid such a numerical 
instability, the Dufort-Frankel method can be used, which uses terms from two time steps rather than just one.

If we apply a central differencing scheme for the time derivative and the Dufort-Frankel method for the second-order spatial 
derivative, the discretized diffusion equation becomes:

\begin{equation}
\frac{ u_j^{n+1} -  u_j^{n-1}}{2 \Delta t} \sim D \frac{u_{j+1}^n - u_j^{n+1} - u_j^{n-1} + u_{j-1}^n}{(\Delta x)^2}. 
\end{equation}

In order to solve this equation numerically, we isolate $u_j^{n+1}$ algebraically, thus solving for $u$ at $t=(n+1)\Delta t$ using
the previously calculated values of $u$ at $t=n \Delta t$ and $t=(n-1)\Delta t$. See Figure \ref{figDF} for a pictorial representation.

Using values averaged over both space and time allows the Laplacian terms 
to be more stable than with simpler, fully explicit differencing methods such as the Euler method. However, where second derivatives 
in two different coordinates appear, in the full 2D anisotropic equations, a time-averaged Laplacian is used in place of the Dufort-Frankel method
because it is easier to implement.

\begin{equation}
\frac{ u_j^{n+1}-u_j^{n-1}}{2 \Delta t} = D \frac{1}{2} \left( \frac{u_{j+1}^n - 2u_j^{n} + u_{j-1}^n}{(\Delta x)^2} + \frac{u_{j+1}^{n-1} - 2u_j^{n-1} + u_{j-1}^{n-1}}{(\Delta x)^2}  \right)
\end{equation}

The trade-off for ease of implementation of the time averaged Laplacian scheme is reduced stability. 
Simulations using this method
have a smaller range of parameters that exhibit stable solutions as compared to the Dufort-Frankel
method used in the constrained 2D models. Furthermore, using 2D arrays
to store the 2D information of each tensor component increases the computing time several fold. Of the three geometries
developed in this work, the time-averaged Laplacian models of 2D systems are by far the most computationally expensive.

For a full derivation of how the dynamical equations become discretized, see Appendix \ref{AppendixDiscretization}.

\section{Precision of the Dufort-Frankel method} 

\begin{figure}[h]
    \centering
    \includegraphics[width=0.75\textwidth]{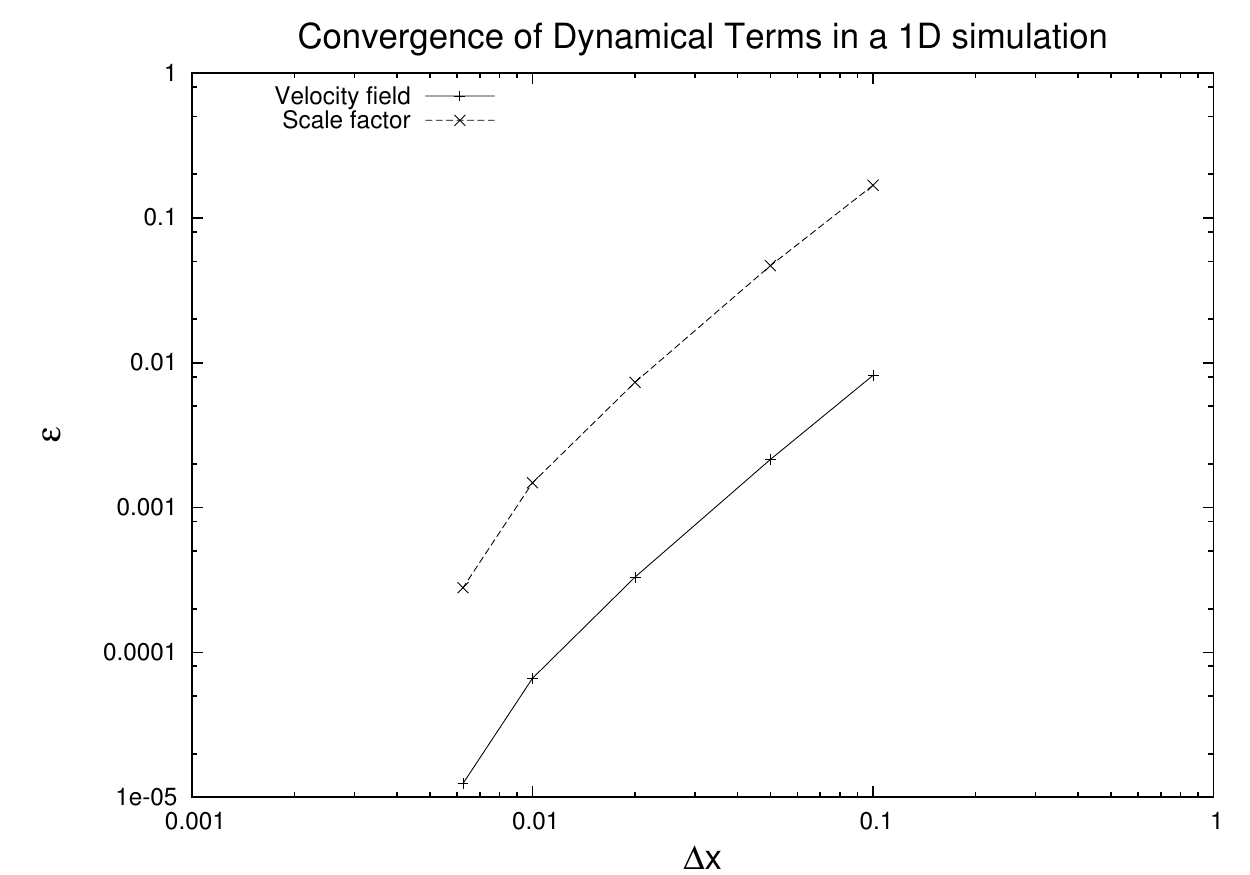}
    \caption{Convergence of a 1D simulation. Here, $\epsilon \sim (\Delta x)^{2.2}$.} \label{figConvergenceTests1D}
\end{figure}

\begin{figure}[h]
    \centering
    \includegraphics[width=0.75\textwidth]{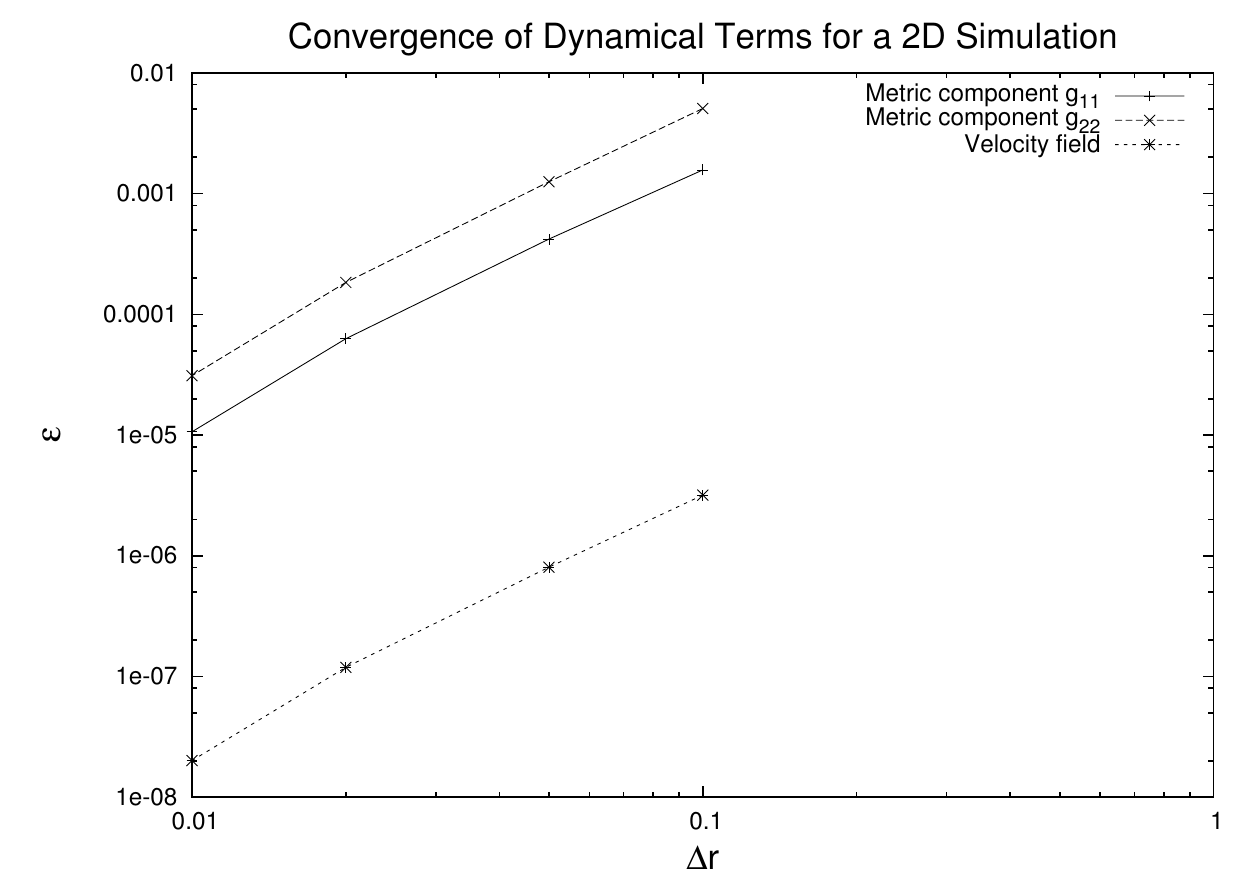}
    \caption{Convergence of a 2D simulation. Here, $\epsilon \sim (\Delta x)^{2.2}$.} \label{figConvergenceTests2D}
\end{figure}

To analyze the precision of the numerical methods used in these simulations, a convergence test can be performed which 
tests the behaviour of the system as the spatial and temporal increments are reduced. For a well-behaved
system, the simulation should have convergent values as $\Delta x$ and $\Delta t$ approach zero. 
It is important to note that using a standard analytical error analysis involving Taylor expansions of the 
derivative terms in the equations would not easily capture the nonlinear feedback effects from first-order
terms in the equations. 

The dynamcial equations used in these convergence tests can be found in Chapters \ref{chapter1DSimulations} and
\ref{chapter2DConstrainedSimulations}. They are not included here for brevity.

In Figures \ref{figConvergenceTests1D} and \ref{figConvergenceTests2D}, convergence tests for 
1D and 2D simulations using the Dufort-Frankel method model are shown. The graphs were generated by running the same 
simulation for increasingly precise time and spatial steps while maintaining the condition $\Delta t = (\Delta x)^2$.
The values used to calculate absolute error were all taken at $r=r_{max}/2$ so that error values can be
compared consistently across rescaled spatial grids. 
The error was calculated relative to the highest precision values, and the correlation between error and
precision was determined using a linear fit of the logarithms of the data sets.

In each case, the trend is for functional values to rapidly approach the highest precision 
value. This is a good indicator that numerical simulations using the Dufort-Frankel 
method are convergent. 

Moreover, these data can be used to determine appropriate spatial and temporal resolutions when running
simulations. This is necessary because a trade-off must be made between precision and computing time; with
more computational power, the simulations can be more precise for the same amount of computing
time. Hence, a balance must be found between reasonable computing time and the precision
of the simulation. In the 1D simulations, $\Delta t = 10^{-5}$ and $\Delta x = 0.05$ were used.
Throughout the 2D simulations, $\Delta t=10^{-4}$ and $\Delta r=0.1$ were used.
Given the hardware available for running the simulations, this meant run times could be kept between
a few minutes to a half hour for 1D simulations of several million time steps, and 2D simulations 
of $750000$ time steps.

The coupling constants and initial conditions used in the convergence tests were chosen to be
representative of those in the simulations presented in Chapters 4, 5 and 6. Thus, for the 1D tests:

\vspace{3mm}
\begin{center}
\begin{tabular}{r l}
\hline
Geometric reaction-diffusion coupling & $\kappa = 20.0$ \\
Growth tensor coupling & $\kappa_1 = 10.0$ \\
Velocity diffusion coupling & $c = 10.0$ \\
$f(t=0,x) = $ & $1.0$ \\
$v(t=0,x) = $ & $[1.0 + \exp (-0.2(x-\frac{1}{2}x_{max}))]^{-1}$ \\ 
\hline 
\end{tabular}
\end{center}
\vspace{3mm}

\noi For the 2D convergence tests:

\vspace{3mm}
\begin{center}
\begin{tabular}{r l}
\hline
Ricci flow coupling & $\kappa = 0.75$ \\
Growth tensor coupling & $\kappa_1 = 0.75$ \\
Velocity diffusion coupling & $c = 0.01$ \\
$f(t=0,r) = $ & $5.0 \exp (-3.0(r-\frac{1}{2}r_{max})^2) + 2.0$ \\
$g(t=0,r) = $ & $2.0 \exp (-3.0(r-\frac{1}{2}r_{max})^2) + 2.0$ \\ 
$v^1(t=0,r) = $ & $0.01/[1.0 + \exp(-5.0(r-\frac{1}{2}r_{max}))]$ \\
             & $-0.01/[1.0 + \exp(\frac{5}{2}r_{max})]$ \\ 
\hline 
\end{tabular}
\end{center}
\vspace{3mm}

The form of the initial velocity field is chosen based on data from root growth experiments. The details
of this are discussed further in Section \ref{section1DIC}. 

Performing convergence tests for the time averaged Laplacian method was not practical because of the 
long run times of these simulations even at $10^5$ time steps. Instead, knowing that the Dufort-Frankel method
is convergent, simulations using the 
time-averaged Laplacian method were compared against simulations done using the Dufort-Frankel method
using the same initial conditions. That is to say,
the time averaged Laplacian simulations had the same results as Dufort-Frankel simulations for the
same initial conditions.

\section{Accuracy of the Dufort-Frankel method: \\ Comparison to an exact solution}

Comparing a numerical solution against an exact solution is possibly the best test of the accuracy of a 
given numerical method. Central to the simulations is the question of Ricci flow, however the problem 
of numerical Ricci flow is not widely studied, so a set of exact solutions had to be found first.

For a 2D conformally flat coordinate system in Cartesian coordinates, the line element is

$$ ds^2 = f(x,y)(dx^2 + dy^2)$$
\noi and the Ricci tensor is:
$$R_{11} = R_{22} = \frac{1}{2f^2} \left[ f\frac{\partial ^2 f}{\partial x^2} + f\frac{\partial ^2 f}{\partial y^2}
- \left(\frac{\partial f}{\partial x} \right)^2 - \left(\frac{\partial f}{\partial y} \right)^2 \right ] $$

For Ricci flow, we seek a solution to the equation

\begin{equation}
\frac{\partial g_{ik}}{\partial t} = \kappa R_{ik}.
\end{equation}

\noi The sign convention used here is different than in Equation \ref{eqnRicciFlow}, but is still mathematically consistent 
with the equations studied in later chapters.

If we impose a constraint for circular symmetry and transform to polar coordinates where $x^2 + y^2 = r^2$, 
the solution for the metric satisfying Ricci flow is

$$ f(r,\theta) = \frac{1}{r^2 + e^{nt}}.$$

\noi where $n$ is a real number.

The Ricci tensor for this metric is

$$R_{ik} = - \frac{2 e^{nt}}{[r^2 + e^{nt}]^2},$$

\noi and the time derivative of the metric is 

$$\partial g_{ik}/ \partial t = - \frac{n e^{nt}}{[r^2 + e^{nt}]^2}.$$

\noi Therefore 
$$\frac{\partial g_{ik}}{\partial t} = \frac{n}{2} R_{ik}. $$

To test the numerical solution, we write the Ricci tensor in polar coordinates and use the Dufort-Frankel method
on the diffusive term:

$$ R_{ik} = \frac{1}{2r^2f^2} \left[\left( f\frac{\partial ^2 f}{\partial r^2} - \left(\frac{\partial f}{\partial r} \right)^2\right) r^2   + r f\frac{\partial f}{\partial r}   \right ]. $$

\begin{figure}[h]
    \centering
    \includegraphics[width=0.75\textwidth]{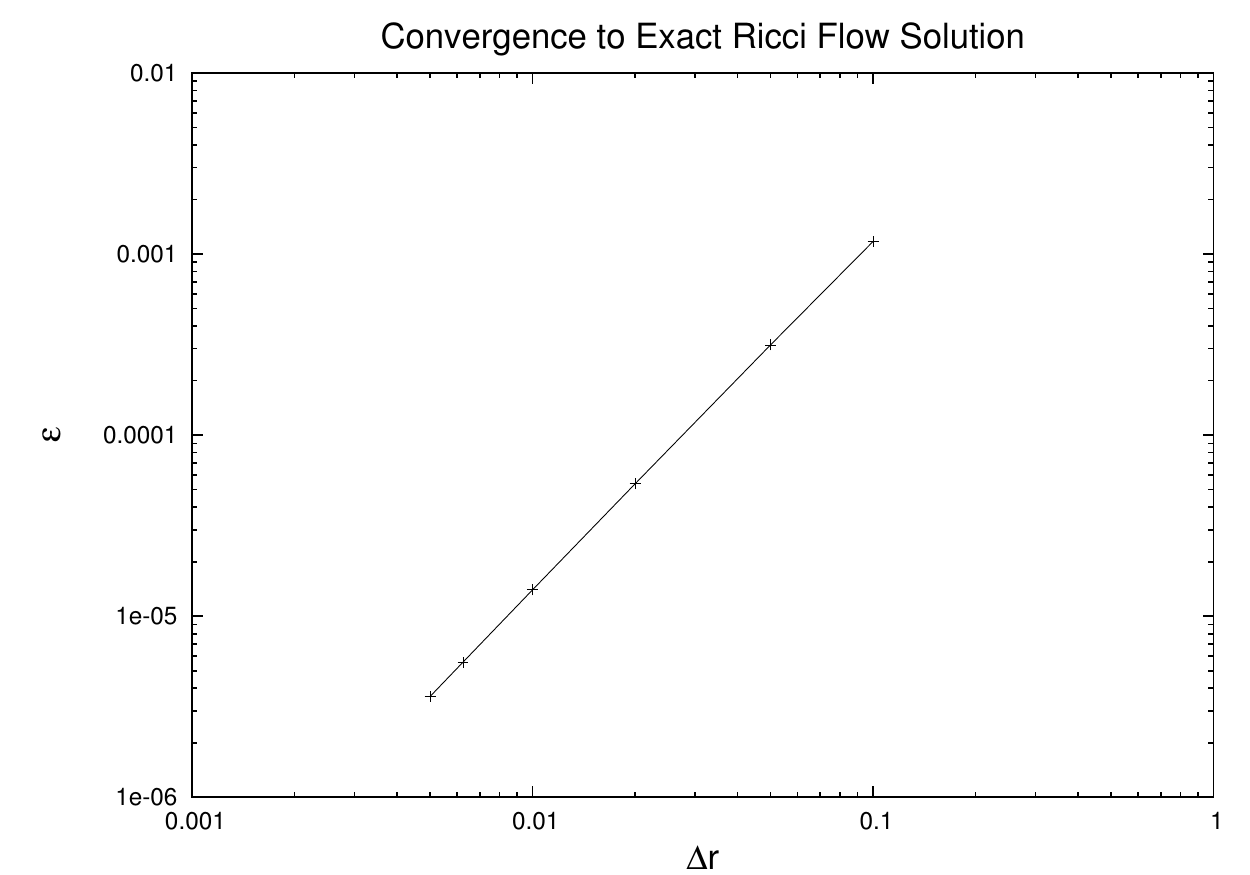}
    \caption{Convergence of numerical Ricci flow solutions to an exact solution using the Dufort-Frankel method. Here, $\epsilon \sim (\Delta r)^{1.9}$.} \label{figExactSolutionConvergence}
\end{figure}

The accuracy itself is tested by decreasing the spatial step size $\Delta r$. The results of these tests are shown in Figure
\ref{figExactSolutionConvergence}. Here, $\epsilon \sim (\Delta r)^{1.9}$ indicating that the Dufort-Frankel method 
accurately represents the diffusive dynamics of Ricci flow.

\section{Coordinate singularities}

In the 2D models of tissue growth, a plane polar coordinate system will be used. The details of this are
discussed in Chapter \ref{chapter2DConstrainedSimulations}, but we note here that 
a significant difference between the 1D and 2D equations is the presence of $1/r$ and $1/r^2$ terms in the 2D case.
This is a direct result of using polar coordinates and causes discretized operators with this behaviour to become singular 
at the origin. These types of singularities can, for the most part, be managed within the numerical simulations 
since their behaviour is known and can be anticipated. This includes imposing appropriate boundary conditions on the
dynamical terms and any other terms experiencing this behaviour (growth tensor components, connection
coefficients, Ricci tensor components, etc), and studying the region near the origin using Taylor series expansions
to ensure that the solutions are in fact regular.

\section{Boundary conditions and initial values}

Nonlinear systems like the one constructed here for plant growth generally have large sensitivities to both
boundary conditions and initial values.

Ideally, both boundary and initial conditions would be set using biological data. However, no data exists
to constrain the boundary behaviour, and very little information exists on specific initial data for the
geometric variables.

Boundary conditions were kept consistent across geometries. For the metric tensor, the first derivative
of the continuous functions defining its components was set to zero (Neumann condition) at the inner and 
outer boundary. For the velocity field 
the inner boundary was always set to zero based on the argument that 
the base of the tissue should have a vanishing velocity (Dirichlet condition). 
At the outer boundary, the velocity field was defined to have a constant first derivative. 
Specifying derivative information rather than functional values at the boundary allows
the absolute values of the functions to change over time in a way that can accommodate a wider range of growth
modes. For instance, specifying a constant velocity field value at the outer boundary would necessarily cause 
the boundary to grow outward at a prescribed rate, rather than be modulated by the dynamics of the whole tissue.
Likewise, specifying the values of the metric components at the boundary would not allow the tissue to grow at 
the boundary based on the physics inside the boundary.

Initial values were chosen based on the available data on velocity fields in 1D plant tissues in ~\cite{SilkSciAm}.
Though not strictly an initial velocity field for a growing root, the data show a monotonically increasing
logistic function for the velocity of cells in a root at some intermediate time of development. This general
behaviour was then extended to the velocity fields in our 2D models. Initial data for the metric is not 
available from experiments. For simplicity, our models start with flat initial metrics for growth models.
In models that test only the metric's response to curvature, we start with Gaussian 
perturbations superimposed onto a flat metric.

\section{Chapter summary: \\ The Dufort-Frankel method for parabolic PDEs}

In choosing the numerical methods required to solve the dynamical equations presented in the previous chapter, the first 
aspects considered were the differential nature of the equations, and whether or not they have conserved quantities. 
Given that the system is most easily expressed in differential form due to the assumption of local interactions, 
and has no obvious conserved quantities because it is an open system, finite differencing schemes are favoured over
finite element methods.

Next, we saw that the dynamical equations have parabolic-type terms due to Ricci flow in the metric and diffusion in the 
velocity field. A strong candidate for the finite differencing scheme was the Dufort-Frankel method, which is semi-implicit
scheme that uses two previous times steps to calculate the next time step. Comparing numerical and exact solutions for Ricci 
flow on a circularly symmetric disk confirmed that the Dufort-Frankel method is accurate, and the absolute error follows 
$\epsilon \sim (\Delta r)^2$. Furthermore, convergence tests showed that the numerical solutions are also well-behaved 
with increasing precision in the time and spatial steps.

A number of other methodological questions were considered as well, including the effects of 
coordinate singularities and boundary conditions.

In the next three chapters, numerical solutions to the equations of plant growth will be developed based on the 
discussion and results presented here. Numerical Ricci flow will provide the basis of the results concerning how the 
metric behaves over time; these are novel results in their own right given how little is known about Ricci flow in 2D
disk-type geometries. The dynamics of disk growth will be further augmented by coupling the velocity field
to the metric. This coupled, nonlinear system of equations will yield different modes of growth that can be 
controlled by selecting appropriate coupling parameters.

\chapter{1D Simulations}
\label{chapter1DSimulations}

We now present results from numerical simulations of 1D growth. This model is a simplification of the full 2D
model discussed up until now. Assuming that organs such as roots and stems can be considered one-dimensional objects,
with the other two dimensions being negligible,
the simulations discussed in this chapter will be applicable to such elongated structures in the macroscopic limit
(i.e. length scales greater than 1mm). 

\section{Dynamical equations of growth in 1D}
\label{section1Dequations}

Studying the 1D model is a natural first step in testing the numerical techniques required to model
the system and for
building intuition about the dynamics of the system. It is not, however, fully analogous to the 2D system. 
The main difference is that Riemannian curvature does not exist for 1D systems, which fundamentally alters 
the mathematics of the system. Without curvature coupling to the metric tensor, the nonlinear behaviour
in the metric tensor equation disappears. 

To retain the nonlinear self-coupling term for the metric evolution, we introduce a term
similar in functional form to 2D Ricci flow. This is consistent with Equation \ref{eqnKardar}, and recovers the
reaction-diffusion behaviour that allows the metric to interact with itself and its derivatives. 
In fact, the 1D nonlinear term is similar in structure to a model of interface growth introduced by 
Kardar, Parisi and Zhang ~\cite{KPZ}. The main difference between the two growth models is that the 1D interface 
growth of a height field accumulates mass by stochastic surface deposition, while in the 1D root model, growth in length
can in principle occur anywhere along the manifold. 
Since the mathematical form of this one dimensional term has self-coupling to the metric and quadratic gradients in the
reaction-diffusion term,
we call it the \textit{geometric reaction-diffusion term} $\mathcal{R}$. Including this term reflects the idea that plant tissues
can grow by depositing new material anywhere throughout their tissue.

In 1D, the reduced metric tensor has only one component $g_{11}$ which can be written as a real-valued function $f$
called the scale factor: 

\begin{equation}
g_{11} = f(\mathbf{x},t)
\end{equation}

\noi where \textbf{x} is the coordinate position relative to the base of the root. The velocity field $v$ 
is also a function of position \textbf{x} and time $t$ and has only one component $v^i = v^1(\mathbf{x},t)$.

Assuming that the root material remains at a constant density throughout its growth, the tensor equations 
for the metric and velocity field evolution (Equations \ref{eqnMetricTensor} and \ref{eqnVelocityField})  
become two coupled PDEs for the scale factor and the material velocity. 
The parameter $x$ is a co-moving coordinate that has a constant coordinate value. This is related directly 
to the way the spatial growth measurements are made. Usually the immature organism is marked with a grid of 
ink spots or pin pricks that are assumed to have no effect on the growth. This is called the \textit{landmark method}
and has been used since the 1930s to measure plant growth (initial experiments measured tobacco leaves ~\cite{avery1933structure}).
More recently, ink jet printing on leaves ~\cite{inkjetLeaves} and advanced optical techniques for roots 
~\cite{rootImagingDoublePeakedGT} ~\cite{newRootDataCurvature} ~\cite{basu2012dblREGRpeak} have allowed for precise
measurements of growth at micrometer scales.

As the organism grows, the physical distance between the markings expands. This is generally a nonhomogeneous process.
The marks are fixed in number (let us say $M$) and their coordinate values 
are also fixed at position $x^i$ where the index $i = 1, 2,... M$. This means that the coordinate separation between 
the marked positions are fixed as well ($\Delta x^i = x^{i+1}-x^i = h^i$) where $h^i$ is a constant numerical 
separation value that depends on how the markings are laid out and enumerated. The growth is then measured by 
the change in the scale factor such that the physical distance of separation is given as 

\begin{equation}
L(t) = \int_{x_1}^{x_2} \sqrt{g_{11}(x,t)} dx = \int_{x_1}^{x_2} \sqrt{f(x,t)} dx.
\label{eqnLength}
\end{equation}

\noi This is exactly the role that the metric is supposed to play for a Riemannian geometry. 
It acts as the mathematical object that turns coordinate separations into physical length separations.

In a 1D geometry, we can express all the geometrical quantities in terms of the scale factor $f(x,t)$. 
The connection coefficient is: 

\begin{equation}
\Gamma_{11}^1 = \frac{1}{2f} \frac{\partial f}{\partial x}.
\end{equation}

The covariant derivative of the velocity is: 

\begin{eqnarray}
\nabla_1 v^1 & = & \frac{\partial v^1}{\partial x} + \Gamma_{11}^1 v^1 \nonumber \\
				& = & \frac{\partial v^1}{\partial x} + \frac{1}{2f} \frac{\partial f}{\partial x} v^1.
\end{eqnarray}

The covariant derivative can be treated as a mixed second rank tensor in order to determine the
curved-space Laplacian, which is needed for the velocity diffusion term:

\begin{equation}
g^{11} \nabla_1 \nabla_1 v^1 = \frac{1}{f} \frac{\partial^2 v^1}{\partial x^2} + \frac{1}{2f^2} \left( \frac{\partial^2 f}{\partial x^2} v^1 + \frac{\partial f}{\partial x} \frac{\partial v^1}{\partial x} - \frac{1}{f} \left( \frac{\partial f}{\partial x} \right)^2 v^1 \right).
\end{equation}

From the covariant derivative, the growth tensor can be calculated as:

\begin{eqnarray}
T_{11} & = &  \nabla_1 v_1 + \nabla_1 v_1 \nonumber \\
		& = & 2 g_{11} \nabla_1 v^1 \nonumber \\
		& = & 2 f \frac{\partial v^1}{\partial x} + \frac{\partial f}{\partial x} v^1 .
\label{eqnGrowthTensor1D}
\end{eqnarray}

Finally, we define the geometric reaction-diffusion term $\mathcal{R}$ which replaces the Riemann tensor $R_{ik}$:

\begin{equation}
\mathcal{R} = \frac{1}{2f} \left[ \frac{\partial^2 f}{\partial x^2} - \frac{1}{f} \left(\frac{\partial f}{\partial x}\right)^2 \right].
\end{equation}

All the contributions to the metric and velocity field time evolution can now be combined for the 1D system, as 
stated in Section \ref{sectionInterplay}. Once again, assuming the density is constant over the time scale during
which the evolution takes place, the equations governing the 1D dynamical system are given by:

\begin{eqnarray}
\label{eqn1Dequations}
\frac{\partial f}{\partial t} & = & \frac{\kappa}{2f} \left[ \frac{\partial^2 f}{\partial x^2} - \frac{1}{f} \left(\frac{\partial f}{\partial x}\right)^2 \right] + \kappa_1 \left( 2 f \frac{\partial v}{\partial x} + v \frac{\partial f}{\partial x} \right) \label{eqn1DScaleFactorPDE} \\
\frac{\partial v}{\partial t} & = & -\frac{1}{2f} \left(\frac{\partial f}{\partial x}\right) v^2 - v \left[ \frac{\partial v}{\partial x} + \frac{1}{2f} \left(\frac{\partial f}{\partial x} \right) v \right] \nonumber \\
& &  + c \left[ \frac{1}{f} \frac{\partial^2 v}{\partial x^2} + \frac{1}{2f^2} \left( \frac{\partial^2 f}{\partial x^2} v + \frac{\partial f}{\partial x} \frac{\partial v}{\partial x} - \frac{1}{f} \left( \frac{\partial f}{\partial x} \right)^2 v \right) \right]  \label{eqn1DVelFieldPDE}
\end{eqnarray}

\noi where we have replaced $v^1$ with $v$.

Before discussing the results obtained from numerical simulations based upon these equations, it should be 
noted that the 1D equations under certain circumstances reduce to other well known equations used to explore 
uni-dimensional systems.

First the velocity equation in the case of flat Euclidean space $(f(x) = 1)$ reduces to the viscous Burger’s equation:

\begin{equation}
\frac{\partial v}{\partial t} = -v \frac{\partial v}{\partial x} + c \frac{\partial^2 v}{\partial x^2},
\end{equation}

\noindent or if $c = 0$, the inviscid Burger’s equation. These equations have been used to model different flow patterns 
arising in systems ranging from turbulent gas dynamics, to urban traffic flow, to large scale cosmological structures.

Equation \ref{eqn1DVelFieldPDE}, the velocity field evolution, is also a generalized 
reaction-diffusion equation of the form:

\begin{equation}
\frac{\partial v}{\partial t} =  -\frac{1}{f} \left(\frac{\partial f}{\partial x}\right) v^2 - \frac{c}{f} \frac{\partial^2 v}{\partial x^2}  + \cdots \nonumber
\end{equation}

The first term is the reaction component, given by nonlinear geodesic flow which is proportional to the connection coefficient. 
Note that the effect of the covariant derivative in the advection term of Equation \ref{eqn1DVelFieldPDE} is to double
the geodesic flow term. The second term
is diffusive flow. The remaining terms in Equation \ref{eqn1DVelFieldPDE} are linear in the velocity or its derivative with respect to $x$.

Equation \ref{eqn1DScaleFactorPDE}, the evolution of the scale factor, reduces to an equation developed by Kardar, Parisi and Zhang \cite{KPZ}
which, as discussed in Section \ref{sectionKPZstructure}, takes the form:

\begin{equation}
\frac{\partial h(t,\mathbf{x})}{\partial t} = \nu \nabla^2 h + \frac{\lambda}{2} (\nabla h)^2 + \cdots  \nonumber
\end{equation}

\noi where $h$ is the dynamical 'height function' that changes due to deposition of material on a two dimensional surface.
Comparing with Equation \ref{eqn1DScaleFactorPDE}, we identify $h(t,\mathbf{x})$ with $f(t,\mathbf{x})$, $\nu$ with $\kappa /2f$ 
and $\lambda /2$ with $-\kappa /f^2$. Recently, the KPZ equation has sparked interest among mathematicians studying nonlinear
PDEs, including a 2014 Fields medal won by Martin Hairer for his work in the area. 

This equation describes the growth of a surface through deposition of material where $h(t,{\bf x})$ is 
a height field measured from some flat horizontal reference surface. Examples include 
the build up of a snowpack during a snow storm, the formation of sand dunes by wind deposition, etc. In our
case the height function is replaced by a scale factor that can evolve as a result of a build up of material
occurring throughout the system. 

The fact that the equations reduce to those discussed above, all of which have been studied extensively, 
provides us with a number of comparisons that can be made with well known results in order to verify 
that the numerical methods we have chosen are valid.

\section{Initial conditions, boundary conditions and coupling parameters for a growing root}
\label{section1DIC}

In the numerical simulations, we evolve the metric tensor and velocity field over time for $5 \times 10^5$ time steps, 
subject to coupling parameters:

\vspace{3mm}
\begin{center}
\begin{tabular}{r l}
\hline
Geometric reaction-diffusion coupling & $\kappa = 20.0$ \\
Growth tensor coupling & $\kappa_1 = (0.0,10.0,20.0,30.0)$ \\
Velocity diffusion coupling & $c = 10.0$ \\
\hline 
\end{tabular}
\end{center}
\vspace{3mm}

\noi where the variation in $\kappa_1$ represents different growth tensor couplings.

The intial conditions are:

\vspace{3mm}
\begin{center}
\begin{tabular}{r l}
\hline
$f(t=0,x) = $ & $1.0$ \\
$v(t=0,x) = $ & $[1.0 + \exp(-0.2(x-\frac{1}{2}x_{max}))]^{-1}$ \\ 
\hline 
\end{tabular}
\end{center}
\vspace{3mm}

The boundary conditions in 1D are: 

\vspace{3mm}
\begin{center}
\begin{tabular}{r c l}
\hline
$\partial_x f(t,x=0)$ & = & 0 \\
$\partial_x f(t,x=x_{max})$ & = & 0 \\
$v(t,x=0)$ & = & $v(t=0,x=0) \sim 0$ \\
$\partial v(t,x=x_{max})/\partial x$ & = & $\phi (t)$ \\
\hline 
\end{tabular}
\end{center}
\vspace{3mm}

The function $\phi (t)$ is equivalent to $\partial v^1/\partial x$ evaluated at $x=x_{max-1}$.

The initial condition on the velocity field was chosen based on data available on relative elemental growth rates of
corn roots, which indicates that the velocity field has a logistic shape ~\cite{SilkSciAm} ~\cite{cornRootGrowth}. 
We approximate the velocity field of the root as a logistic 
function. The functional fit is not a true regression, but is consistent with the form of measured velocity fields 
in roots across many species ~\cite{rootImaging}. The magnitude of the velocity field is scaled so that
the maximum speed is 1.0 (distance unit)/(time unit). Since distance and time units in this model are arbitrary, the magnitude of the velocity 
field can be rescaled without loss of generality.  We preserve the convention found in most root growth literature that
the tip is the origin of the coordinate system, i.e. $x=0$ corresponds to the tip of the growing root with the base
of the root appearing to grow away from the tip.

For the initial condition on the metric tensor, we assumed a simple flat geometry. 
This is analogous to inscribing a pattern of equidistant markers on the root at the beginning of a growth
experiment.

\section{What do the data look like?}

In presenting the information, we choose the total plant length and sectional growth as data most easily
comparable to biological data on root growth (using Equation \ref{eqnLength}). 

The sectional growth in our model is the equivalent of the landmark method discussed earlier in the chapter.
This form of presenting the data allows us to understand where growth occurs in a tissue. In 
plant roots, growth does not occur homogeneously throughout the tissue. The portion of root that grows
the most is called the \textit{elongation zone}, and is a common feature in many species found just behind
the root tip extending for several mm, and ending in the mature root tissue that experiences little to 
no growth once it leaves the elongation zone ~\cite{taiz}. The markings placed on the root tissue are, in our model, represented by 
initially equidistant coordinate positions. As the dynamics of the system evolve, growth occurs nonhomogeneously 
which becomes evident as different segments come to have different lengths.

We also present the metric tensor, 
material velocity, geometric reaction-diffusion term and growth tensor which reflect the dynamics of the system itself.

An additional interpretation of the velocity data can be made by plotting the velocity field as a function of
proper radial distance $r'$ from the origin (tip), where $dr'= \sqrt{g_{11}}dr$. 
This format will be called a \textit{expansion diagram} because it shows how the velocity is distributed 
over the increasing length of the growing root. For the 2D plant
leaf, it is a representation of the velocity of points on the leaf relative to an observer
at the origin of the leaf ($x_0=0$) where the local velocity vanishes. From this point of view the expansion 
diagram is analogous to a Hubble diagram that in cosmology plots the recessional velocity of galaxies with 
respect to the observer. It is interesting to note the analogy between plant growth and the expansion of 
the universe is rather close. The markers used in cosmology are the positions of the galaxies. Each galaxy can have
its own proper motion due to a combination of interactions with other galaxies and its initial conditions.
These proper motions are 'dragged' along with the expansion of the universe, so one of the main problems in 
cosmology is to distinguish between the motion strictly due to expansion and the motion driven by local interactions.
However, since the standard model of our universe is homogeneous and isotropic, the scale factor that measures the physical 
distances between the galaxies depends only on time. Also, both in cosmology and a growing plant tissue, local expansion 
contributes to the overall structure on the largest scales.

Lastly, we can also use the velocity field to construct the expansion scalar $\Theta$. As discussed in Sections
\ref{sectionGrowthWithTensors} and \ref{sectionMechTensors}, growth is a tensor quantity that is
closely related to viscoelastic deformations (expansion, shear and rotation). In the 1D system, given the definitions
in Section \ref{sectionMechTensors}, shear is directly proportional to the growth tensor component $T_{11}$, rotation vanishes, 
and the expansion scalar becomes:

\beq
\Theta = \nabla_1 v^1 = \frac{T_{11}}{2f}.
\label{eqn1Dexpansion}
\eeq

Recall from Section \ref{sectionMassDensityEqns} that the assumption of constant tissue density in space and time
leads to the source field $S(x,t)$ being directly proportional to the expansion scalar $\Theta$. Hence, mapping out the
expansion scalar will also give us information on how the source of material is distributed through the tissue.



\begin{figure}[t]
    \centering
    \includegraphics[width=0.75\textwidth]{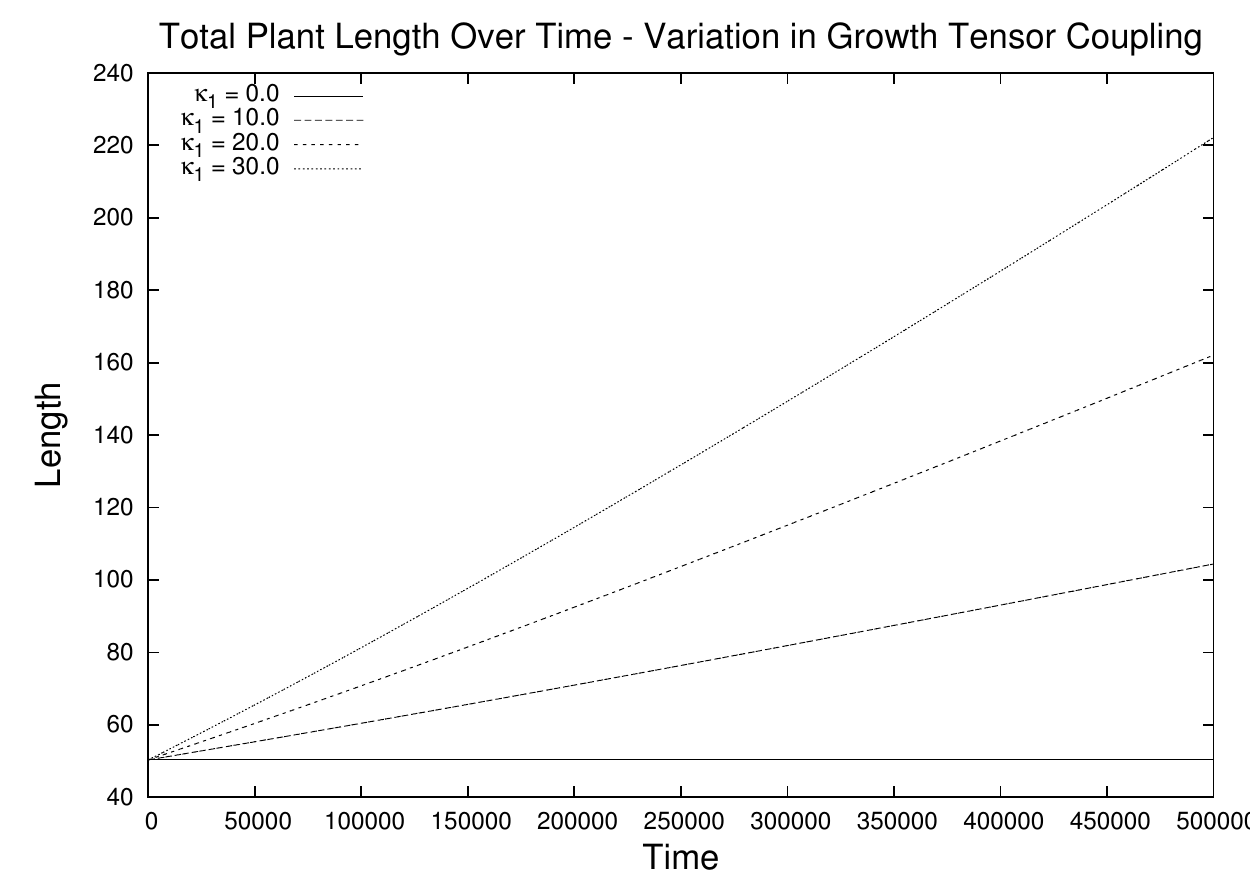}
    \caption{Plant length in 1D for four different growth tensor couplings.}\label{figure1DPlantLength}
\end{figure}

\begin{figure}[h!]
    \centering
    \includegraphics[width=0.75\textwidth]{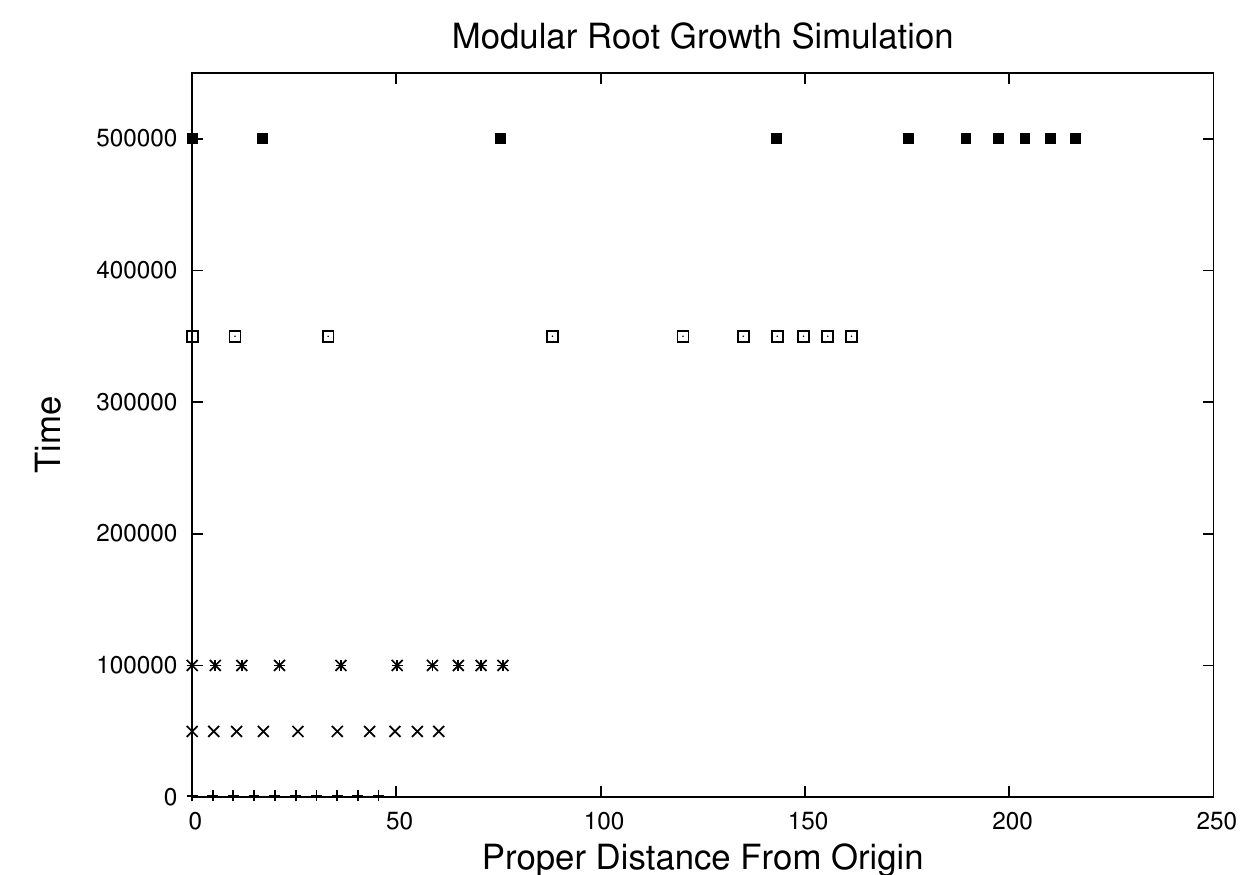}
    \caption{Modular growth in 1D for growth tensor coupling $\kappa_1 = 30.0$. Note that the tip is designated as the origin of the coordinate system.}\label{figure1DModGrowth}
\end{figure}

\section{Results for 1D simulations}

Figure \ref{figure1DPlantLength} shows that the total root length increases in a nearly linear fashion (slightly superlinear).
Depending on how strongly the growth tensor is coupled to the evolution of the metric, the root
may not grow at all (null coupling), or more than quadruple in length (highest coupling).

Figure \ref{figure1DModGrowth} shows the sectional 
growth pattern of the root simulation. Initially, the 'landmarks' are equally
spaced on the spatial domain. Over time, growth is localized near the origin, as can be seen by the 
landmarks near the origin spreading apart from each other, while landmarks further away from the origin
experience little growth relative to their neighbours. This qualitatively resembles the general pattern of root growth ~\cite{taiz}, 
with a clearly recognizable elongation zone behind the growing tip.
For a direct comparison with biological data, see Figure 3 in ~\cite{cornRootGrowth}, and Figures 4 and 7 in ~\cite{SilkSciAm}.

\begin{figure}[h]
    \centering
    \includegraphics[width=0.75\textwidth]{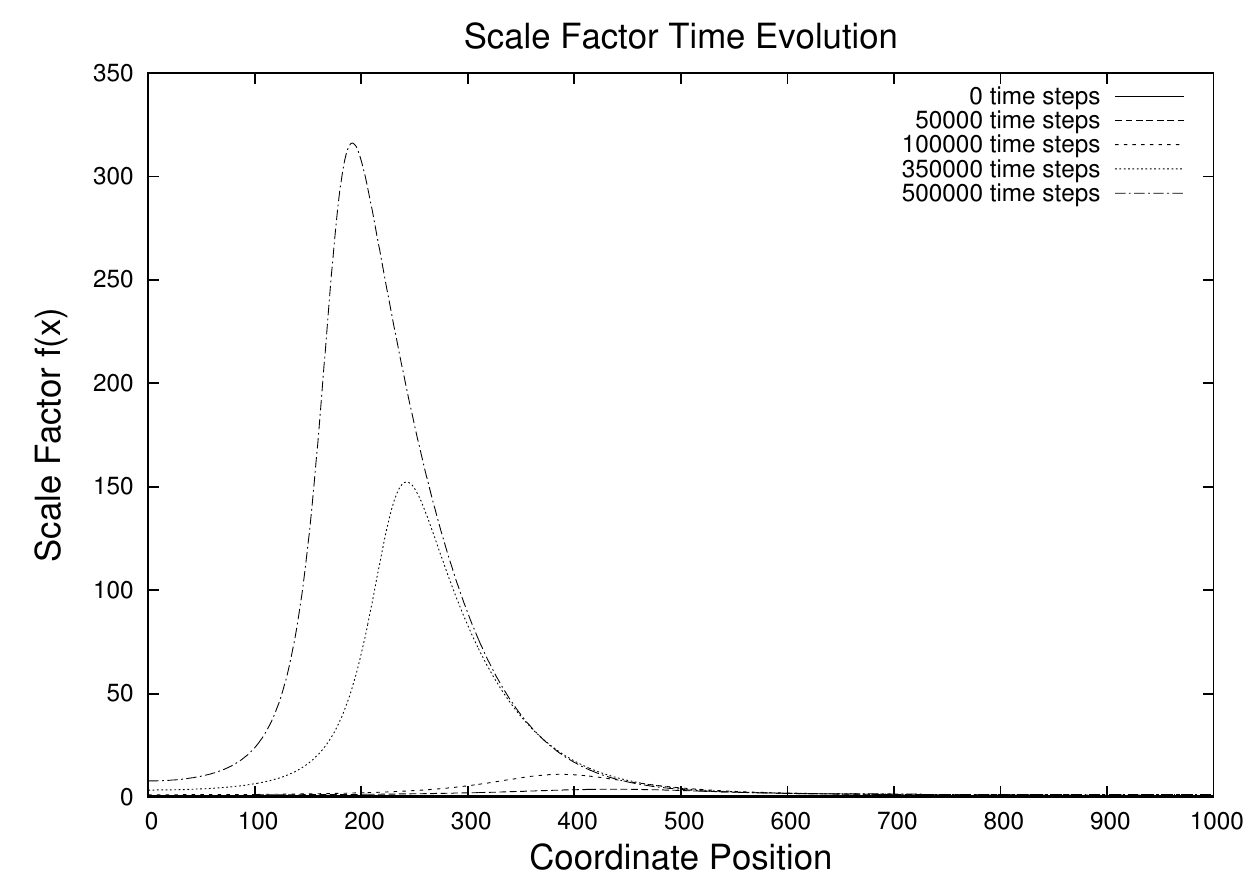}
    \caption{Time evolution of the metric tensor for growth tensor coupling $\kappa_1 = 30.0$.}\label{figure1Dmetric}
\end{figure}

\begin{figure}[h]
    \centering
    \includegraphics[width=0.75\textwidth]{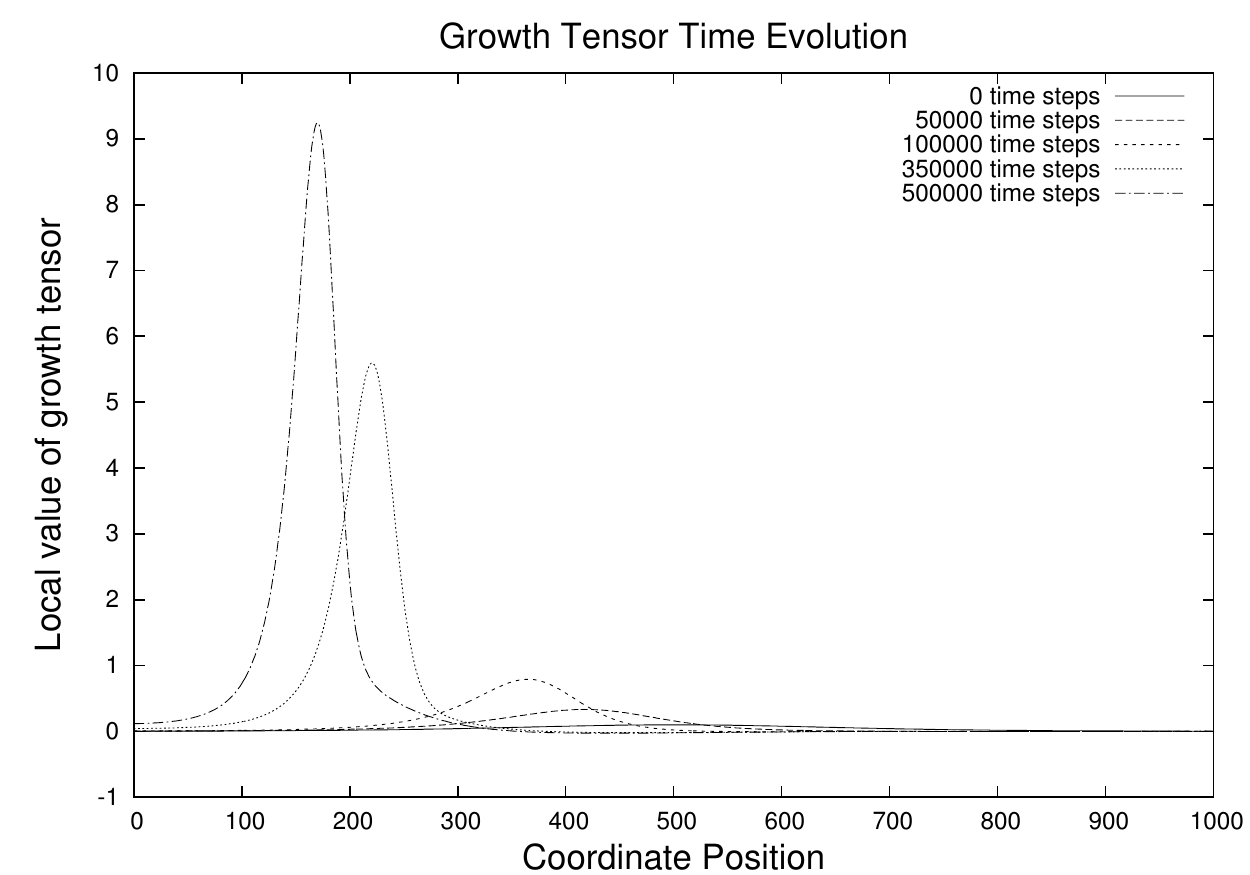}
    \caption{Time evolution of the growth tensor for growth tensor coupling $\kappa_1 = 30.0$.}\label{figure1DGT}
\end{figure}

\begin{figure}[h]
    \centering
    \includegraphics[width=0.75\textwidth]{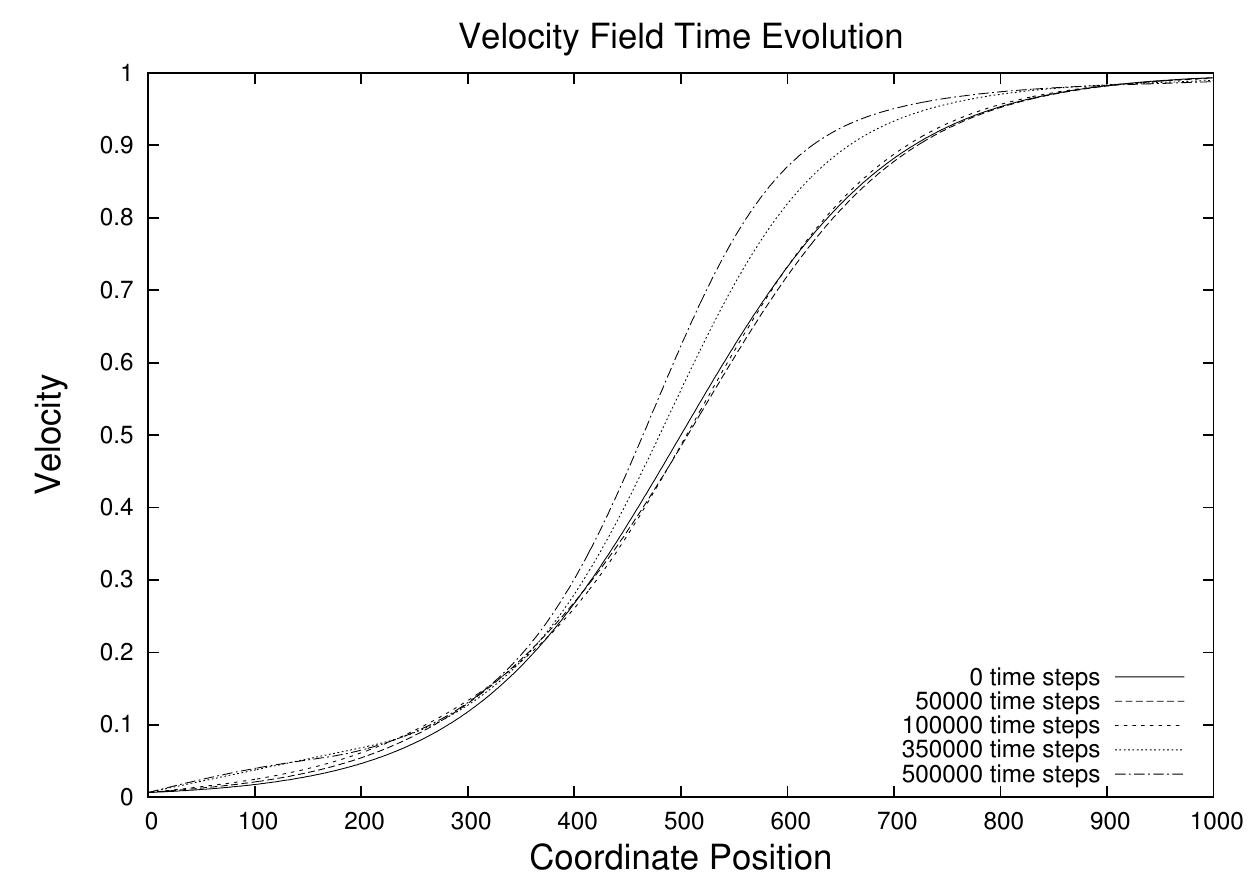}
    \caption{Time evolution of the velocity field for growth tensor coupling $\kappa_1 = 30.0$.}\label{figure1Dvelocity}
\end{figure}

\begin{figure}[h]
    \centering
    \includegraphics[width=0.75\textwidth]{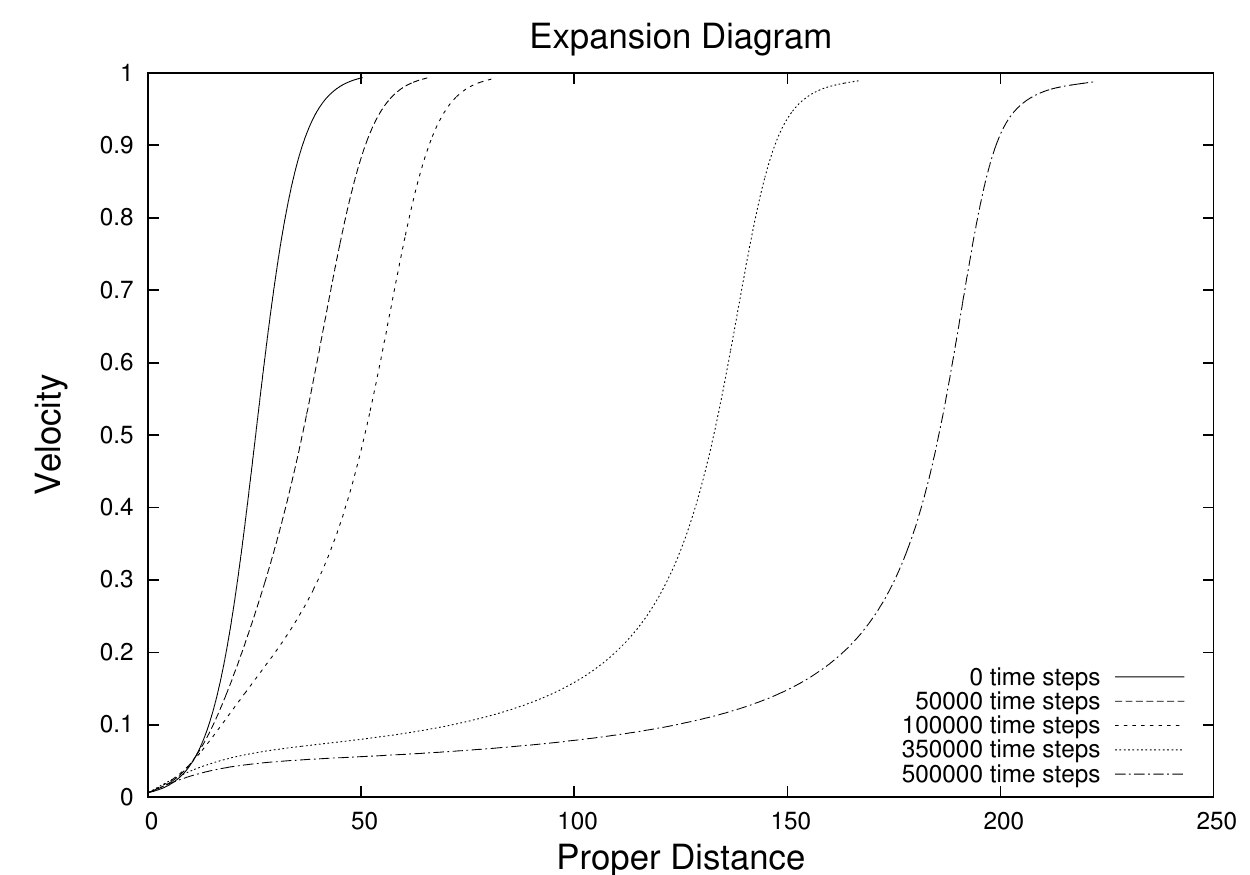}
    \caption{Expansion diagram for growth tensor coupling $\kappa_1 = 30.0$.}\label{figure1Dhubble}
\end{figure}

\begin{figure}[h]
    \centering
    \includegraphics[width=0.75\textwidth]{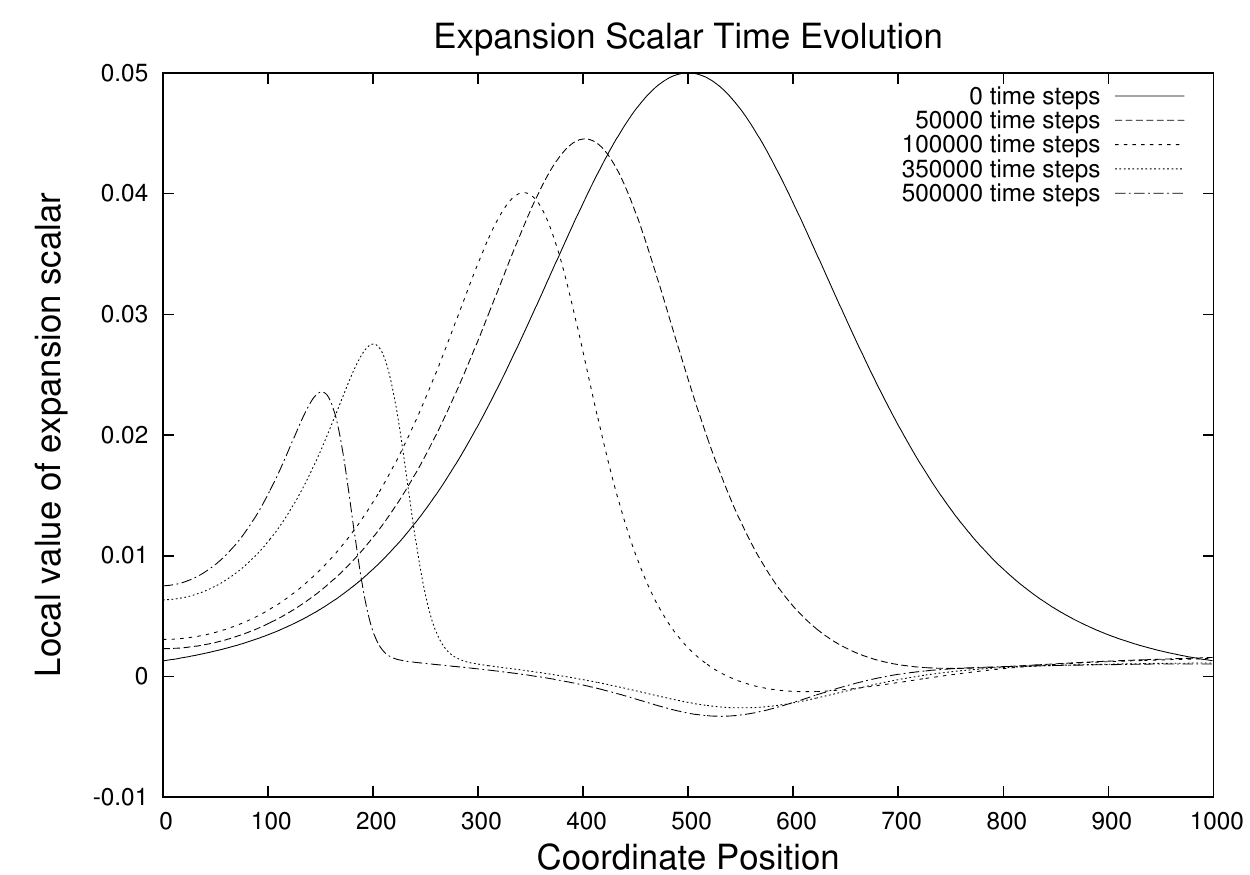}
    \caption{Expansion scalar time evolution for growth tensor coupling $\kappa_1 = 30.0$.}\label{figure1Dexpansion}
\end{figure}

\begin{figure}[h]
    \centering
    \includegraphics[width=0.75\textwidth]{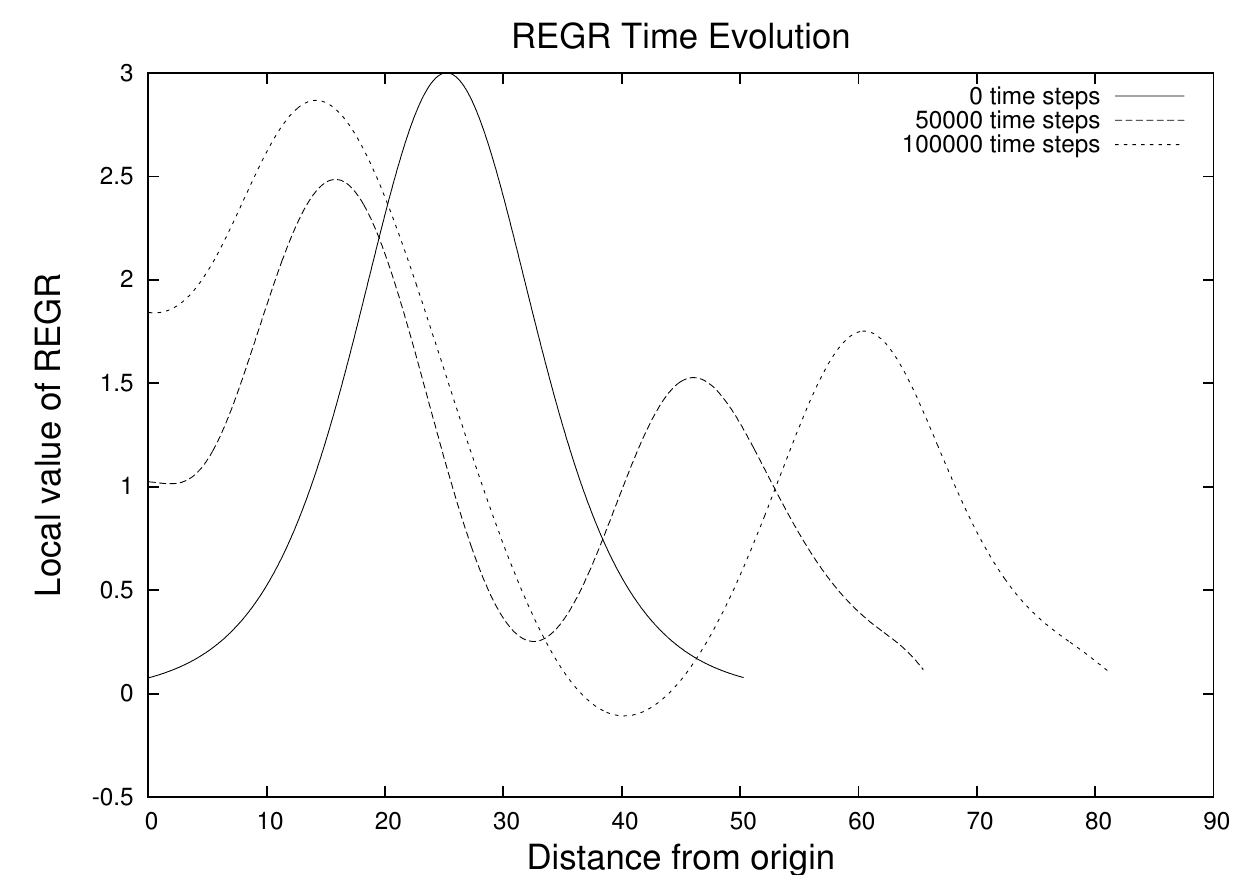}
    \caption{REGR time evolution for growth tensor coupling $\kappa_1 = 30.0$.}\label{figure1Dregr}
\end{figure}

\begin{figure}[h]
    \centering
    \includegraphics[width=0.75\textwidth]{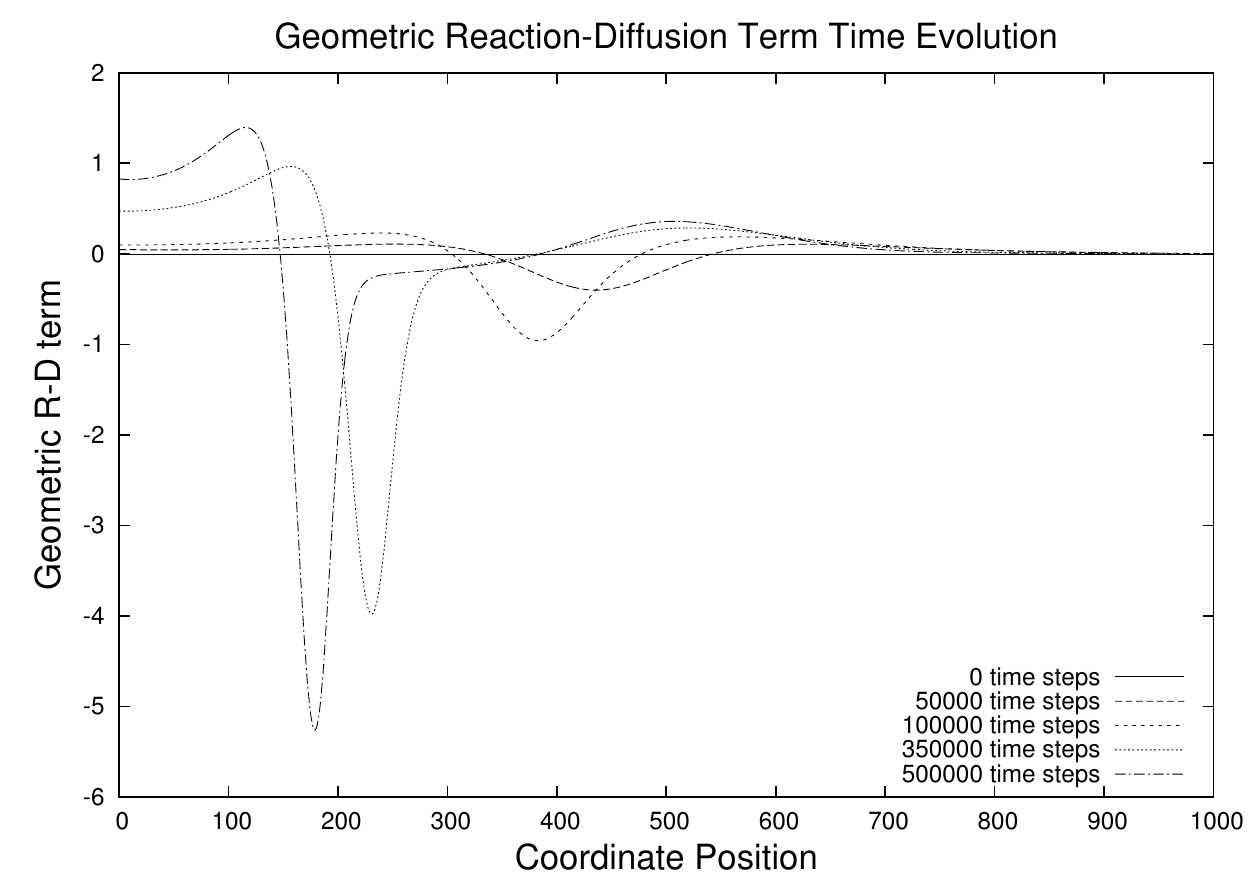}
    \caption{Time evolution of the geometric reaction-diffusion term for growth tensor coupling $\kappa_1 = 30.0$. Note
    that the $5\times 10^5$ timestep is omitted so that earlier time steps may be more clearly discerned.}\label{figure1Dkpz}
\end{figure}

Figures \ref{figure1Dmetric}, \ref{figure1DGT} and \ref{figure1Dvelocity} show that the scale factor, growth tensor
and material velocity are self-consistent with the results 
above. The metric tensor and growth tensor show that distances between neighbouring points increase the most in a region that
corresponds with the elongation zone seen in Figure \ref{figure1DModGrowth}.

The velocity field undergoes nontrivial dynamics due to both viscous Burgers' type behaviour and
contributions from the position-dependent expansion of the root. The steepening of the velocity field is typical
of what happens in Burger's equation, but the diffusive term prevents the flow from forming a discontinuity (shock)
as happens in the inviscid Burger's equation. This type of shock in the root model would in fact be physically impossible
to sustain, since the steepening of the velocity field would eventually lead to a step function, indicating that the
material would have two different and discontinuous velocities in adjoining regions of the tissue. An alternate mathematical
technique to avoid these types of shocks would be to average the velocity field over neighbouring spatial regions. Physically, it is
unclear how this would happen in the tissue; a diffusive mechanism provides a more intuitive way for the tissue to avoid such 
shocks.

The functional form of the initial velocity field is an important factor in the model. Recalling from earlier in the chapter,
the initial velocity field is based on experimental root growth data which shows that roots have a monotonically increasing
logistic velocity field. Most significantly, the gradients of the velocity field determine where growth occurs initially. 
As determined earlier in the chapter, the growth tensor in the 1D geometry is:

\begin{equation}
T_{11} = 2 f \frac{\partial v^1}{\partial x} + \frac{\partial f}{\partial x} v^1 .
\end{equation}

\noi Hence, at early times when the metric is flat or nearly flat, the region 
with the steepest velocity gradient will experience the highest growth rates, causing the scale factor in that region to
expand significantly more than in neighbouring regions. However at late times, the feedback between 
the scale factor and the growth tensor becomes dominant. This
late-time behaviour is interesting in that the dominant term in the growth tensor is no longer the spatial derivative of the velocity
field, but rather the spatial derivative of the scale factor (metric). This can be seen in Figures \ref{figure1Dmetric}, \ref{figure1DGT}
and \ref{figure1Dvelocity} where the peak of the growth tensor is initially small and correlates to the steepest part
of the velocity's logistic shape, but in late times the peak of the growth tensor shifts closer to the origin and is
correlated with the highest values of the scale factor's spatial derivatives. Hence, the geometry of the root has a 
significant impact on where growth occurs in the tissue.

Figure \ref{figure1Dhubble} shows how the velocity field evolves as the length of the root increases. This plot
combines information about the root length as it would be measured in an experiment (the proper length) with the
velocity field of the material that makes up the root. True to the general model of root growth, the base of the root
does not contribute much to the growth, whereas the region behind the elongation zone continues to grow, pushing the root forward
and allowing the tip itself to maintain a nearly constant velocity relative to the base.

Figure \ref{figure1Dexpansion} shows how the expansion scalar evolves over time. As Equation \ref{eqn1Dexpansion} shows,
it is closely related to the growth tensor $T_{11}$, and from Section \ref{sectionMassDensityEqns} we also know the expansion
scalar is directly proportional to the distributed source field $S(x,t)$. We can see that expansion decreases over time and
moves in a wave-like fashion toward the root tip. Given how large the scale factor $f$ gets over time (Figure 
\ref{figure1Dmetric}), this plot
essentially shows that growth has a positive feedback quality to it; at late times, even a small expansion leads to 
large overall growth because the geometric space itself has expanded. This plot is also consistent with the dynamics of the
growth tensor discussed above, where at early times $T_{11}$ is dominated by velocity field gradients, while at late
times is dominated by scale factor gradients.

Another interesting aspect of the expansion scalar (or equivalently in our model, the source term) is that 
it is related to the geometric expansion of the root. For an initially constant scale factor, Equation 
\ref{eqn1DScaleFactorPDE} reduces to

\beq
\frac{\partial f}{\partial t} \sim \kappa_1 T_{11}.
\eeq

\noi If we divide both sides by $2f$ and use Equation \ref{eqn1Dexpansion}:

\beq
\frac{1}{2f} \frac{\partial f}{\partial t} \sim \kappa_1 \frac{T_{11}}{2f} = \kappa_1 \Theta.
\label{eqnFlatExpansion}
\eeq

\noi Hence, the expansion scalar gives us some information on where the scale factor, and hence the geometry, is expanding.

Interestingly, the relation in Equation \ref{eqnFlatExpansion} is often used in tissue growth literature to express the
relative elemental growth rate (see for example ~\cite{newRootDataCurvature}) such that

\beq
REGR = \frac{1}{f} \frac{\partial f}{\partial t}.
\label{eqnREGRgeneral}
\eeq

Clearly, the REGR is proportional to the expansion scalar $\Theta$
only in the absence of a geometric reaction-diffusion term $\mathcal{R}$. Indeed, the standard definition of REGR in
plant growth dynamics literature equates REGR with the expansion scalar. In the model developed here, the time evolution of the scale factor
also depends on the diffusive behaviour of the geometric reaction-diffusion term. Hence, in a model of plant growth
that includes diffusive coupling to the geometry, the REGR and expansion scalar are not equivalent. 
The REGR can be obtained by dividing Equation \ref{eqn1DScaleFactorPDE} by the scale factor $f$ and using the
definition from Equation \ref{eqn1Dexpansion}:

\beq
REGR = \kappa \frac{\mathcal{R}}{f}  + 2 \kappa_1 \Theta.
\label{eqnREGR}
\eeq

Recalling that $\mathcal{R}$ is similar to the deposition term in the KPZ equation (Section \ref{section1Dequations}), 
this result states that growth is due to both a deposition of material throughout the tissue as well as expansion of the tissue,
which is also consistent with the distributed source term introduced in the continuity equation (Section \ref{sectionMassDensityEqns}).
Furthermore, REGR can be understood as a true geometric measure of growth when calculated from the metric tensor as in Equation 
\ref{eqnREGR}. In fact, the same measure of geometric growth is used in cosmology to describe the expansion of the universe; 
in that context it is known as the Hubble parameter $H = \dot{S}/{S}$ where $S(t)$ is a scale factor representing the 'radius' 
of the universe ~\cite{dinverno}.

The time evolution of the REGR using proper distances is shown in Figure \ref{figure1Dregr}, in a format that allows for direct
comparison to the results of Figure 7 in ~\cite{rootImagingDoublePeakedGT} and Figure 4 in ~\cite{basu2012dblREGRpeak}, 
where the dissipative, double peaked behaviour of the REGR 
near the root tip is consistent with the results of our simulation. If we had assumed that only the gradient of the velocity field
contributes to the REGR, then Figures \ref{figure1Dvelocity} and \ref{figure1Dhubble} would imply a very different behaviour
than is seen in Figure \ref{figure1Dregr}. The difference is due to the effects of the 1D geometry, both in the geometric reaction-diffusion
term, and the covariant derivative used to calculate the expansion scalar. In the case of this simulation, the REGR has nearly
equal contributions from the expansion scalar and geometric reaction-diffusion term, but with opposing signs and slightly misaligned
peak values, leading to the double-peaked behaviour. It is interesting to note that new, high-resolution imaging techniques of root growth have 
indeed shown that the REGR is not merely the gradient of the velocity, but that it exibits a double-peaked, time-varying behaviour not fully correlated 
with just the velocity field structure (Figure 5 in ~\cite{rootImagingDoublePeakedGT} and Figure 4 in ~\cite{basu2012dblREGRpeak}). 
To our knowledge, a hypothesis explaining these results is still lacking, however, our theory provides a mechanism to extend the 
definition of the REGR that could in principle be used to explain time-varying REGRs with behaviours not 
solely linked to the gradient of the velocity field.
 
Let us now look at the time evolution of the geometric reaction-diffusion term $\mathcal{R}$ in Figure \ref{figure1Dkpz}
The most important feature is the emergence of domains in the geometry that alternate between positive and negative values
of the geometric reaction-diffusion term. This hints at the possibility that in 2D, domains of positive and negative curvature 
can emerge from growth dynamics, yielding the ruffled and buckled geometries characteristic of non-flat leaves.

Another feature of the dynamics of the geometric reaction-diffusion term and other variables is that there is a wavelike motion of the peaks 
and troughs along the length of the organism, including an increase in the magnitude of the peaks and troughs over time. 
This behaviour occurs in many other systems whose dynamics is 
governed by reaction-diffusion equations where patterns in the field variables form and shift over time.

The results shown above are a small subset of the parameter sets tested overall. They are, however, representative
of the main features that came to light. Namely, the coupling to the growth tensor $\kappa_1$ dominates the dynamics of the
simulation. In terms of the coupling parameters in the equations, we find that the same variance in geometric 
reaction-diffusion coupling $\kappa$ or velocity diffusion $c$ 
(i.e. $0.0-30.0$ in coupling strength) does not produce as dramatic differences in final root length and the
associated dynamics in the metric tensor and velocity field. 

Another important aspect of the parameter space is that in the absence of geometric reaction-diffusion coupling ($\kappa = 0$) 
the system is unstable and the growth is unrestricted. 
If we look at the evolution of an intially flat metric under this condition we have, from Equation \ref{eqn1DScaleFactorPDE}:

\begin{equation}
\frac{\partial f}{\partial t} = 2 \kappa_1 \frac{\partial v}{\partial x} f,
\end{equation}

\noi which has an exponential solution of the form:

\begin{equation}
f(x,t) \propto \exp \left[ \int \left( 2 \kappa_1 \frac{\partial v(x,t)}{\partial x} \right) dt \right].
\end{equation}

Moreover, as $f$ increases exponentially, the velocity field is driven even more quickly to a shock front. This can be seen in 
Equation \ref{eqn1DVelFieldPDE} where velocity diffusion is inversely proportional to $f^3$. Hence, as $f$ increases,
the velocity field is less and less capable of dissipating. 

Thus, in the absence of a geometric reaction-diffusion term, the growth of the metric is initially exponential, with 
the fastest growth occurring where the spatial derivative of $v$ is largest. At the same time, the exponential growth of the metric
causes a shock front in the velocity field by shutting down the diffusion term where it is needed most. 
This seems to imply that a \textit{geometric} dissipative term is essential in distributing the “energy” associated with 
growth between elongation and buckling of the metric.

\section{Conclusions on 1D results}
\label{section1DConclusion}

The 1D results presented here serve as a proof of concept for the dynamical equations of plant growth. 
The results show qualitative similarities to real biological systems (roots) and demonstrate that the dynamics of a model
based on nonlinear coupled tensor equations can produce biologically relevant global properties. For a comparison to
biological data, see Figure 3 in ~\cite{cornRootGrowth} which shows the modular growth of a corn root; Figures 7 and 8 in ~\cite{SilkSciAm} 
showing time-continuous root growth data and the corresponding root velocity field and REGR data;
Figures 4,5 and 7 showing the REGR distribution in three different root specimens, the time variation of REGR
at short time scales, and the REGR time evolution in a root over 120min in ~\cite{rootImagingDoublePeakedGT};
Figure 4 in ~\cite{basu2012dblREGRpeak} showing a double-peaked REGR time evolution over 5 hours in two different plant species.

In the 1D model, growth is initiated by gradients in the velocity field, which is initially a logistic function mimicking
the velocity field of the primary root in many plant species. Gradients in the velocity field provide the growth tensor with 
its initial form. The growth tensor then causes significant changes in the metric tensor. This in turn affects
the transport of material through the connection coefficients found in the velocity equation. As seen in analyzing the REGR,
growth occurs both through material expansion via a distributed source of material, and the deposition of material throughout 
the tissue.

To our knowledge, no other root growth model has predicted the temporal behaviour of the velocity field, metric tensor or any 
of the other tensor quantities presented here. Furthermore, our theory provides a mechanism to extend the definition
of the REGR that could in principle be used to explain time-varying REGRs with behaviours not solely linked to the gradient of the velocity field, as
observed in ~\cite{rootImagingDoublePeakedGT} and ~\cite{basu2012dblREGRpeak}. This is because the REGR in our model 
includes a geometric reaction-diffusion (deposition) term in addition to a covariantly defined expansion scalar in the REGR.

One major limitation in producing these simulations is that how the velocity field of a primary root changes over long 
periods of time (i.e. over a quadrupling in length scale) is not well documented in biological literature. 
The measurement is difficult with such large root tip displacements, and corresponds to time scales on the 
order of several days ~\cite{beemsterBaskin}.
Indeed, the initial velocity field used here represents some intermediate stage of growth and would
presumably be different during different stages of development (which appears as a prediction of velocity field dynamics
in our model). Also, the physiology of a plant changes with age, 
which in this model could be represented by time-varying coupling constants that make the plant more or less 
susceptible to growth.

Lacking sufficient experimental data to constrain all these parameters in our dynamical model, the initial conditions for the metric
tensor and velocity field are based on available static and short-time data 
that shows a logistic velocity profile in the primary root of many species ~\cite{SilkSciAm} ~\cite{cornRootGrowth} ~\cite{rootImaging}. 
The coupling constants for the geometric reaction-diffusion term, growth tensor and velocity diffusion are 
currently static and arbitrary. 

In the next chapters, 2D models representing simplified leaf shapes will be developed. Modeling of the 2D system is necessary to 
encompass the effect of curvature flow on the metric tensor. This needs to be built up, first from 2D systems
exhibiting symmetries like circular symmetry that constrain and simplify the system. A thorough investigation of the
parameter space of the system is also needed, including the space of geometries produced by the dynamics. 

Comparison of 2D systems to experiment can begin to be more quantitative. The metric tensor from a 2D numerical 
simulation of our model can be compared against the metric from plastic thin-sheet experiments ~\cite{thinSheetShapes}
and models ~\cite{AudolyBoudaoud}. Comparing the metric of a mature plant leaf is also possible. The more difficult problem
will be to compare the growth dynamics between model and experiment since this would require a non-destructive way of
probing leaf tissue geometry as it grows over time.

\clearpage

\chapter{2D Simulations with Circular Symmetry}
\label{chapter2DConstrainedSimulations}

\section{Why circular symmetry?}

The next step we take is to build a 2D growth model that represents growing disks. A disk with circular symmetry
reduces the complexity of the tensor equations and boundary conditions required to model its growth, thereby simplifying the numerical simulations as well.
It is a toy model because of its simple boundary shape and angular symmetry, but can still exhibit the non-zero curvatures 
found in plant leaves. Also, a circularly symmetric 2D geometry can be thought of as a 1D model generalized through
rotational symmetry to two dimensions.

Geometries with circular symmetries do mimic simple plant leaf shapes like the lotus leaf, nasturtium leaf, lily pad and the cap of
the algae \textit{Acetabularia}. They can also be extended to other shapes parameterized in $(r,\theta)$ coordinates such as ellipses, lobes and cusps.

Another significant advantage of modeling growing disks is the ability to compare numerical simulations against experimental data on
deformations of thin gel disks with circular symmetry. The physics of thin disk deformations have been studied both experimentally 
~\cite{thinSheetShapes} and theoretically ~\cite{AudolyBoudaoud}, in large part to understand the physical and geometric basis of 
biological shapes like plant leaves. However these studies concentrate on objects whose total mass is fixed so that the deformations result 
from externally impose forces rather than the addition of material.

In building these first 2D models, we will examine two cases in particular. The first will be to study Ricci flow in disk 
geometries without the influence of a velocity field. This is important because Ricci flows have not been studied widely
using numerical simulations or in 2D disks. This will also give us an idea of how the metric behaves when coupled only
to its own curvature, thus separating out the influence of geometry in our dynamical model.

The second case that will be studied is that of a growing disk. This will involve the full dynamical equations developed in
Chapter \ref{chapterEquations}. The results of both the Ricci flow and dynamical growth simulations will be compared and 
discussed in the context of thin disk models of plant tissues.

\section{Dynamical equations of growth with circular symmetry}

The critical step in deriving the dynamical equations for a growing disk is to cast the components of the metric
tensor and velocity field as continuous functions of the radius $r$. Indeed, a similar approach was taken in the 1D
case, and stems directly from the principles of Riemannian geometry ~\cite{hawkingBook}. It allows the coupled tensor 
equations for plant growth to be restated
as a set of nonlinear coupled PDEs, which can then be simulated numerically.

Imposing circular symmetry necessarily removes any angular dependence, hence the metric components and velocity field 
vary only in $r$. Symmetry also requires the off diagonal terms to vanish since the condition $d\theta = -d\theta$ is required
by isotropy. Hence $g_{12}=g_{21}=0$ so that distances of the form $g_{12}drd\theta$ vanish. However, $g_{22}$ does not 
vanish since purely angular lengths are of the form $g_{22}d\theta^2$, thus allowing $(d\theta)^2 = (-d\theta)^2$.

The metric components are therefore:

\begin{equation}
g_{ik} = 
\begin{pmatrix}
f(r) & 0 \\
0 & r^2g(r) \\
\end{pmatrix}.
\end{equation}

Equivalently, 

\begin{eqnarray}
g_{11} &=& g_{rr} = f(r) \nonumber \\ 
g_{22} &=& g_{\theta \theta} = r^2 g(r). \nonumber \\
g_{12} &=& g_{21} = g_{r \theta} = 0 \nonumber \\
\end{eqnarray}

The velocity field components reduce to 

\begin{eqnarray}
v^1 &=& v(r) \nonumber \\
v^2 &=& 0. \nonumber \\
\end{eqnarray}

This is equivalent to stating that the velocity vector $v^i$ has no $\theta$ component, and is consistent with isotropy
imposing the $d\theta = -d\theta$ condition ($v^2 = -v^2 = 0$).

As discussed in Chapter \ref{chapterEquations}, the full tensor equations for plant growth under the assumptions of constant 
tissue density are:

\begin{eqnarray}
\frac{\partial g_{ik}}{\partial t} &=& - \kappa R_{ik} + \kappa_1 T_{ik} \\[10pt]
\frac{\partial v^i}{\partial t} &=& -\Gamma_{jk}^i v^j v^k - v^k \nabla_k v^i + c g^{jk} \nabla_j \nabla_k v^i 
\end{eqnarray}

\noi where $T_{ik}$ is the growth tensor.

Writing the metric and velocity field components as continuous functions of $r$ allows all the 
parts of the dynamical equations to also be cast as functions of $r$. 

The connection coefficients are:
\begin{eqnarray}
\Gamma_{11}^1 & = & \frac{f_{,r}}{2f} \nonumber \\
\Gamma_{11}^2 & = & 0 \nonumber \\
\Gamma_{12}^1 & = & 0 \nonumber \\
\Gamma_{12}^2 & = & \frac{g_{,r}r + 2g}{2gr} \nonumber \\
\Gamma_{22}^1 & = & -\frac{g_{,r}r^2 + 2gr}{2f} \nonumber \\
\Gamma_{22}^2 & = & 0.
\end{eqnarray}

\noi where $f_{,r} = \partial f/ \partial r$.

The scalar curvature function (also known as the Ricci scalar) is:
\begin{equation}
R = g^{lm}R_{lm} = - \frac{1}{4d^2} \left( 2g_{,rr}fg - f(g_{,r})^2 - f_{,r}gg_{,r} + \frac{1}{r} (4fgg_{,r} - 2f_{,r}g^2) \right) \label{eqnScalarCurvatureConstrained}
\end{equation}

\noi where $d = -fg$.

In 2D, the Ricci tensor can be expressed as 

\begin{equation}
R_{ik} = R g_{ik},
\end{equation}

\noi so the metric evolution becomes: 
\begin{eqnarray}
\frac{\partial g_{11}}{\partial t} \rightarrow \frac{\partial f}{\partial t} & = & - \kappa Rf + \kappa_1 T_{11} \label{eqng11}\\[10pt]
\frac{\partial g_{22}}{\partial t} \rightarrow \frac{\partial g}{\partial t} & = & - \kappa Rg + \kappa_1 \frac{T_{22}}{r^2} \label{eqng22}
\end{eqnarray}

The general expression for the growth tensor is:
\begin{eqnarray}
T_{ik} & = & \nabla_i v_k + \nabla_k v_i \nonumber \\
       & = & g_{kl} \nabla_i v^l + g_{ip} \nabla_k v^p
\end{eqnarray}

The specific expressions are:
\begin{eqnarray}
T_{11} & = & 2f(v_{,r}^1 + \Gamma_{11}^1 v^1)  \nonumber \\
T_{22} & = & 2gr^2 \Gamma_{12}^2 v^1   \nonumber\\
T_{12} & = & 0. \label{eqnGTcomponents}
\end{eqnarray}

The velocity equations become:
\begin{eqnarray}
\frac{\partial v^1}{\partial t} & = & -\Gamma_{11}^1 v^1 v^1 - v^1 [v_{,r}^1 + \Gamma_{11}^1 v^1] + c g^{11} \left( \frac{\partial^2 v^1}{\partial r^2} + \Gamma_{11,r}^1 v^1 + \Gamma_{11}^1 v_{,r}^1 \right) \nonumber \\ 
           &   & + c g^{22} \left( - \Gamma_{22}^1 v_{,r}^1 + \Gamma_{22}^1 (\Gamma_{12}^2 - \Gamma_{11}^1) v^1 \right) \label{eqnv1}\\
\frac{\partial v^2}{\partial t} &=& 0 \label{eqnv2}. 
\end{eqnarray}

Hence, by introducing continuous functions of $r$ to represent the metric and velocity field components, we can
view the coupled tensor equations as nonlinear coupled PDEs. The additional constraint of null $g_{12}$ and $g_{21}$ terms 
means that the dynamics are not affected by any mixed-derivative terms. Meanwhile, the imposed angular symmetry removes
any derivatives with respect to $\theta$, leaving the equations dependent only on $r$ and derivatives in the radial direction.

\section{Numerical methods}

The set of coupled PDEs we are left to consider are Equations \ref{eqng11}, \ref{eqng22} and \ref{eqnv1}. Most importantly, 
the time evolution of $g_{22}$ and $v^1$ contain parabolic (diffusive) terms that can be stabilized by implementing
the Dufort-Frankel method, similarly to how the parabolic terms in the 1D equations were handled.

Having more nonlinear terms than the 1D equations, the 2D evolution equations are expected to be less numerically stable 
than their 1D counterparts. This manifests itself in requiring smaller time steps and smaller coupling coefficients in the 2D models, which 
effectively slows down the simulation. Simulations of the 2D dynamics will be more computationally expensive than the 1D dynamics.

The boundary conditions in the isotropic 2D simulations are: 

\vspace{3mm}
\begin{center}
\begin{tabular}{r c l}
\hline
$\partial_r f(t,r=0)$ & = & 0 \\
$\partial_r f(t,r=r_{max})$ & = & 0 \\
$\partial_r g(t,r=0)$ & = & 0 \\
$\partial_r g(t,r=r_{max})$ & = & 0 \\
$v^1(t,r=0)$ & = & 0 \\
$\partial v^1(t,r=r_{max})/\partial r$ & = &  $\phi (t)$ \\
\hline 
\end{tabular}
\end{center}

The function $\phi (t)$ is equivalent to $\partial v^1/\partial r$ evaluated at $r=r_{max-1}$.

\section{What do the data look like?}

The data generated by the 2D simulations is not unlike the 1D data set. Information about the state of the metric, 
velocity field, growth tensor, connection coefficients and scalar curvature is gathered at prescribed time slices.
The bigger problem is how to represent this data in a meaningful graphical way.

Ideally, the data would be represented visually as a growing disk with ruffling and buckling developing over time, in a way
that strongly resembles a time-lapse video of a plant leaf growing over time. However, the problem of interpreting 
geometric data in this way is a mathematically difficult problem. It is called the \textit{embedding problem}, which can 
be summarized as the problem of representing a $N$ dimensional geometry in a $M$ dimensional space where $N<M$. In the case
of plant growth considered here, we would seek a solution of the embedding problem of a non-flat 2D manifold embedded 
in a flat 3D space. However, the general solution to the embedding problem when $M = N+1$ remains an open problem in geometry. 
Moreover, no general solution exists for embedding a 2D surface of non-constant curvature in a flat 3D space 
~\cite{blackHoleEmbedding}, this despite the fact that many leaves found in nature have negatively curved surfaces
and remain comfortably embedded in 3D space. Computational methods do exist for visualizing these types of surfaces, however
the techniques remain nontrivial and would generally require a finite element representation of the data rather than the
finite differencing data used here ~\cite{genesAndGeoModel} ~\cite{PrusinkiewiczConstraints}.

Lacking a general method to visualize the data generated by the simulations, what are other possible ways to interpret 
the data? One option is 3D printing each time slice, for which there are a number of methods to explore: hinged plastic 
components ~\cite{3Ddress}, programmed thin disks ~\cite{thinSheetShapes} and even crochet ~\cite{crochet}. 
The advantages of physically representing the 
data in this way is that it is highly realistic and intuitive. The disadvantage is that these representations would
be sparse in time and miss the dynamical aspect of the simulations. 

Another interesting consideration is how similar the geometric data will be to the actual shape of the plant tissue.
As a soft material conforms to a non-flat metric, the material must bend and stretch to accommodate its new shape.
Studies of non-Euclidean plates have indicated that it is the interplay of a target metric with the properties of the material
(such as elasticity and plasticity) that determine its final shape ~\cite{NEPmechanics}. It is hypothesized that growing plant tissues
experience these effects, but a full understanding of the interplay of growth and embedding remains an open question 
~\cite{embeddingAndElasticity}.

In the current work, data representing the dynamics of the system is presented in a relatively raw format. For the
circularly symmetric simulations, the metric components, velocity field, expansion diagram, growth tensor components and modular
growth are graphed as functions of the coordinate distance $r$ along the $\theta=0$ direction, 
so are similar in format to the 1D data. 

We can also determine the deformation tensors for expansion, shear and rotation. In the 2D isotropic case, the expansion is:

\beq
\Theta = \frac{T_{11}}{2f} + \frac{T_{22}}{2gr^2}
\label{eqn2DisoExpansion}
\eeq

\noi where $T_{11}$ and $T_{22}$ are the growth tensor components from Equation \ref{eqnGTcomponents}. The shear tensor 
can be defined in its trace-free form:

\beq
\sigma_{ab} = \frac{1}{2} (\nabla_a v_b + \nabla_b v_a) - g_{ab} \Theta.
\eeq

\noi This form of the shear tensor is useful given that we are dealing with expanding surfaces; subtracting
the trace eliminates the effect of expansion on the shear. The components of the trace-free shear tensor are then:

\begin{eqnarray}
\sigma_{11} & = & -f \frac{T_{22}}{2gr^2}    \nonumber \\
\sigma_{22} & = & -g \frac{T_{11}}{2f}  
\label{eqn2DisoShear}
\end{eqnarray}

\noi and the rotation is null.

Calculating the deformation tensors is useful in that these represent the physical changes the tissue goes through as
it grows. These quantities could be measured using techniques similar to those in ~\cite{inkjetLeaves} where a regular grid
of dots is printed on a growing leaf. Over time, the grid becomes deformed, and could presumably be analyzed to yield the
deformations that have resulted from growth. Also, under the assumption of constant tissue density in space and time, the expansion
scalar is directly proportional to the distributed source field (Section \ref{sectionMassDensityEqns}), 
providing yet another way to characterize tissue growth empirically. 

\begin{figure}[ht]
    \centering
        \subfloat{
                \includegraphics[width=0.3\textwidth]{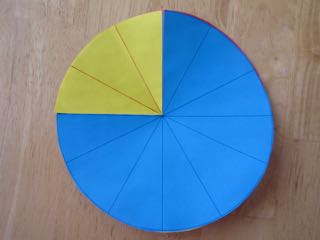}
        }        
        \subfloat{
                \includegraphics[width=0.3\textwidth]{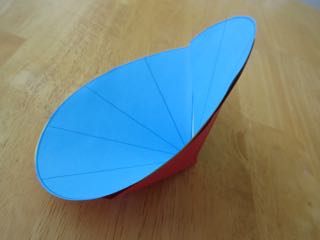}
        }
        \subfloat{
                \includegraphics[width=0.3\textwidth]{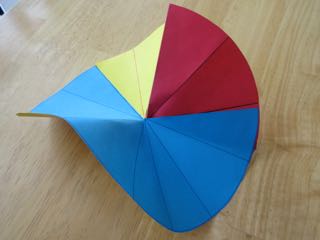}
        }
        \caption{From left to right, geometries with flat, positive and negative curvatures.}\label{figureCurvaturePhotos} 
\end{figure} 

Data specific to 2D simulations also exists and requires additional plots. The perimeter $C$ of the growing disk
can be plotted relative to the proper radial distance $r'$. 
This is entirely analogous to the plots in ~\cite{thinSheetShapes}
and is chosen so that comparisons to existing experiments on expanding thin disks can be readily made. Indeed,
plotting the perimeter as a function of radial distance is a fairly intuitive way of knowing whether a surface
is positively or negatively curved. A basic result of Gauss' \textit{Theorema Egregium} is that surfaces
of positive curvature will have a perimeter $C<2\pi r'$ while surfaces of negative curvature will have $C > 2\pi r'$.
In Figure \ref{figureCurvaturePhotos}, this is represented by a positive curvature disk that has a deficit perimeter when wedges 
of the disk are removed, and the negative curvature disk that has excess perimeter when wedges are added. 

An analogy of the relative elemental growth rate (REGR) can also be defined in 2D. In the 1D case, the REGR was
the rate of change of the scale factor normalized by the local length scale. In 2D, we must work with area instead.
One way to define area is using the product of the two scale factors $f$ and $g$

\beq
dA = \sqrt{f}dr \times \sqrt{g}rd\theta = \sqrt{fg} \, rdrd\theta.
\eeq

\noi The corresponding measure of elemental growth is then

\beq
REGR = \frac{1}{A} \frac{\partial A}{\partial t}.
\eeq

\noi Using the expressions for the time evolution of the metric components (Equation \ref{eqng11} and \ref{eqng22}), we obtain

\beq
REGR = - \kappa R + \kappa_1 \Theta
\label{eqn2Dregr}
\eeq

\noi where $R$ is the scalar curvature (Equation \ref{eqnScalarCurvatureConstrained}) and $\Theta$ is the expansion scalar
(Equation \ref{eqn2DisoExpansion}). Hence, the REGR equation shows that growth of a disk is not just the expansion of the space, 
but also a result of curvature dissipation. When we consider that the scalar curvature is a KPZ-type function generalized to 2D
space, the notion that growth is both expansion and deposition again emerges, in analogy to the results of Chapter \ref{chapter1DSimulations}.

Measurements of the REGR for a 2D leaf or petal could be made using existing experimental methods ~\cite{inkjetLeaves}. By imprinting
a growing tissue with a grid of dots (the landmark method from Chapter \ref{chapter1DSimulations}), the area inside each segment of tissue
could be measured over time, thereby allowing the REGR to be calculated.
 
Finally, the curvature evolution of the disk is represented as a 2D heat map. Regions of positive and negative
curvature are mapped to the disk. Regions of positive curvature, when embedded in flat 3D space, 
will generally look like portions of a sphere, while areas of negative curvature will be buckled and ruffled.  
Two versions of the heat map will be presented. One is a disk that changes size over time, which has the radial
growth information overlayed with curvature information. Another format is the mapping the curvature data
onto the unit disk. This does not mean that the disk does not grow; rather, it is the curvature information
separated from growth information. In these cases, the radial growth graphs are shown following the curvature maps.

\section{Numerical Simulation: Ricci flow on the disk}

The first 2D simulation we can look at is the evolution of the metric in response to its own curvature, i.e. pure
Ricci flow. The equation using the Ricci tensor $R_{ik}$ is:

\begin{equation}
\frac{\partial g_{ik}}{\partial t} = - \kappa R_{ik}.
\end{equation}

Just as before, we can replace the metric $g_{ik}$ with its components $g_{11} = f(r)$ and $g_{22} = r^2 g(r)$, and the Ricci
tensor with $Rg_{ik}$, where $R$ is given by \ref{eqnScalarCurvatureConstrained}. The evolution equation becomes:

\begin{eqnarray}
\frac{\partial g_{11}}{\partial t} \rightarrow \frac{\partial f}{\partial t} & = & - \kappa Rf \label{eqnRFg11}\\[10pt]
\frac{\partial g_{22}}{\partial t} \rightarrow \frac{\partial g}{\partial t} & = & - \kappa Rg. \label{eqnRFg22}
\end{eqnarray}

We choose the coupling constant and initial conditions as:

\vspace{3mm}
\begin{center}
\begin{tabular} {r l}
\hline
Ricci flow coupling & $\kappa = 0.5$ \\
$f(t=0,r) = $ & $1.0$ \\
$g(t=0,r) = $ & $0.5\exp(-2.0(r-\frac{1}{2}r_{max})^2) + 1.0$ \\ 
\hline 
\end{tabular}
\end{center}
\vspace{3mm}

\begin{figure}[h]
    \centering
		\subfloat{
                \includegraphics[width=0.5\textwidth]{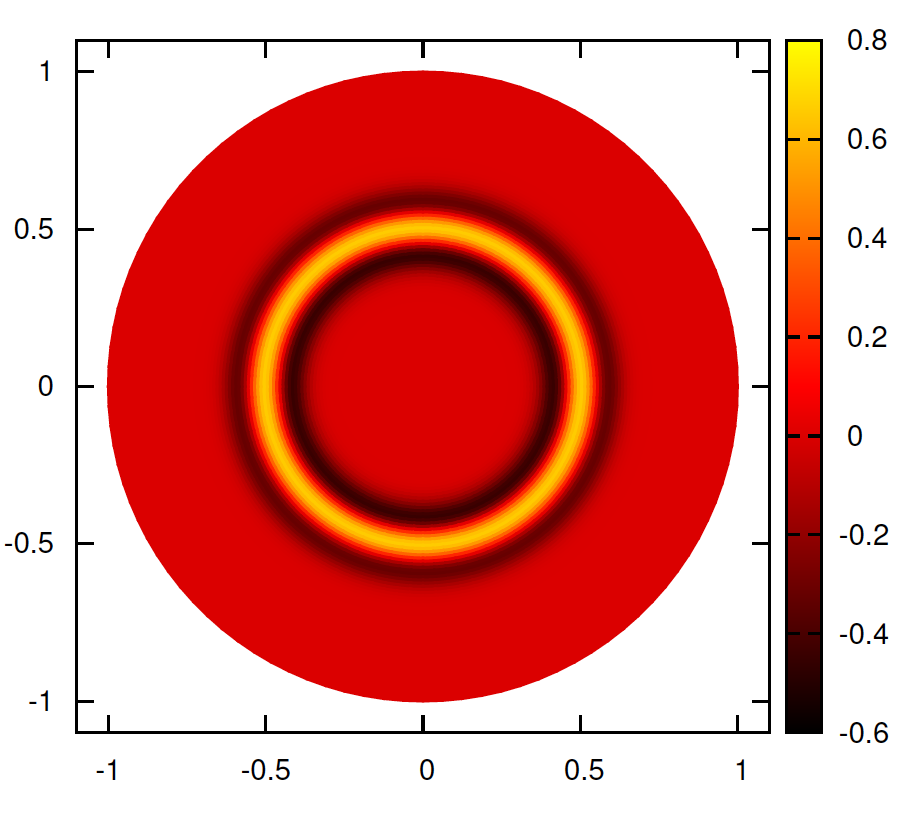}
        }

        \subfloat{
                \includegraphics[width=\textwidth]{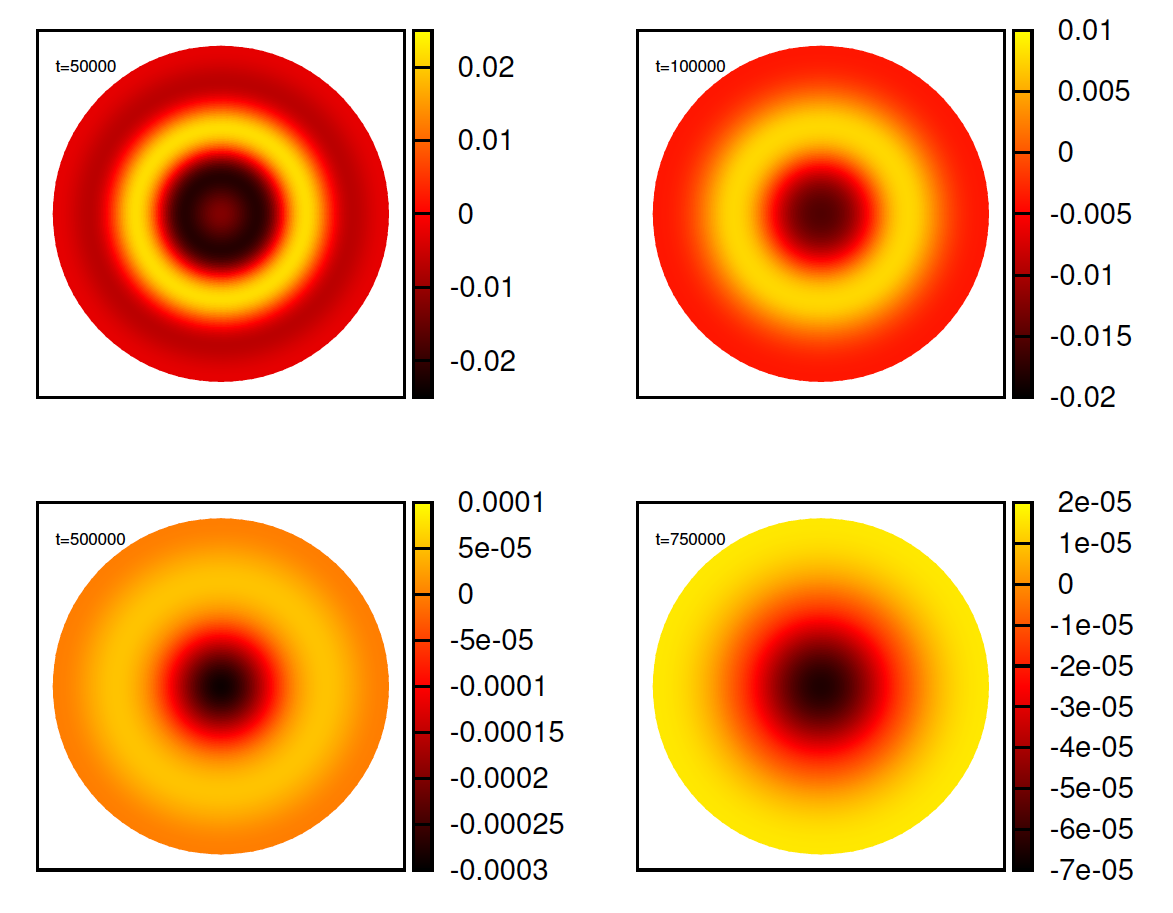}
        }
        \caption{Time evolution of the scalar curvature under Ricci flow.}\label{figureRFCurvatureConstrained} 
\end{figure}

\begin{figure}[h]
    \centering
        \subfloat{
                \includegraphics[width=0.7\textwidth]{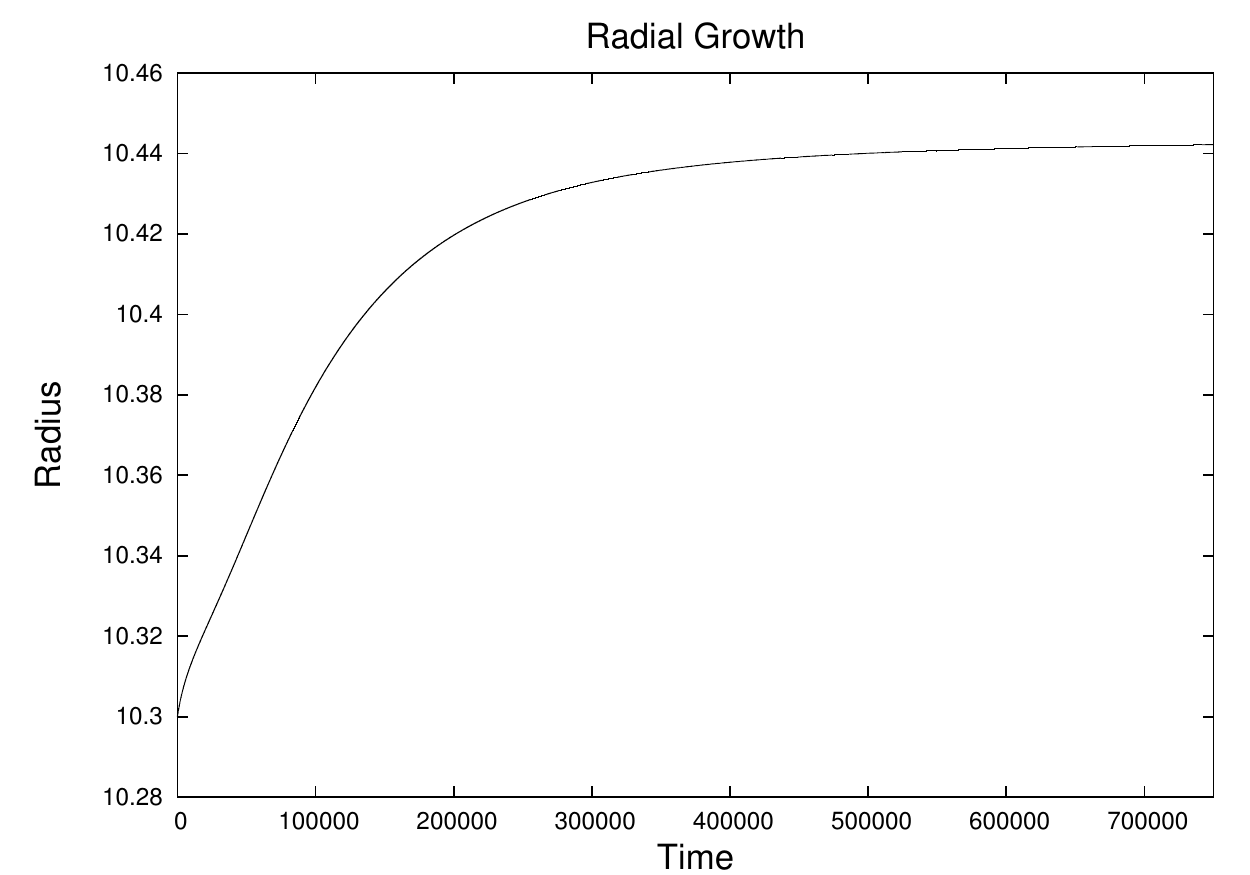}
        }
        
        \subfloat{
                \includegraphics[width=0.48\textwidth]{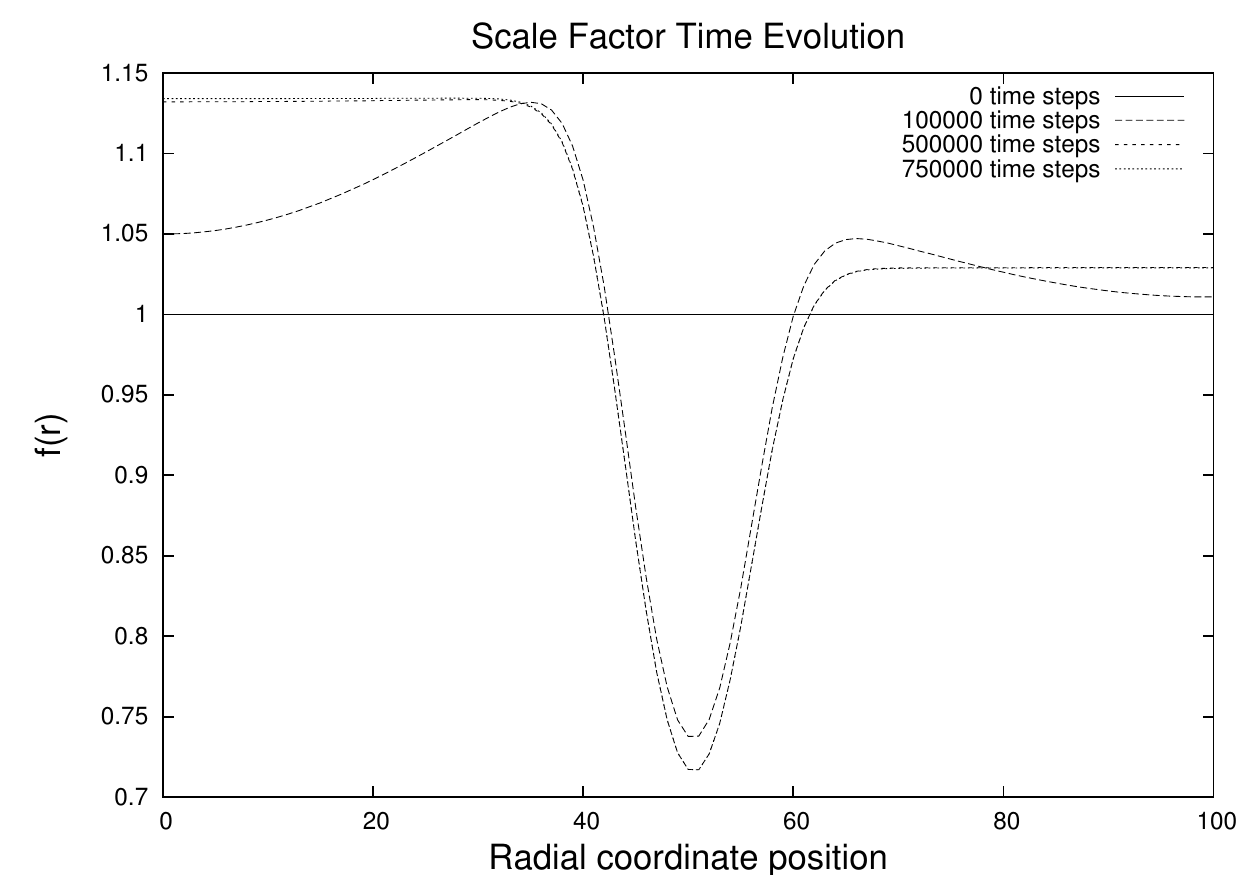}
        }
        \subfloat{
                \includegraphics[width=0.48\textwidth]{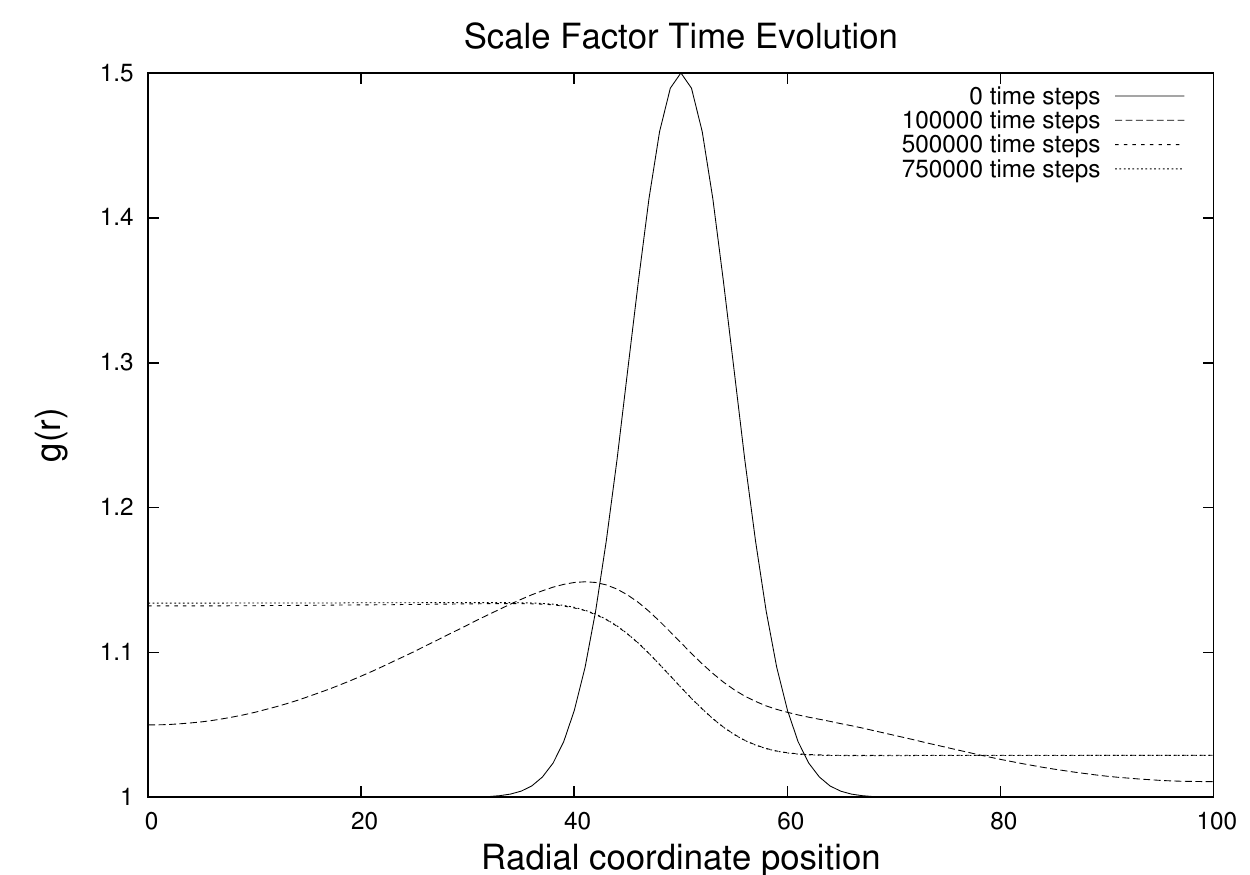}
        }
        \caption{Dynamics of the metric under Ricci flow.}\label{figureRFDynamicsConstrained} 
\end{figure}

\noi and the results are shown in Figures \ref{figureRFCurvatureConstrained} and \ref{figureRFDynamicsConstrained}.

Since Ricci flow on a disk has not been widely studied, it is important to understand the dynamics of this curvature coupling
to the metric before introducing coupling to the velocity field. There are several important points to make about these results.

First, the radius of the disk increases over time, indicating that growth occurs due to Ricci flow alone as anticipated by 
Equation \ref{eqn2Dregr}. The stagnation of
growth is related to curvature values dropping to nearly zero at late times, which itself is due to the dissipation term 
in $\partial g_{22}/ \partial t$. This result also highlights the numerical 
stability of the code, which does not pick up oscillations when curvature values hover near zero.

Second, it is important to note that the curvature tends toward zero at late times, but the components of the metric 
are not trivial. A flat metric can be easily achieved if the components are constant (i.e. do not vary with radius).
However, this is not the case at late times in this simulation where both $g_{11}$ and $g_{22}$ reach 
asymptotic solutions that clearly vary with radius.

\section{Numerical Simulation: The growing disk}

We now move on to study an example of growth driven by the velocity field and growth tensor coupling. 
Coupling to the Ricci tensor remains, and as will be discussed, is a key part of the dynamics of growth.

Starting with a flat disk and a radial logistic velocity field (a generalization of the one used in 1D simulations),
the disk evolves under Equations \ref{eqng11}, \ref{eqng22} and \ref{eqnv1}. The coupling constants and
initial conditions are:

\vspace{3mm}
\begin{center}
\begin{tabular}{r l}
\hline
Ricci flow coupling & $\kappa = 7.0$ \\
Growth tensor coupling & $\kappa_1 = 5.0$ \\
Velocity diffusion coupling & $c = 0.05$ \\
$f(t=0,r) = $ & $1.0$ \\
$g(t=0,r) = $ & $1.0$ \\ 
$v^1(t=0,r) = $ & $0.01/[1.0 + \exp(-5.0(r-\frac{1}{2}r_{max}))]$ \\
             & $-0.01/[1.0 + \exp(\frac{5}{2}r_{max})]$ \\ 
\hline 
\end{tabular}
\end{center}
\vspace{3mm}

\begin{figure}[h]
    \centering
	\includegraphics[width=\textwidth]{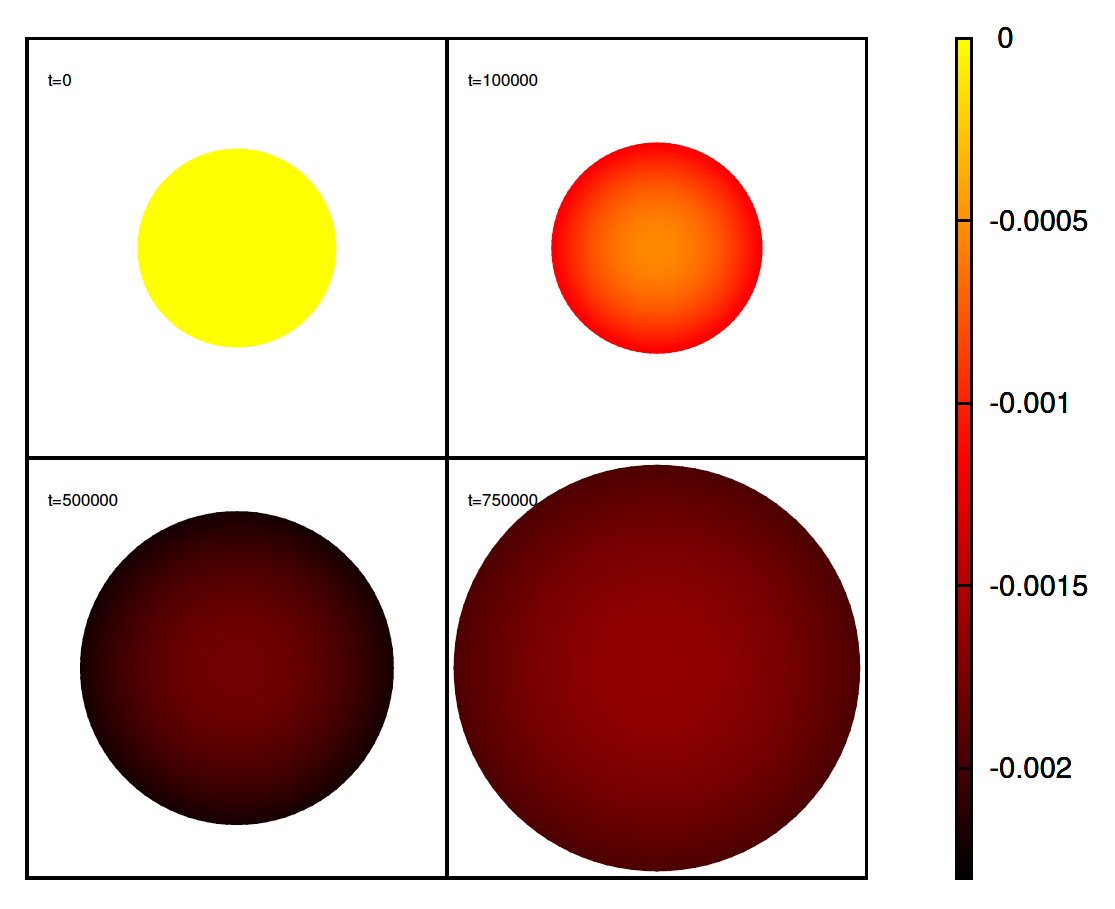}
    \caption{Time evolution of the scalar curvature of a growing disk.}\label{figureGDcurvature} 
\end{figure}

\begin{figure}[t]
    \centering
    \includegraphics[width=0.7\textwidth]{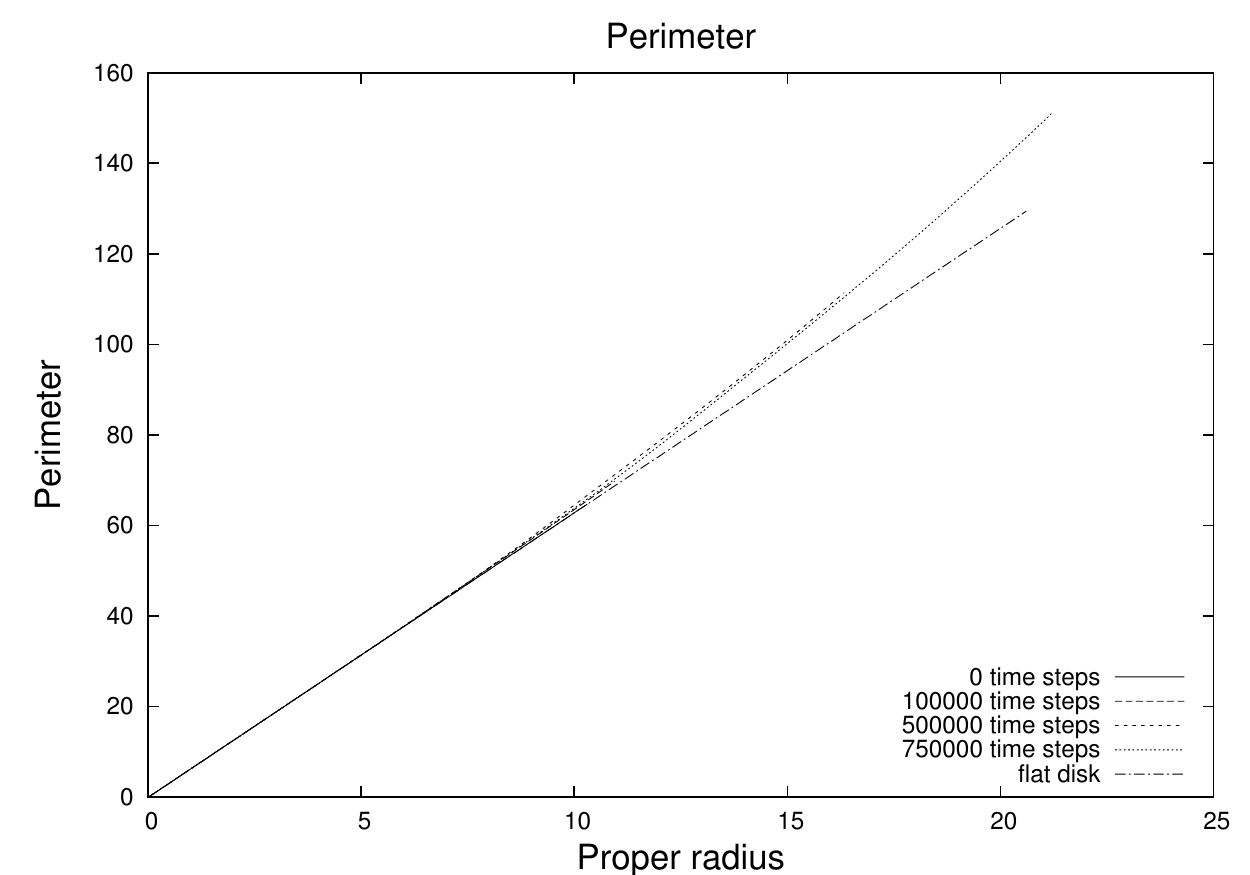}
    \caption{Perimeter of a growing disk over time.}\label{figureGDperimeter} 
\end{figure}

\begin{figure}
    \centering
    \includegraphics[width=0.7\textwidth]{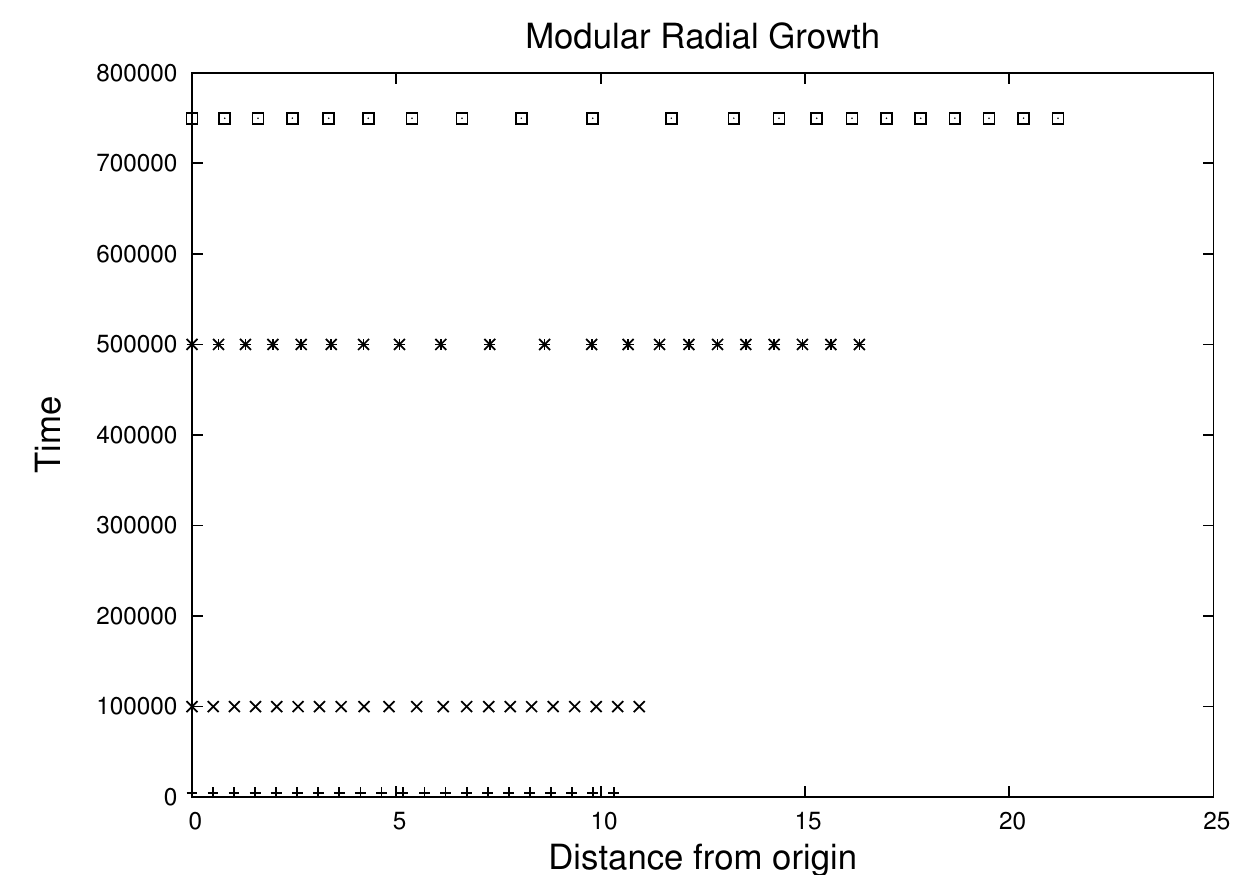}
    \caption{Modular radial growth of a growing disk.}\label{figureGDmodGrowth} 
\end{figure}

\begin{figure}[h]
    \centering
    \subfloat{\includegraphics[width=0.7\textwidth]{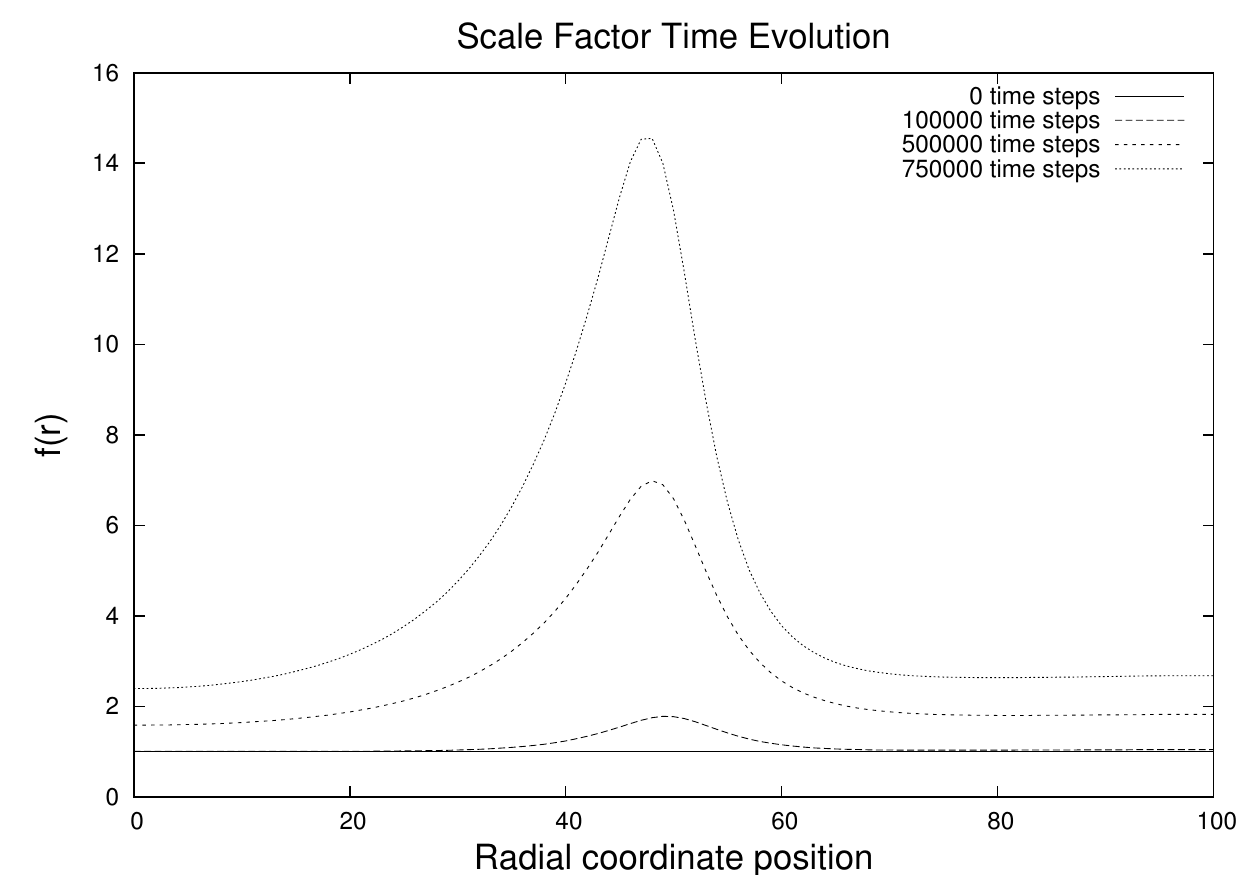} 
    }
    
    \subfloat{\includegraphics[width=0.7\textwidth]{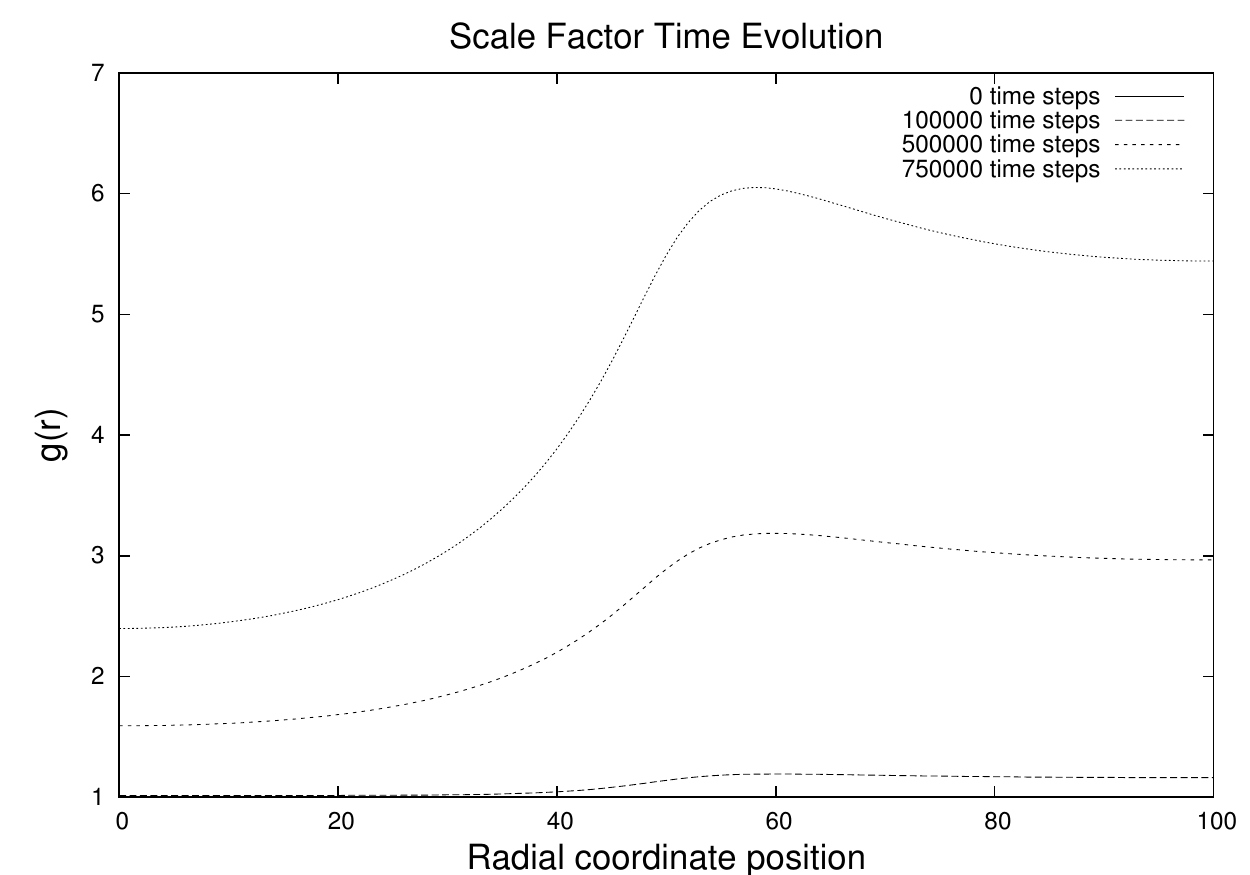}}
    \caption{Time evolution of the metric components of a growing disk.}\label{figureGDmetric} 
\end{figure}

\begin{figure}[h]
    \centering
    \subfloat{\includegraphics[width=0.7\textwidth]{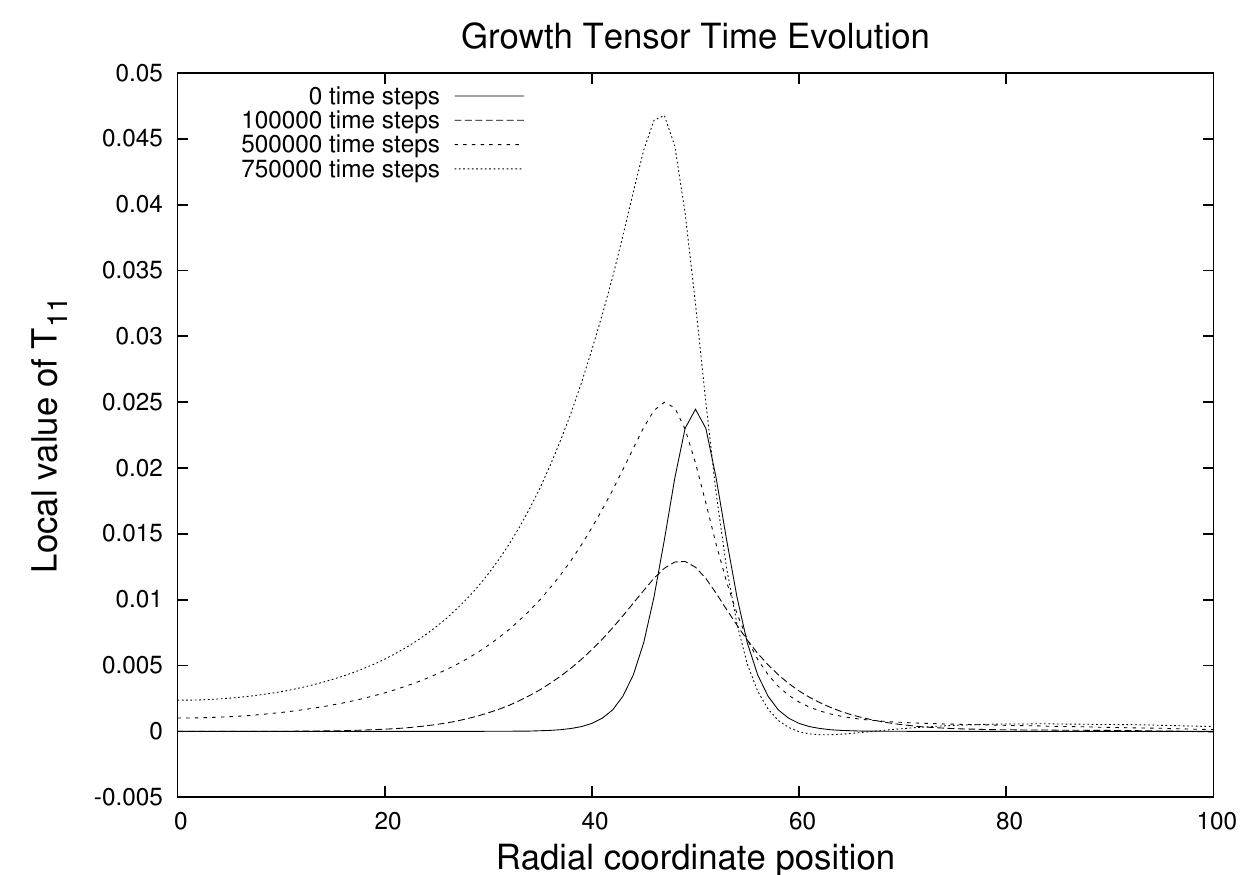}
    }

    \subfloat{\includegraphics[width=0.7\textwidth]{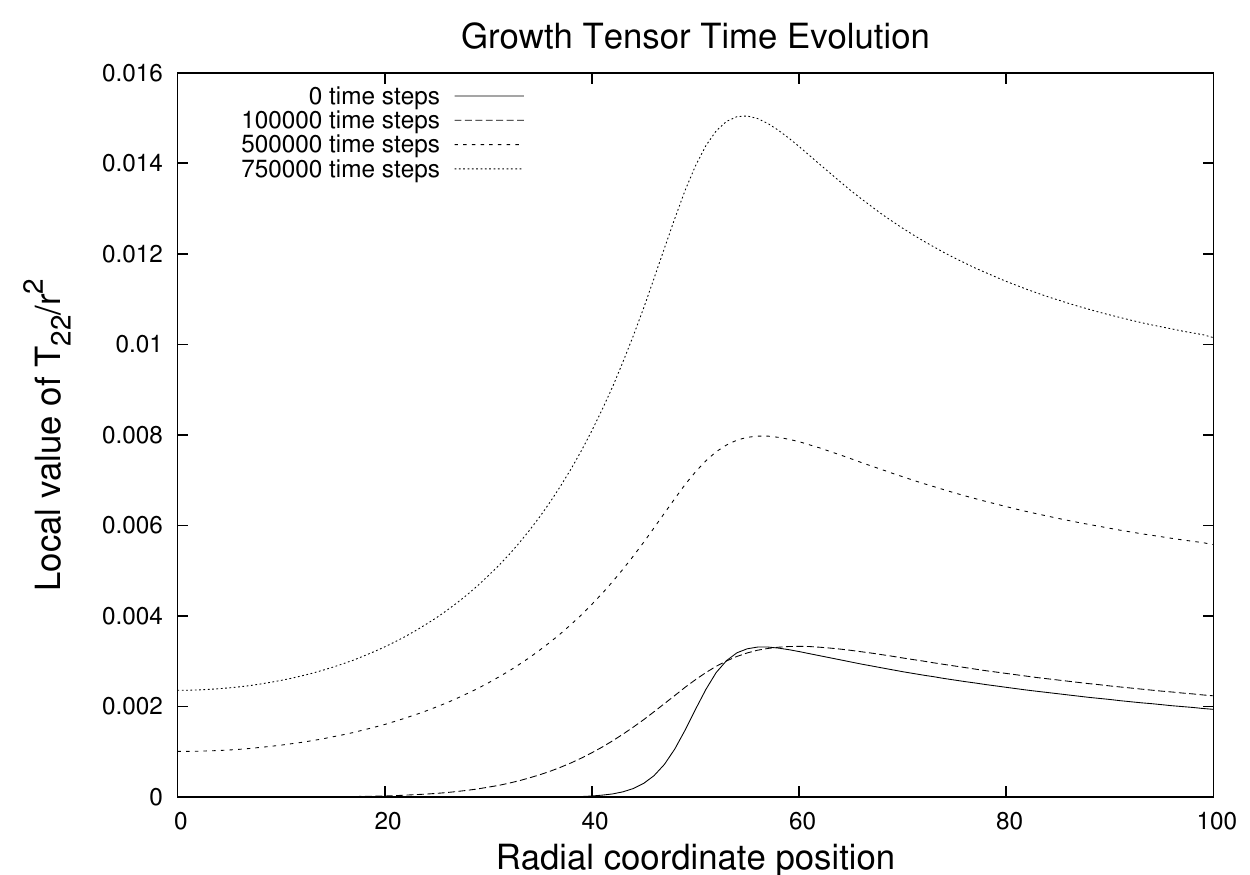} 
    }   
    \caption{Time evolution of the growth tensor components of a growing disk.}\label{figureGDgts} 
\end{figure}

\begin{figure}[h]
    \centering
    \subfloat{\includegraphics[width=0.7\textwidth]{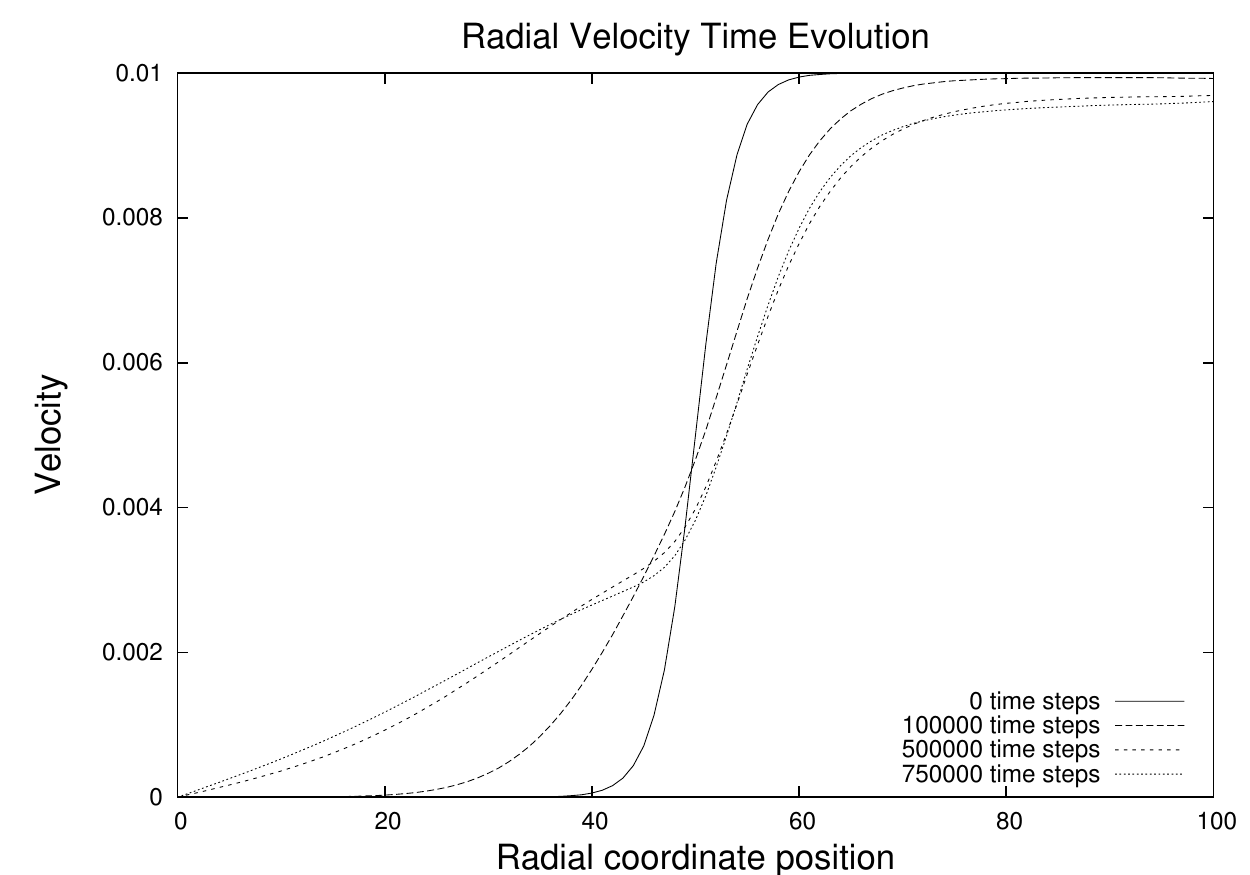}
    }
    
    \subfloat{\includegraphics[width=0.7\textwidth]{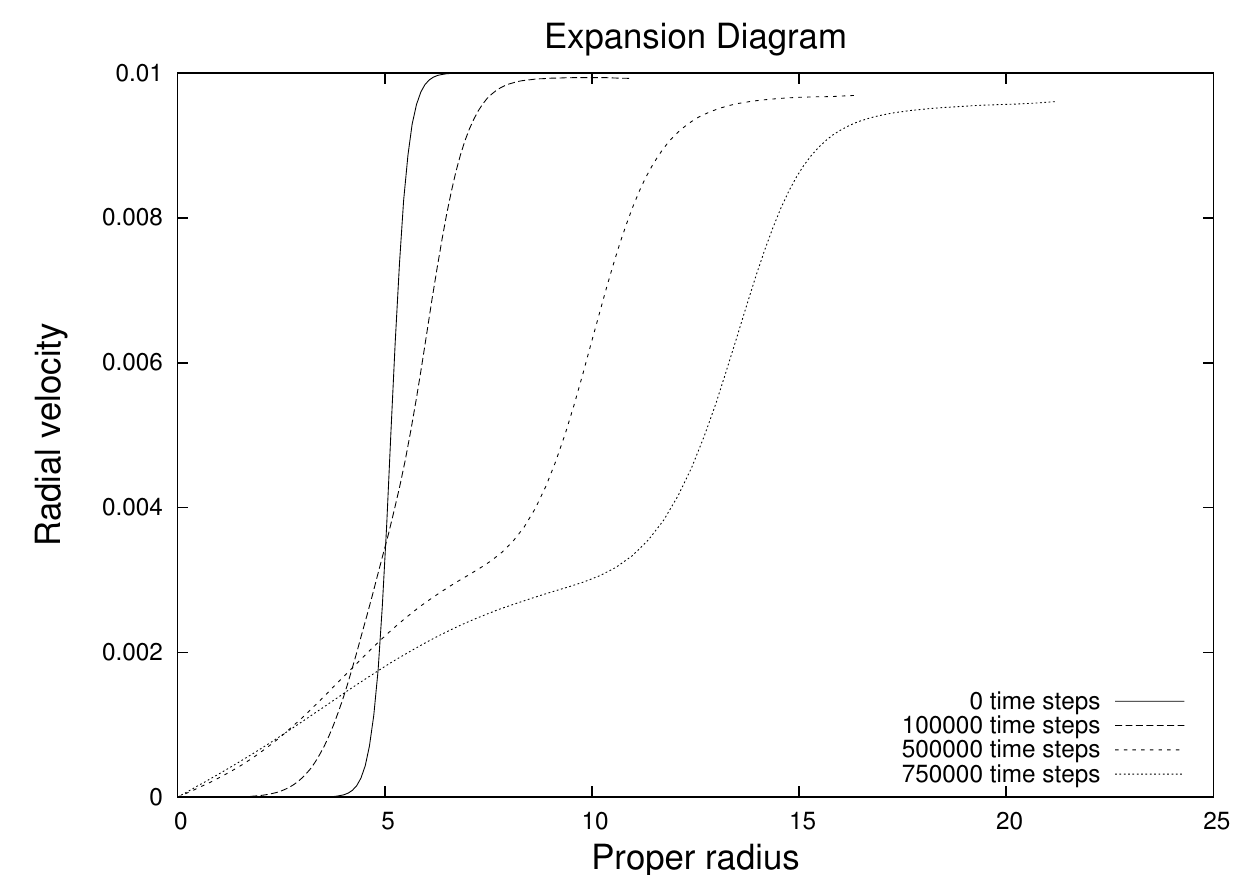}
    }
    \caption{Time evolution of the velocity field of a growing disk.}\label{figureGDvelField} 
\end{figure}

\begin{figure}[h]
    \centering
    \subfloat{\includegraphics[width=0.7\textwidth]{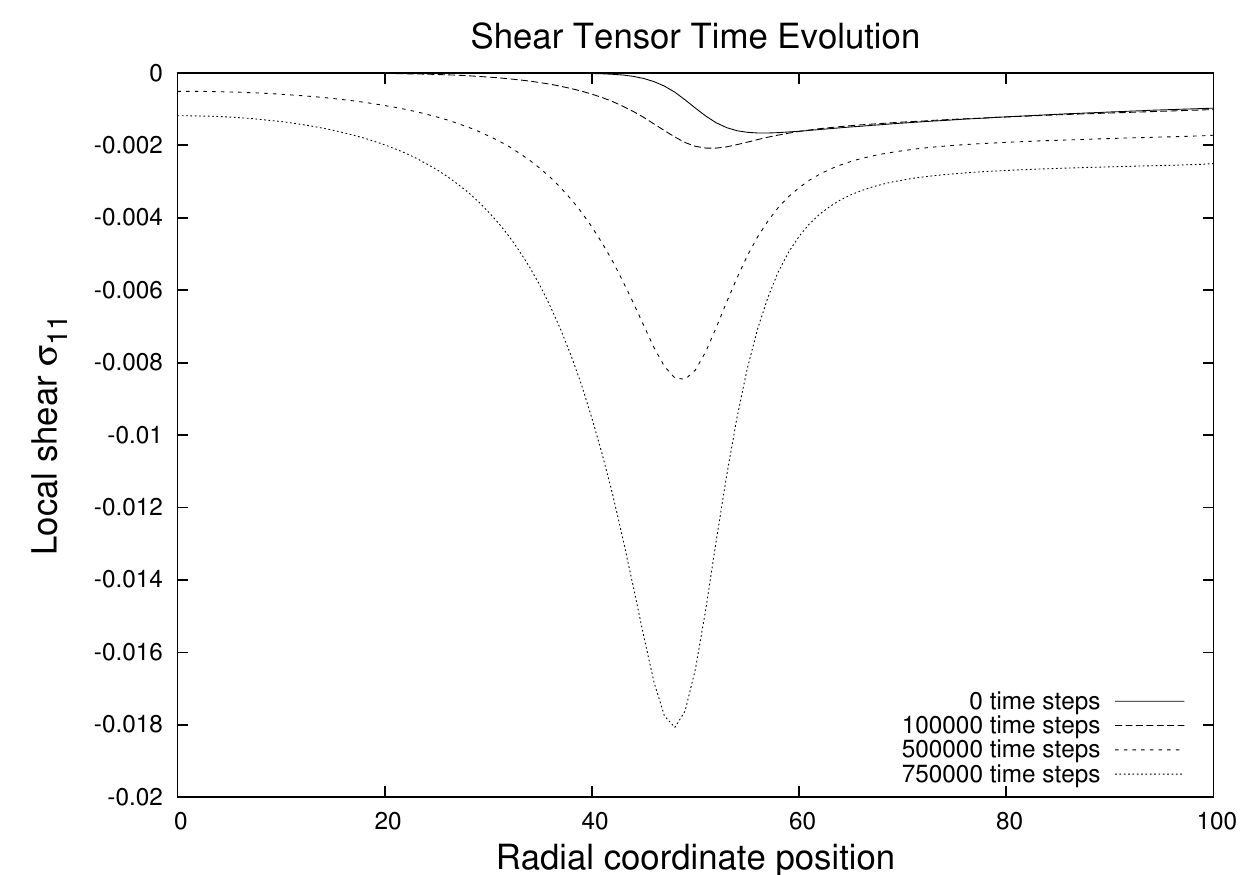}
    }
    
    \subfloat{\includegraphics[width=0.7\textwidth]{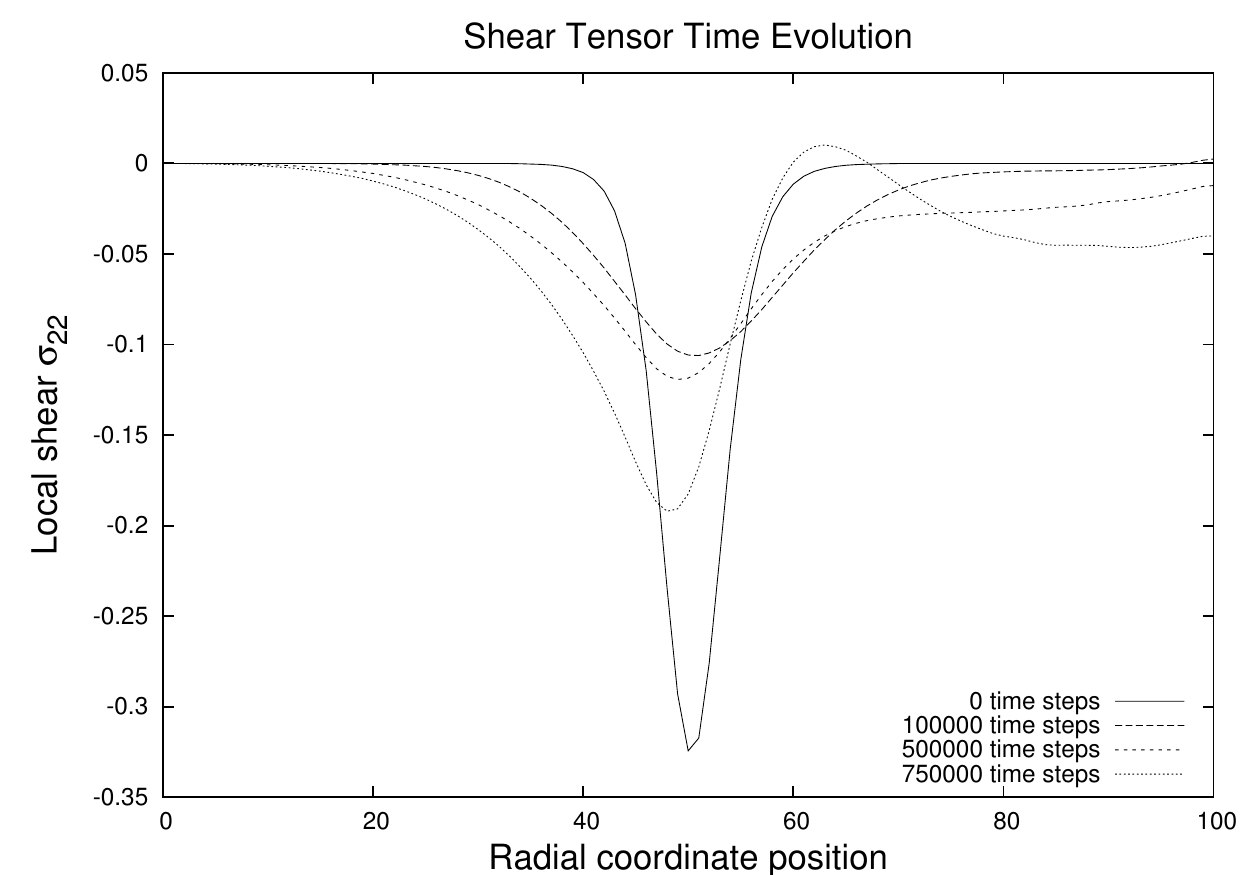}
    }
    \caption{Time evolution of the shear tensor components of a growing disk.}\label{figureGDshear} 
\end{figure}

\begin{figure}[h]
    \centering
    \includegraphics[width=0.7\textwidth]{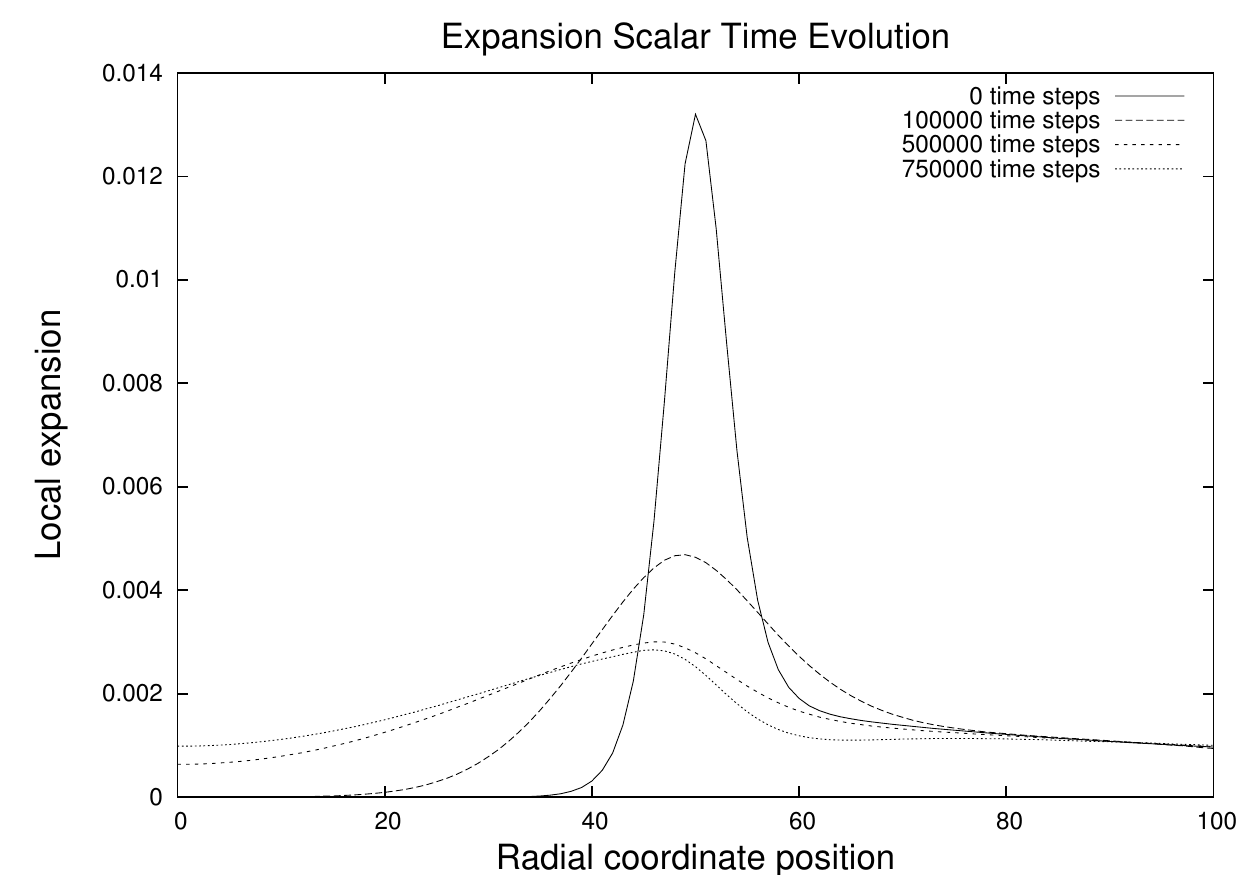}
    \caption{Time evolution of the expansion scalar of a growing disk.}\label{figureGDexpTensor} 
\end{figure}

\begin{figure}[h]
    \centering
    \includegraphics[width=0.7\textwidth]{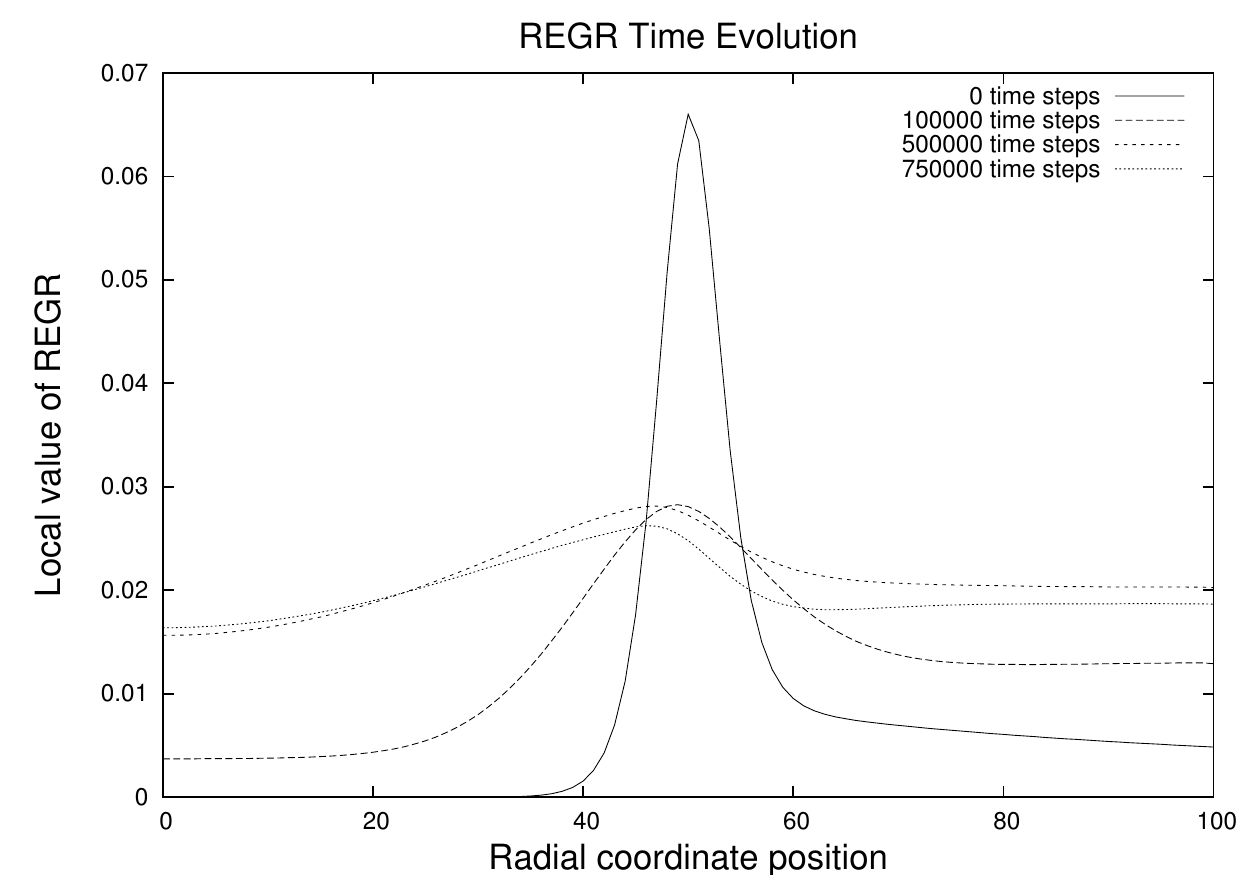}
    \caption{Time evolution of the REGR for a growing disk.}\label{figureGDregr} 
\end{figure}

\noi and the results are shown in Figures \ref{figureGDcurvature} to \ref{figureGDregr}.

The disk evolves in several notable ways. First, the radius of the disk roughly doubles (Figures \ref{figureGDperimeter} and \ref{figureGDmodGrowth}), 
demonstrating significant growth over what is physiologically a relatively short span of time. The disk growth is approximately linear in time, 
so it is likely only a portion of the larger sigmoidal pattern of growth that a real leaf would experience
over its entire development.

The graphs of perimeter (Figure \ref{figureGDperimeter}) and curvature (Figure \ref{figureGDcurvature})
show that ruffling on the edges will occur at late times for this growth pattern 
due to the globally negative curvature values and an excess of perimeter compared to a flat disk. Correspondingly, Figures 
\ref{figureGDmodGrowth} and \ref{figureGDmetric} all show that the disk grows more at the outer edge than 
at the center, again indicating the emergence of negative curvature.

Hence, the data indicate that under the conditions of this simulation, a growing, buckled surface will 
develop from an initially flat disk.

\section{Numerical Simulation: Modeling the growing cap of \textit{Acetabularia}}
\label{sectionAceta}

A very interesting organism on which to test the assumptions of our model is the algae \textit{Acetabularia} (Figure \ref{figureAcetaPhoto}).
The algae is single-celled, but grows to centimeter length scales. Moreover, the algae has a cap structure it uses
in its reproductive cycle that starts as positively curved, then grows to a flat circular shape, and finally becomes 
negatively curved. Along with this, it is known that growth occurs in an annulus around the central stem. ~\cite{acetabularia}

{\it Acetabularia} has a property that is rare in our understanding of dynamical Riemannian geometries. The growth of its cap 
begins from the top of the stem with a positive curvature everywhere.
As it matures its curvature lessens and eventually becomes flat.  However it then continues to evolve until the cap's
curvature is completely negative.  This is unusual since flat space it often a stable fixed point in for dynamical Riemannian curvatures.
For example in pure Ricci flow, in two-dimensions, the Ricci tensor vanishes only if the Riemann tensor vanishes therefore 
when the manifold is flat $R_{ik} = 0$ implies $\partial g_{ik}/\partial t = 0$.  If such a fixed point is reached asymptotically 
then the curvature cannot not change its sign.  Similarly if the energy momentum tensor of general relativity obeys certain
``positive energy conditions'' (generally the energy density of matter and fields in the space-time is greater than the momentum
density which is itself positive), the signature of constant curvature space times cannot change without becoming singular in its 
evolution ~\cite{bergmann2013topological}. 

Of course one can engineer a flat thin disk into a positively or negatively curved surface with the appropriate tools, or change
the topology of the universe from positive curvature to flat by adding negative energy (Dark Energy), but it would seem that 
some how some plants are able to do this quite naturally without requiring external manipulation or bizarre physical conditions.    
Thus {\it Acetabularia} presents a challenge to any biological models to explain how its cap changes the signature of its
curvature through the growth process. 

As a first attempt one can ask if the coupling of Ricci flow to the growth tensor is capable of changing the sign of the 
curvature if the surface has a spatially constant but time dependent curvature.  
One can then ask what the necessary conditions are on the velocity field to drive the surface past zero curvature.
As will be proven below, \textit{a constant curvature space cannot change signature, even when driven by a growth tensor.}

The simplest non-flat surfaces to describe are those with constant curvature.  If the curvature is positive the surface is
that of a sphere, if it is zero it is a flat plane and if it is negative the surface can be described by a hyperbolic paraboloid.
Constant curvature surfaces have constant Gaussian curvature
$$ K = \frac{1}{\rho_1 \rho_2}. $$
If $\rho_1$ and $\rho_2$ are the radii of curvatures of the principle directions of the surface, the surface has positive
curvature of the radii are measured form the same side of the surface, if one measures the radii from two different sides then
the curvature is negative and if one of the radii is infinite then the surface has zero curvature and is flat.

A space of constant Riemannian curvature is isotropic and homogeneous and the line element in two dimensions can be 
written as:

\beq
ds^2 = \frac{1}{1-Kr^2} dr^2 + r^2 \theta^2.
\eeq

A space of constant curvature has the following relation between the metric and the Riemann 
tensor:
$$ R_{1212} = K(g_{11}g_{22} - (g_{12})^2) $$
and therefore 
$$ R_{ik} = K g_{ik}.$$
Suppose it is assumed that the curvature is a spatial constant but can evolve over time.
Then $K= K(t)$.

Given the line element above we can identify the metric coefficients;
$$ f = \frac{1}{1-Kr^2} \qquad \qquad {\rm and} \qquad \qquad g = 1. $$
Therefore 
$$ R_{11} = \frac{K}{1-Kr^2} \qquad \qquad {\rm and} \qquad \qquad R_{22} = K r^2. $$
With a time dependent $K$ the time derivatives of the metric coefficients are 
$$ \frac{\partial f}{\partial t} = \frac{\dot{K} r^2}{(1-Kr^2)^2} \qquad \qquad {\rm and} \qquad \qquad \frac{\partial g}{\partial t} = 0 .$$
Given the metric above connection coefficients are:
$$ \Gamma^1_{11} = \frac{Kr}{1-Kr^2}, \qquad  \Gamma^2_{12} = \frac{1}{r} \qquad \Gamma^1_{22} = r(Kr^2 -1) $$
and these lead to:
$$ T_{11} = 2f(\frac{\partial v^1}{\partial r} + \Gamma^1_{11} v^1)$$
and
$$ T_{22} = 2r^2 \Gamma^2_{12} v^1 .$$

Since the time derivative of $g$ vanishes, the metric evolution equation leads to an expression for $v^1$ in terms of the radial coordinate:
\begin{eqnarray}
 \frac{\partial g}{\partial t} = & 0   = & -\kappa R_{22} + \kappa _1 T_{22} \nonumber \\
 &   = &-\kappa Kr^2 + 2 \kappa _1 r^2 \left(\frac{1}{r} \right) v^1  \nonumber \end{eqnarray}
and this leads to :
$$ v^1 = \frac{\kappa}{2 \kappa _1} K r.$$
Therefore the velocity field grows linearly with the radius $r$.
The other metric evolution equation is:
\begin{eqnarray}
 \frac{\partial f}{\partial t} &   = & -\kappa R_{11} + \kappa _1 T_{11} \nonumber \\
 &   = &-\kappa \left (\frac{K}{1-Kr^2} \right) + \kappa_1 \left(\frac{2}{(1-Kr^2)}\right) \left[ +\frac{\kappa}{2 \kappa _1} K + \frac{Kr}{1-Kr^2} \left(
-\frac{\kappa}{2 \kappa _1} K r \right ) \right. ] \nonumber
\end{eqnarray}
The first two terms cancel exactly and writing the explicit expression for $\partial f/\partial t$ leads to 
$$ \frac{\dot{K} r^2}{(1-Kr^2)^2}  =   \kappa \left(\frac{K^2 r^2}{(1-Kr^2)^2} \right ) $$ 
which reduces to:
$$ \frac{d K}{dt} =  \kappa K^2.$$
Using separation of variables and assuming that the constant curvature at $t=0$ is $K=K_0$ the explicit time dependence
for $K(t)$ is
$$ K(t) = \frac{K_0}{(1-\kappa K_0 t)}. $$
Therefore the explicit velocity field required to maintain a time dependent spatially constant curvature should be
$$ v^1 = \frac{1}{2} \frac{\kappa}{\kappa_1} \left(\frac{K_0}{1- \kappa K_0 t} \right ) r. $$
If $K$ is initially positive then a sign change cannot occur without the formation of a singularity. If $K<0$ then 
the curvature becomes flat asymptotically and if $K=0$ no evolution occurs. This result is similar to that 
found for cosmological spacetimes filled with positive energy.

From this discussion we can infer two things about the initial conditions on a model of \textit{Acetabularia} development.
The first is that if the sign of the curvature is to change, the idea of constant curvature must be abandoned and a more general approach 
needs to be taken.  However the cap of the {\it Acetabularia} is close to constant curvature so we will chose as
initial conditions for the metric a functional form for the coefficients that is close to that for the constant
curvature case. Indeed, closer examination of the the cap structure
at early times reveals a concentration of positive curvature near the stem, and a flatter but still positively
curved surface toward the outer edge of the cap (Figure 7 in ~\cite{acetabularia}). In the simulation, we choose 
$f(t=0,r) = 1.0 + 0.25[1+\exp(-(r-\frac{1}{2}r_{max})]^{-1}$ and $g(t=0,r) = 1.0$, which allows the metric to have
globally positive curvature, but also spatial variability such that the inner regions of the disk begin with higher
positive curvature.

The second inference is that the velocity field 
cannot be linear if it is to drive a sign change from positive to negative curvature. Indeed, significantly more material needs to 
be transfered to the outer regions of the tissue in order to flatten out and then buckle the surface. 
Furthermore, the cap appears to have a localized annular area where most of the growth occurs (Figure 4 in ~\cite{acetabularia}), which hints at a 
velocity field that is not linear but rather has a steep gradient across part of the radius that drives the 
growth in this way. These behaviours were also seen in the previous simulation of a flat disk developing negative curvature over time. 
We therefore initiate the simulation with a monotonically increasing sigmoidal velocity field.

In summary, to simulate the cap changing curvature we use the following parameters:

\vspace{3mm}
\begin{center}
\begin{tabular}{r l}
\hline
Ricci flow coupling & $\kappa = 1.0$ \\
Growth tensor coupling & $\kappa_1 = 1.0$ \\
Velocity diffusion coupling & $c = 0.01$ \\
$f(t=0,r) = $ & $1.0 + 0.25[1+\exp(-(r-\frac{1}{2}r_{max})]^{-1}$ \\
$g(t=0,r) = $ & $1.0$ \\ 
$v^1(t=0,r) = $ & $0.01/[1.0 + \exp(-5.0(r-\frac{1}{2}r_{max}))]$ \\
             & $-0.01/[1.0 + \exp(\frac{5}{2}r_{max})]$ \\ 
\hline 
\end{tabular}
\end{center}
\vspace{3mm}

Compared to the flat disk simulation in the previous section, the coupling parameters are smaller but of the same
magnitude, and rather than an initially flat metric, we choose a metric that is globally positive but not constant.
The velocity field is the same as the previous simulation.

\begin{figure}[h]
    \centering
        \subfloat{ 
            \includegraphics[width=0.7\textwidth]{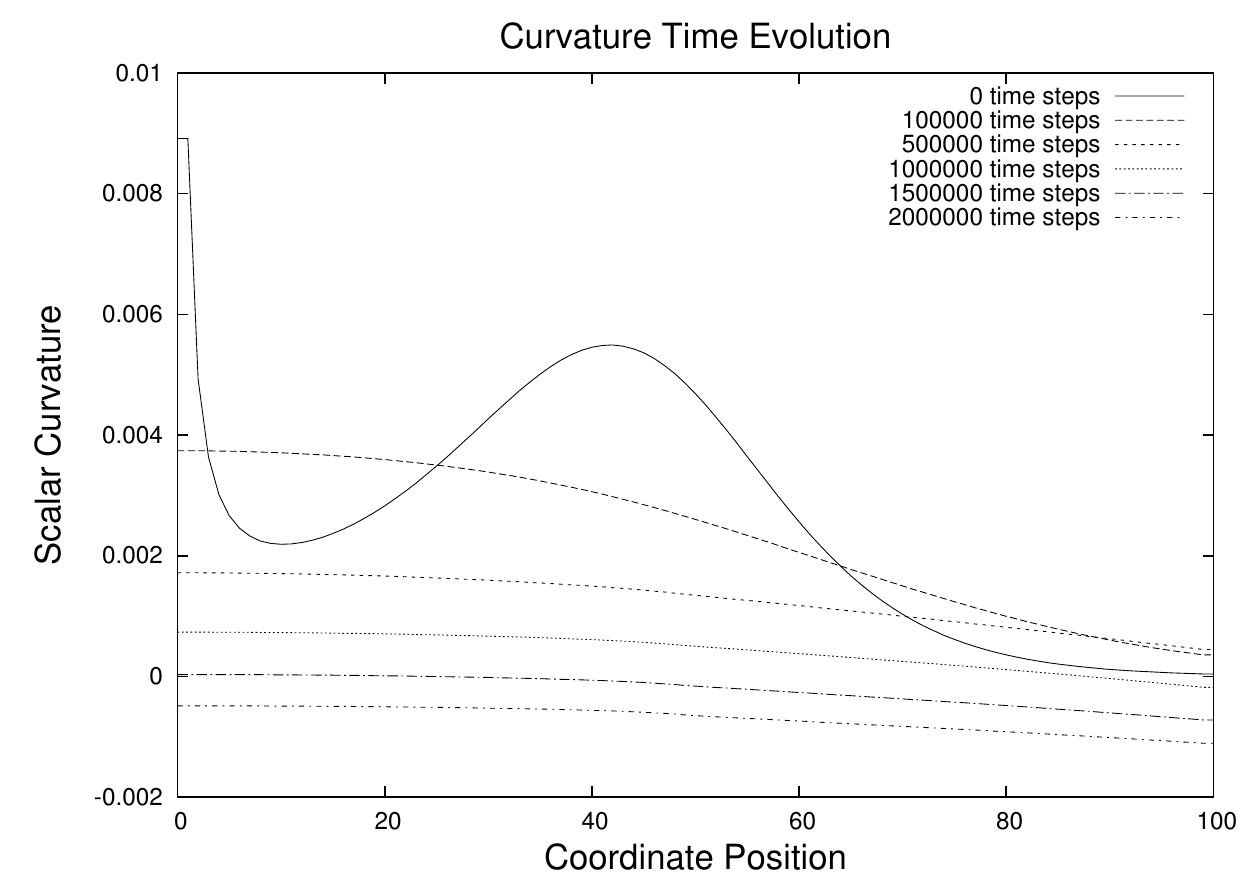}
        }

        \subfloat{\includegraphics[width=0.7\textwidth]{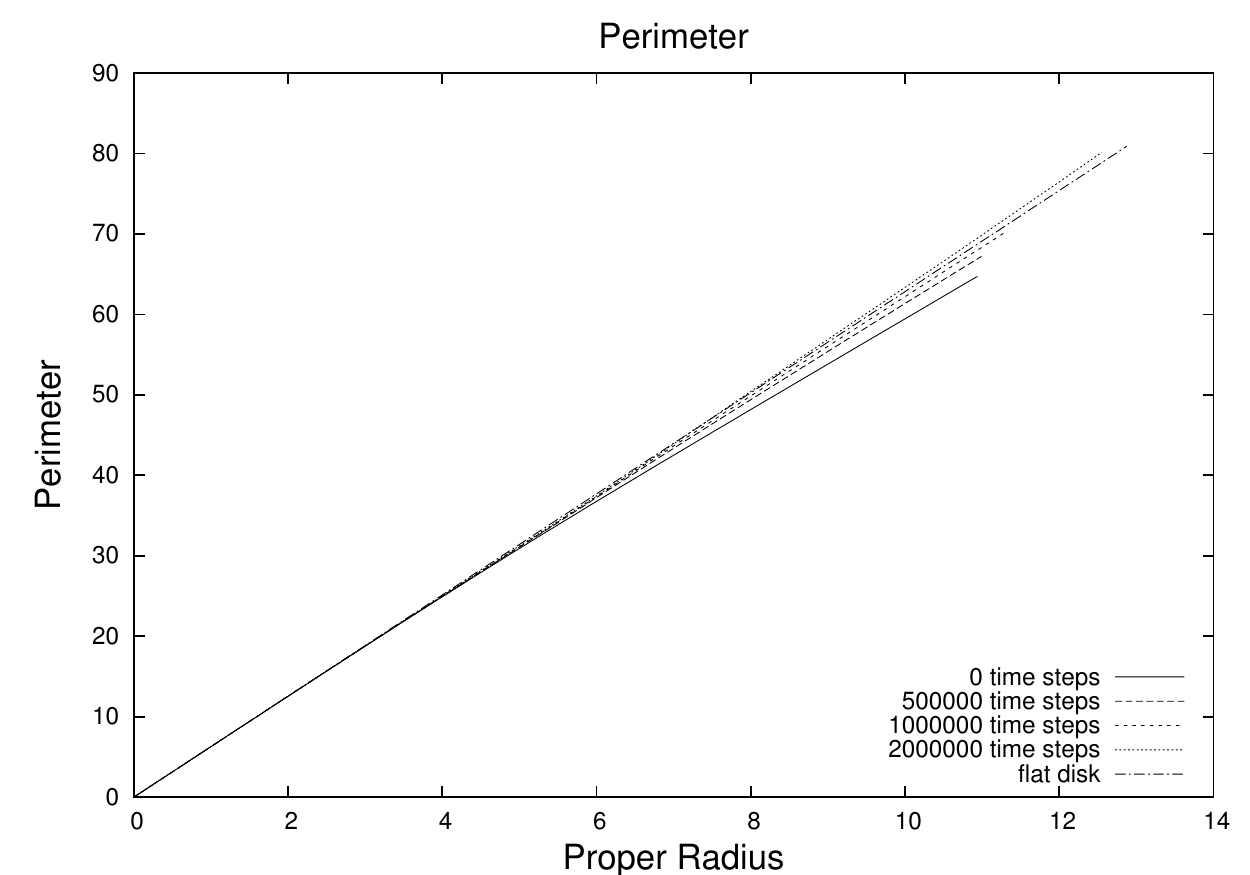}
        }
        \caption{Time evolution of the scalar curvature and perimeter of a disk with changing curvature signature.}\label{figureAcetaCurvature} 
\end{figure}

\begin{figure}[h]
    \centering
        \subfloat{\includegraphics[width=0.7\textwidth]{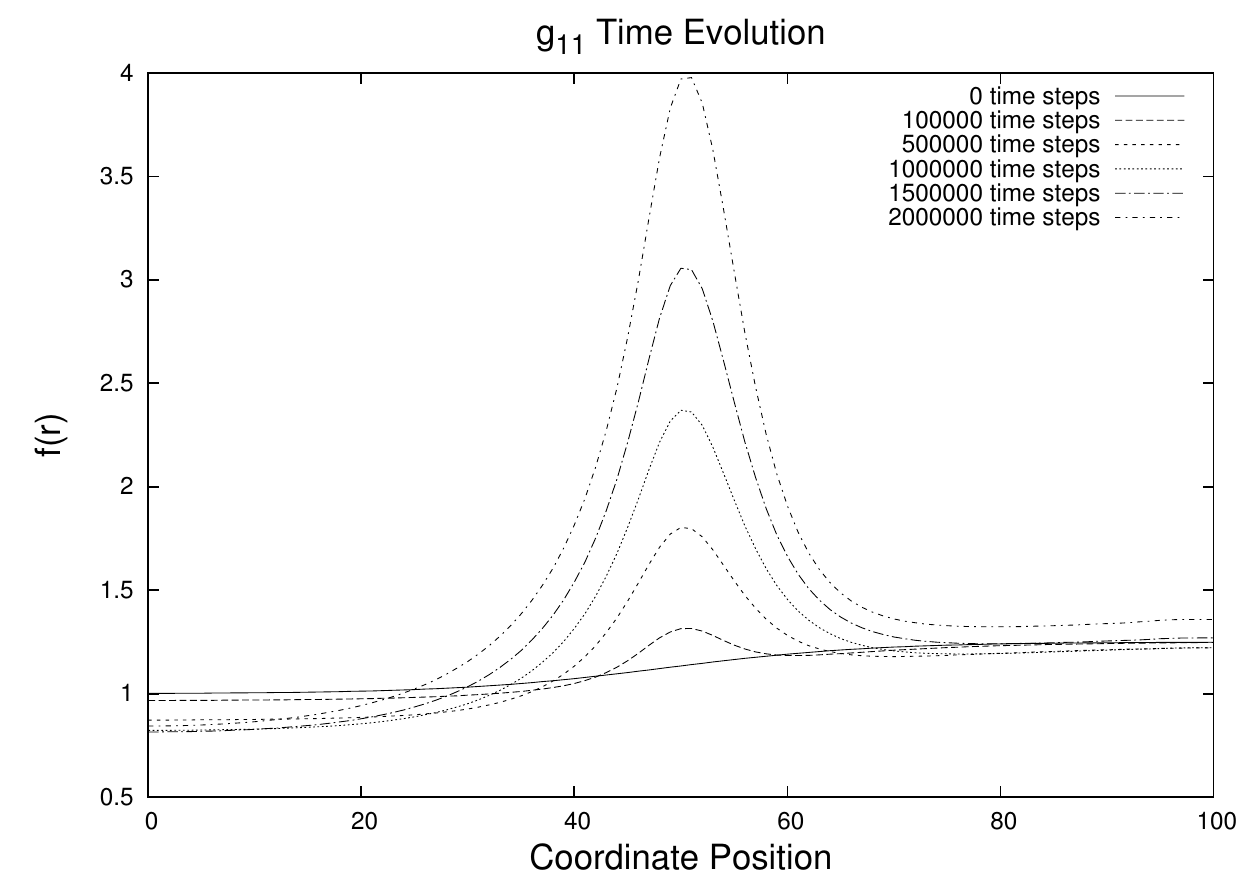}
        }

        \subfloat{\includegraphics[width=0.7\textwidth]{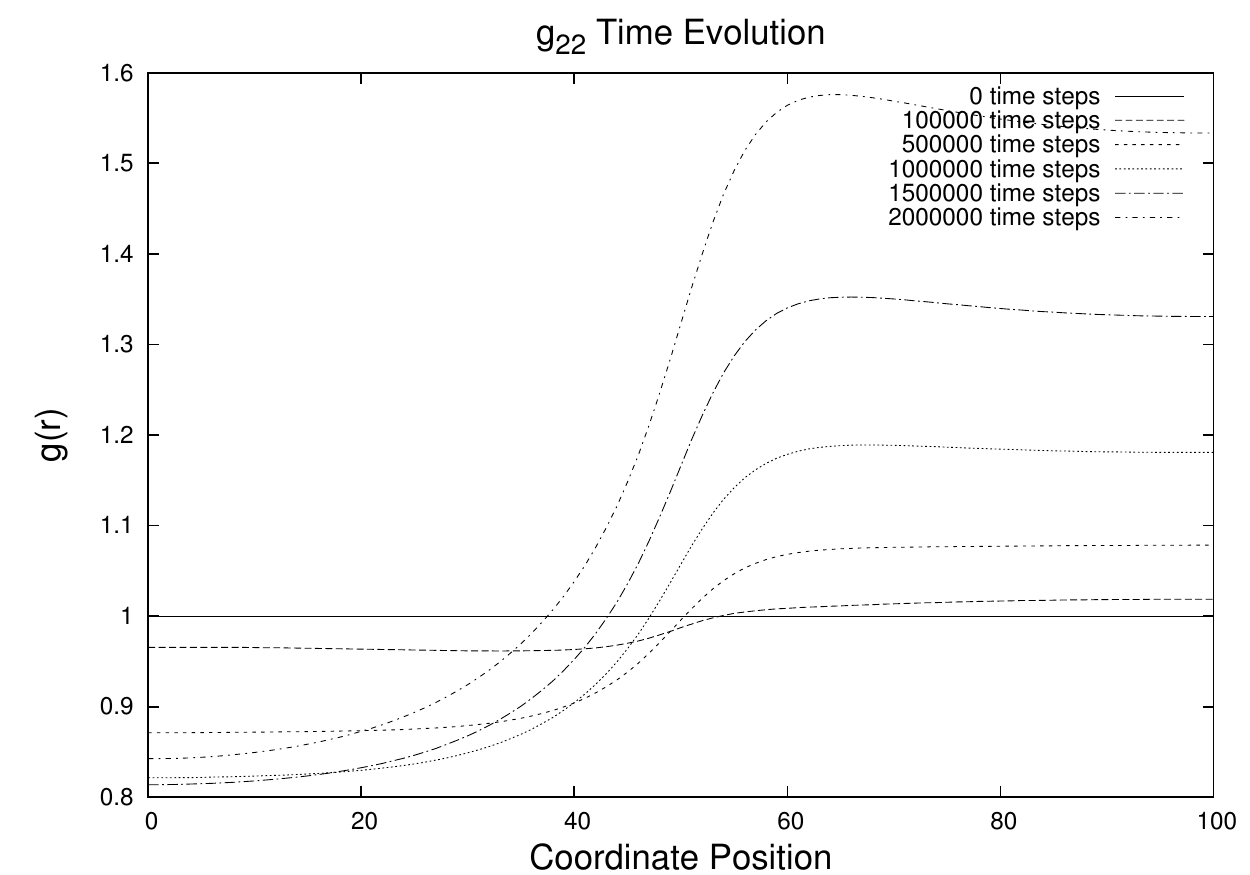}
        }
        \caption{Time evolution of the metric components of a disk with changing curvature signature.}\label{figureAcetaMetric} 
\end{figure}

\begin{figure}[h]
    \centering
        \subfloat{\includegraphics[width=0.7\textwidth]{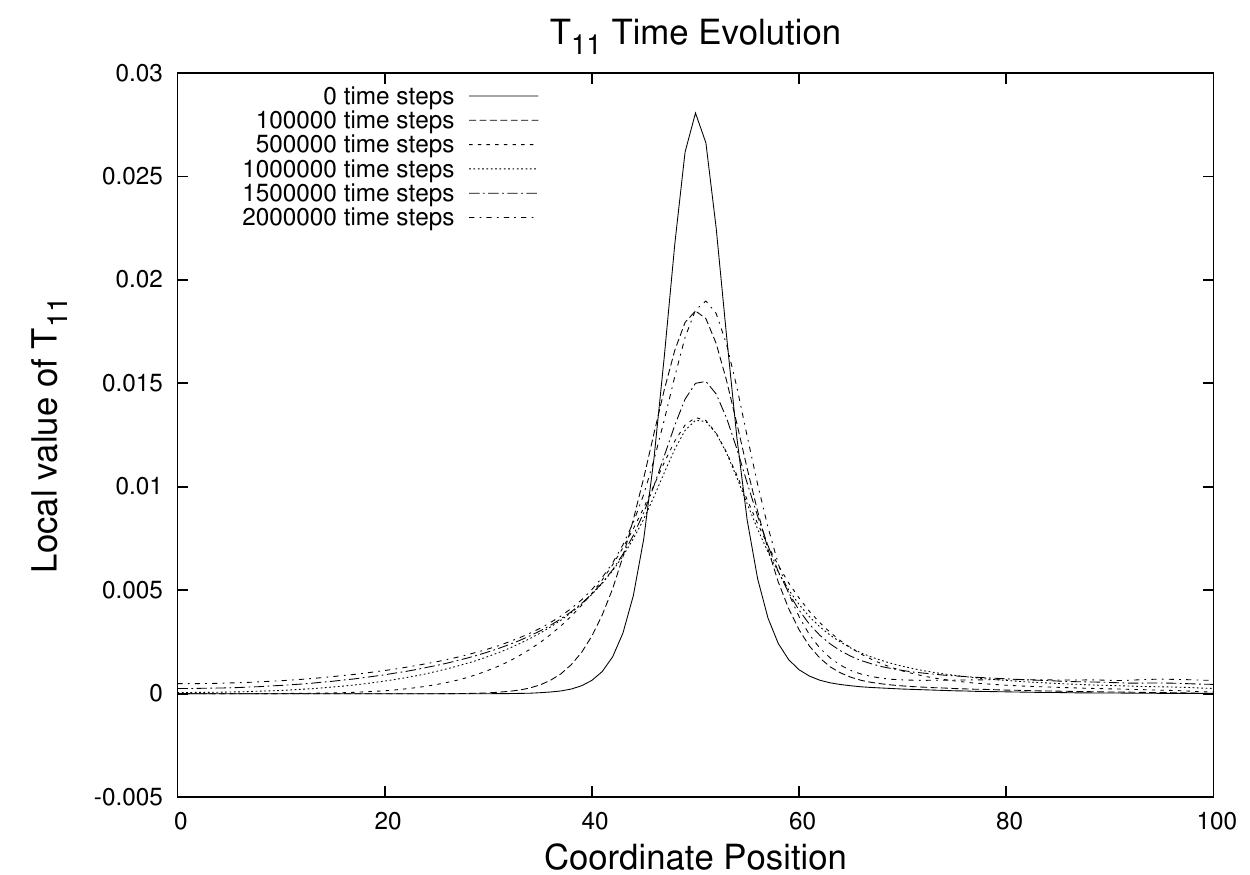}
        }

        \subfloat{\includegraphics[width=0.7\textwidth]{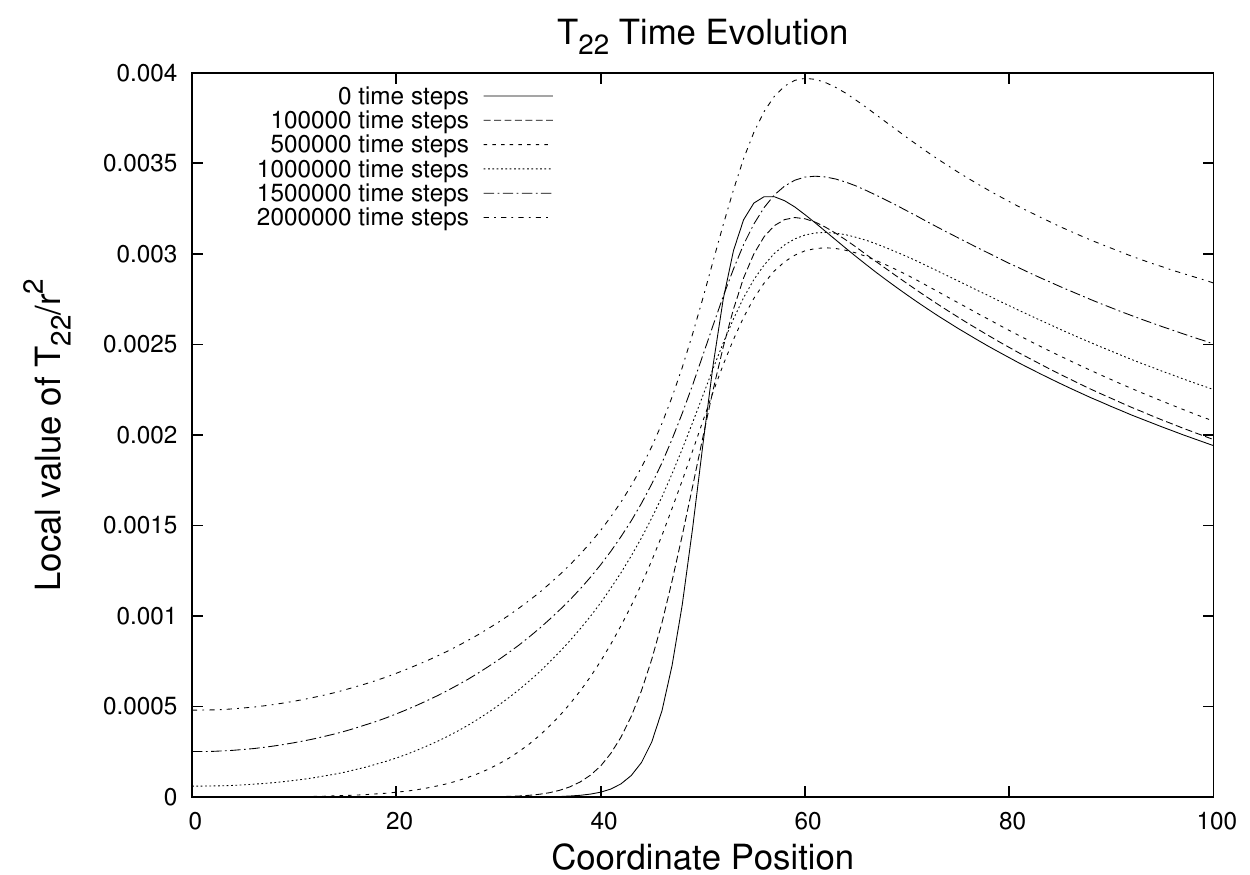}
        }
        \caption{Time evolution of the growth tensor components of a disk with changing curvature signature.}\label{figureAcetaGTs} 
\end{figure}

\begin{figure}[h]
    \centering
    \includegraphics[width=0.7\textwidth]{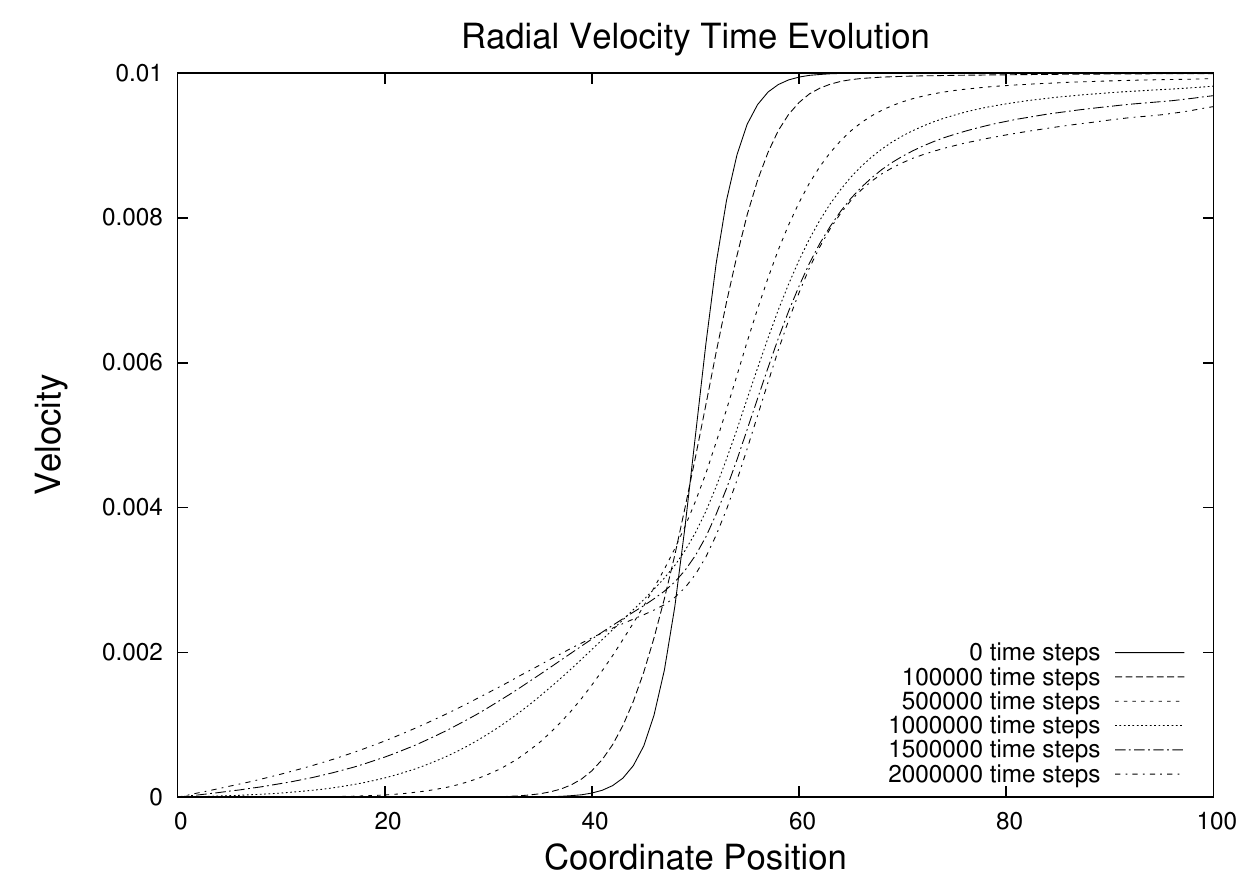}
    \caption{Time evolution of the velocity field of a disk with changing curvature signature.}\label{figureAcetaVelocity} 
\end{figure}

The results can be found in Figures \ref{figureAcetaCurvature} to \ref{figureAcetaVelocity}. 
The curvature transitions from positive to negative values 
seamlessly. Also, the radial metric component increases in a localized ring around the center, while 
the angular metric component increases more at the outer boundary than near the center; both of these
behaviours are consistent with biological observations. For brevity, we do not show the deformation 
tensors given the strong similarities
between this simulation and the previous one for a growing disk that develops negative curvature.

\section{Analysis of 2D simulations with circular symmetry}

From the three simulations presented above, several key observations emerge.

First is the role of Ricci flow in disk growth. Ricci flow allows areas of high curvature to dissipate, which
contributes to the growth of the disk as the metric expands (and in certain areas contracts) to accommodate a 
more uniform distribution of curvature. Another important observation is how the velocity field influences growth. 
As is clear from the two simulations, Ricci flow drives the growth of the disk for some time, but a non-constant velocity
field is required to keep growth going.

With an initial logistic velocity field,
the growth tensor generally has its highest values where the logistic function is steepest. Hence growth is 
localized, causing the metric to expand much more wherever the growth tensor has its largest values. 
This can be seen in the correspondence between the highest values of the growth tensor components (Figure \ref{figureGDgts})
with the steepest parts of the velocity field (Figure \ref{figureGDvelField}). 
Without Ricci flow, these regions of fast growth lead to the development of a shock front in the metric 
where growth builds up in one area while neighbouring regions remain stagnant. 

At the tissue level, such drastic differences in growth rates between neighbouring regions cause strain. 
In Figures \ref{figureGDshear} and \ref{figureGDexpTensor}, the values of the
deformation tensors are shown over time. All the deformations are most pronounced in the area close to the steepest
part of the velocity field, and hence the fastest growth. Here, expansion dissipates over time, as does the $\sigma_{22}$
component of the shear tensor, but the $\sigma_{11}$ shear component appears to keep increasing over time. 
Closer inspection of the same figures also shows that both shear components increase in magnitude at the outer
boundary of the disk over time, even though the expansion scalar remains unchanged in this area at late times.
These results suggest that although the dynamics are initially driven by the steep velocity field
in the mid-region of the disk, the outer edge certainly begins to experience deformations due to the 
changing geometry of the disk. Note how this is also correlated with the metric tensor components steadily
increasing over time at the outer boundary in Figure \ref{figureGDmetric}, and with the 
REGR becoming almost uniform everywhere on the disk over time (Figure \ref{figureGDregr}), 
despite the growth tensor components having localized maxima in the mid-radius regions. 

The deformation tensors and REGR also exhibit wave-like motion of the peaks over time. This shows how the underlying
reaction-diffusion dynamics affect the global properties of the tissue. Also, similarly to the 1D case, growth 
has a positive feedback behaviour, with the expansion scalar dissipating over time, yet generating large
growth tensor values due to the geometric effects of an expanding metric. The REGR evolution also shows that
the globally negative curvature of the disk allows growth to become almost equally distributed over the area
of the disk rather than highly localized as the growth tensor dynamics alone would suggest.

The localization of deformations on the disk also suggests that if a disk or tissue experiences rapid growth, 
it must also have a mechanism to dissipate both
its curvature and velocity field. In this phenomenological model, Ricci flow and velocity diffusion 
provide such a mechanism. When modeled numerically, this dependency manifests itself as a lower threshold 
on the Ricci flow and velocity diffusion coupling terms for a given value of growth tensor coupling.
The lower threshold rises when growth tensor coupling is increased.

Recall from Section \ref{sectionMassDensityEqns} that the assumption of constant tissue density in space and time
leads to the source field $S(r,\theta,t)$ being directly proportional to the expansion scalar $\Theta$. Hence, plotting the
expansion scalar as in Figure \ref{figureGDexpTensor} will also give us information on how the source of material 
is distributed through the tissue. Throughout the simulation of disk growth and \textit{Acetabularia} development, 
the highest values of $\Theta$ are localized in an annular region of the disk, again consistent with the region of highest
growth rates observed in \textit{Acetabularia} measurements. A second feature of note in the simulation is that 
expansion at late times also influences the regions near the origin, indicating that the source of material becomes
more uniformly distributed over time compared to the initial configuration. How these dynamics could be orchestrated by
a single-celled organism is a truly intriguing question.

Lacking extensive data on the velocity field of a growing leaf at mm scales and larger, a circularly 
symmetric logistic function was assumed 
as an initial condition. The logistic function has certain characteristics that strongly influence
the growth pattern of the disk. Having a steep increase at half the radial distance to the edge means that 
growth will occur mostly in a ring around the origin, as seen in the metric components of both growing disks
(Figures \ref{figureGDmetric} and \ref{figureAcetaMetric}). Adopting this initial condition on the velocity field
is also consistent with the mathematical result in Section \ref{sectionAceta} that constant curvature metrics driven by linear
velocity fields cannot experience a change of curvature signature.

The logistic function increases monotonically with radius. In Equation \ref{eqnGTcomponents}, we see that $T_{22}$ 
(the angular component  of the growth tensor) is proportional to the value of $v^1$. Hence, it can be expected that a monotonically 
increasing velocity field will lead to $g(r)$ also having monotonically increasing values over time 
(Equation \ref{eqng22}). This behaviour has strong implications for the final shape of the disk because the circumference of the 
disk is given by $C = 2 \pi \sqrt{g_{22}} = 2 \pi r \sqrt{g(r)}$. If $g(r)$ is greater than unity and
increases with radius, then $C>2 \pi r$ for every value of $r$, and must therefore represent a ruffled, 
negatively curved surface (Figure \ref{figureGDperimeter}).

An interesting question is what initial conditions might generate a growing disk with positive curvature.
One might start with $0<g(r)<1$ to force $C<2 \pi r'$ for all values of $r$, however this is somewhat
contrived. A monotonically decreasing velocity field would produce the desired effect on $T_{22}$, but
would generate only shrinking solutions rather than growing ones. Lastly, there may be an initial 
configuration of both metric components that leads to a positive curvature solution.

\section{Conclusions on 2D simulations with circular symmetry}

The data presented here clearly indicate that under the conditions of this simulation, a growing, ruffled surface will 
emerge from an initially flat disk due to the development of a globally negative curvature on the disk.
The monotonically increasing velocity field plays a major role in this behaviour, as does curvature coupling of the
metric to its Ricci tensor. Comparing Figure \ref{figureGDperimeter} with the results in ~\cite{thinSheetShapes}
shows that the curvature patterns in our model of disk growth would be physically realizable shapes
when embedded in flat 3D space.

A biological model for this type of growth can be found in \textit{Acetabularia}. Two morphologically significant
results emerge from our simulation. The first is that a positively curved disk can evolve to a negatively curved
disk when driven by a logistic velocity field, thereby reproducing the curvature evolution of the \textit{Acetabularia} 
cap. The second is that an annulus of growth is observed in the model, which is consistent with the pattern of 
growth seen in ~\cite{acetabularia}.

Building the simplest 2D model of disk growth is the bridge to the next level of complexity: 
simulations of 2D disks with $\theta$-dependent quantities. These are discussed in the next chapter. 

Looking back on the similarities between the 1D and 2D dynamical equations, it is possible to draw
comparisons between the two. Both react to an initial logistic velocity field by exhibiting an elongation zone
where the velocity field is steepest, and hence the growth tensor is largest. Both require 
dissipation terms in the metric and velocity fields to prevent shocks from building up as a 
result of localized growth. Both grow as a result of deposition (metric self-coupling)
and material expansion.

The differences between the 1D and 2D models ultimately stem from their 
different dimensionalities. Even under similar initial conditions, in particular the logistic 
velocity field, the two systems behave differently. The 2D model has more degrees of freedom in how 
the metric can evolve, and has two components of the growth tensor (as opposed to one in 1D). This causes
the feedback between the velocity field and metric to occur differently than in the 1D case. For instance, 
the REGR in 2D does not evolve a double-peak behaviour, even though it has contributions from both the curvature 
and expansion.

These 2D simulations are also the first to have curvature coupling terms. The 1D simulations
required a functionally similar term to mimic the effect of curvature since intrinsic curvature cannot 
be defined with only one spatial dimension. As seen
in the 2D isotropic simulations, it is possible to look at the effects of curvature coupling and velocity 
field coupling separately and together to determine their relative roles in the dynamics of growth. With the
aid of physical deformation tensors, the REGR, perimeter and modular growth calculations, the dynamics of 
the geometry and velocity field can be interpreted in terms of measurable quantities. To our knowledge, these
are the first such calculations made for a 2D disk growth model.

\chapter{2D Simulations with Anisotropies}
\label{chapter2DFullSimulations}

\section{Breaking circular symmetry}
In this chapter, simulations of growth involving angular anisotropies will be discussed. 
Circular symmetry can be broken by introducing $\theta$-dependent initial conditions
on the metric or the velocity field.

As in the previous chapter on 2D circularly symmetric solutions, we will proceed in a similar
progression of simulations. First, Ricci flow on the disk with angular anisotropies will be studied in order to understand
the dynamics of the disk when it is coupled only to its own curvature. Building on these results,
we will then introduce a circularly symmetric velocity field to the anisotropic disks. This will allow
us to see the first effects of angular anisotropies in the full 2D dynamical equations. Lastly,
an initially flat disk will be evolved under the influence of an anisotropic velocity field in a model
that results in elongation of the disk along an axis as well as nontrivial curvature patterns. 

\section{Dynamical equations of growth in anisotropic polar coordinates}

The equations for growth with angular anisotropy are derived in a similar way to 
the isotropic 2D equations. A fully general 2D metric in polar coordinates can be
written as: 

\begin{equation}
g_{ik} = 
\begin{pmatrix}
f(r,\theta) & rh(r,\theta) \\
rh(r,\theta) & r^2g(r,\theta) \\
\end{pmatrix}
\end{equation}

The connection coefficients are:
\begin{eqnarray}
\Gamma_{11}^1 & = & \frac{(2hh_{,r} - f_{,r}g)r + 2h^2 - f_{,\theta}h}{2dr} \nonumber \\
\Gamma_{11}^2 & = & -\frac{(2fh_{,r} - f_{,r}h)r + 2fh - ff_{,\theta}}{2dr^2} \nonumber \\
\Gamma_{12}^1 & = & \frac{g_{,r}hr + 2gh - f_{,\theta}g}{2d} \nonumber \\
\Gamma_{12}^2 & = & -\frac{fg_{,r}r - f_{,\theta}h + 2fg}{2dr} \nonumber \\
\Gamma_{22}^1 & = & \frac{gg_{,r}r^2 + (-2gh_{,\theta} + g_{,\theta}h + 2g^2)r}{2d} \nonumber \\
\Gamma_{22}^2 & = & -\frac{g_{,r}hr - 2hh_{,\theta} + 2gh + fg_{,\theta}}{2d}
\end{eqnarray}

\noi where $d=h^2-fg$.

The only unique Riemann tensor component is:
\begin{eqnarray}
R_{1212} & = & \frac{1}{4d} [ (2g_{,r}hh_{,r} - 2g_{,rr}d - f(g_{,r})^2 - f_{,r}gg_{,r})r^2 \nonumber \\
&  & + ((-4hh_{,\theta} + 4gh + 2fg_{,\theta})h_{,r} + 4dh_{,r\theta} + 2f_{,r}gh_{,\theta} - 6g_{,r}h^2 \nonumber \\
& & + (f_{,\theta}g_{,r} - f_{,r}g_{,\theta})h + 4fgg_{,r} - 2f_{,r}g^2)r \nonumber \\
&  & + (2f_{,\theta}h - 4fg)h_{,\theta} - 2df_{,\theta\theta} + (2fg_{,\theta} + 2f_{,\theta}g)h - ff_{,\theta}g_{,\theta} - f_{,\theta}^2g ]\nonumber \\
\end{eqnarray}

The Riemann tensor in 2D has only one component and is related to the scalar curvature $R$ by:

\begin{equation}
R_{1212} = R[g_{11}g_{22} - (g_{12})^2].
\end{equation}

Since the Ricci tensor can be expressed as $R_{ik} = R g_{ik}$, the metric evolution becomes: 
\begin{eqnarray}
\frac{\partial g_{11}}{\partial t} \rightarrow \frac{\partial f}{\partial t} & = & - \kappa Rf + \kappa_1 T_{11} \nonumber \\[10pt]
\frac{\partial g_{12}}{\partial t} \rightarrow \frac{\partial h}{\partial t} & = & - \kappa Rh + \kappa_1 \frac{T_{12}}{r} \nonumber \\[10pt]
\frac{\partial g_{22}}{\partial t} \rightarrow \frac{\partial g}{\partial t} & = & - \kappa Rg + \kappa_1 \frac{T_{22}}{r^2} \nonumber \\
\label{eqnMetric}
\end{eqnarray}

\noindent The general expression for the growth tensor is:
\begin{eqnarray}
T_{ik} & = & \nabla_i v_k + \nabla_k v_i \nonumber \\
       & = & g_{kl} \nabla_i v^l + g_{ip} \nabla_k v^p
\end{eqnarray}

\noi since $\nabla_i g_{kl} = 0$. 

\noindent The specific expressions are:
\begin{eqnarray}
T_{11} & = & 2f(v_{,r}^1 + \Gamma_{11}^1 v^1 + \Gamma_{12}^1 v^2) + 2hr(v_{,r}^2 + \Gamma_{11}^2 v^1 + \Gamma_{12}^2 v^2)   \nonumber \\
T_{12} & = & hr(v_{,r}^1 + \Gamma_{11}^1 v^1 + \Gamma_{12}^1 v^2) + gr^2(v_{,r}^2 + \Gamma_{11}^2 v^1 + \Gamma_{12}^2 v^2)   \nonumber \\
       & & + f(v_{,\theta}^1 + \Gamma_{21}^1 v^1 + \Gamma_{22}^1 v^2) + hr(v_{,\theta}^2 + \Gamma_{21}^2 v^1 + \Gamma_{22}^2 v^2) \nonumber \\
T_{22} & = & 2hr(v_{,\theta}^1 + \Gamma_{21}^1 v^1 + \Gamma_{22}^1 v^2) + 2gr^2(v_{,\theta}^2 + \Gamma_{12}^2 v^1 + \Gamma_{22}^2 v^2)   \nonumber\\
\end{eqnarray}

\noindent The velocity equations are:
\begin{eqnarray}
\frac{\partial v^1}{\partial t} & = & -\Gamma_{11}^1 v^1 v^1 - 2 \Gamma_{12}^1 v^1 v^2 - \Gamma_{22}^1 v^2 v^2 - v^1 \nabla_1 v^1 - v^2 \nabla_2 v^1 + c(g^{11} \nabla_1 \nabla_1 v^1 + g^{22} \nabla_2 \nabla_2 v^1 ) \nonumber \\
           & = & -\Gamma_{11}^1 v^1 v^1 - 2 \Gamma_{12}^1 v^1 v^2 - \Gamma_{22}^1 v^2 v^2 - v^1 [v_{,r}^1 + \Gamma_{11}^1 v^1 + \Gamma_{12}^1 v^2 ] - v^2 [v_{,\theta}^1 + \Gamma_{12}^1 v^1 + \Gamma_{22}^1 v^2 ] \nonumber \\ 
           &   & + c g^{11} \left( \frac{\partial^2 v^1}{\partial r^2} + \Gamma_{11,r}^1 v^1 + \Gamma_{12,r}^1 v^2 + \Gamma_{11}^1 v_{,r}^1 + 2 \Gamma_{12}^1 v_{,r}^2 - \Gamma_{11}^2 v_{,\theta}^1 + (\Gamma_{12}^1 \Gamma_{12}^2 - \Gamma_{11}^2 \Gamma_{22}^1) v^2 \right) \nonumber \\ 
           &   & + c g^{22} \left( \frac{\partial^2 v^1}{\partial \theta^2} + \Gamma_{12,\theta}^1 v^1 + \Gamma_{22,\theta}^1 v^2 + 2\Gamma_{12}^1 v_{,\theta}^1 + 2\Gamma_{22}^1 v_{,\theta}^2 - \Gamma_{22}^1 v_{,r}^1 - \Gamma_{22}^2 v_{,\theta}^1 \right) \nonumber \\
           &   & + c g^{22} \left( \Gamma_{12}^1 (\Gamma_{12}^1 - \Gamma_{22}^2) + \Gamma_{22}^1 (\Gamma_{12}^2 - \Gamma_{11}^1 ) \right) v^1 \label{eqnV1}
\end{eqnarray}

\begin{eqnarray}
\frac{\partial v^2}{\partial t} & = & -\Gamma_{11}^2 v^1 v^1 - 2 \Gamma_{12}^2 v^1 v^2 - \Gamma_{22}^2 v^2 v^2 - v^1 \nabla_1 v^2 - v^2 \nabla_2 v^2 + c(g^{11} \nabla_1 \nabla_1 v^2 + g^{22} \nabla_2 \nabla_2 v^2 )  \nonumber \\
          & = & -\Gamma_{11}^2 v^1 v^1 - 2 \Gamma_{12}^2 v^1 v^2 - \Gamma_{22}^2 v^2 v^2 - v^1 [v_{,r}^2 + \Gamma_{11}^2 v^1 + \Gamma_{12}^2 v^2 ] - v^2 [v_{,\theta}^2 + \Gamma_{12}^2 v^1 + \Gamma_{22}^2 v^2 ] \nonumber \\
          &   & + c g^{11} \left( \frac{\partial^2 v^2}{\partial r^2} + \Gamma_{11,r}^2 v^1 + \Gamma_{12,r}^2 v^2 + 2 \Gamma_{11}^2 v_{,r}^1 + 2 \Gamma_{12}^2 v_{,r}^2 - \Gamma_{11}^1 v_{,r}^2 - \Gamma_{11}^2 v_{,\theta}^2 \nonumber \right) \\
          &   & + c g^{11}(\Gamma_{11}^2 (\Gamma_{12}^1 - \Gamma_{22}^2) + \Gamma_{12}^2 (\Gamma_{12}^2 - \Gamma_{11}^1) ) v^2  \nonumber \\ 
          &   & + c g^{22} \left( \frac{\partial^2 v^2}{\partial \theta^2} + \Gamma_{12,\theta}^2 v^1 + \Gamma_{22,\theta}^2 v^2 + 2\Gamma_{12}^2 v_{,\theta}^1 - \Gamma_{22}^2 v_{,\theta}^2 - \Gamma_{22}^1 v_{,r}^2 \right) \nonumber \\
          &   & + c g^{22}(\Gamma_{12}^2 \Gamma_{12}^1 - \Gamma_{22}^1 \Gamma_{11}^2 ) v^1 \label{eqnV2}
\end{eqnarray}

\section{Numerical methods}

Again, the coupled tensor equations are now written as a set of nonlinear coupled PDEs. We seek to model 
Equations \ref{eqnMetric}, \ref{eqnV1} and \ref{eqnV2}.

The equations now include more nonlinear terms due to the presence of the angular velocity, off-diagonal metric terms
and the full set of connection coefficients. The equations now also contain diffusion terms in both $r$ and $\theta$. 
To handle the additional
complexity of the $\theta$-dependent equations, we implement a simpler, fully explicit finite differencing scheme. 
For the second-derivative terms appearing in the scalar curvature, a time averaged 
scheme for the Laplacian terms is used as discussed in Section \ref{sectionDFmethod}.

The boundary conditions in the anisotropic 2D simulations now require conditions on both the radial and
angular coordinates. The most appropriate condition for angular coordinate is a periodic boundary since
the solutions can be mapped to the unit circle.

\vspace{3mm}
\begin{center}
\begin{tabular}{r c l}
\hline
$\partial_r f(t,r=0,\theta)$ & = & 0 \\
$\partial_r f(t,r=r_{max},\theta)$ & = & 0 \\
$\partial_r g(t,r=0,\theta)$ & = & 0 \\
$\partial_r g(t,r=r_{max},\theta)$ & = & 0 \\
$\partial_r h(t,r=0,\theta)$ & = & 0 \\
$\partial_r h(t,r=r_{max},\theta)$ & = & 0 \\
$f(t,r,\theta=0)$ & = & $f(t,r,\theta=2\pi)$ \\
$g(t,r,\theta=0)$ & = & $g(t,r,\theta=2\pi)$ \\
$h(t,r,\theta=0)$ & = & $h(t,r,\theta=2\pi)$ \\
$v^1(t,r=0,\theta)$ & = & 0 \\
$\partial v^1(t,r=r_{max},\theta)/\partial r$ & = & $\phi (t,\theta)$ \\
$v^1(t,r,\theta=0)$ & = & $v^1(t,r,\theta=2\pi)$ \\
\hline 
\end{tabular}
\end{center}

The function $\phi (t,\theta)$ is equivalent to $\partial v^1/\partial r$ evaluated at $r=r_{max-1}$.

\section{Numerical Simulation: Ricci flow with angular anisotropies}
\label{sectionRF4gauss}

To understand the effect of pure Ricci flow on a metric with angular anisotropies, we study a metric with
$4$ Gaussian perturbations arranged symmetrically on the metric. These are similar to the
Gaussian used in the isotropic 2D case, but are now modulated by a Gaussian in the $\theta$ coordinate.

The $4$-Gaussian simulation has parameters:
\vspace{3mm}
\begin{center}
\begin{tabular} {r l}
\hline
Ricci flow coupling & $\kappa = 0.5$ \\[10pt]
$f(t=0,r) = $ & $5.0\exp(-2.0(r-\frac{1}{2}r_{max})^2) $\\
			& $\times [\exp(-2.0(\theta-\frac{1}{8}\theta_{max})^2) + \exp(-2.0(\theta-\frac{3}{8}\theta_{max})^2)$ \\
			& $+ (\exp(-2.0(\theta-\frac{5}{8}\theta_{max})^2) + \exp(-2.0(\theta-\frac{7}{8}\theta_{max})^2)] + 2.0$ \\[10pt]
$g(t=0,r) = $ & $2.0\exp(-2.0(r-\frac{1}{2}r_{max})^2) $\\ 
			& $\times [\exp(-2.0(\theta-\frac{1}{8}\theta_{max})^2) + \exp(-2.0(\theta-\frac{3}{8}\theta_{max})^2)$ \\ 
			& $+(\exp(-2.0(\theta-\frac{5}{8}\theta_{max})^2) + \exp(-2.0(\theta-\frac{7}{8}\theta_{max})^2)] + 2.0$ \\[10pt]
$h(t=0,r) = $ & $0.5\exp(-2.0(r-\frac{1}{2}r_{max})^2) $\\
			& $\times [0.5\exp(-2.0(\theta-\frac{1}{8}\theta_{max})^2) + 0.5\exp(-2.0(\theta-\frac{3}{8}\theta_{max})^2)$ \\
			& $+(0.5\exp(-2.0(\theta-\frac{5}{8}\theta_{max})^2) + 0.5\exp(-2.0(\theta-\frac{7}{8}\theta_{max})^2)] + 1.0$ \\
\hline 
\end{tabular}
\end{center}
\vspace{3mm}

Figures \ref{figureRF4gaussCurvature} to \ref{figureRF4gaussMetric2} show the results of these simulations.

Figure \ref{figureRF4radius} indicates that the regions with high initial curvature will shrink radially over time, while regions
that are initially relatively flat will grow radially. Just like for the isotropic 2D simulations, 
as the curvature dissipates, growth stagnates and the metric approaches a nontrivial asymptotic 
configuration, as shown in Figures \ref{figureRF4gaussMetric4} and \ref{figureRF4gaussMetric2}. 

An interesting difference in the curvature dynamics compared to the isotropic 2D simulations
is the final configuration of the scalar curvature. In the isotropic simulations, the center
of the disk had negative curvature and the edges were positive. In the anisotropic simulations,
the center is positive while the edges have negative curvature. The cause of this difference
is not known, but is clearly a result of introducing angular anisotropy into the system.

\section{Numerical Simulation: Anisotropic growth of a disk}
\label{sectionAniDisk}

We can also study an initially flat metric under the influence of an anisotropic velocity field.
Here, the velocity field is a logistic function just as in previous simulations, only this
time modulated by a $\theta$-dependent Gaussian function centered on $\theta=\pi$:

\vspace{3mm}
\begin{center}
\begin{tabular}{r l}
\hline
Ricci flow coupling & $\kappa = 0.75$ \\
Growth tensor coupling & $\kappa_1 = 0.75$ \\
Velocity diffusion coupling & $c = 0.005$ \\
\hline 
\end{tabular}
\end{center}
\vspace{3mm}

\begin{center}
\begin{tabular}{r c l}
\hline
$f(t=0,r)$ & = &  $1.0$ \\
$g(t=0,r)$ & = &  $1.0$ \\ 
$h(t=0,r)$ & = &  $0.0$ \\ 
$v^1(t=0,r)$ & = &  $(0.01/[1.0 + \exp(-10.0(r-\frac{1}{2}r_{max}))] - 0.01/[1.0 + \exp(5.0r_{max}))])$ \\ 
			 & &  $\times \exp(-0.1(\theta-\frac{1}{2}\theta_{max})^2)$ \\
$v^2(t=0,r)$ & = &  $0.0$ \\ 
\hline 
\end{tabular}
\end{center}
\vspace{3mm}

Figures \ref{figureAniDiskCurvature} to \ref{figureAniDiskRotationPi2} show the results of these simulations.

The results of growth driven by a radially anisotropic velocity field show that
regions of buckling and ruffling can emerge from an initially flat disk. Highest growth rates at 
$\theta = \pi$ correspond to the largest values of the growth tensor and velocity field. These
also become the areas of highest curvature. Indeed, the growth tensor for this simulation not only
drives elongation of the disk, but also where and how much curvature the disk develops.

\section{Analysis of 2D simulations with anisotropy}

The results of the two simulations presented in this chapter can be interpreted both geometrically and physically.

\begin{figure}[t]
    \centering
		\subfloat{
            \includegraphics[width=0.5\textwidth]{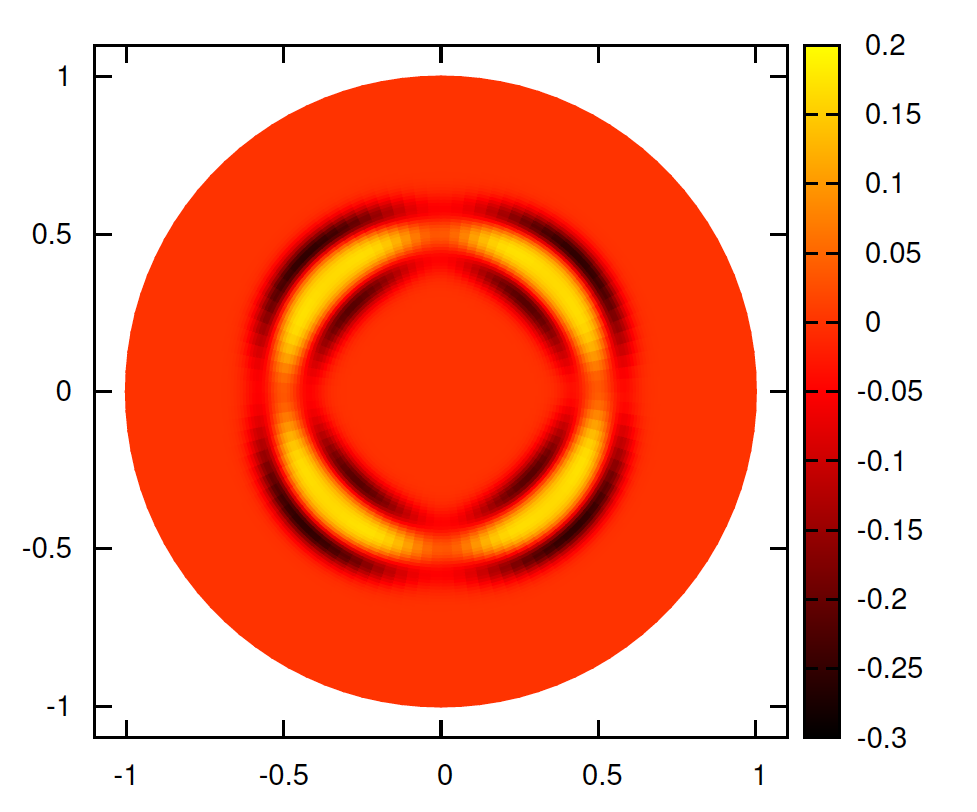}
        }

        \subfloat{                
            \includegraphics[width=\textwidth]{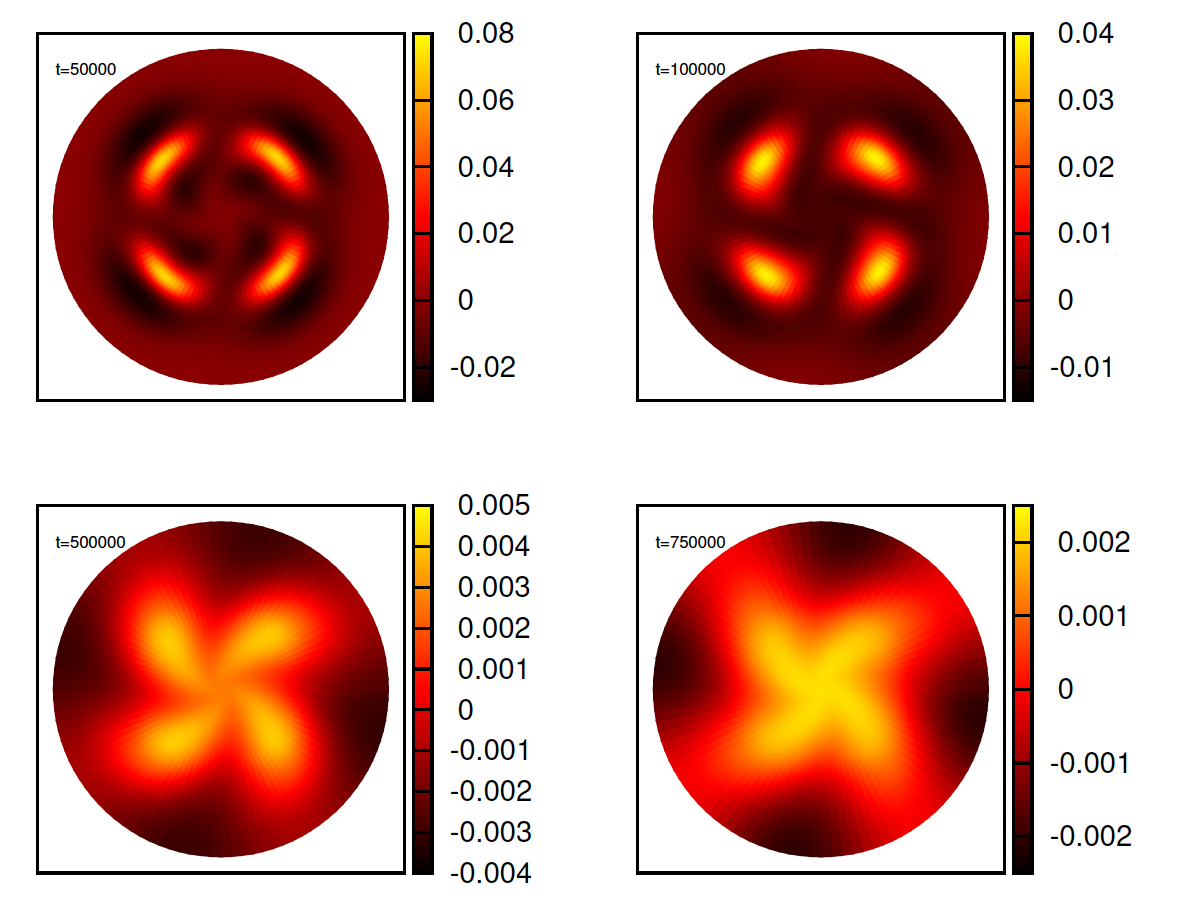}
        }
        \caption{Time evolution of the scalar curvature under Ricci flow for a $4$-gaussian initial metric.}\label{figureRF4gaussCurvature} 
\end{figure}

\begin{figure}
	\centering
	\includegraphics[width=0.7\textwidth]{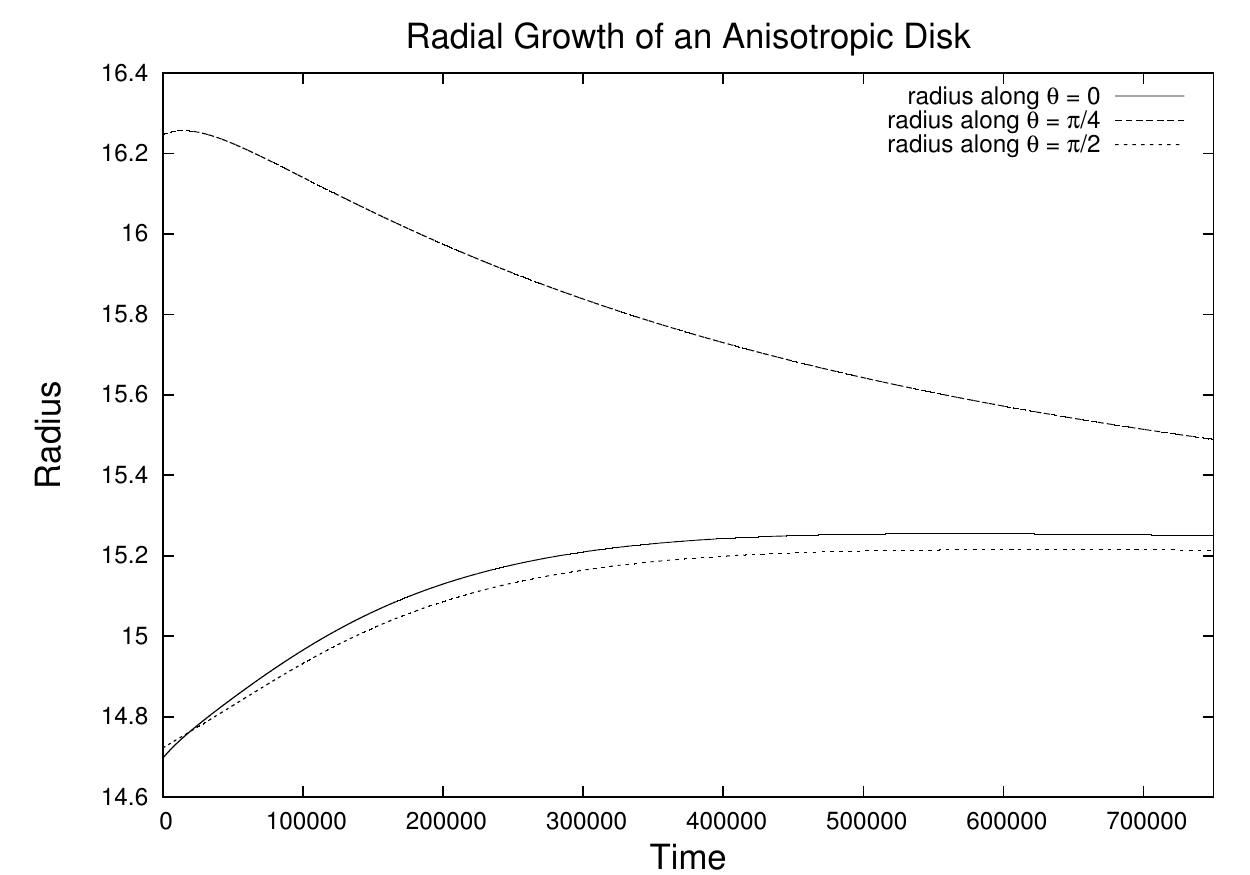}
	\caption{Radial growth of a $4$-gaussian initial metric under Ricci flow along different $\theta$ directions.}
	\label{figureRF4radius}
\end{figure}

\begin{figure}
    \centering
        \subfloat{
                \includegraphics[width=0.6\textwidth]{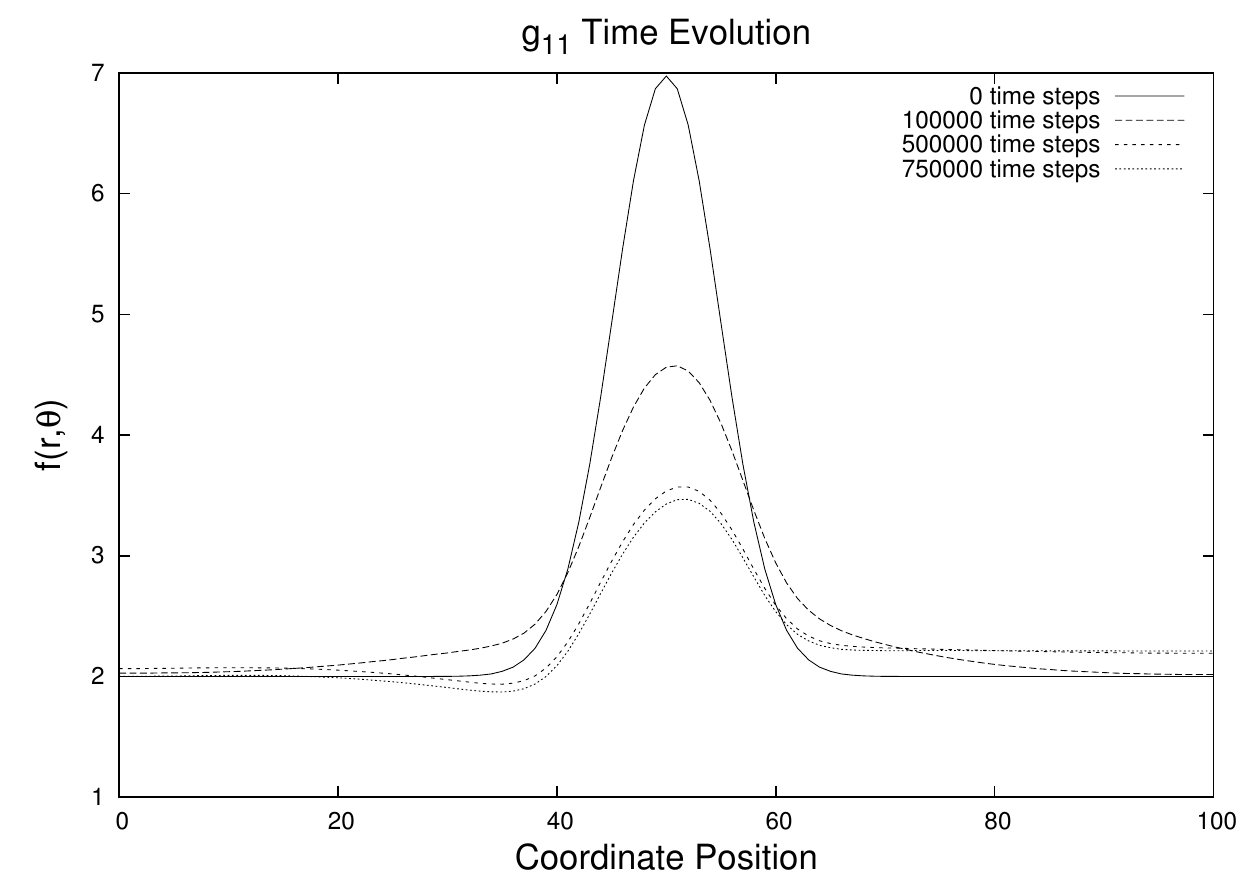}
        }

        \subfloat{
                \includegraphics[width=0.6\textwidth]{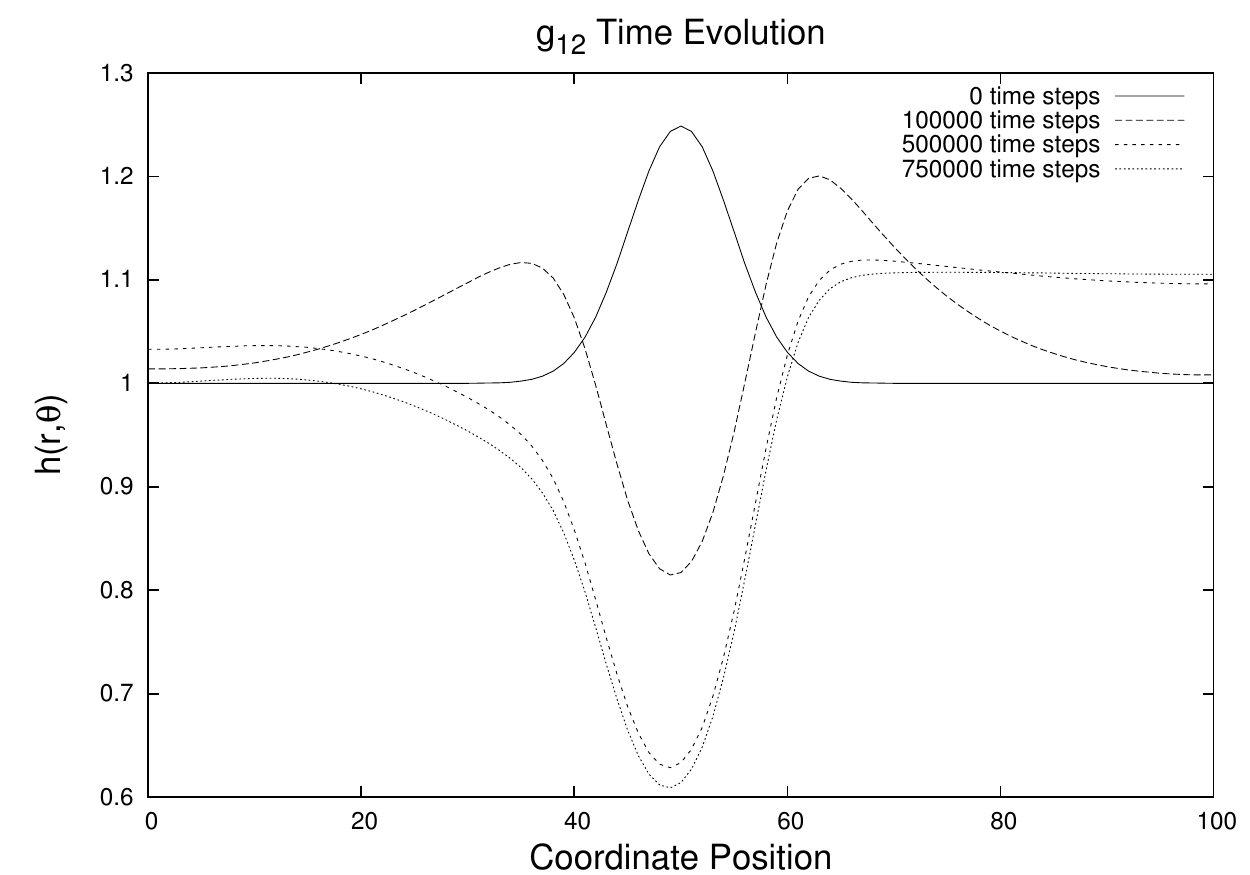}
        }

        \subfloat{
                \includegraphics[width=0.6\textwidth]{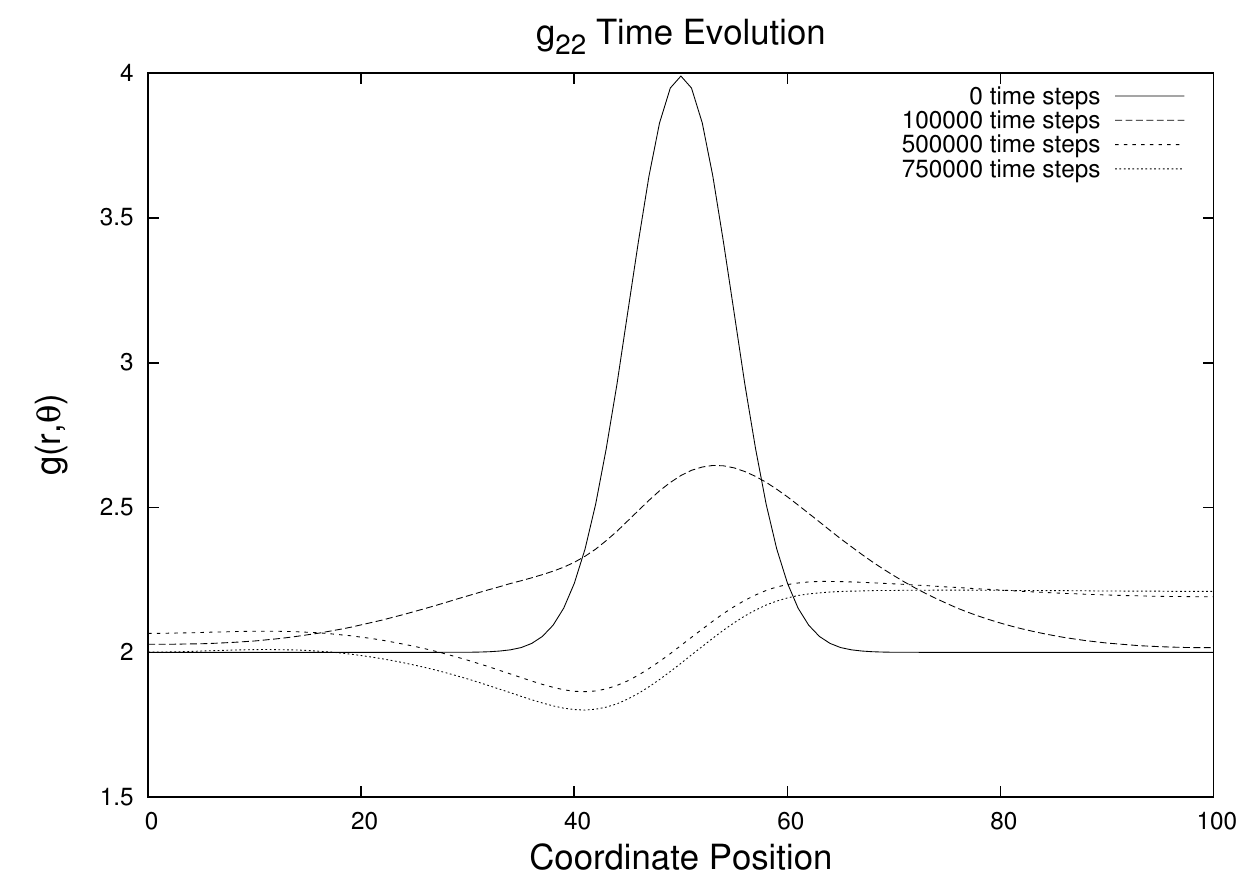}
        }
        \caption{Dynamics for a $4$-gaussian initial metric under Ricci flow at $\theta=\pi/4$.}\label{figureRF4gaussMetric4} 
\end{figure}

\begin{figure}
    \centering
        \subfloat{
                \includegraphics[width=0.6\textwidth]{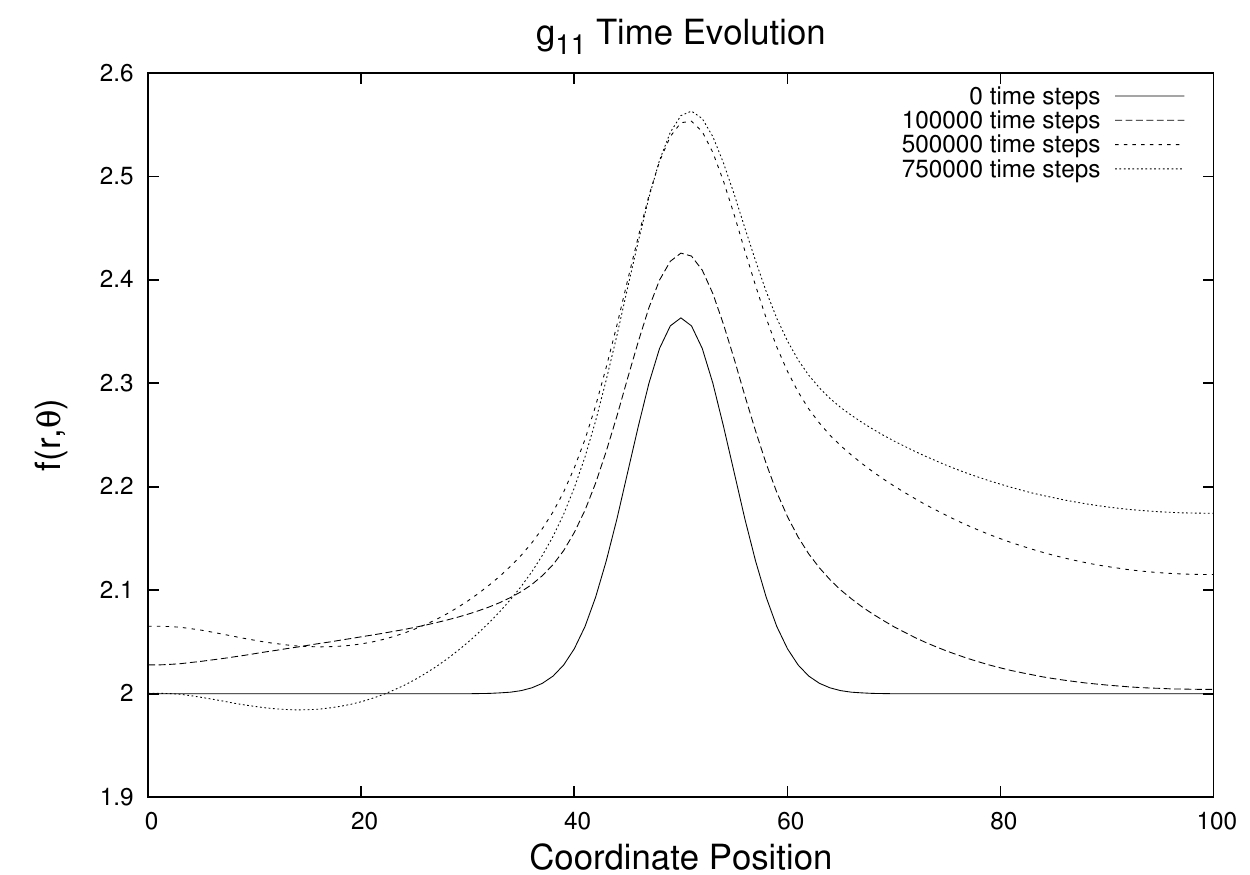}
        }

        \subfloat{
                \includegraphics[width=0.6\textwidth]{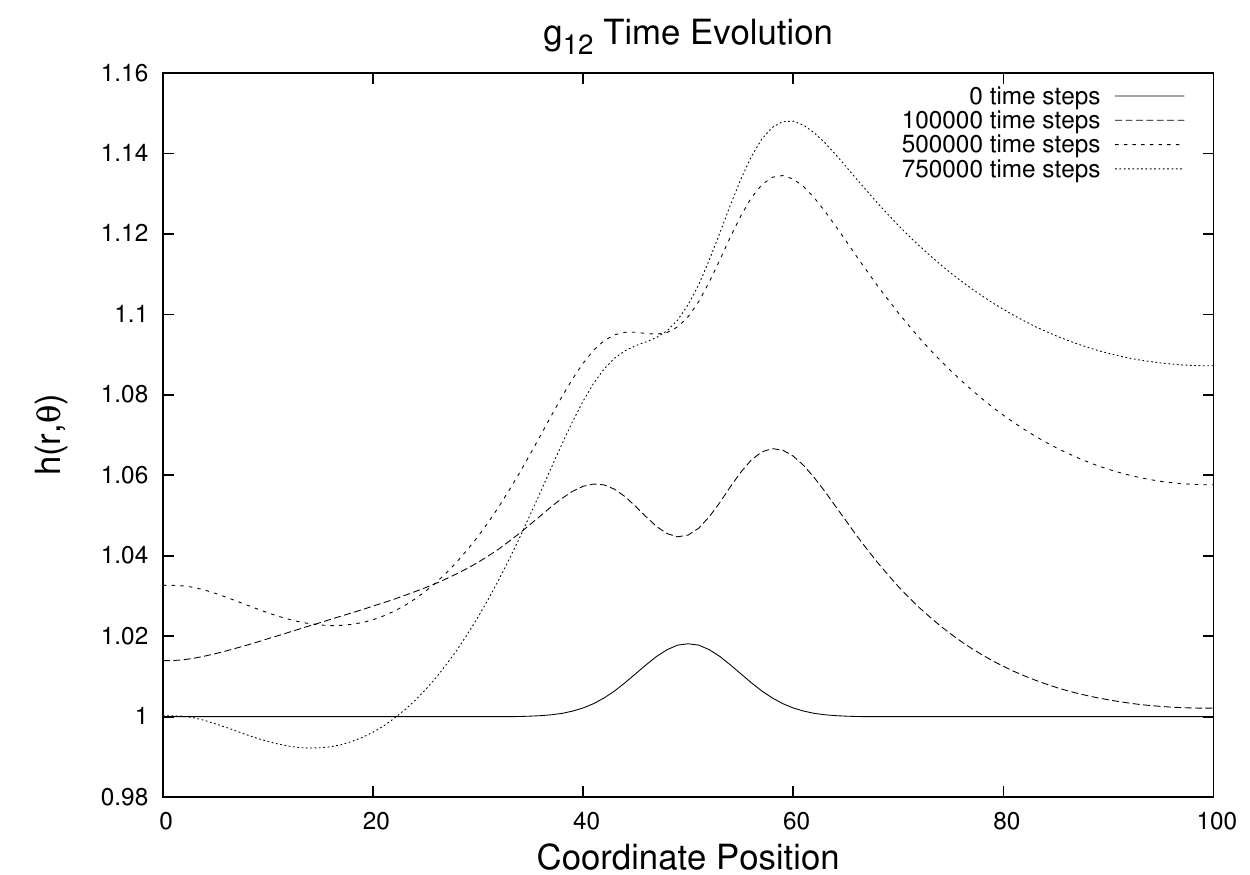}
        }

        \subfloat{
                \includegraphics[width=0.6\textwidth]{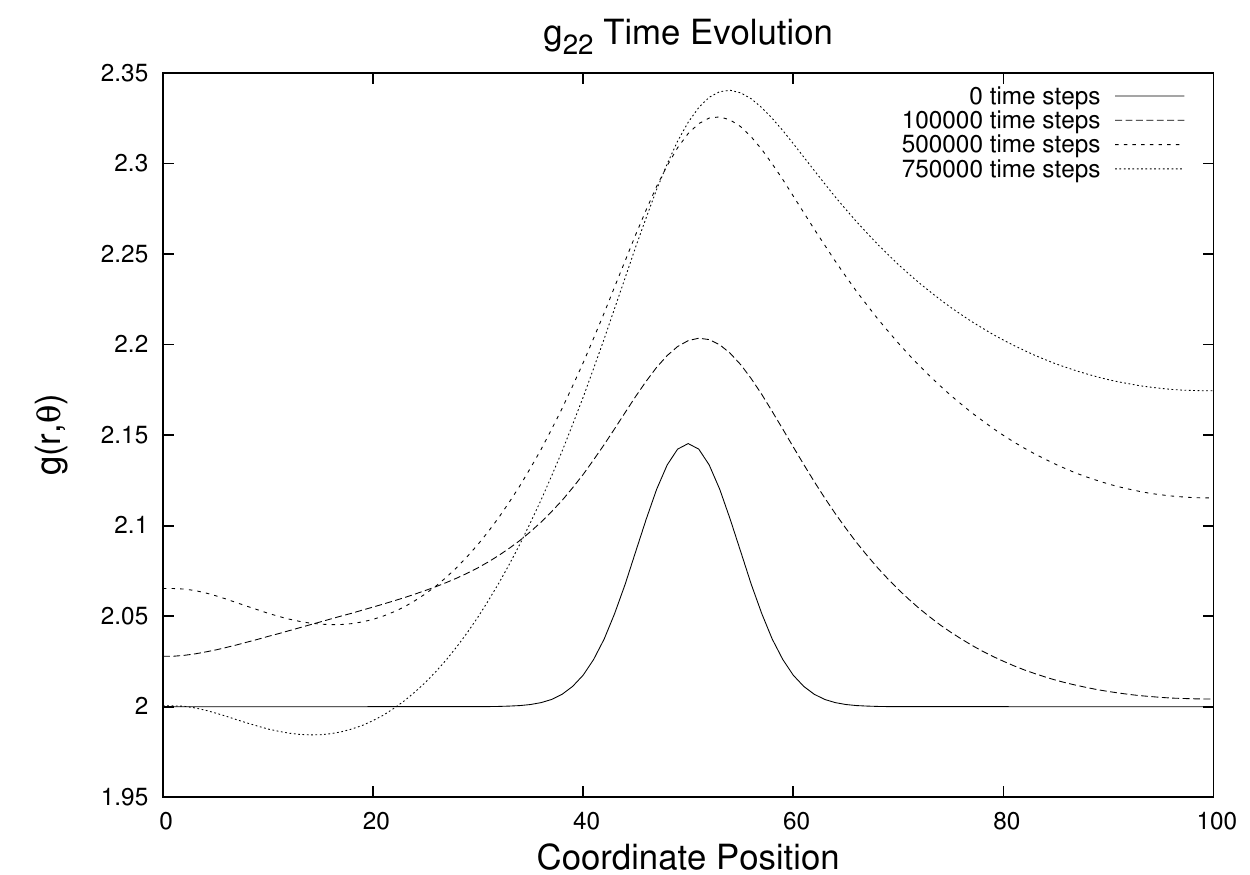}
        }
        \caption{Dynamics for a $4$-gaussian initial metric under Ricci flow at $\theta=\pi/2$.}\label{figureRF4gaussMetric2} 
\end{figure}

\begin{figure}
    \centering
	\includegraphics[width=\textwidth]{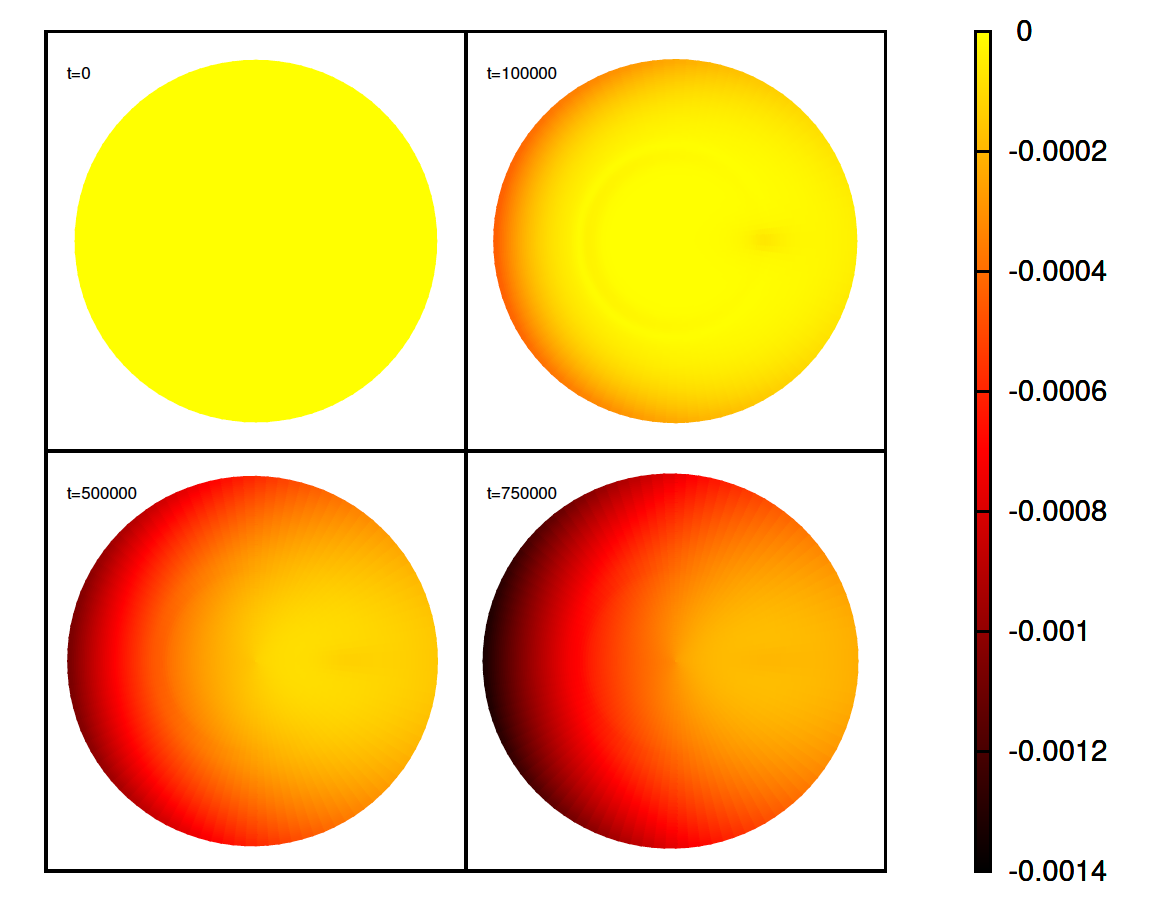}
    \caption{Time evolution of the scalar curvature for an initially flat metric coupled to an anisotropic velocity field.}\label{figureAniDiskCurvature} 
\end{figure}

\begin{figure}
	\centering
	\includegraphics[width=0.7\textwidth]{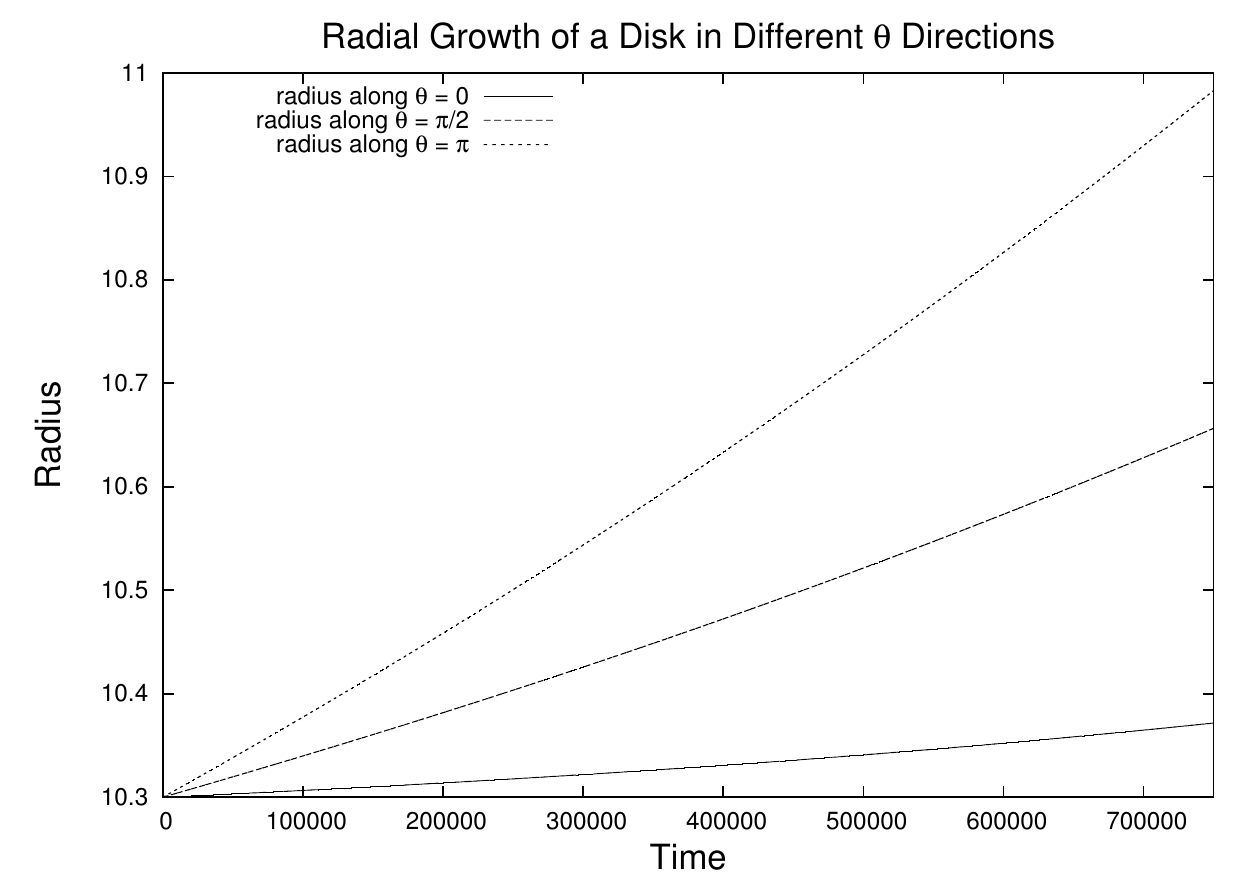}
	\caption{Radial growth for an initially flat metric coupled to an anisotropic velocity field along different $\theta$ directions.}
	\label{figureAniDiskRadius}
\end{figure}

\begin{figure}
    \centering
        \subfloat{
                \includegraphics[width=0.6\textwidth]{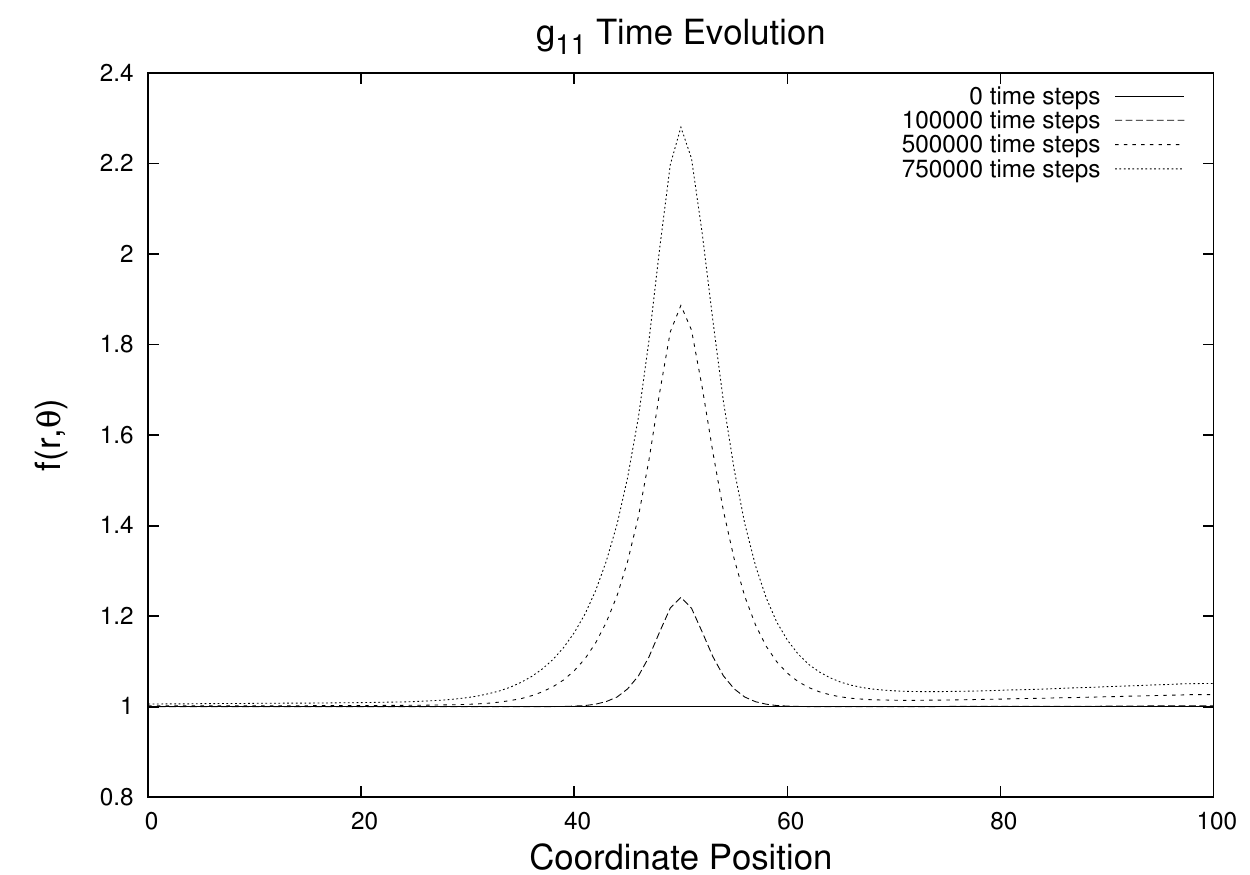}
        }

        \subfloat{
                \includegraphics[width=0.6\textwidth]{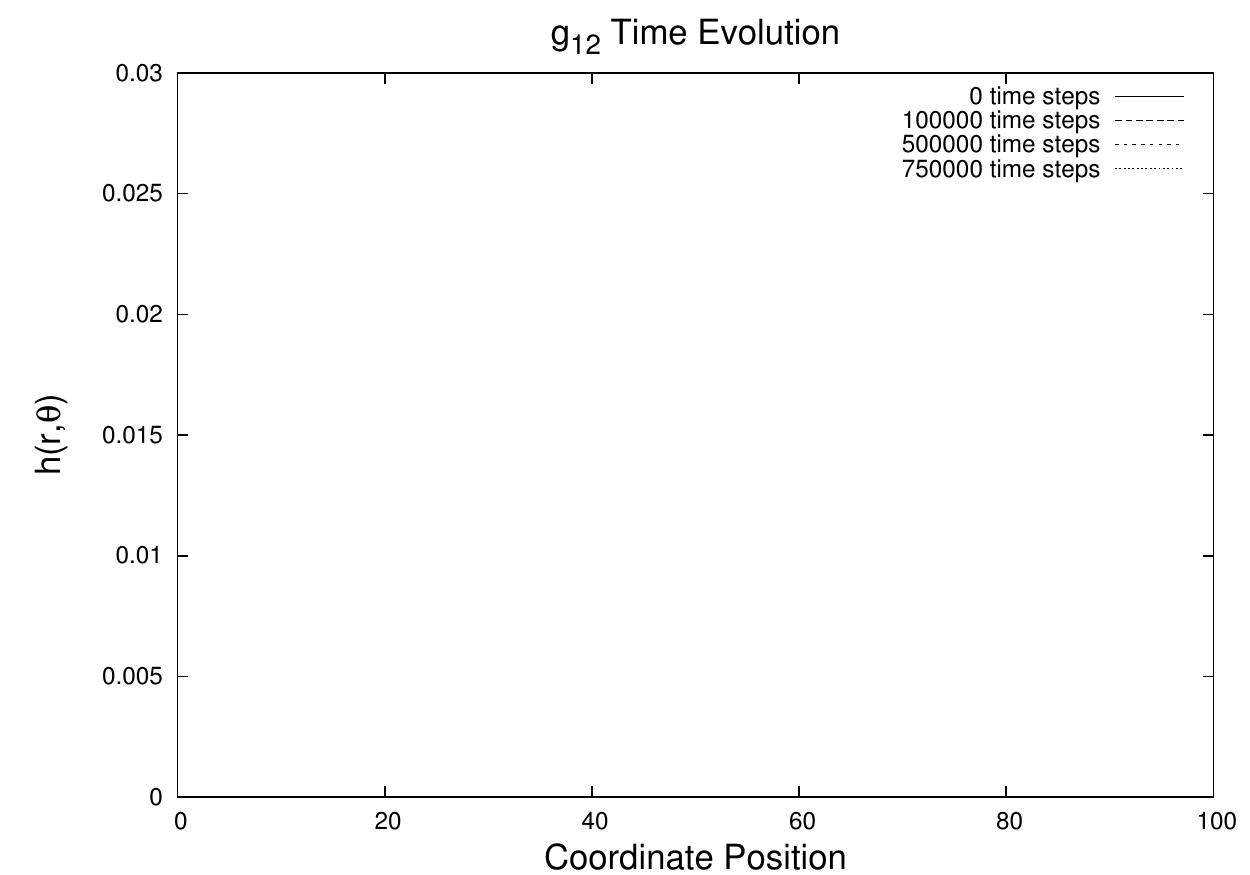}
        }

        \subfloat{
                \includegraphics[width=0.6\textwidth]{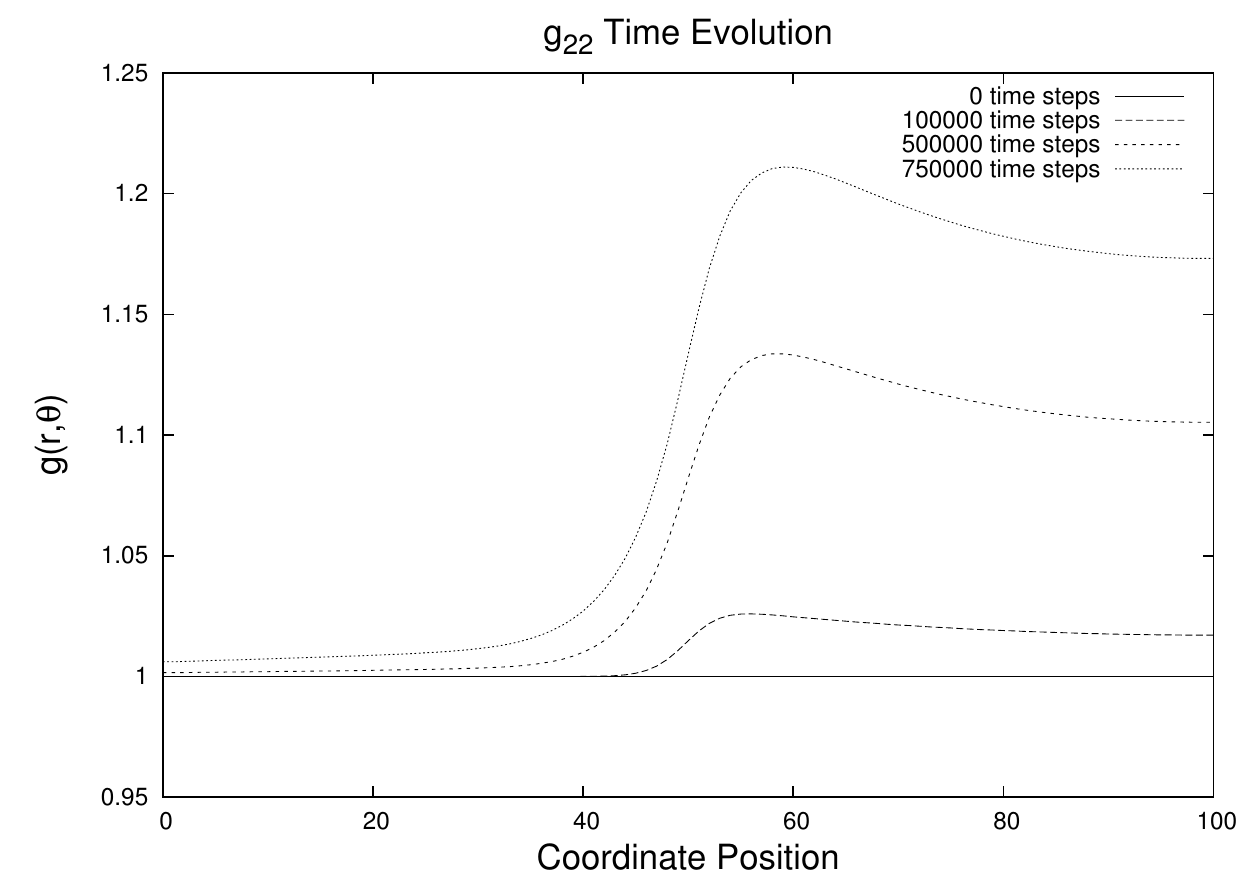}
        }
        \caption{Dynamics of the metric at $\theta=\pi$ for an initially flat metric coupled to an anisotropic velocity field.}\label{figureAniDiskMetricPi} 
\end{figure}

\begin{figure}
    \centering
        \subfloat{
                \includegraphics[width=0.6\textwidth]{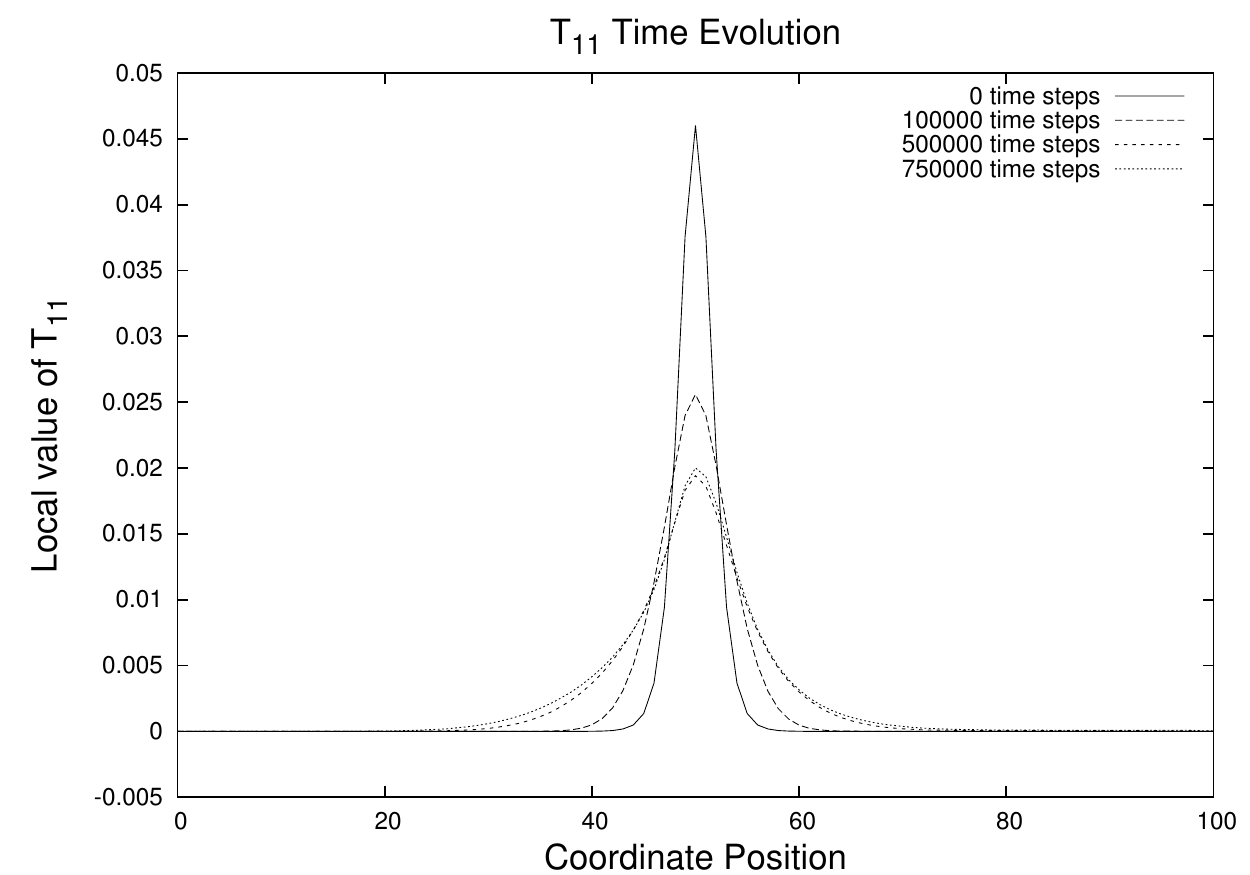}
        }

        \subfloat{
                \includegraphics[width=0.6\textwidth]{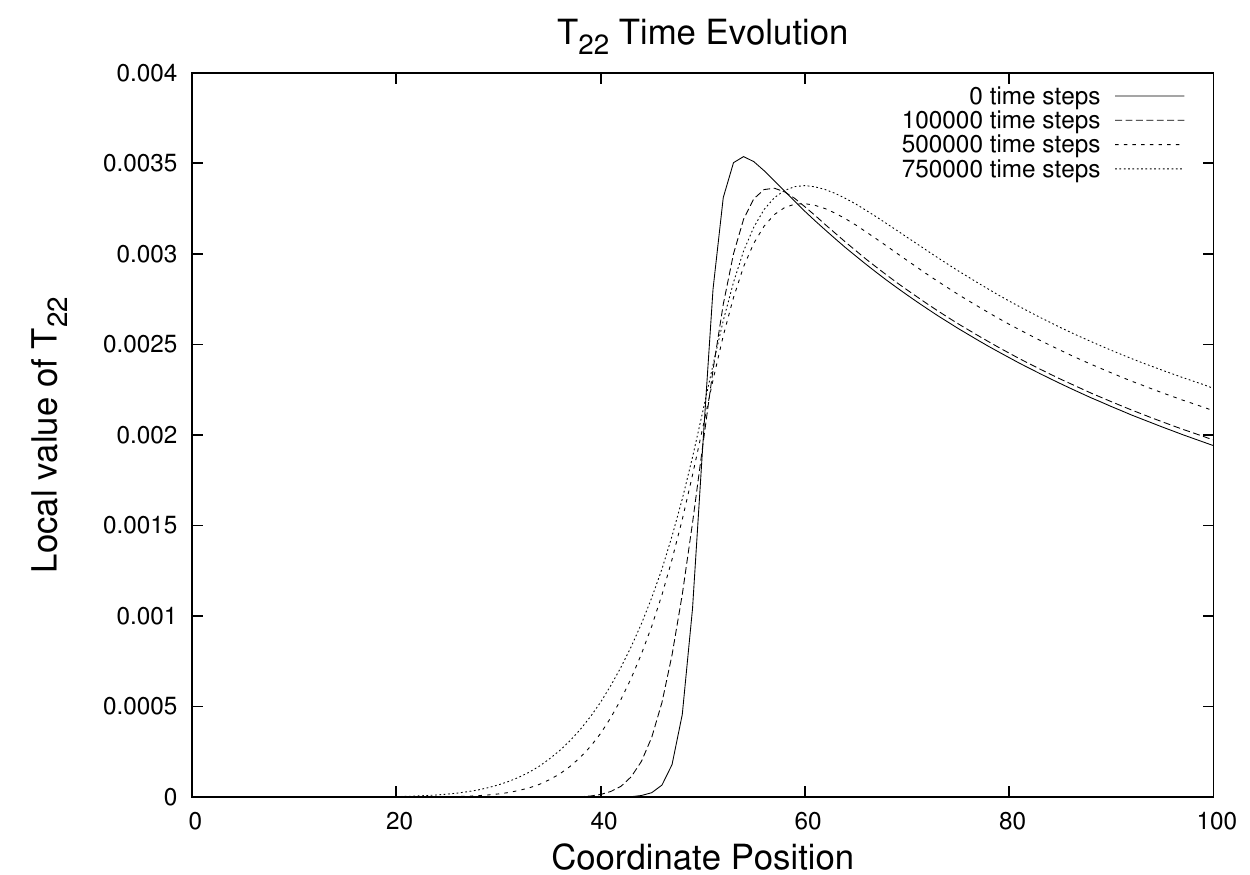}
        }

        \subfloat{
                \includegraphics[width=0.6\textwidth]{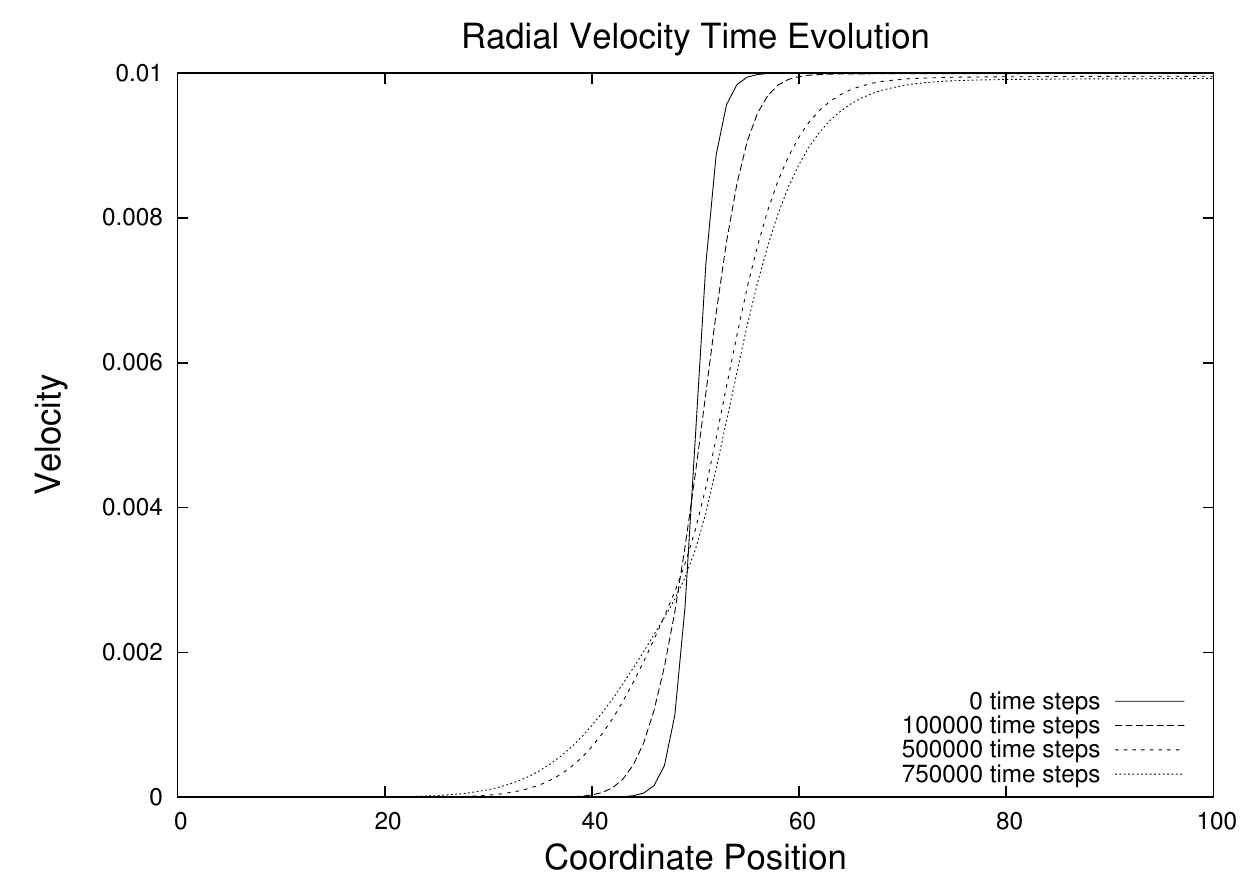}
        }
        \caption{Growth tensor components and radial velocity field at $\theta=\pi$ for an initially flat metric coupled to an anisotropic velocity field.}\label{figureAniDiskGTvelPi} 
\end{figure}

The Ricci flow simulation shows that curvature can cause both expansion and contraction of the manifold depending
on the local distribution of curvature (Figure \ref{figureRF4radius}). Globally, the curvature still dissipates over 
time, flattening the manifold (Figure \ref{figureRF4gaussCurvature}).

The expansion, rotation and shear tensors, as well as the REGR, give insight into the physical development of the
tissue. As before, the REGR gives a measure of local growth. In a general 2D manifold, the area is now

\beq
A = \sqrt{\det(g_{ik})}drd\theta = \sqrt{fg-h^2}rdrd\theta,
\eeq

\noi but the expression for REGR reduces to the same one as in the 2D isotropic case:

\beq
REGR = - \kappa R + \kappa_1 \Theta.
\label{eqn2DregrGeneral}
\eeq

\noi As before, $R$ is the scalar curvature and $\Theta=\nabla_i v^i$ is the expansion scalar.

\begin{figure}
    \centering
        \subfloat{
                \includegraphics[width=0.6\textwidth]{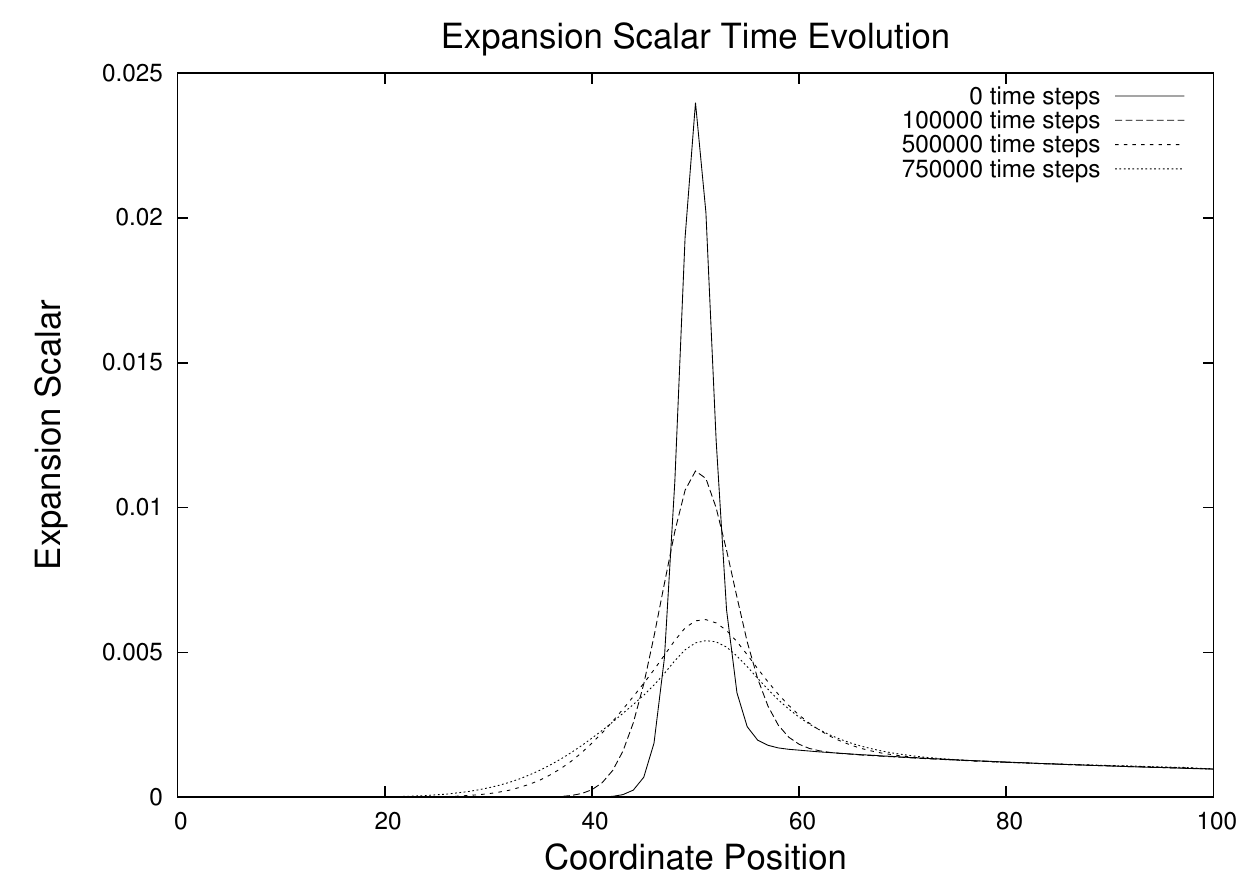}
        }

        \subfloat{
                \includegraphics[width=0.6\textwidth]{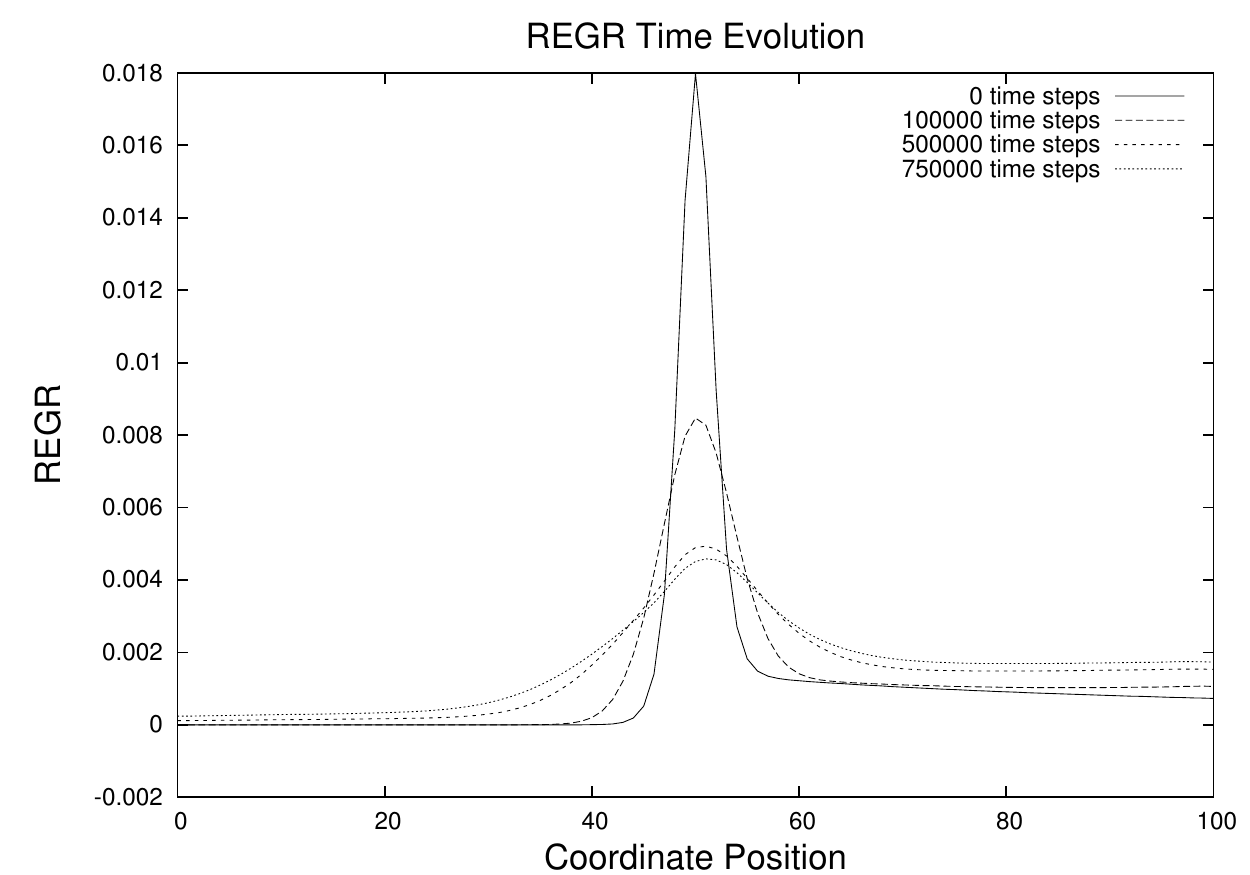}
        }
        \caption{Expansion scalar and REGR at $\theta=\pi$ for an initially flat metric coupled to an anisotropic velocity field.}\label{figureAniDiskThetaREGRpi} 
\end{figure}

As can be seen in Figures \ref{figureAniDiskThetaREGRpi} and \ref{figureAniDiskThetaREGRpi2}, growth has
contributions from both the expansion scalar and the curvature term, just as in the 2D isotropic case. 
The REGR at the outer is edge is clearly higher than at the center in both plots, indicating that a
ruffled structure is emerging. Also, the magnitude of REGR along the outer edge at $\theta=\pi$ is
roughly twice that at $\theta=\pi/2$, showing that ruffling will be more pronounced along the $\theta=\pi$ direction
and will taper off in parts of the disk that are not growing as fast. This contribution is
also clearly related to the anisotropic curvature of the disk, since the expansion scalars do not increase
at the outer edge over time.

\begin{figure}
    \centering
        \subfloat{
                \includegraphics[width=0.6\textwidth]{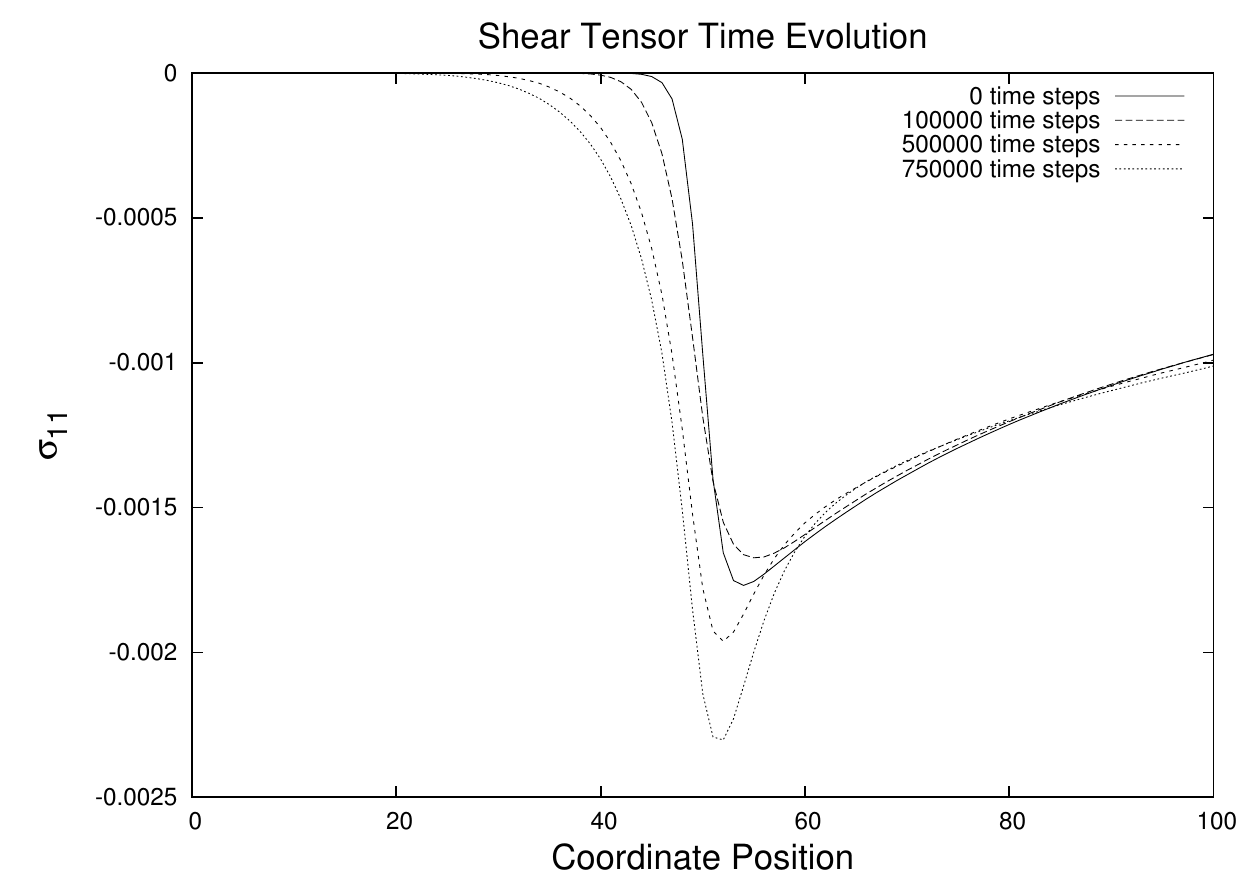}
        }

        \subfloat{
                \includegraphics[width=0.6\textwidth]{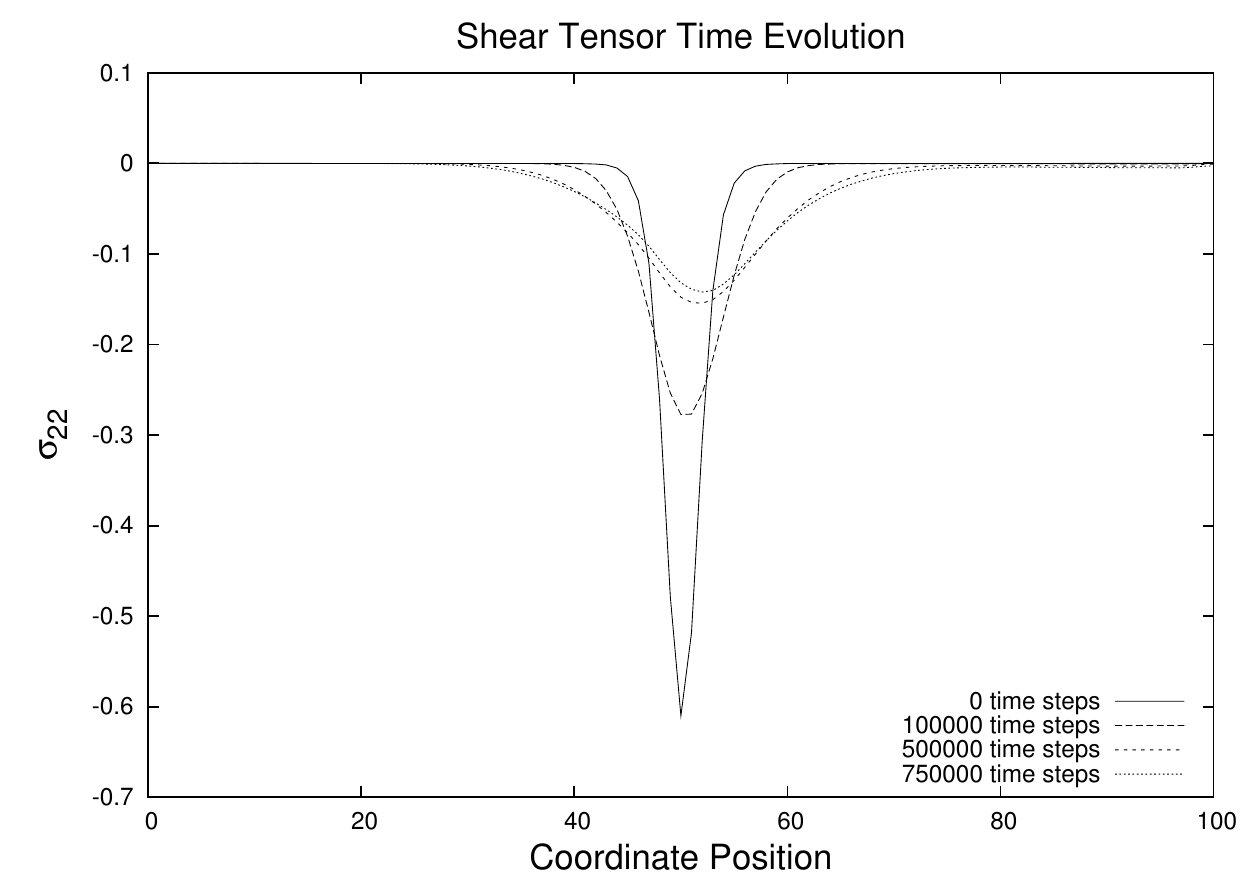}
        }

        \subfloat{
                \includegraphics[width=0.6\textwidth]{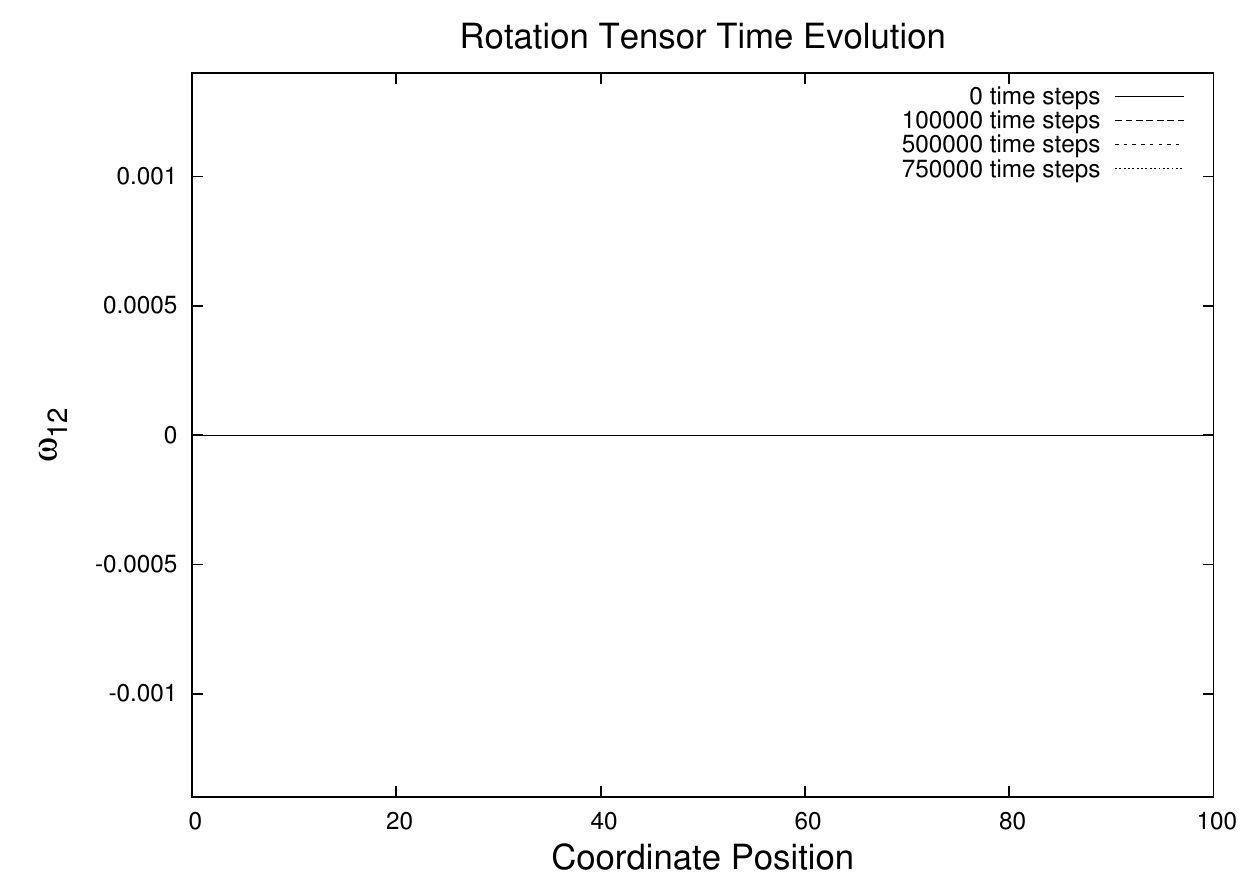}
        }
        \caption{Shear and rotation tensor components at $\theta=\pi$ for an initially flat metric coupled to an anisotropic velocity field.}\label{figureAniDiskDefTensorsPi} 
\end{figure}

\begin{figure}
    \centering
        \subfloat{
                \includegraphics[width=0.6\textwidth]{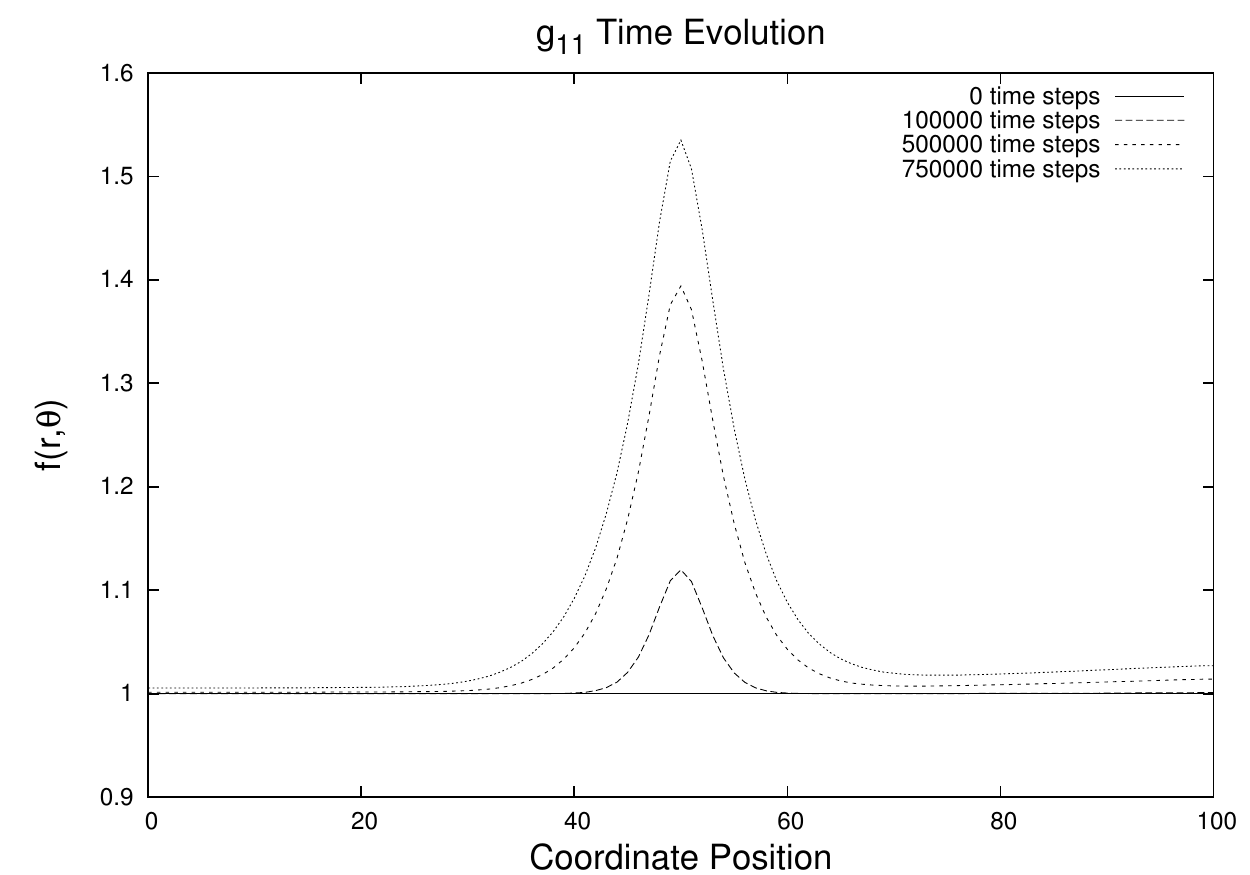}
        }

        \subfloat{
                \includegraphics[width=0.6\textwidth]{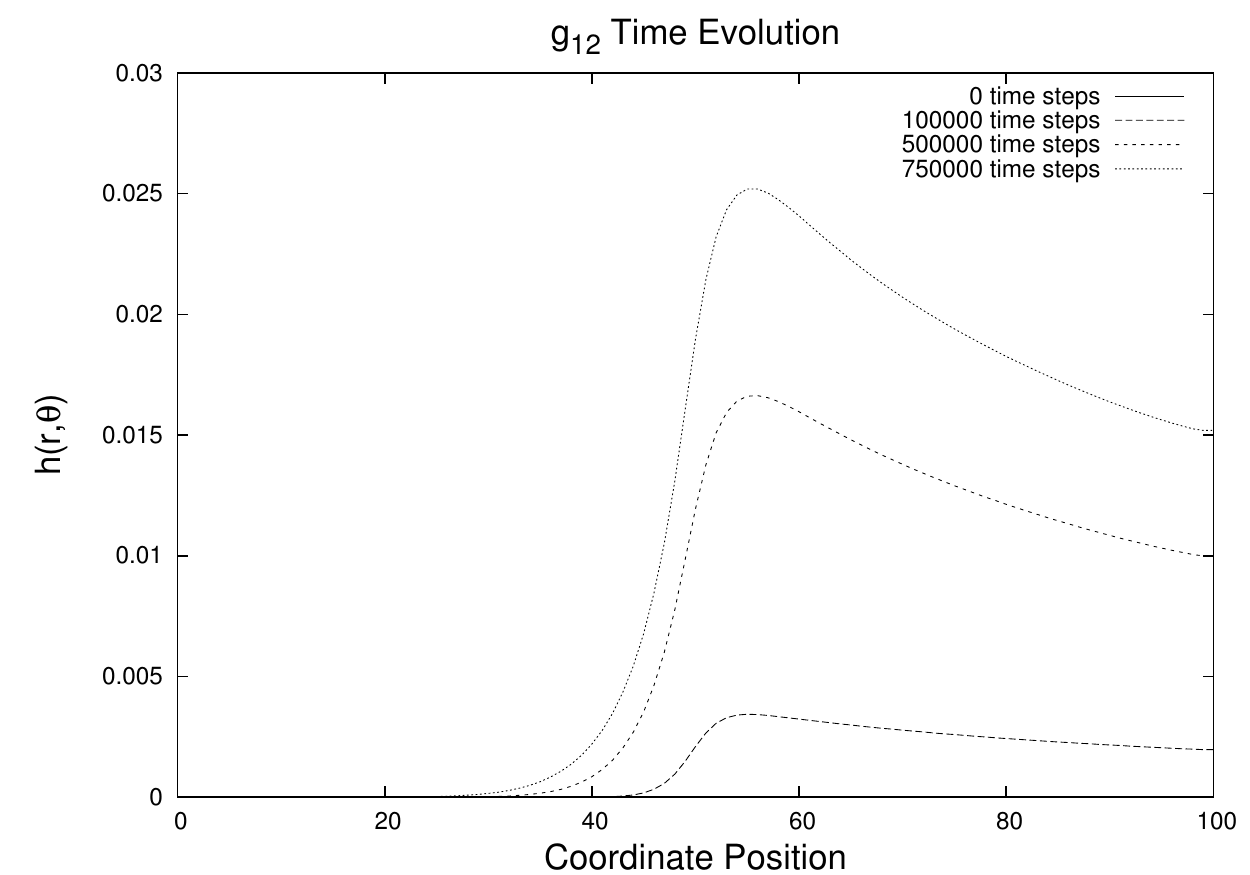}
        }

        \subfloat{
                \includegraphics[width=0.6\textwidth]{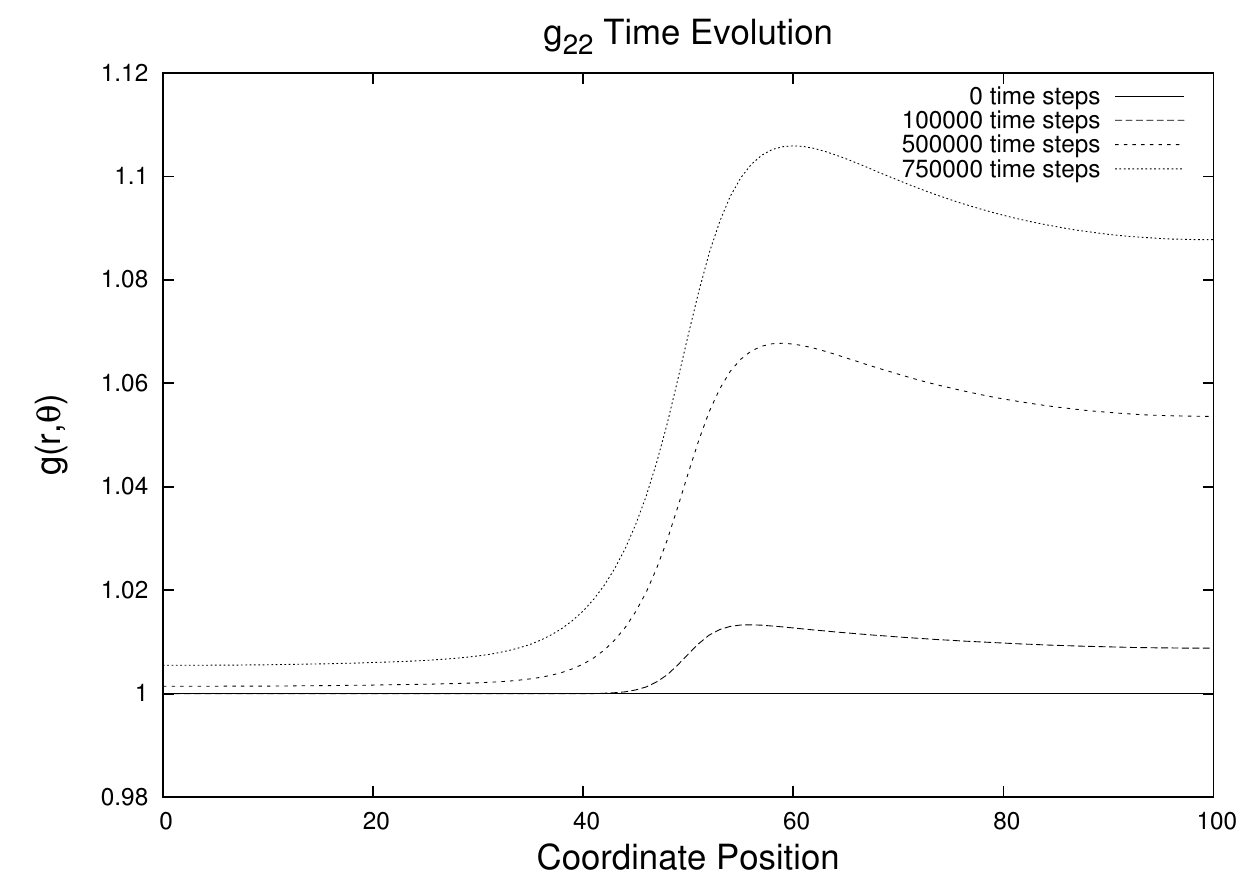}
        }
        \caption{Dynamics of the metric at $\theta=\pi/2$ for an initially flat metric coupled to an anisotropic velocity field.}\label{figureAniDiskMetricPi2} 
\end{figure}

\begin{figure}
    \centering
        \subfloat{
                \includegraphics[width=0.6\textwidth]{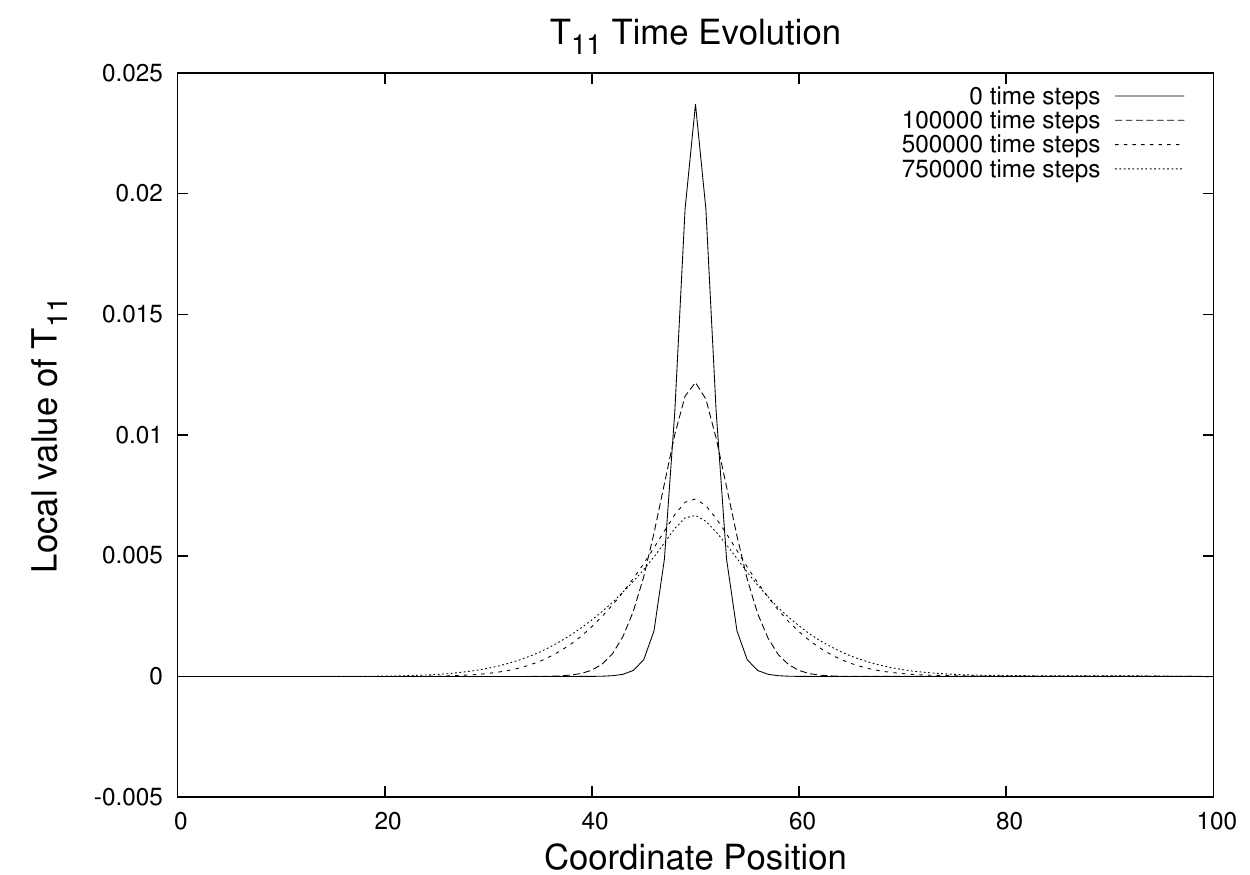}
        }

        \subfloat{
                \includegraphics[width=0.6\textwidth]{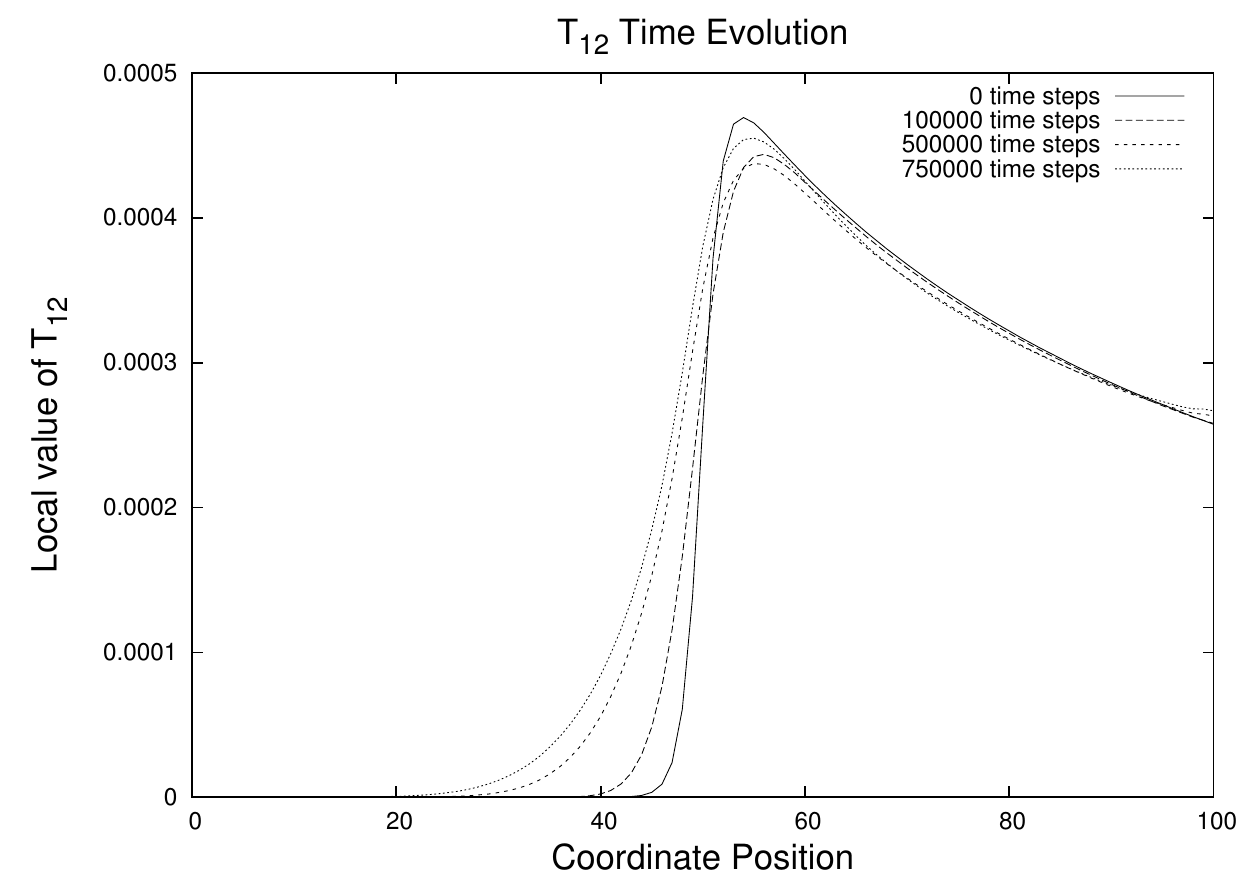}
        }

        \subfloat{
                \includegraphics[width=0.6\textwidth]{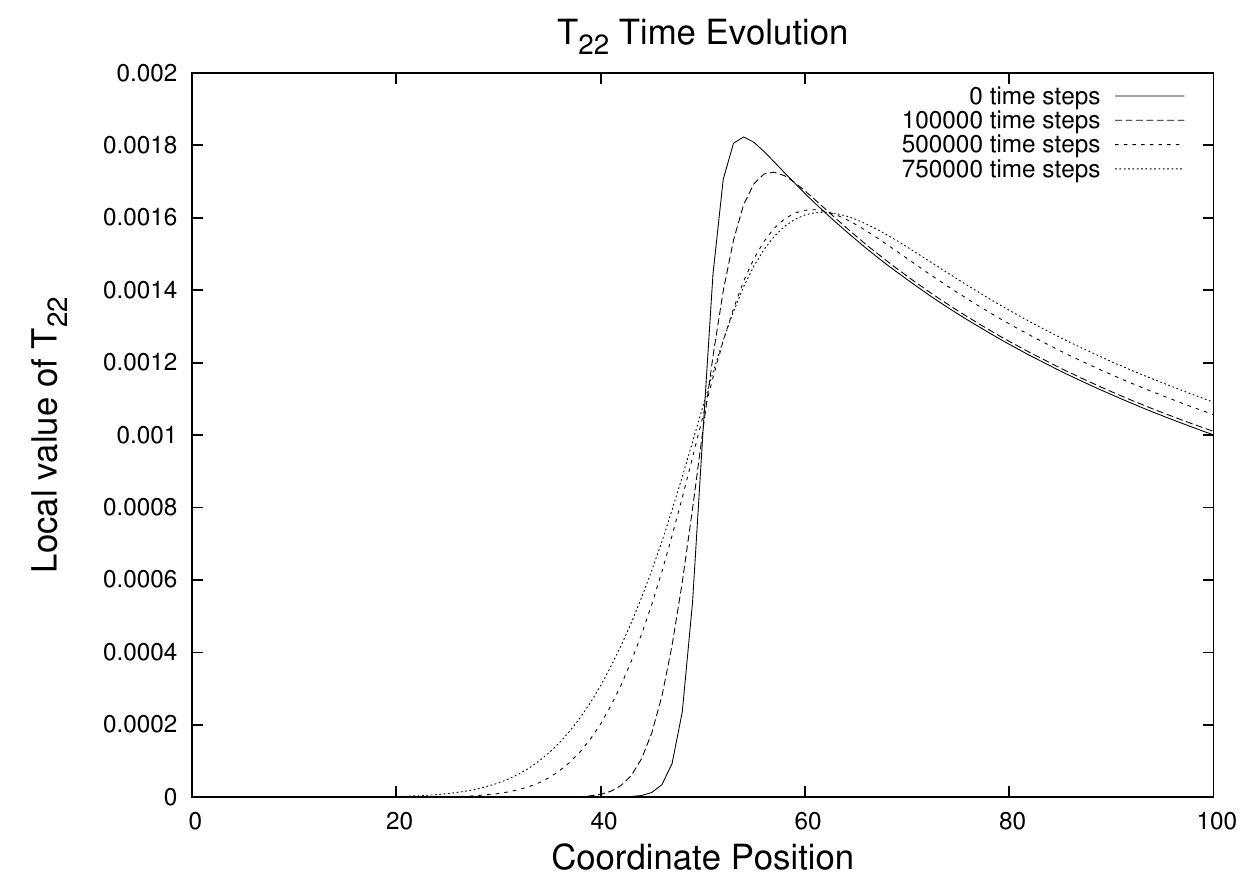}
        }
        \caption{Growth tensor components at $\theta=\pi/2$ for an initially flat metric coupled to an anisotropic velocity field.}\label{figureAniDiskGTPi2} 
\end{figure}

\begin{figure}
    \centering
        \subfloat{
                \includegraphics[width=0.6\textwidth]{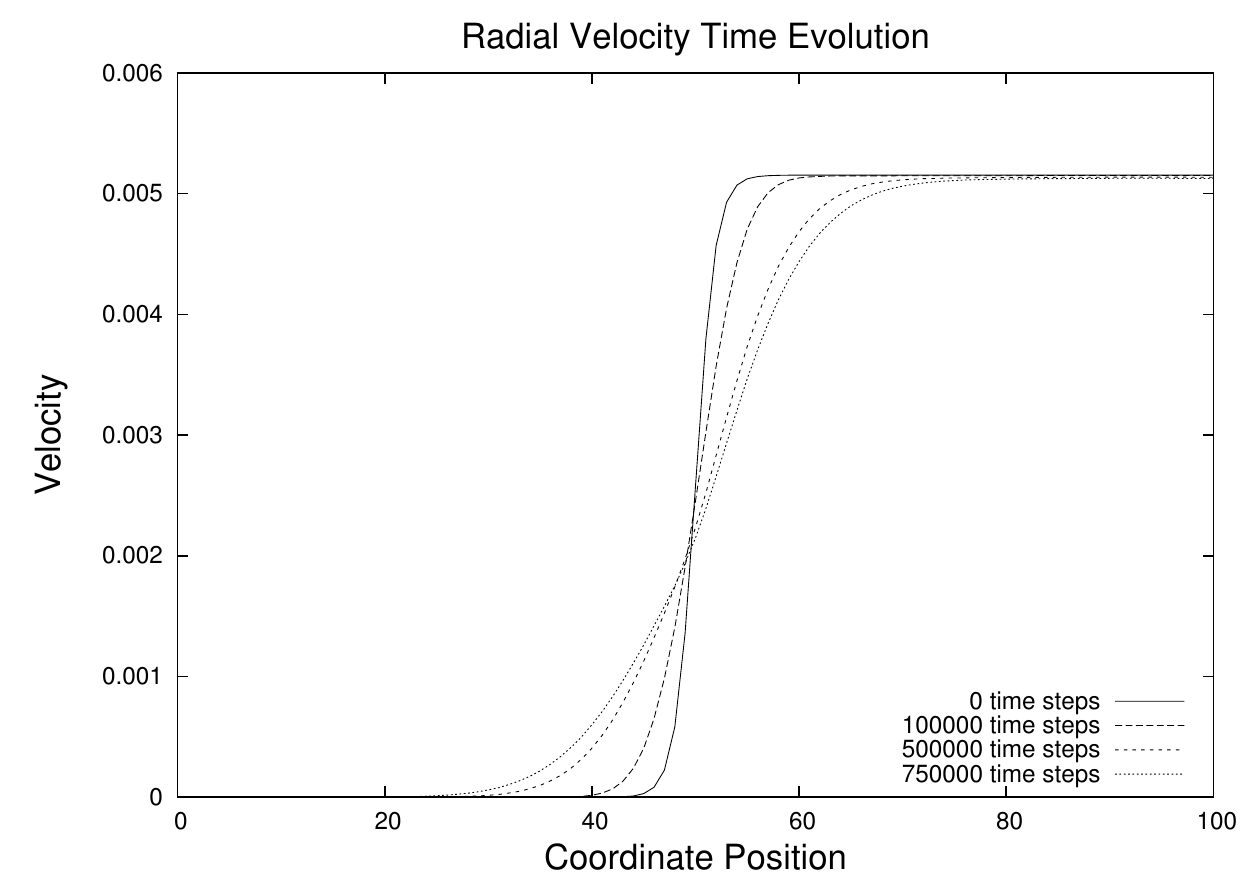}
        }

        \subfloat{
                \includegraphics[width=0.6\textwidth]{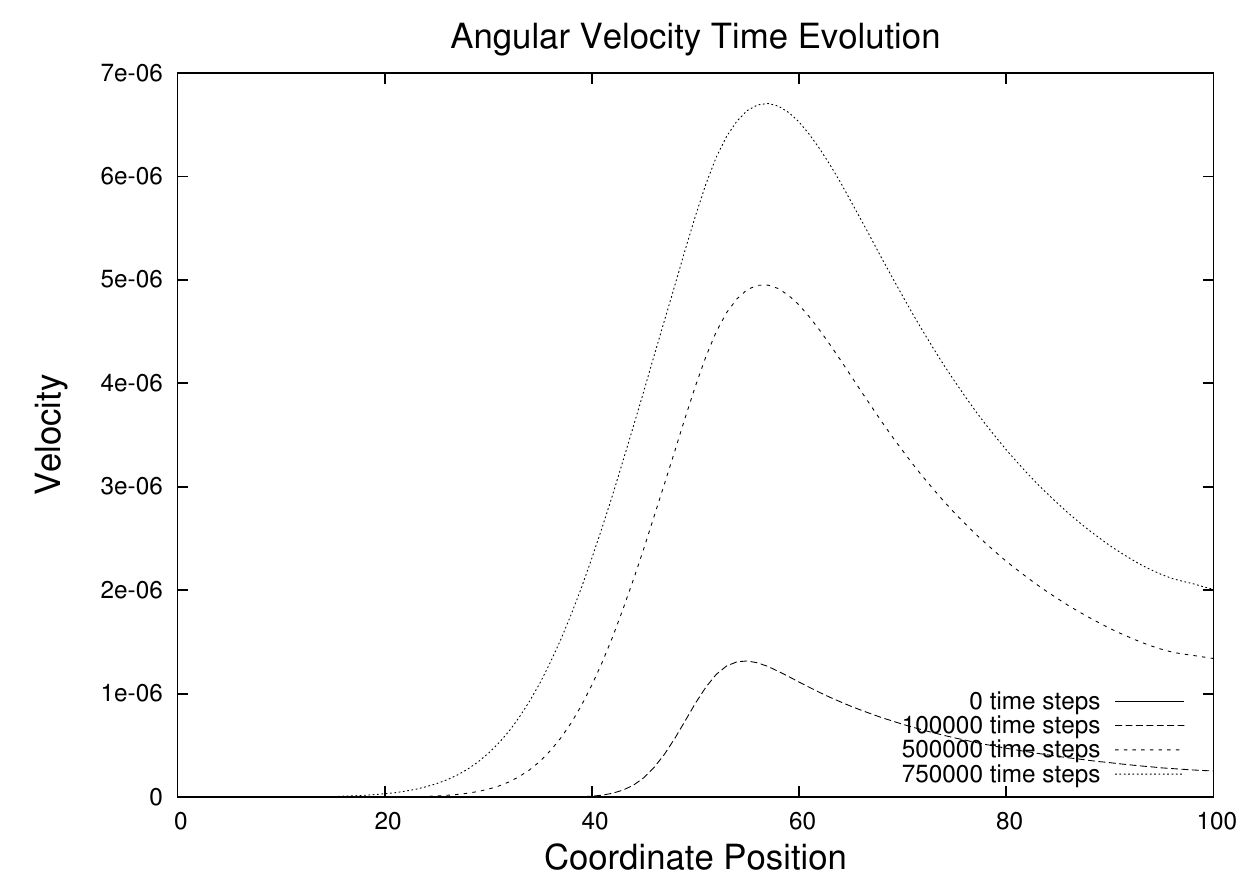}
        }
        \caption{Velocity field components at $\theta=\pi/2$ for an initially flat metric coupled to an anisotropic velocity field.}\label{figureAniDiskVelPi2} 
\end{figure}

\begin{figure}
    \centering
        \subfloat{
                \includegraphics[width=0.6\textwidth]{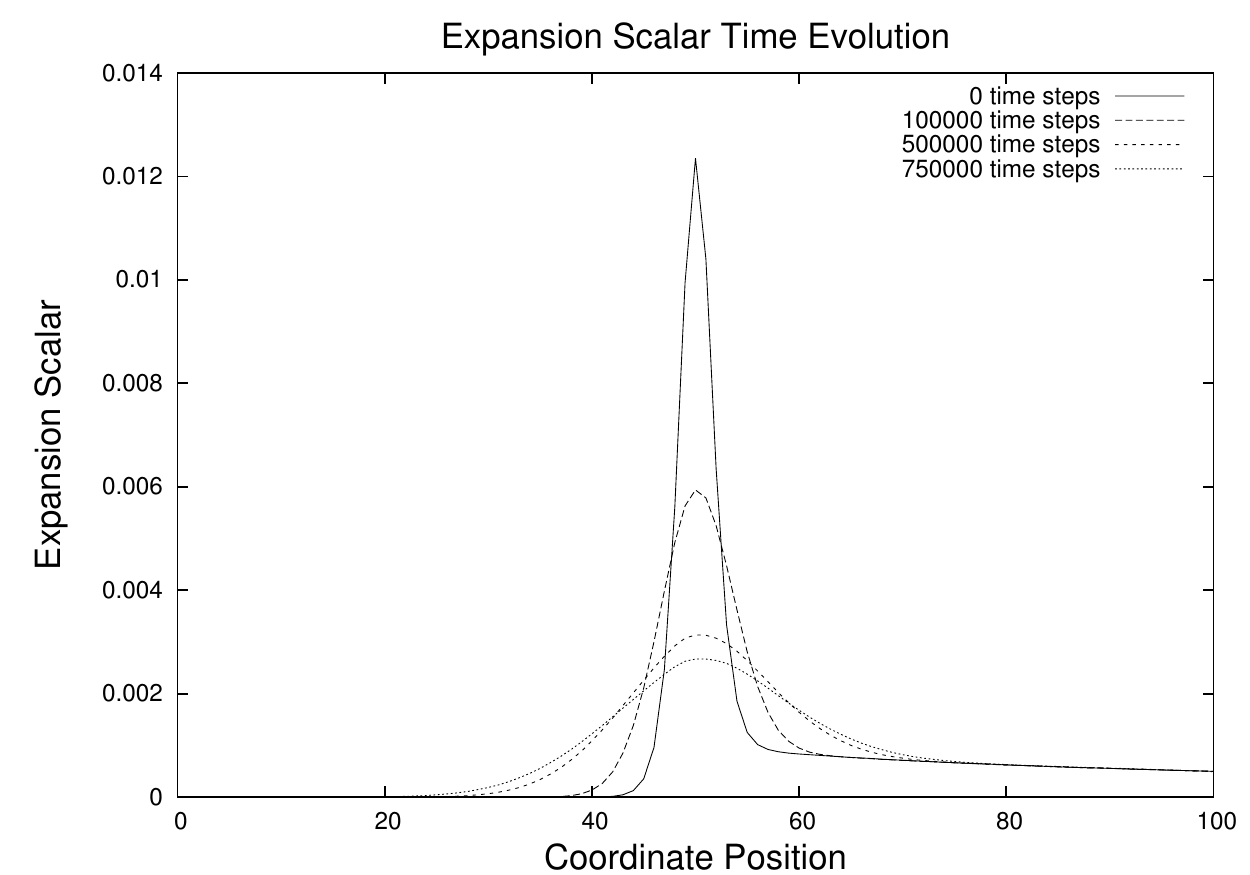}
        }

        \subfloat{
                \includegraphics[width=0.6\textwidth]{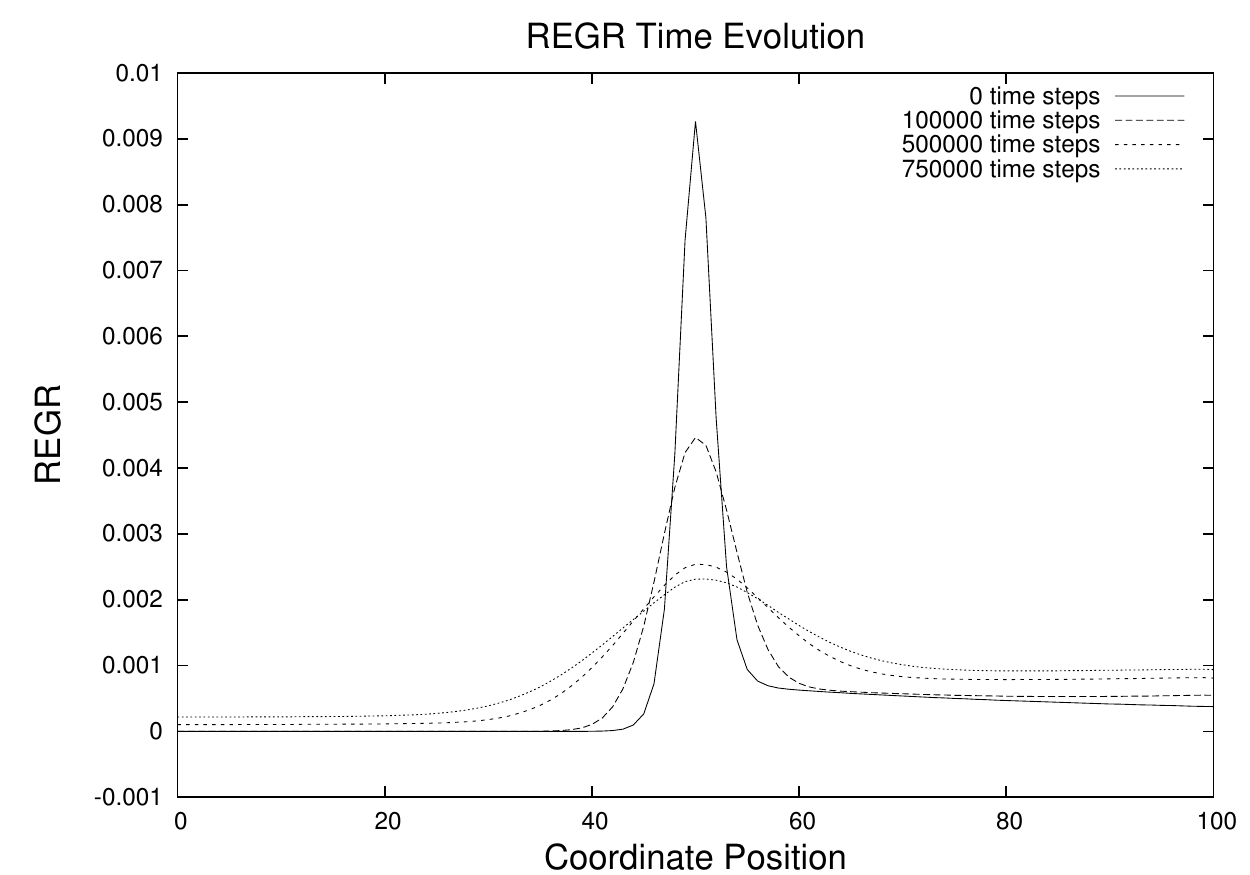}
        }
        \caption{Expansion scalar and REGR at $\theta=\pi/2$ for an initially flat metric coupled to an anisotropic velocity field.}\label{figureAniDiskThetaREGRpi2} 
\end{figure}

\begin{figure}
    \centering
        \subfloat{
                \includegraphics[width=0.6\textwidth]{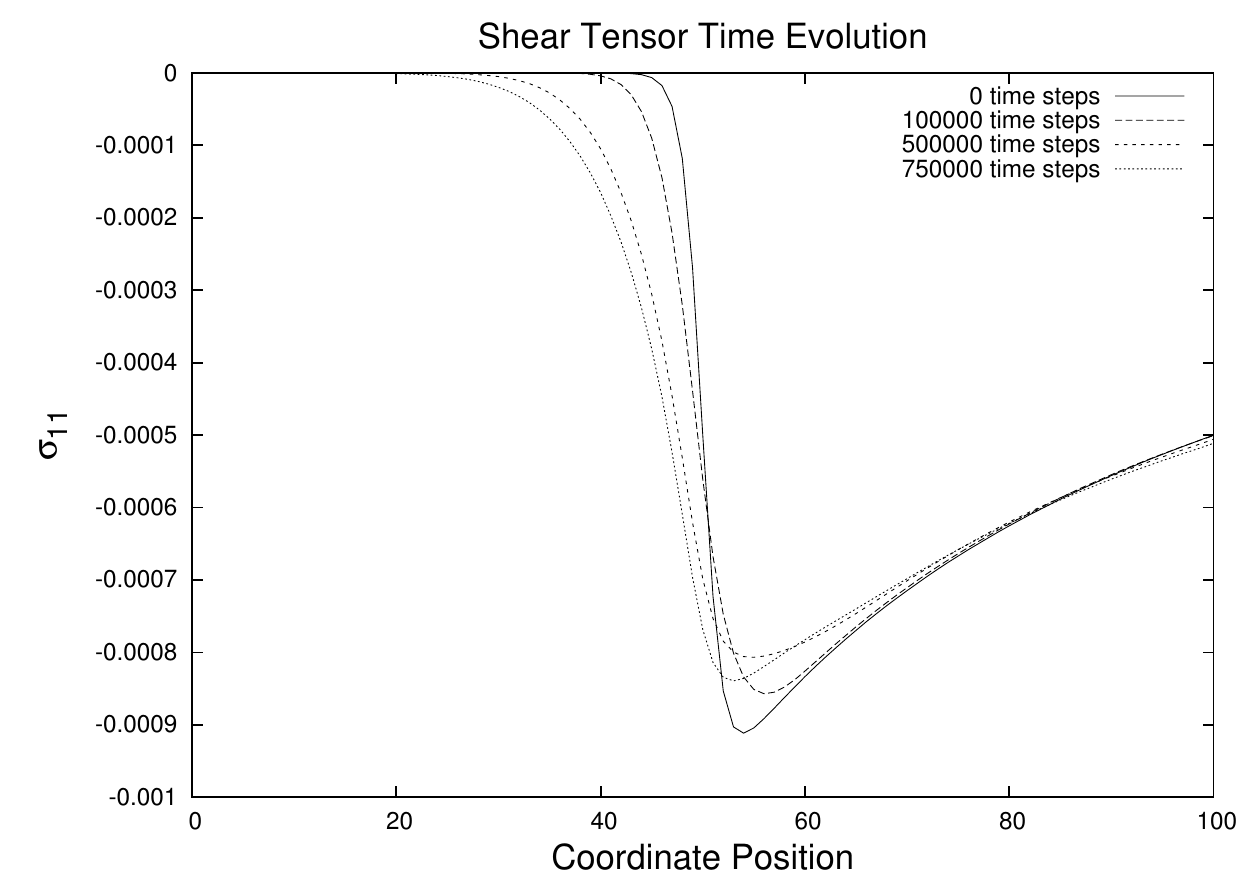}
        }

        \subfloat{
                \includegraphics[width=0.6\textwidth]{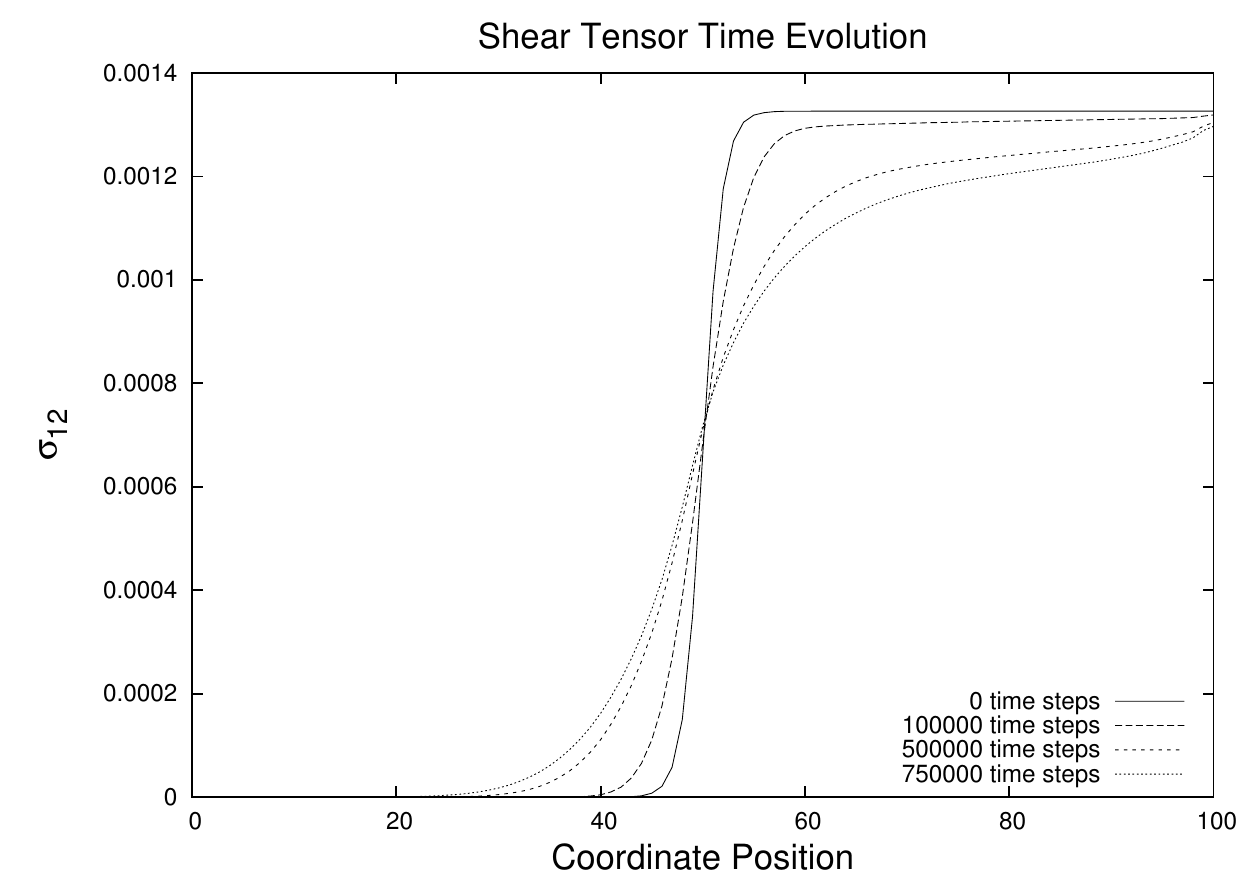}
        }

        \subfloat{
                \includegraphics[width=0.6\textwidth]{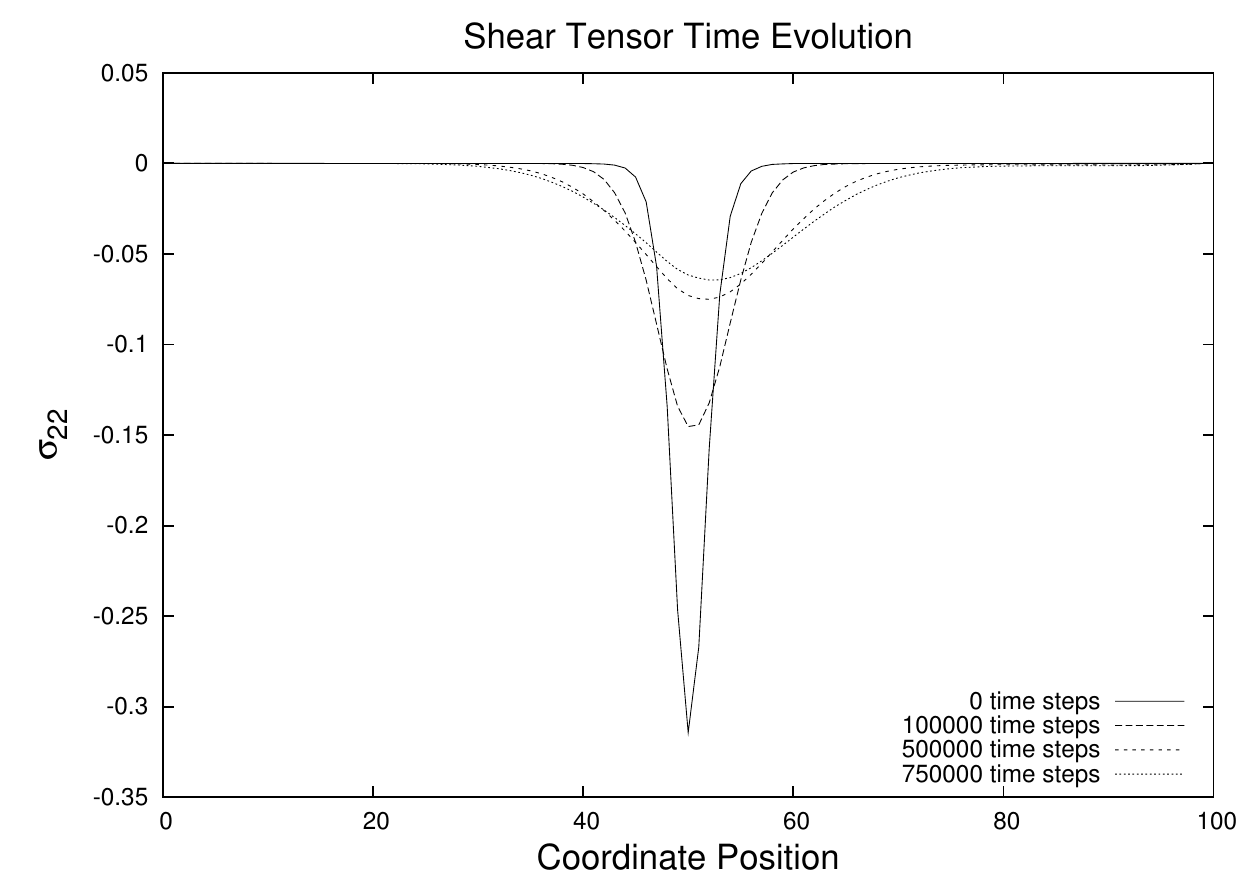}
        }
        \caption{Shear tensor components at $\theta=\pi/2$ for an initially flat metric coupled to an anisotropic velocity field.}\label{figureAniDiskShearPi2} 
\end{figure}

\begin{figure}
    \centering
    \includegraphics[width=0.6\textwidth]{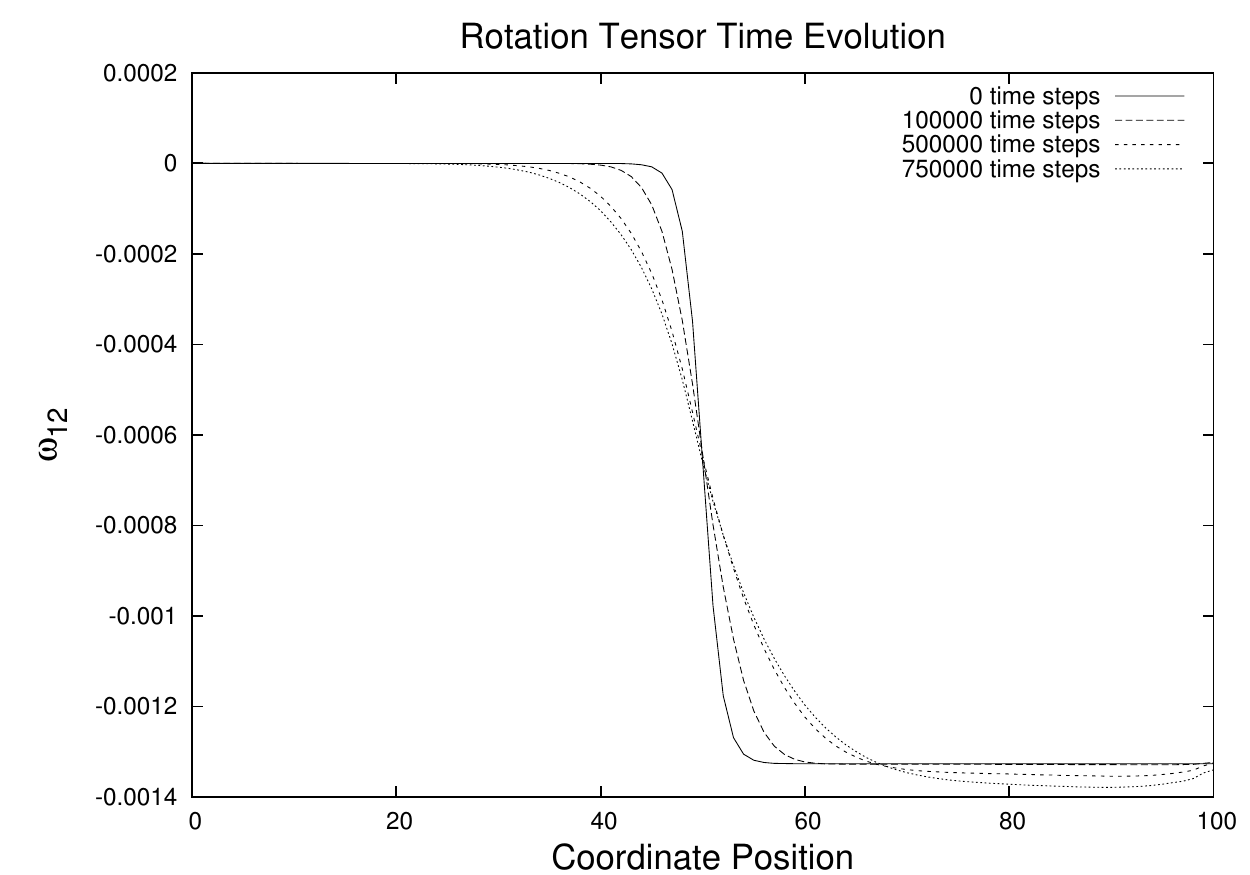}
    \caption{Rotation tensor component at $\theta=\pi/2$ for an initially flat metric coupled to an anisotropic velocity field.}\label{figureAniDiskRotationPi2} 
\end{figure}

Unlike the isotropic case in the previous chapter, rotation is now possible. This is a direct consequence of 
having an angular velocity develop over time (Figure \ref{figureAniDiskVelPi2}). Moreover, the angular velocity varies 
in both the radial and angular directions; indeed, the angular velocity and rotation tensor are null at $\theta = \pi$
but are clearly non-zero along $\theta = \pi/2$. This also implies that the shear tensor components will have more 
structure than in the isotropic case (Figures \ref{figureAniDiskDefTensorsPi} and \ref{figureAniDiskShearPi2}). 

The dynamics of the initially flat disk show that growth of the metric is driven by the gradients of the initial
velocity field, a result consistent with the simulations in 1D and 2D isotropic cases studied in previous chapters.
Overall, the deformations in the anisotropic case are richer than those of previous simulations, yet
follow the same broad patterns. Deformations are all localized near the steepest gradient of the velocity field,
and with the expansion of the metric, these effects tend to dissipate even as the manifold continues to grow. 
The expansion scalar (or equivalently in our model, the source field) has similar dynamics to the disk growth 
and \textit{Acetabularia} models, indicating that growth will be localized in a deformed annulus around the center 
of the anisotropic disk.

The results for these simulations are limited due to the less stable numerical methods used to achieve them. However
there is strong indication that, just as in the isotropic 2D simulations, growth and
curvature feed back onto each other to produce non-trivial geometries. In these models,
growth and buckling occur in regions where the velocity field and growth tensor are largest.

Similarly to the isotropic 2D simulations, a lower threshold for velocity diffusion and
Ricci flow coupling exists for a given growth tensor coupling. Using the less stable finite differencing scheme
in this version puts a strong upper limit on the growth tensor coupling. Again, dissipative terms are 
required to prevent large differences in growth rates between neighbouring regions that would otherwise
excessively strain the tissue.

\section{Conclusions on 2D simulations in anisotropic polar coordinates}

The data presented in this chapter are a prediction of how anisotropic growth may 
occur in a leaf. Different radial distances along different values of $\theta$ indicate
that the embedded shape of such geometries would be elongated along the $\theta=\pi$ axis. 
Variations in scalar curvature
suggest the tissue would be ruffled or buckled near the growing edge, but flat near the base. 
Under the influence of an anisotropic velocity
field, the simulations show that the disk becomes elongated with nontrivial curvature patterns
throughout the disk. To our knowledge, no experimental data in the plant biology or thin disk 
communities exist to verify these simulated results.


\chapter{Conclusions and Future Work}
\label{chapterConclusions}

\section{Conclusions on 1D and 2D Simulations}

The results presented for 1D and 2D simulations show that biologically relevant growth dynamics 
occur for a system where the metric, velocity field and mass density are coupled in a way that represents the 
biology and physics of macroscopic plant leaf growth. 

In the 1D case, several features of primary root growth emerge from the mathematical model. 
Amongst these is a pronounced elongation zone in the root tip, and a double peaked REGR that
emphasized the contribution of geometric growth in the model in addition to the effects of a 
logistic-type velocity field. It was also found that a minimum  
coupling to dissipation terms in the scale factor and velocity field terms is required to prevent shocks from 
occurring as neighbouring regions experience different rates of growth.

In the 2D simulations, growth occurs both due to Ricci flow and coupling to the growth tensor. 
Similarly to the 1D case, a minimum amount of coupling to the Ricci tensor and velocity diffusion 
terms is needed to counteract the effect of localized growth due to the growth tensor. 

Additionally, the 2D simulations show that for a monotonically increasing logistic velocity field, 
the tissue can transition from globally positive to globally negative curvature. 
Simulations of a flat disk and an initially positively curved surface show that with a logistic 
velocity field, the perimeter of the leaf grows outward, becomes ruffled, and has an annular 
growth zone, just as in the curvature evolution of the \textit{Acetabularia} cap. 
Comparing these results to thin disk experiments strongly suggests 
these are also physically realizable shapes. It is equally conceivable that the elongated and locally
curved disks resulting from anisotropic velocity fields are also physically realizable shapes 
that correspond to leaf shapes found in nature. 

Using a logistic velocity field as an initial condition
in the 2D models was an extension of 1D velocity field profile in roots. In the case of \textit{Acetabularia},
this velocity field profile allows the annular growth pattern to arise, and is consistent with the mathematical
finding that constant curvature metrics driven by linear velocity fields cannot change the sign of their curvature
(Section \ref{sectionAceta}). 

An important feature of growth that emerges from this model is the dual contributions to the relative elemental
growth rate (REGR) from a deposition term and an expansion term. As seen early on, the expansion term can be linked
back to a distributed source function that drives the influx of material to the tissue. The deposition term, however, 
comes from a geometric KPZ reaction-diffusion term in one dimension, and the Ricci flow in two dimensions.

The correspondence of the KPZ term with Ricci flow is striking 
since they originate in very different contexts. The KPZ term has its origins in a surface growth model, which we 
generalize to a one dimensional geometric model where growth can occur not just at the surface, but throughout the tissue. 
Ricci flow is a process that allows a highly curved geometry to evolve continuously to a smoothed geometry. For a 
two dimensional tissue, it is what allows the tissue's curvature to change over time.
The mathematical similarity between the KPZ equation and Ricci flow resides in 
quadratic gradient terms appearing as reaction-diffusion processes in both one and two dimensions. Indeed, one
can think of Ricci flow as a generalization of the KPZ equation to higher dimensions, under the constraints 
of being a tensor of the same valence and symmetry as the metric tensor. On the other hand, one can start with Ricci
flow in two dimensions and recover a geometric reaction-diffusion term for a one dimensional system that is consistent
with the assumptions of local dynamical interactions governing open, driven systems like the KPZ equations.

Characterizing the dynamics of these growth patterns is significant. Geometry is known to be correlated 
with nonhomogeneous growth processes, but an 
understanding of the dynamic feedback between curvature and growth has been lacking. On the theoretical side,
this required developing numerical techniques to study Ricci flow applied to disk geometries, then coupling 
the geometric flow to a velocity field driven by a source of material. Improving this macroscopic and
dynamic model of plant growth will require input from the 
sophisticated plant imaging techniques coming to the fore in experimental biology 
~\cite{basu2012dblREGRpeak} ~\cite{rootImagingDoublePeakedGT} ~\cite{inkjetLeaves}, allowing us to 
expand our understanding of the biophysics happening at large scales.

\section{Future Directions and Open Questions}

The geometric information calculated from the simulations of 1D and 2D growth provides 
a unique understanding of the physics of a growing plant leaf. However, this information 
is not easy to interpret and there are many questions left to answer. 

The next step in interpreting the data will be to embed the 2D data in flat 3D space. 
This will allow for a more intuitive way to visualize the curved structures resulting 
from growth. The mathematics to solve the embedding problem are not trivial, requiring 
solutions to another set of coupled nonlinear PDEs ~\cite{blackHoleEmbedding}.

Another interesting problem would be to look at the geodesics of the metric as it grows. 
Are the geodesics linked to stress patterns and venation? A related question is how geodesics 
evolve under the influence of Ricci flow and growth. For this question, using the Raychaudhuri 
equations may be an interesting way forward, not least because these are questions usually 
only posed by cosmologists when studying spacetime dynamics ~\cite{Raychaudhuri}.

For the simulations themselves, much more is left to explore. Symmetric, noncircular boundary shapes 
could be imposed, thereby simulating the shapes of budded plant species whose leaf shape 
is determined by space constraints within the bud ~\cite{buddedLeaves}. 

Other aspects of the mathematics and physics of the model remain open questions. For example, 
can the full 2D equations be efficiently evolved in a conformally flat space? Such a 
formulation of the equations would allow the use of more stable finite differencing models, 
like in the constrained 2D model. Another geometric question pertains to extending this model
to 3D tissues such as those found in animal organs, or diseases like cancer. 
The mathematics of curvature flows in three spatial dimensions would be more
complex than in the 1D and 2D models studied here, but the approach of coupling
curvature and growth would still be a valid one to study. Just as in the lower dimensional models,
nonhomogenous growth will again lead to stresses in the tissue as well as curved geometries 
that could alter cell behaviour through mechanotransduction.

Gathering more biological data will also be immensely useful for mapping out the solution 
space of the equations. Constraints on the parameters and initial conditions of the system 
that come directly from experiment would help to decide which numerical solutions are biologically 
relevant. For this, data on the velocity field evolution, macroscopic growth patterns (including 
measurements of deformation tensors and REGR in 2D tissues), and 
mechanical properties of leaf tissues would be required. Luckily, the last 5 to 10 years have
seen a resurgence in collecting high-resolution digitized data on root and leaf growth that
can be applied to many different species of plants 
~\cite{basu2012dblREGRpeak} ~\cite{rootImagingDoublePeakedGT} ~\cite{inkjetLeaves}. 

As mentioned in Chapter 4, the shape of a physical non-Euclidean disk in a 3D space is governed by 
both a target metric (i.e. pure geometry) as well as the physical properties of the material that 
forms the disk. The question then arises: what is the interplay of geometry with the material 
properties of the plant tissue as it grows? How do tissue properties evolve over the lifetime 
of the plant? Does this influence how the geometry of the leaf is 'interpreted' in flat 3D space? 
Experiments on thin disks suggest this would indeed be an important aspect of how a plant leaf 
forms its curved shape since a disk that grows in a non-uniform way must minimize its stretching 
and bending energies when its target metric is not flat ~\cite{NEPmechanics}.

The question of uniqueness of individual plants also arises, especially since the equations presented 
in this work are deterministic. In the current model, individuation could occur by selecting heterogeneous
initial conditions or coupling constants. Since the system is nonlinear, it is possible that 
this would be one way to generate significant variations in the final shapes of the plant leaves. 
Another method to introduce variability in the simulations would be to add a stochastic noise term 
to the dynamical equations themselves. This approach is taken by Kardar, Parisi and Zhang ~\cite{kardar} 
in their model of crystal growth. Again, the nonlinearity of the equations combined with a noise
term could allow the system to more fully explore its solution space, allowing it to choose among
several energetically equivalent, yet morphologically distinct configurations.

Lastly, readers with a plant biology background may ask, \textit{where is the auxin?} Or for that matter,
any of a number of biochemical processes that enable plant growth. In plant systems, 
auxin is a hormone that causes growth wherever it is present, and indeed, this phenomenological model 
does not explicitly address where growth comes from. However, the velocity field, which largely dictates 
the form of the growth tensor, is an initial condition that has a profound influence on where growth 
occurs in the system. Additionally, the coupling to the growth tensor is a free parameter in this model, 
which also affects how receptive the system is to growth; in a more complicated model, this coupling 
constant could instead be a function of space and time as well. Hence, it is possible to incorporate 
the effects of biochemical and genetic processes into this phenomenological model through the 
growth tensor and its coupling strength.

Indeed, plant growth operates on multiple scales, from genetic encoding to tissue level mechanics. As
the authors who studied \textit{Acetabularia} caps state: 'In sum, morphogenesis in \textit{A. acetabulum} may result partly from
biophysical constraints arising indirectly from the general growth process, rather than from the action of genes
that code for specific shapes' ~\cite{acetabularia}. Integrating the levels of complexity in biological growth remains a vibrant field
of study that will require expertise from many disciplines of science over the years to come.

\appendix
\chapter{Discretization of the Dynamical Equations}
\label{AppendixDiscretization}

In this section, the details of how the dynamical equations used in the models of plant
growth become discretized are discussed. 

In general, we begin with identifying the diffusive terms in order to apply
the Dufort-Frankel method. The remaining differential terms are discretized using spatial central differencing
on a single time step.

\section{1D discretization}

In the 1D case, we start with two equations, one for the scale factor time evolution $g_{11} = f$ and the other
for the velocity field time evolution $v^i = v^1 = v$. These are the same as Equations \ref{eqn1DScaleFactorPDE} and 
\ref{eqn1DVelFieldPDE} in Chapter \ref{chapter1DSimulations}.

\begin{eqnarray}
\frac{\partial f}{\partial t} & = & \frac{\kappa}{2f} \left[ \frac{\partial^2 f}{\partial x^2} - \frac{1}{f} \left(\frac{\partial f}{\partial x}\right)^2 \right] + \kappa_1 \left( 2 f \frac{\partial v}{\partial x} + v \frac{\partial f}{\partial x} \right)  \nonumber \\
\frac{\partial v}{\partial t} & = & -\frac{1}{2f} \left(\frac{\partial f}{\partial x}\right) v^2 - v \left[ \frac{\partial v}{\partial x} + \frac{1}{2f} \left(\frac{\partial f}{\partial x} \right) v \right] \\
& &  + c \left[ \frac{1}{f} \frac{\partial^2 v}{\partial x^2} + \frac{1}{2f^2} \left( \frac{\partial^2 f}{\partial x^2} v + \frac{\partial f}{\partial x} \frac{\partial v}{\partial x} - \frac{1}{f} \left( \frac{\partial f}{\partial x} \right)^2 v \right) \right] . 
\end{eqnarray}

Both equations have a diffusive term as well as a series of nonlinear differential terms $N$ which can
be summarized in the general form:

\begin{equation}
\frac{\partial u}{\partial t} = D \frac{\partial^2 u}{\partial x^2} + N \left( u, \frac{\partial u}{\partial x}, w, \frac{\partial w}{\partial x}\right).
\label{eqnGeneralForm}
\end{equation}

In order to discretize this type of PDE, we take the following approach:

\begin{itemize}
\item Apply the Dufort-Frankel discretization to the second-order spatial derivative.
\item Apply a central differencing to the first-order time derivatives.
\item Apply a central differencing to any first-order spatial derivatives.
\item Evaluate functional values at $(x=j \Delta x, t=n \Delta t$).
\end{itemize}

For the scale factor, the discretized function will be:

\begin{eqnarray}
\frac{ f_j^{n+1}-f_j^{n-1}}{2 \Delta t} & \sim & \frac{\kappa}{2f_j^n} \left( \frac{f_{j+1}^n - f_j^{n+1} - f_j^{n-1} + f_{j-1}^n}{(\Delta x)^2} \right) - \frac{\kappa}{2(f_j^n)^2} \left(\frac{ f_{j+1}^{n} -  f_{j-1}^{n}}{2 \Delta x} \right)^2  \nonumber \\
      & & + \kappa_1 \left( 2 f_j^n \frac{ v_{j+1}^{n} -  v_{j-1}^{n}}{2 \Delta x} + v_j^n \frac{ f_{j+1}^{n} -  f_{j-1}^{n}}{2 \Delta x} \right).
\end{eqnarray}

A similar discretization of the velocity equation can also be done. Straightforward algebra then allows for isolating $f_j^{n+1}$ and $v_j^{n+1}$, 
which are the values of the scale factor and velocity field at the future time step $t=(n+1)\Delta t$ and spatial step $x=j\Delta x$.

When implementing this algorithm, the discretized functional values of $f$ and $v$ can be stored in arrays and a sequential calculation of
$f_j^{n+1}$ and $v_j^{n+1}$ is then performed using the existing functional values at $t=n\Delta t$ and $t=(n-1)\Delta t$.

\section{2D isotropic discretization}

For the 2D isotropic equations, the discretization method is entirely analogous to the one presented above for 1D systems. 

The equations to solve are now: 

\begin{eqnarray}
\frac{\partial g_{11}}{\partial t} \rightarrow \frac{\partial f}{\partial t} & = & - \kappa Rf + cT_{11}  \nonumber \\[10pt]
\frac{\partial g_{22}}{\partial t} \rightarrow \frac{\partial g}{\partial t} & = & - \kappa Rg + c\frac{T_{22}}{r^2}  \nonumber \\
\frac{\partial v^1}{\partial t} & = & -\Gamma_{11}^1 v^1 v^1 - 2 \Gamma_{12}^1 v^1 v^2 - \Gamma_{22}^1 v^2 v^2 - v^1 \nabla_1 v^1 - v^2 \nabla_2 v^1 + c(g^{11} \nabla_1 \nabla_1 v^1 + g^{22} \nabla_2 \nabla_2 v^1 ) \nonumber \\
           & = & -\Gamma_{11}^1 v^1 v^1 - v^1 [v_{,r}^1 + \Gamma_{11}^1 v^1] + c g^{11} \left( \frac{\partial^2 v^1}{\partial r^2} + \Gamma_{11,r}^1 v^1 + \Gamma_{11}^1 v_{,r}^1 \right)  \nonumber \\ 
           &   & + c g^{22} \left( - \Gamma_{22}^1 v_{,r}^1 + \Gamma_{22}^1 (\Gamma_{12}^2 - \Gamma_{11}^1) v^1 \right)  \nonumber \\
\end{eqnarray}

\noi where the scalar curvature $R$ is given by:

\begin{equation}
R = - \frac{1}{4d^2} \left( 2g_{,rr}fg - f(g_{,r})^2 - f_{,r}gg_{,r} + \frac{1}{r} (4fgg_{,r} - 2f_{,r}g^2) \right) \label{eqnScalarCurvatureConstrainedAppendix}
\end{equation}

\noi As before, $g_{,r} = \partial g/ \partial r$ and  $g_{,rr} = \partial^2 g/ \partial r^2$.

There are more terms than before, but the equations are still of the form \ref{eqnGeneralForm}, so we apply the same algorithm as for 
the 1D equations. In fact, the evolution equation for 
$g_{11}$ is no longer diffusive, since $R$ contains second spatial derivatives of $g_{22}$ only.

\section{2D anisotropic discretization}

The full 2D equation set as presented in Chapter \ref{chapter2DFullSimulations} require a slightly modified discretization method
because of the mixed second derivatives found in the Riemann tensor:

\begin{eqnarray}
R_{1212} & = & \frac{1}{4d} [ (2g_{,r}hh_{,r} - 2g_{,rr}d - f(g_{,r})^2 - f_{,r}gg_{,r})r^2 \nonumber \\
&  & + ((-4hh_{,\theta} + 4gh + 2fg_{,\theta})h_{,r} + 4dh_{,r\theta} + 2f_{,r}gh_{,\theta} - 6g_{,r}h^2 \nonumber \\
& & + (f_{,\theta}g_{,r} - f_{,r}g_{,\theta})h + 4fgg_{,r} - 2f_{,r}g^2)r \nonumber \\
&  & + (2f_{,\theta}h - 4fg)h_{,\theta} - 2df_{,\theta\theta} + (2fg_{,\theta} + 2f_{,\theta}g)h - ff_{,\theta}g_{,\theta} - f_{,\theta}^2g ]. \nonumber \\
\end{eqnarray} 

Additionally, there are significantly more terms to calculate in this set of equations. To make the numerical simulation 
easier to implement, and therefore less prone to programmer error, we use a time-averaged Laplacian in place of the Dufort-Frankel
discretization as discussed in Chapter \ref{chapterNumericalMethods}:

\begin{equation}
\frac{ u_j^{n+1}-u_j^{n-1}}{2 \Delta t} = D \frac{1}{2} \left( \frac{u_{j+1}^n - 2u_j^{n} + u_{j-1}^n}{(\Delta x)^2} + \frac{u_{j+1}^{n-1} - 2u_j^{n-1} + u_{j-1}^{n-1}}{(\Delta x)^2}  \right).
\end{equation}

Aside from this change, the algorithm to discretize the equations is the same as before. The goal is to algebraically isolate
the discretized differential equations for the future time step values $f_j^{n+1}$, $h_j^{n+1}$, $g_j^{n+1}$ representing the
metric components, and similarly for the two velocity components $(v^1)_j^{n+1}$ and $(v^2)_j^{n+1}$.

Another complicating factor in the full 2D simulations is a second spatial dimension, the angle $\theta$, that must be discretized.
In implementing this numerical simulation, we must therefore use a 2D array for each functional value, as well as for any calculated
values such as connection coefficients and the curvature tensor. The use of 2D arrays instead of 1D arrays as before significantly 
increases computing time for these simulations.

\chapter{Supplementary numerical simulations of 2D anisotropic growth}
\label{AppendixMore2Ddata}

\section{The interaction of an initially anisotropic metric with a circularly symmetric velocity field}

Anisotropic growth can be studied by introducing a circularly symmetric
velocity field to the anisotropic initial metrics studied in Chapter \ref{chapter2DFullSimulations}. 

For a $2$-Gaussian initial metric:
\vspace{3mm}
\begin{center}
\begin{tabular}{r l}
\hline
Ricci flow coupling & $\kappa = 0.75$ \\
Growth tensor coupling & $\kappa_1 = 0.75$ \\
Velocity diffusion coupling & $c = 0.01$ \\
\hline
\end{tabular}
\end{center}
\vspace{3mm}

\begin{center}
\begin{tabular}{r c l}
\hline
$f(t=0,r)$ & = & $5.0\exp(-2.0(r-\frac{1}{2}r_{max})^2) \times [\exp(-2.0(\theta-\frac{1}{4}\theta_{max})^2) $ \\ 
			& & $+ \exp(-2.0(\theta-\frac{3}{4}\theta_{max})^2)] + 2.0$ \\[10pt]
$g(t=0,r)$ & = & $2.0\exp(-2.0(r-\frac{1}{2}r_{max})^2) \times [\exp(-2.0(\theta-\frac{1}{4}\theta_{max})^2) $ \\ 
			& & $+ \exp(-2.0(\theta-\frac{3}{4}\theta_{max})^2)] + 2.0$ \\[10pt]
$h(t=0,r)$ & = & $0.5\exp(-2.0(r-\frac{1}{2}r_{max})^2) \times [0.5\exp(-2.0(\theta-\frac{1}{4}\theta_{max})^2) $ \\ 
			& & $+ 0.5\exp(-2.0(\theta-\frac{3}{4}\theta_{max})^2)] + 1.0$ \\[10pt]
$v^1(t=0,r)$ & = & $0.01/[1.0 + \exp(-5.0(r-\frac{1}{2}r_{max}))] - 0.01/[1.0 + \exp(\frac{5}{2}r_{max})]$ \\[10pt]
$v^2(t=0,r)$ & = & $0.0$ \\ 
\hline 
\end{tabular}
\end{center}
\vspace{3mm}

For a $4$-Gaussian initial metric:

\vspace{3mm}
\begin{center}
\begin{tabular}{r l}
\hline
Ricci flow coupling & $\kappa = 0.75$ \\
Growth tensor coupling & $\kappa_1 = 0.75$ \\
Velocity diffusion coupling & $c = 0.01$ \\
\hline
\end{tabular}
\end{center}
\vspace{3mm}

\vspace{3mm}
\begin{center}
\begin{tabular}{r c l}
\hline
$f(t=0,r)$ & = & $5.0\exp(-2.0(r-\frac{1}{2}r_{max})^2) $\\
			& & $\times [\exp(-2.0(\theta-\frac{1}{8}\theta_{max})^2) + \exp(-2.0(\theta-\frac{3}{8}\theta_{max})^2)$ \\
			& & $+ (\exp(-2.0(\theta-\frac{5}{8}\theta_{max})^2) + \exp(-2.0(\theta-\frac{7}{8}\theta_{max})^2)] + 2.0$ \\[10pt]
$g(t=0,r)$ & = & $2.0\exp(-2.0(r-\frac{1}{2}r_{max})^2) $\\ 
			& & $\times [\exp(-2.0(\theta-\frac{1}{8}\theta_{max})^2) + \exp(-2.0(\theta-\frac{3}{8}\theta_{max})^2)$ \\ 
			& & $+(\exp(-2.0(\theta-\frac{5}{8}\theta_{max})^2) + \exp(-2.0(\theta-\frac{7}{8}\theta_{max})^2)] + 2.0$ \\[10pt]
$h(t=0,r)$ & = & $0.5\exp(-2.0(r-\frac{1}{2}r_{max})^2) $\\
			& & $\times [0.5\exp(-2.0(\theta-\frac{1}{8}\theta_{max})^2) + 0.5\exp(-2.0(\theta-\frac{3}{8}\theta_{max})^2)$ \\
			& & $+(0.5\exp(-2.0(\theta-\frac{5}{8}\theta_{max})^2) + 0.5\exp(-2.0(\theta-\frac{7}{8}\theta_{max})^2)] + 1.0$ \\[10pt]
$v^1(t=0,r)$ & = & $0.01/[1.0 + \exp(-5.0(r-\frac{1}{2}r_{max}))] - 0.01/[1.0 + \exp(\frac{5}{2}r_{max})]$ \\[10pt] 
$v^2(t=0,r)$ & = & $0.0$ \\ 
\hline 
\end{tabular}
\end{center}
\vspace{3mm}

\begin{figure}[t]
    \centering
		\subfloat{
            \includegraphics[width=0.5\textwidth]{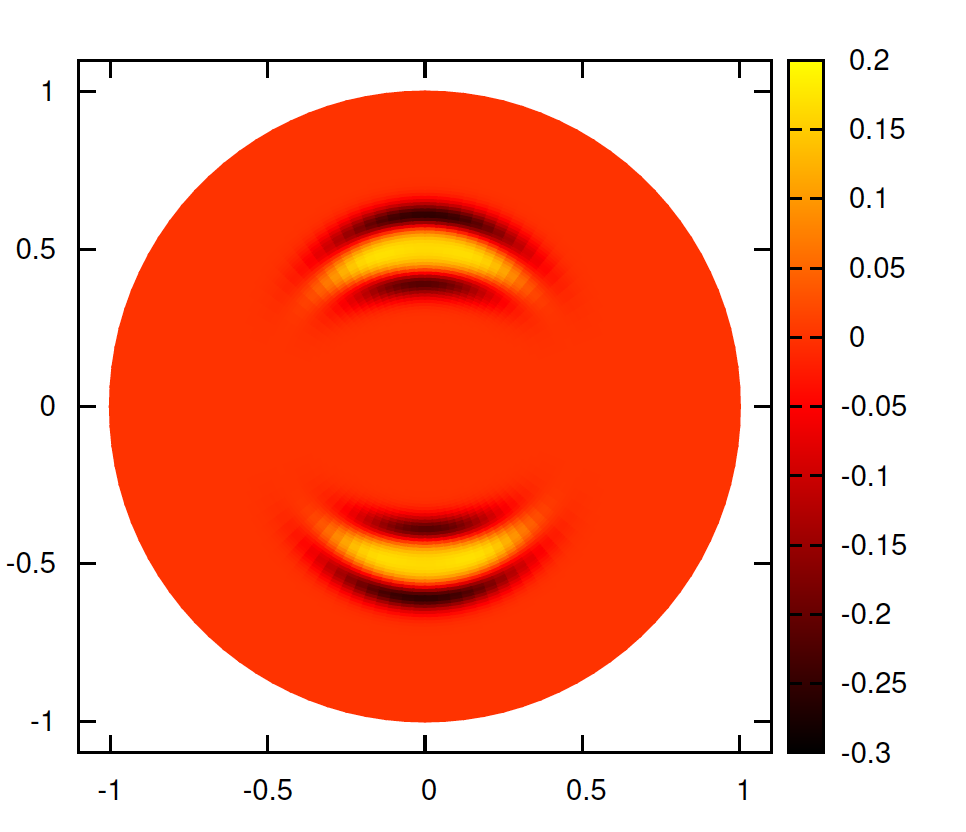}
        }

        \subfloat{                
            \includegraphics[width=\textwidth]{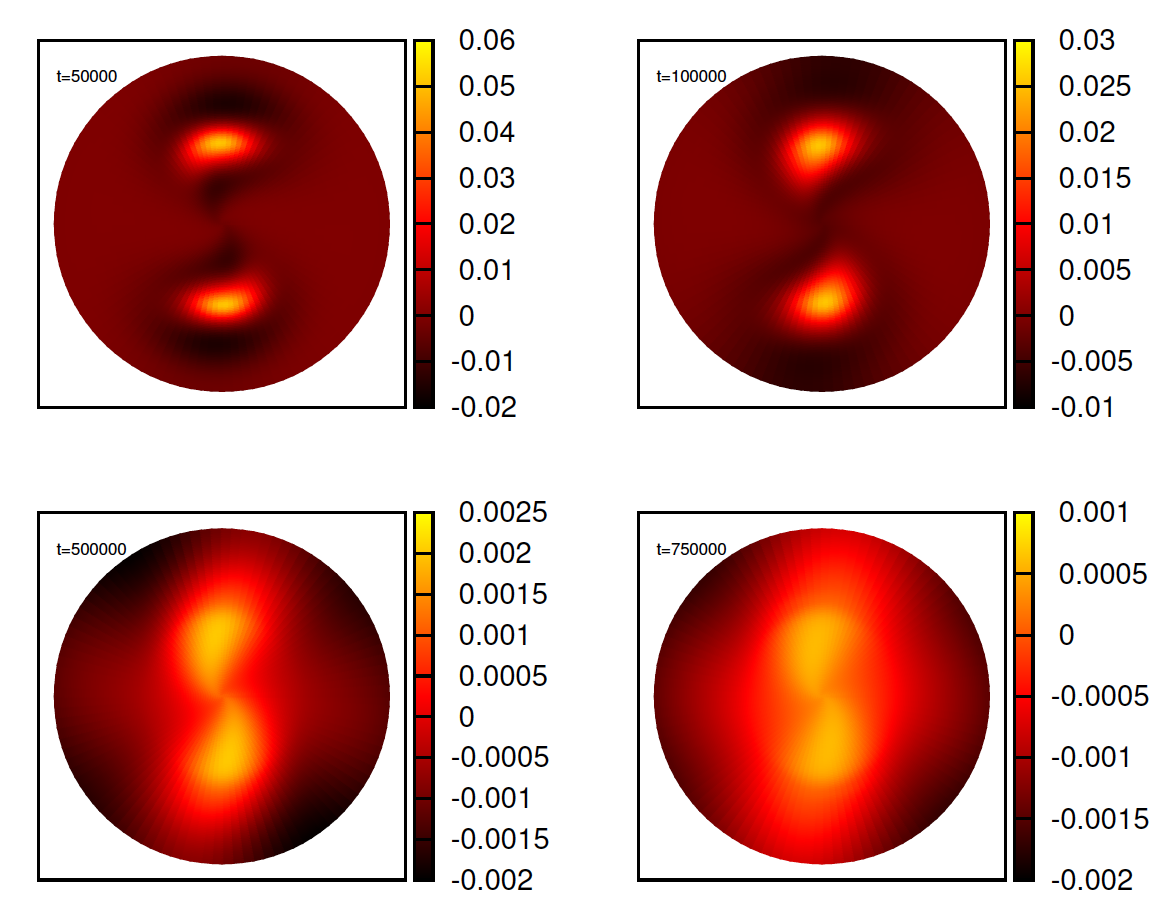}
        }
        \caption{Time evolution of the scalar curvature for a $2$-gaussian initial metric and a circularly symmetric velocity field.}\label{figure2gaussCurvature} 
\end{figure}

\begin{figure}
	\centering
	\includegraphics[width=0.9\textwidth]{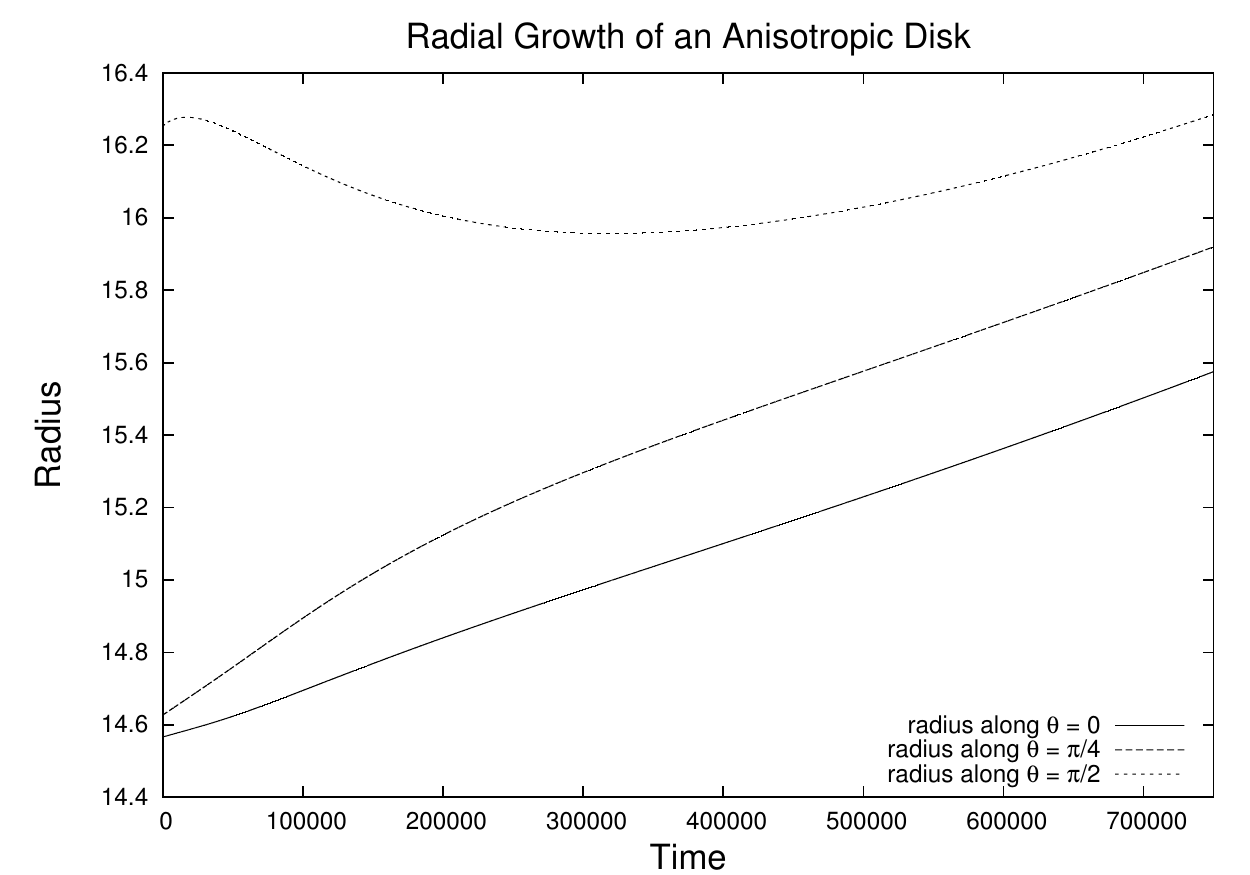}
	\caption{Radial growth of a $2$-gaussian initial metric and a circularly symmetric velocity field along different $\theta$ directions.}
	\label{figure2gaussRadius}
\end{figure}

\begin{figure}
    \centering
        \subfloat{
                \includegraphics[width=0.6\textwidth]{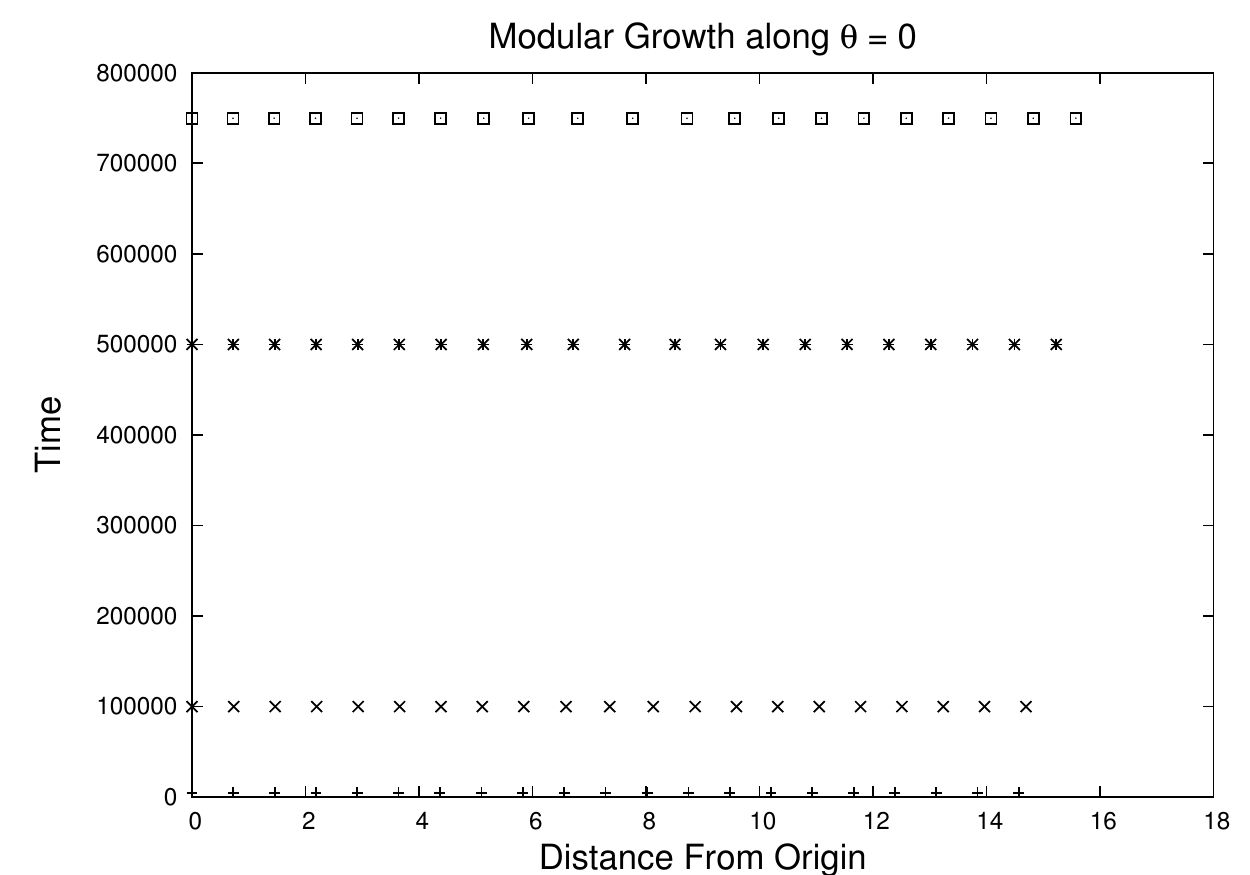}
        }

        \subfloat{
                \includegraphics[width=0.6\textwidth]{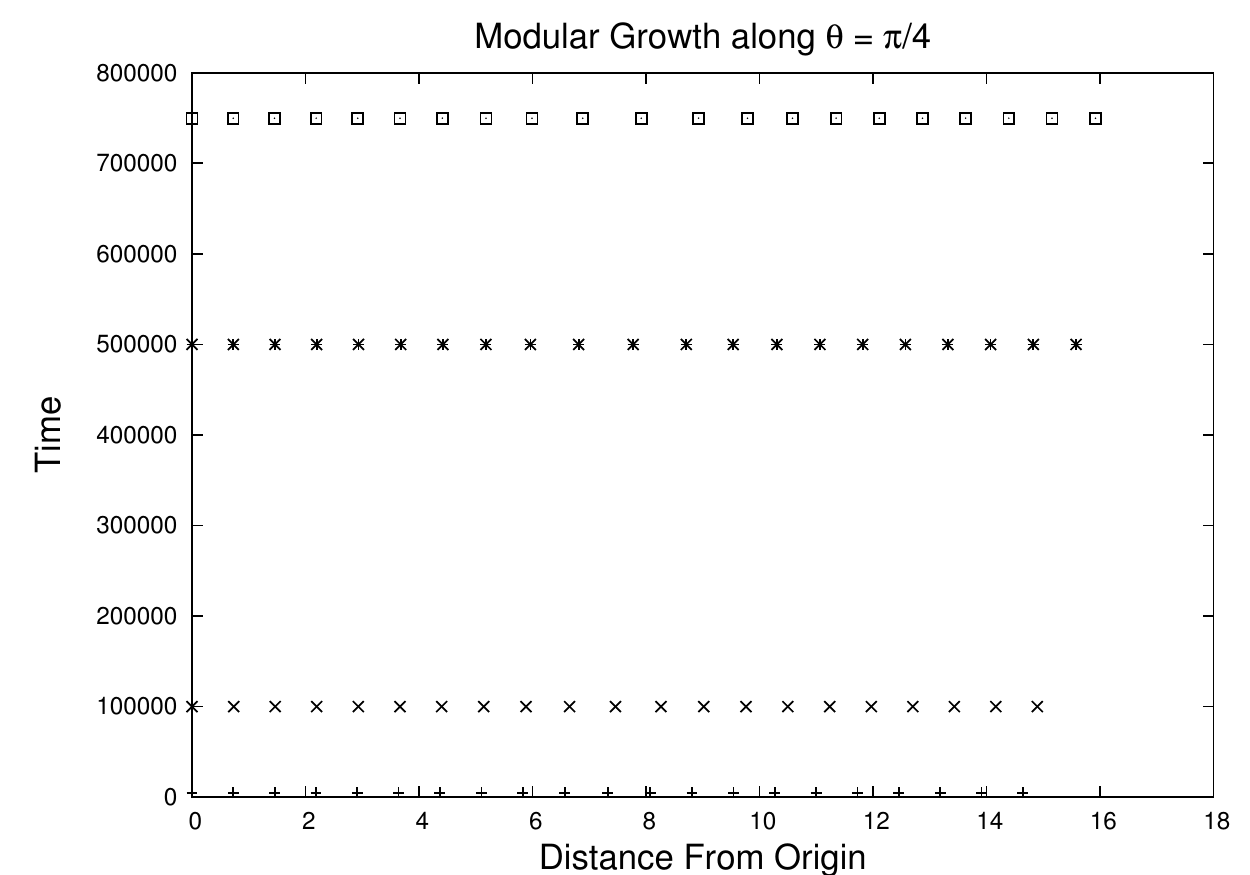}
        }

        \subfloat{
                \includegraphics[width=0.6\textwidth]{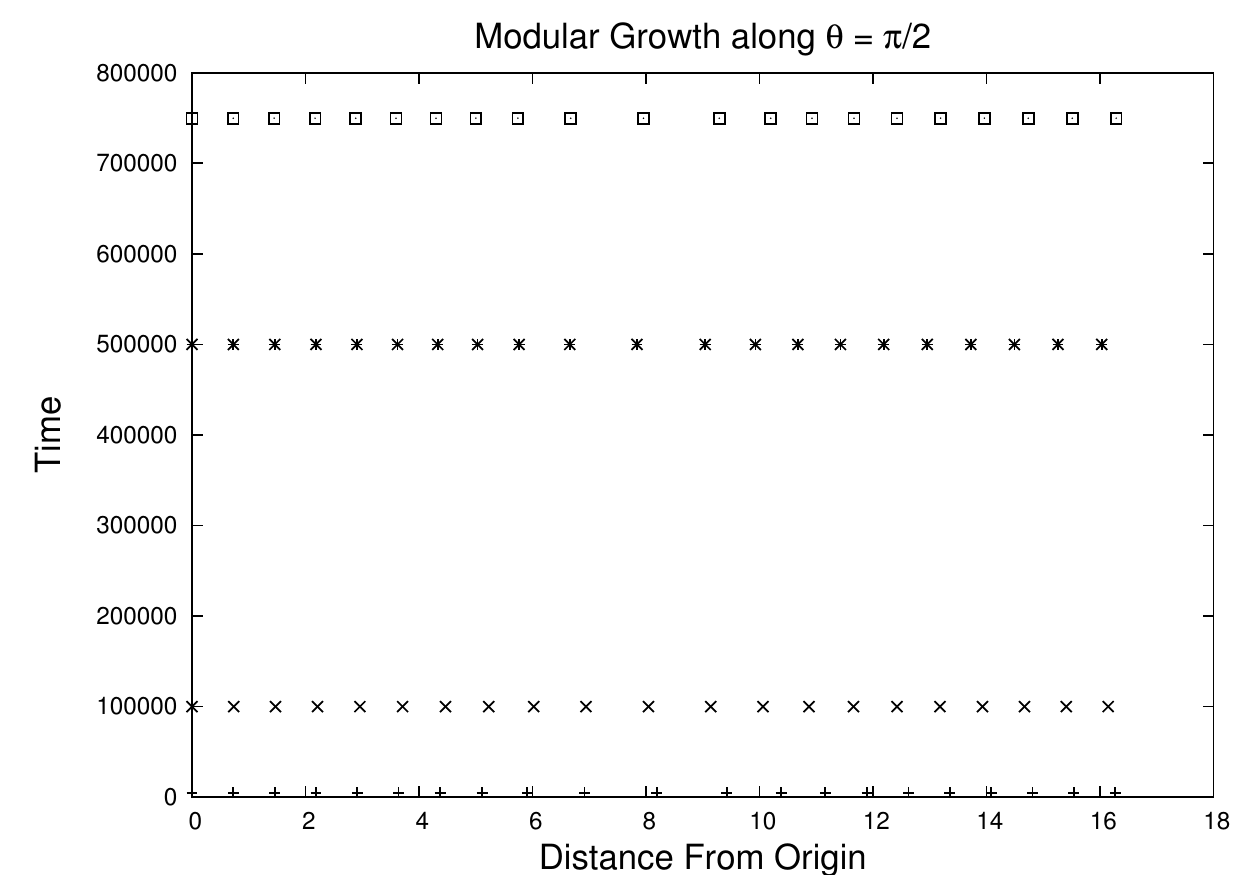}
        }
        \caption{Modular growth along different $\theta$ directions for a $2$-gaussian initial metric and a circularly symmetric velocity field.}\label{figure2gaussModular} 
\end{figure}

\begin{figure}
    \centering
        \subfloat{
                \includegraphics[width=0.4\textwidth]{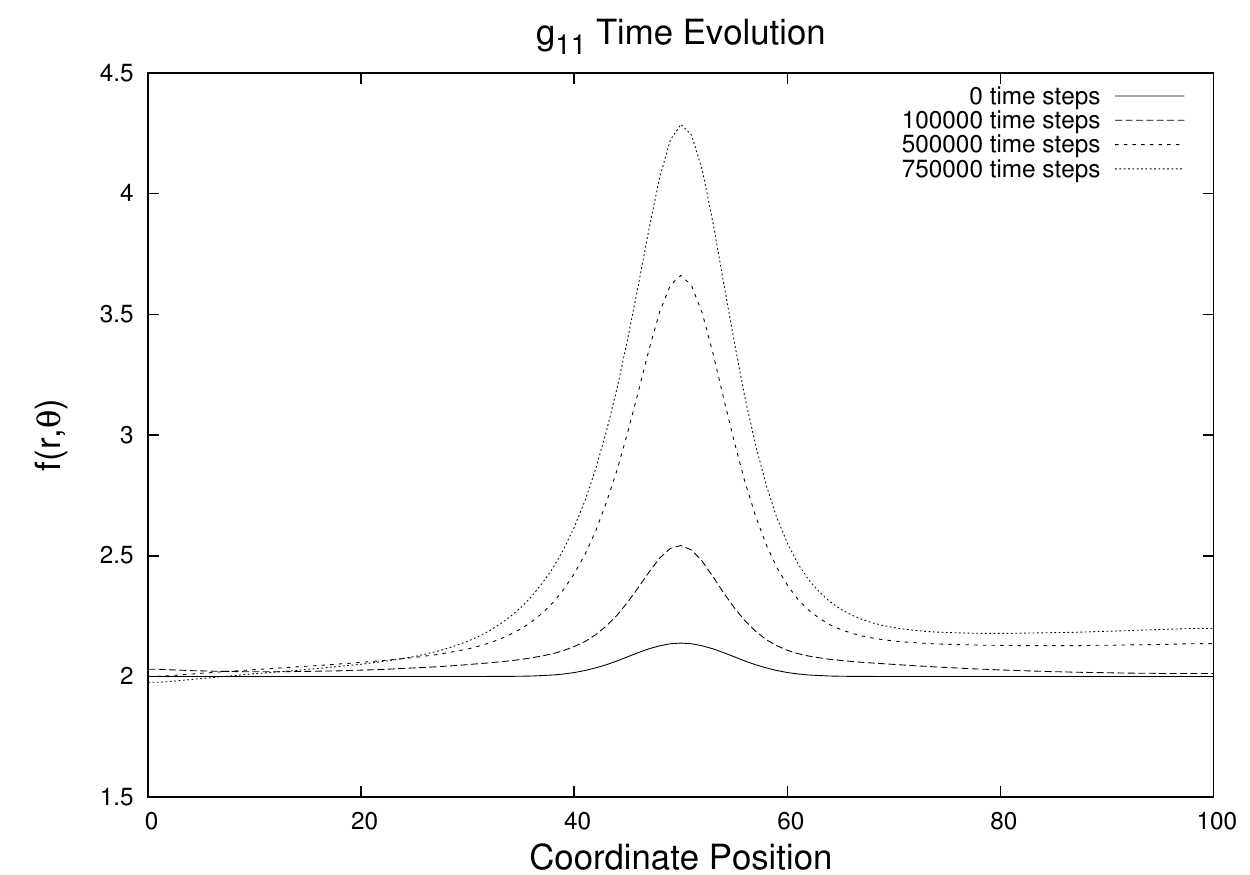}
        }
        \subfloat{
                \includegraphics[width=0.4\textwidth]{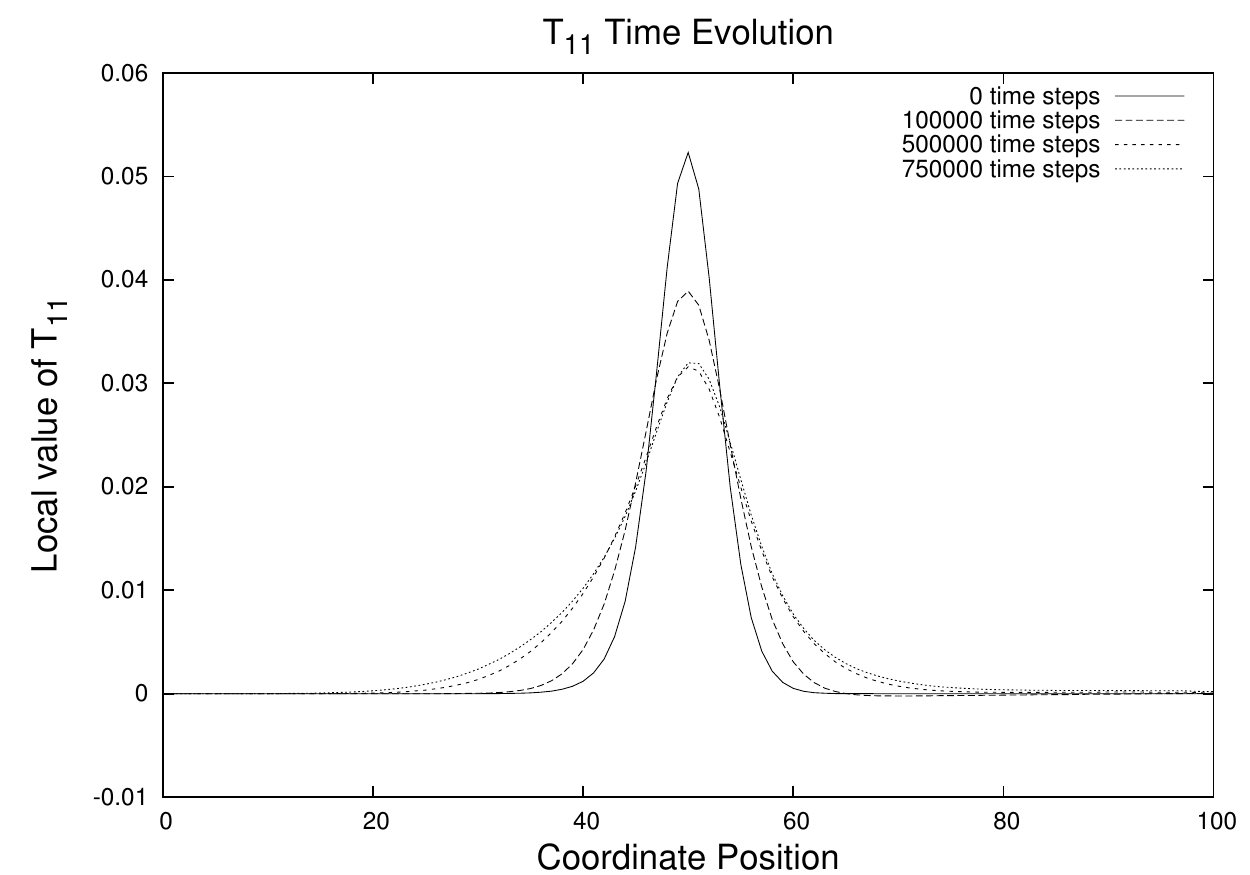}
        }

        \subfloat{
                \includegraphics[width=0.4\textwidth]{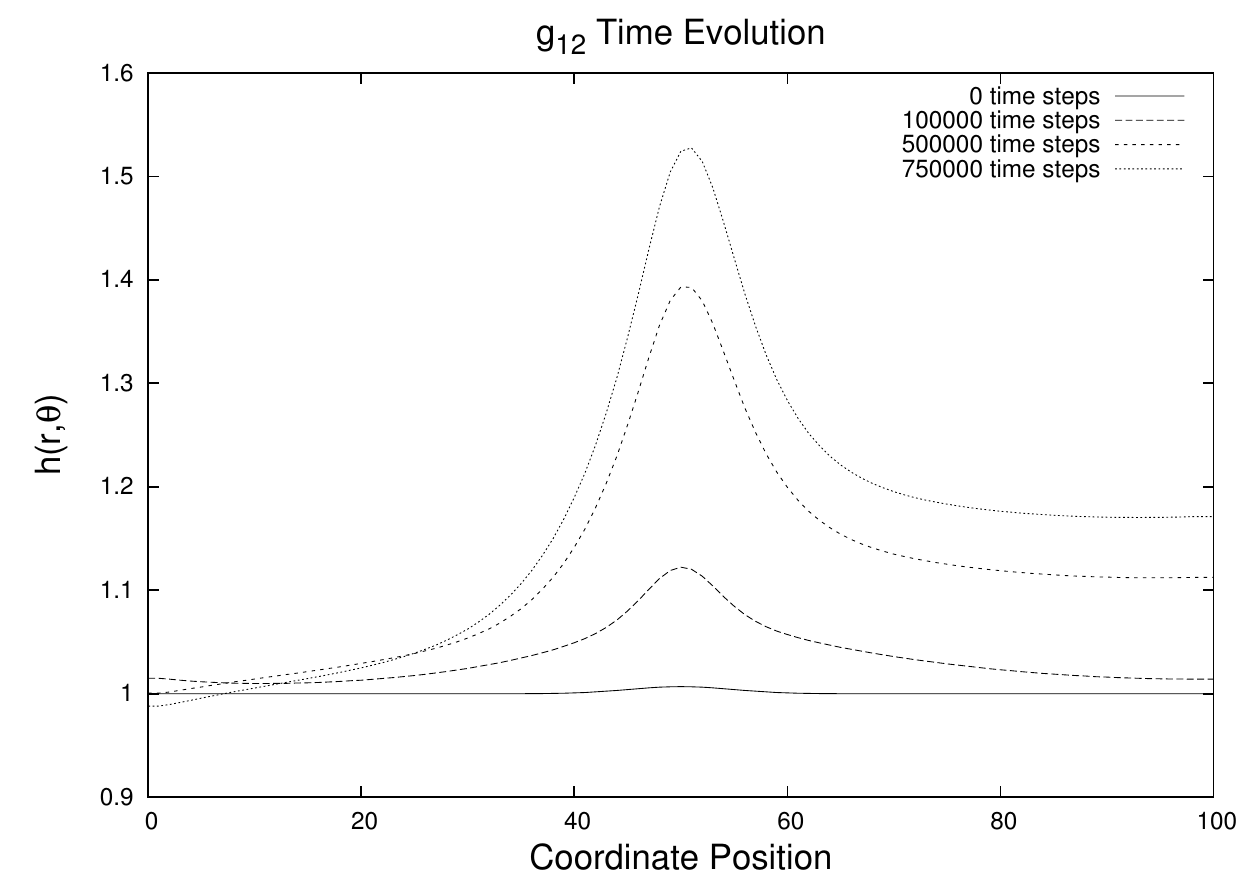}
        }
        \subfloat{
                \includegraphics[width=0.4\textwidth]{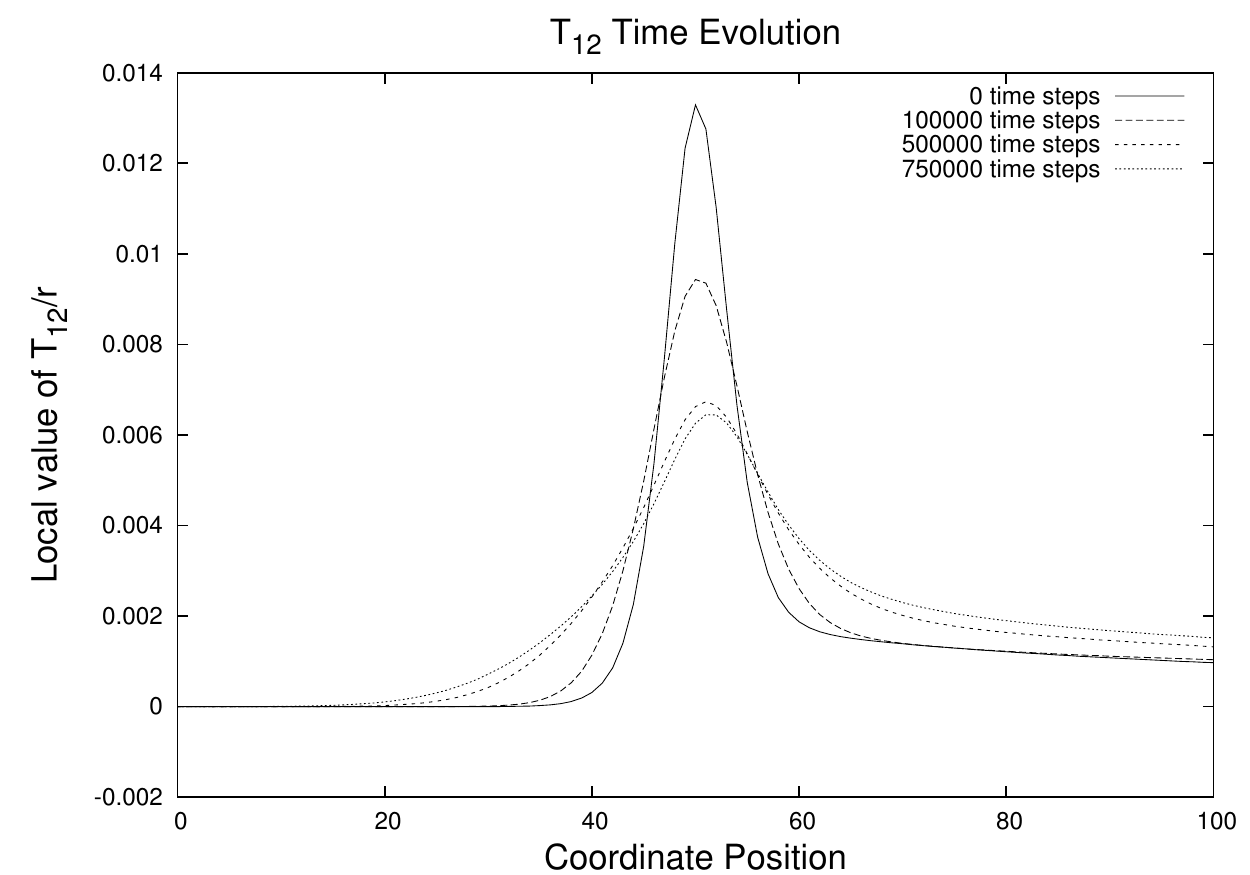}
        }

        \subfloat{
                \includegraphics[width=0.4\textwidth]{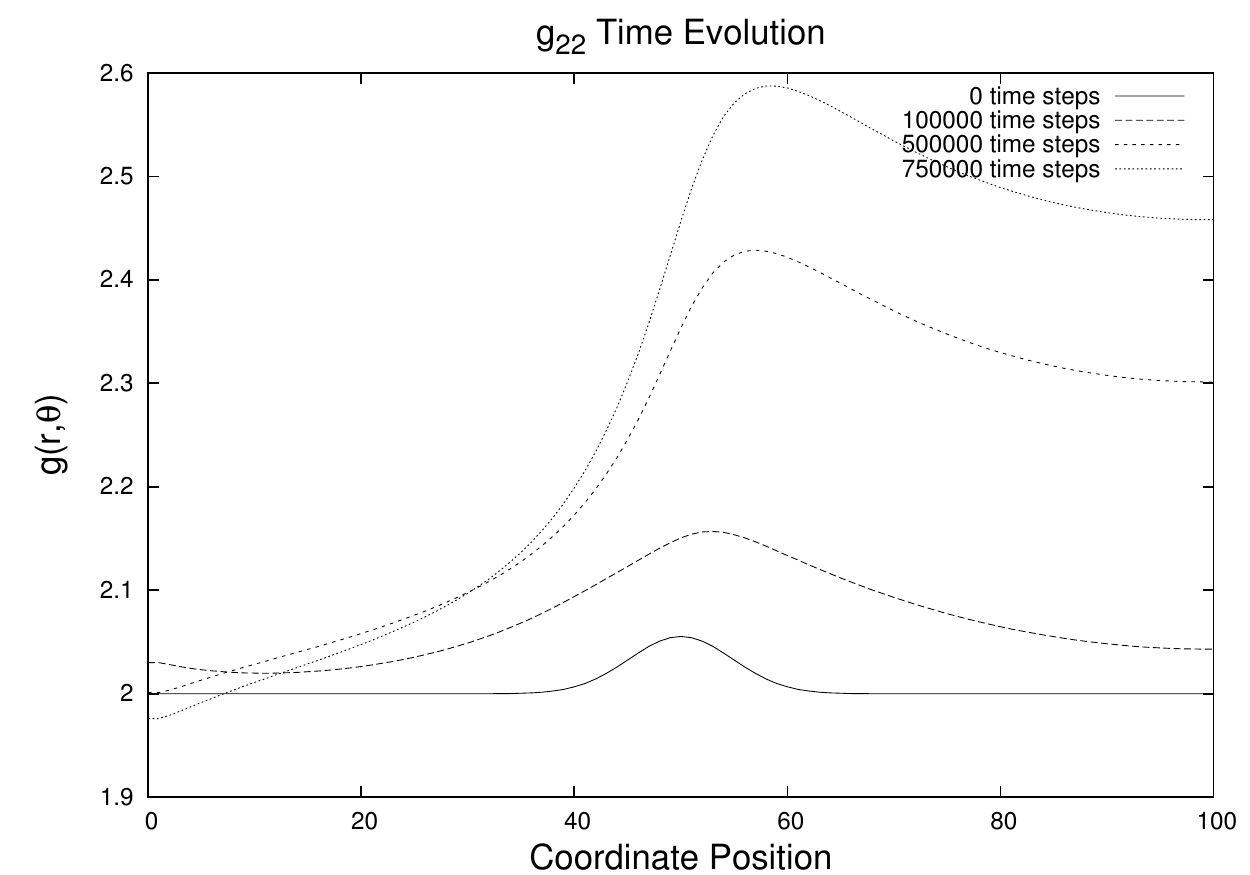}
        }
	   \subfloat{
                \includegraphics[width=0.4\textwidth]{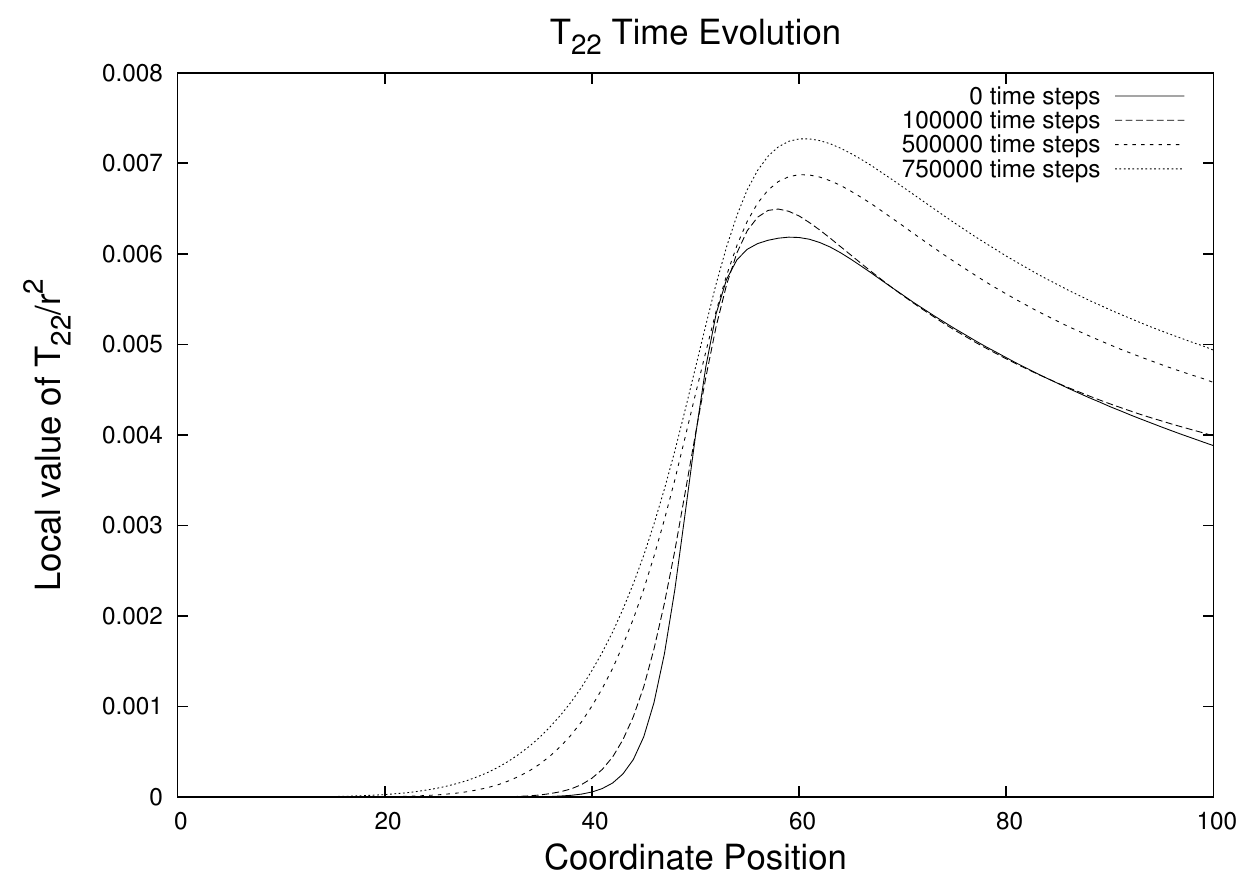}
        }
        
        \subfloat{
                \includegraphics[width=0.4\textwidth]{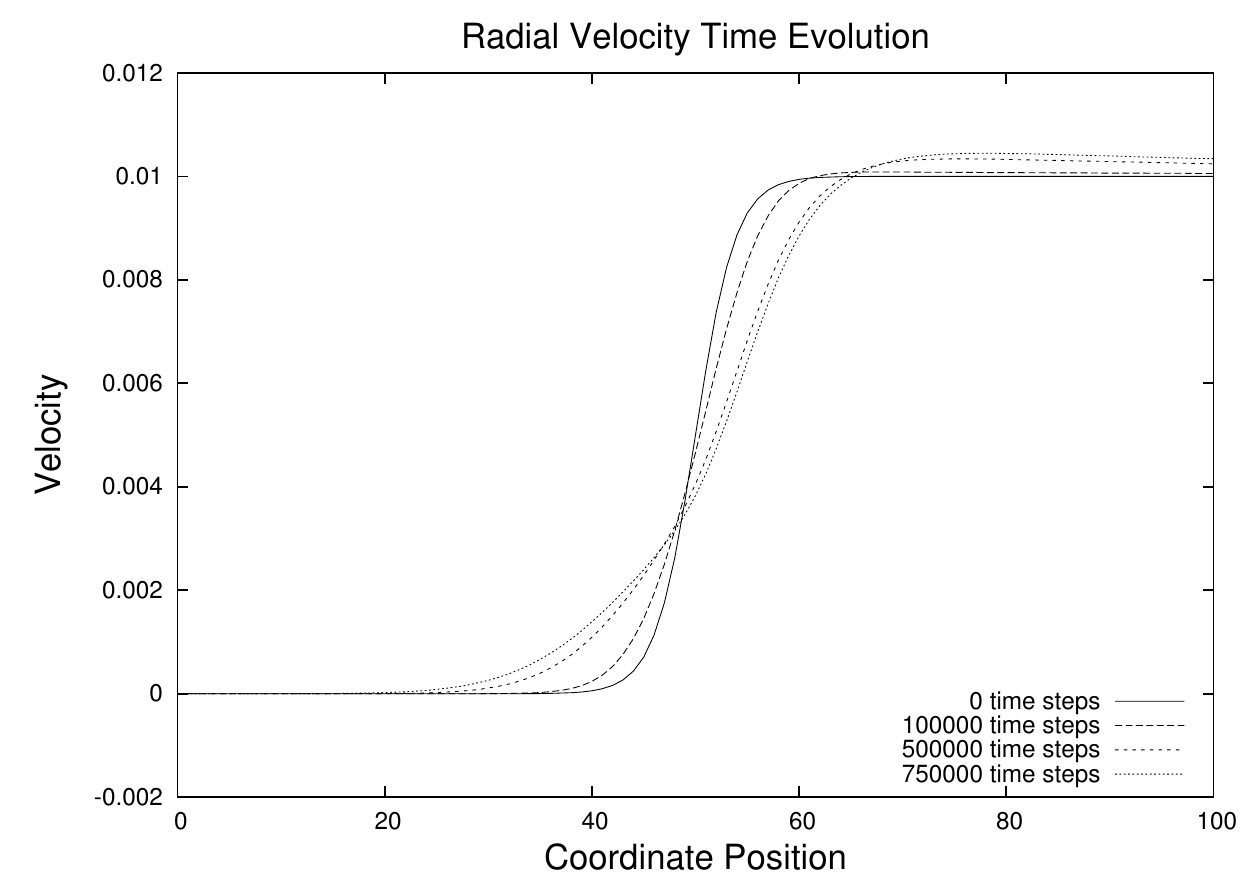}
        }
        \subfloat{
                \includegraphics[width=0.4\textwidth]{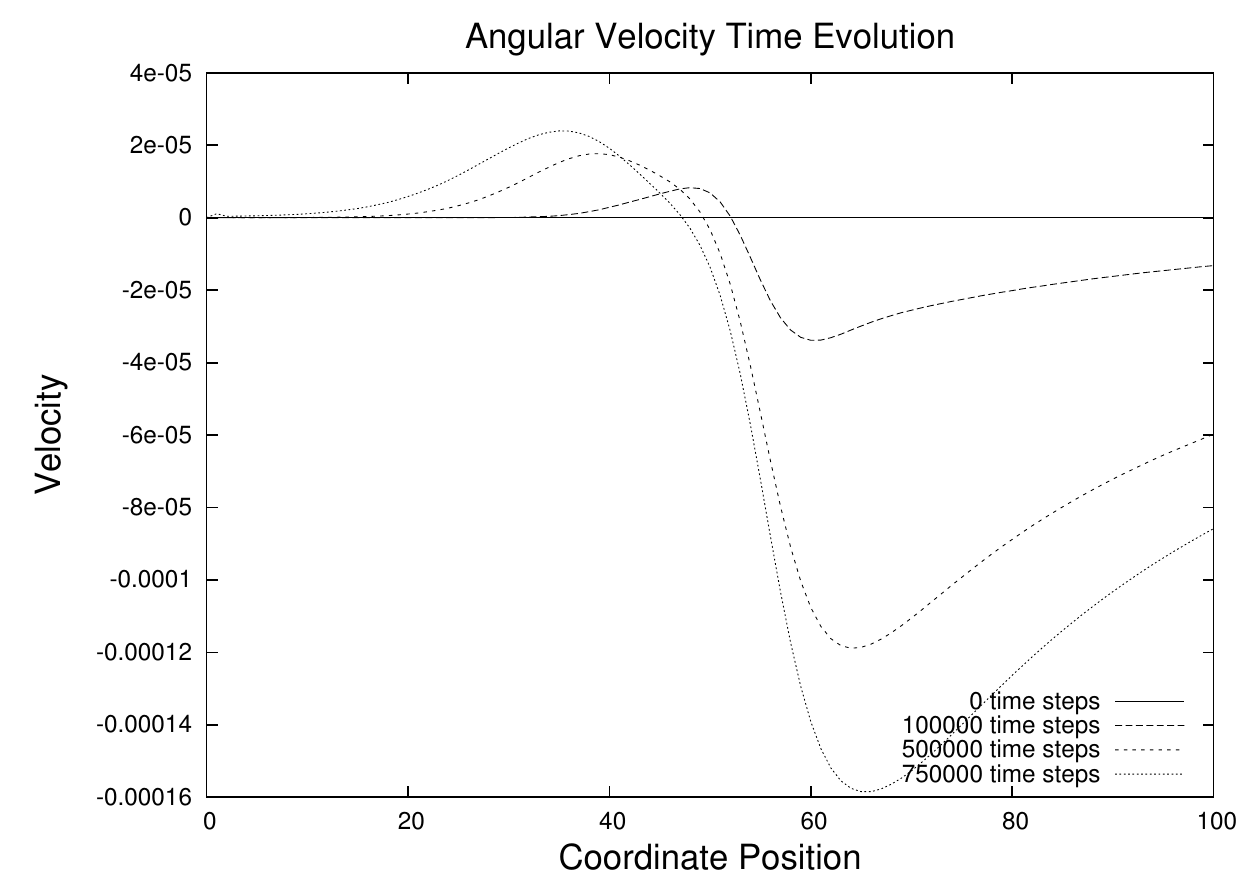}
        }
        \caption{Dynamics of the metric and velocity field at $\theta=\pi/4$ for a $2$-gaussian initial metric and a circularly symmetric velocity field.}\label{figure2gaussDynamics4} 
\end{figure}

\begin{figure}
    \centering
        \subfloat{
                \includegraphics[width=0.4\textwidth]{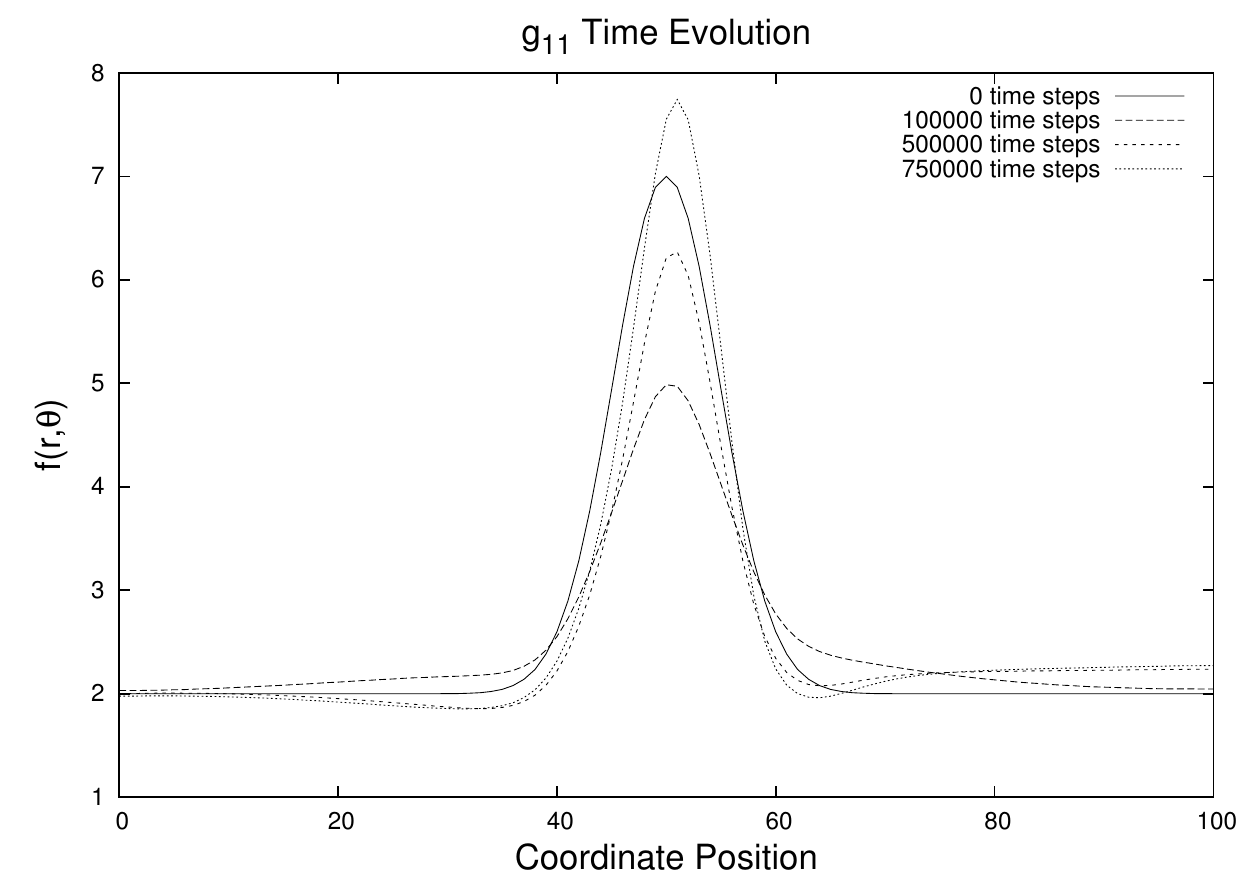}
        }
        \subfloat{
                \includegraphics[width=0.4\textwidth]{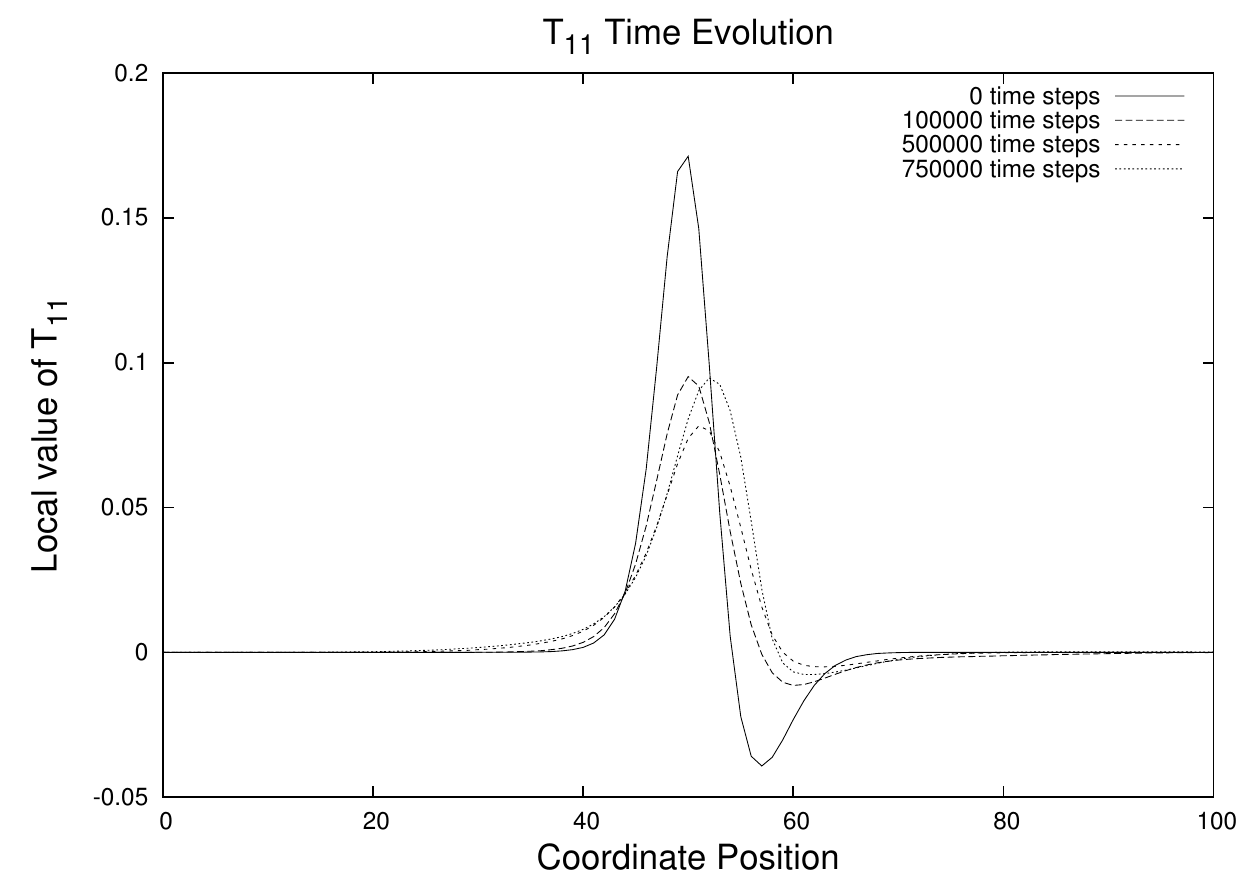}
        }

        \subfloat{
                \includegraphics[width=0.4\textwidth]{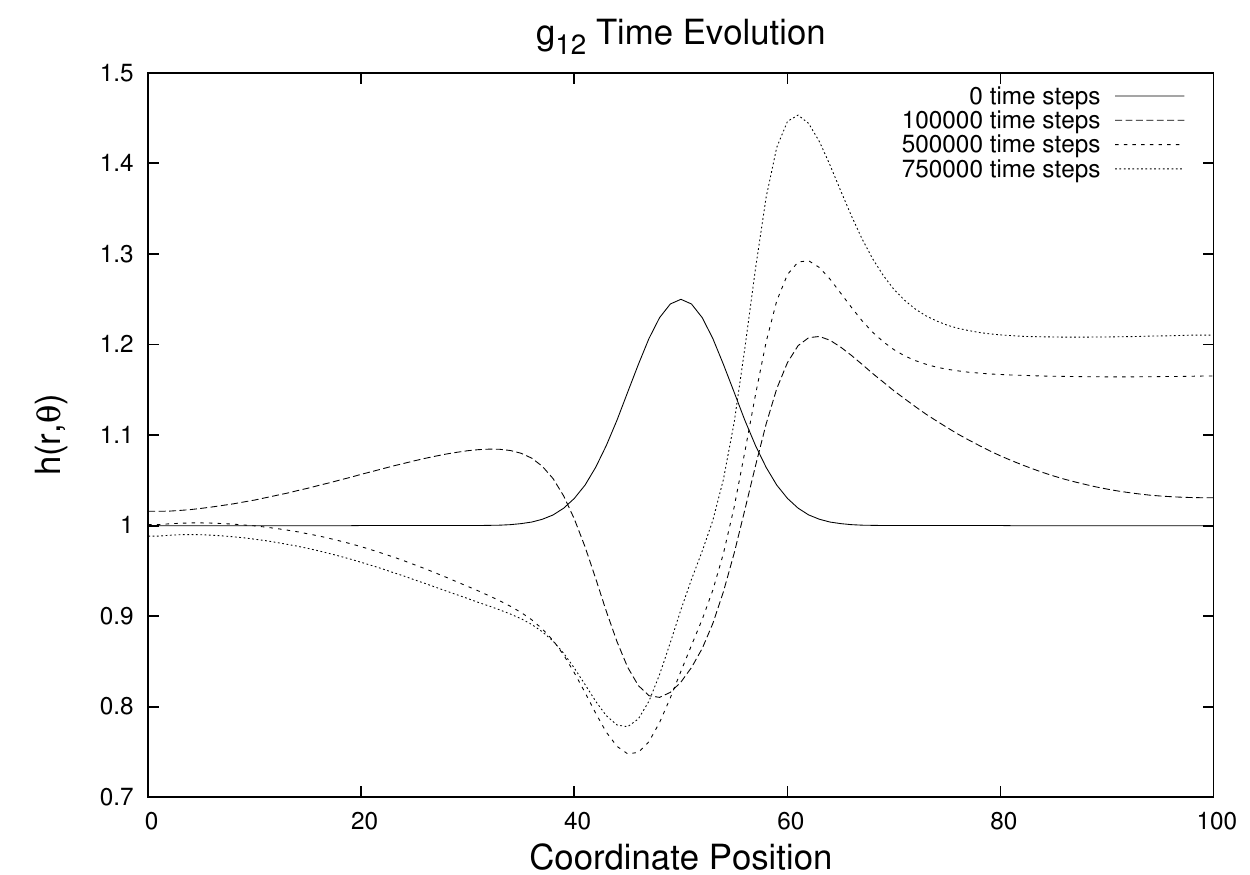}
        }
        \subfloat{
                \includegraphics[width=0.4\textwidth]{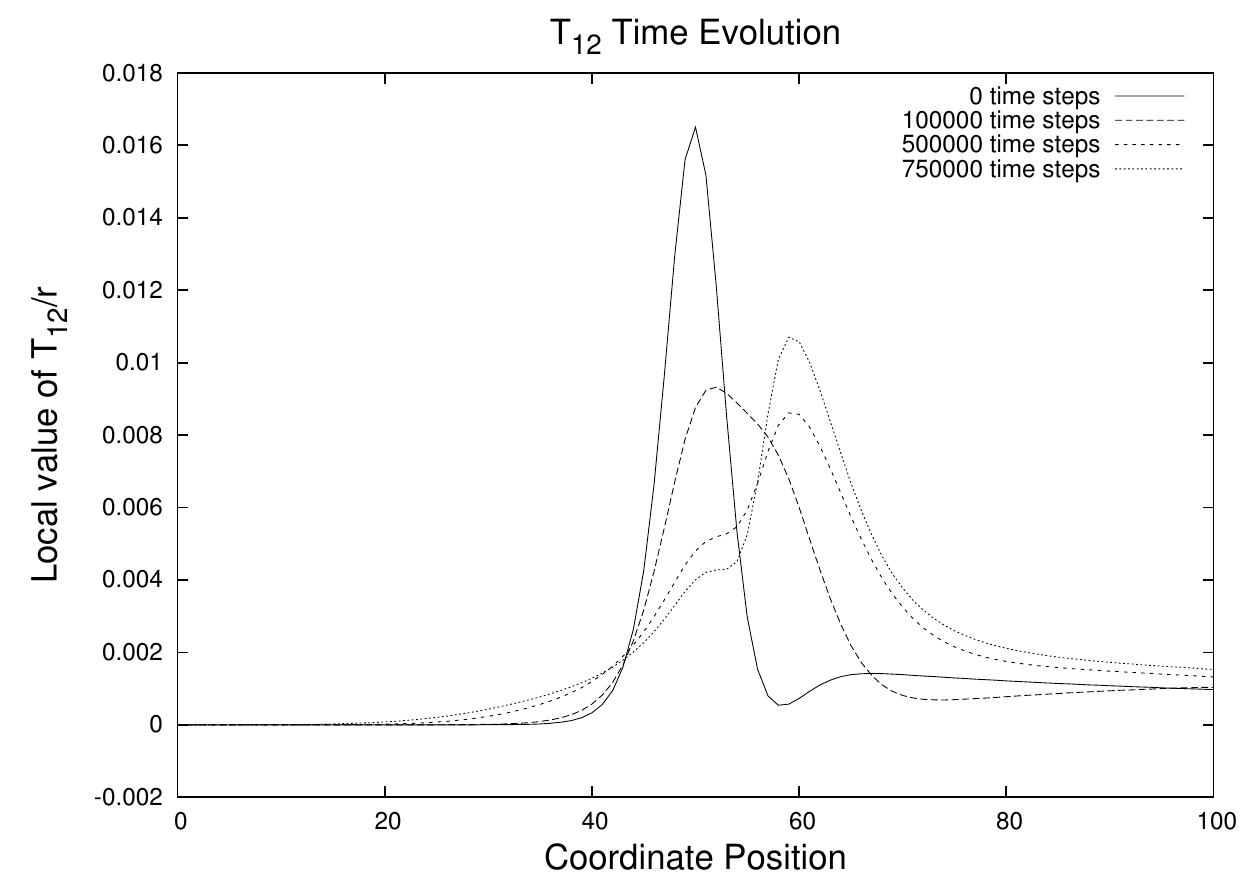}
        }

        \subfloat{
                \includegraphics[width=0.4\textwidth]{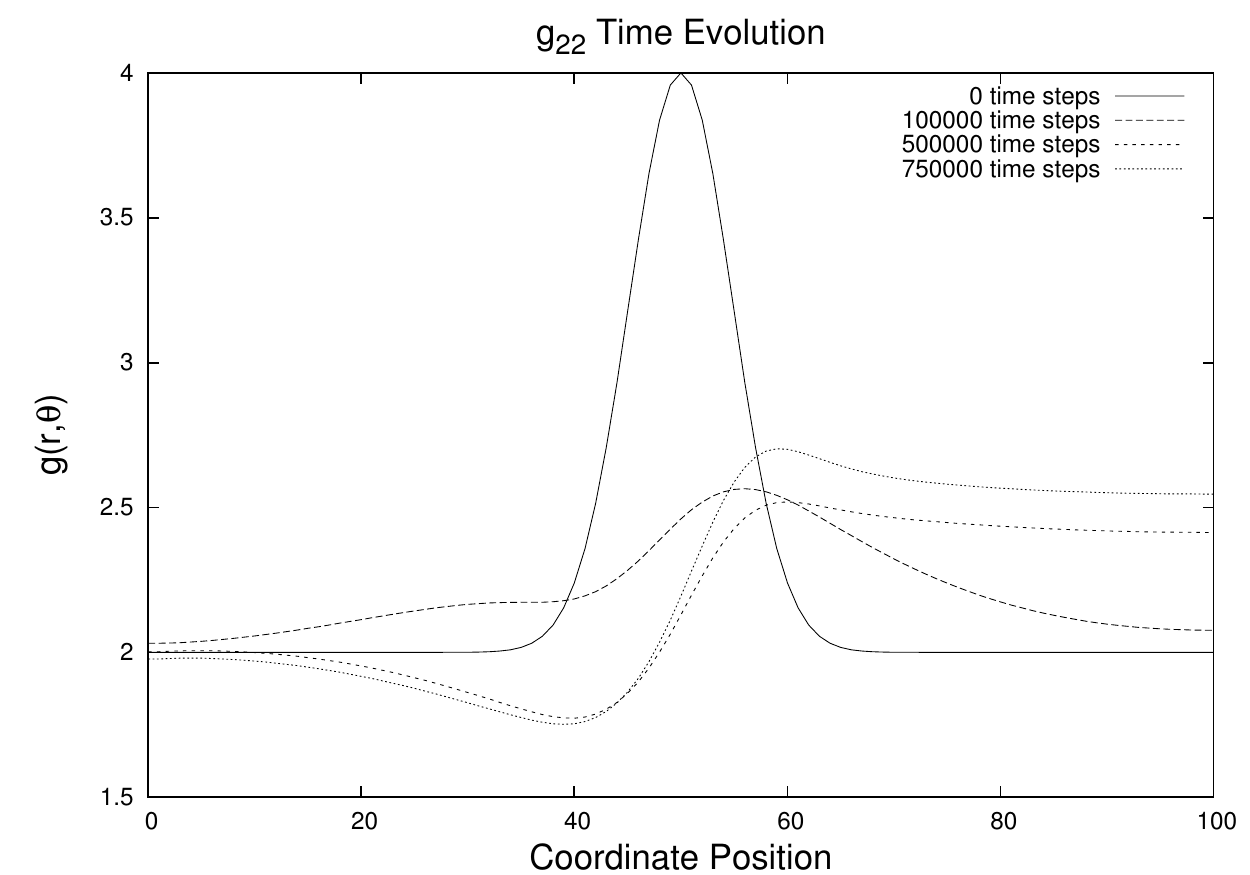}
        }
        \subfloat{
                \includegraphics[width=0.4\textwidth]{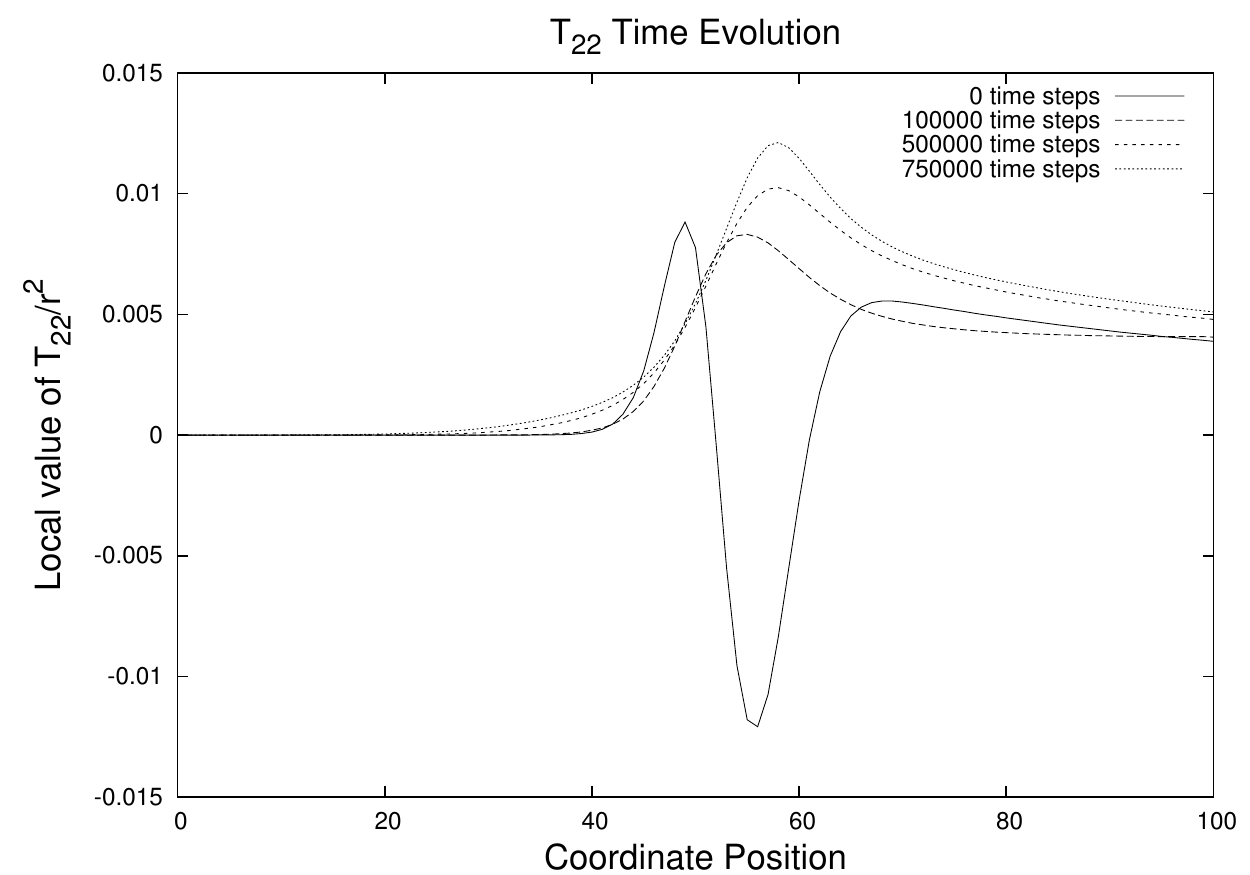}
        }

        \subfloat{
                \includegraphics[width=0.4\textwidth]{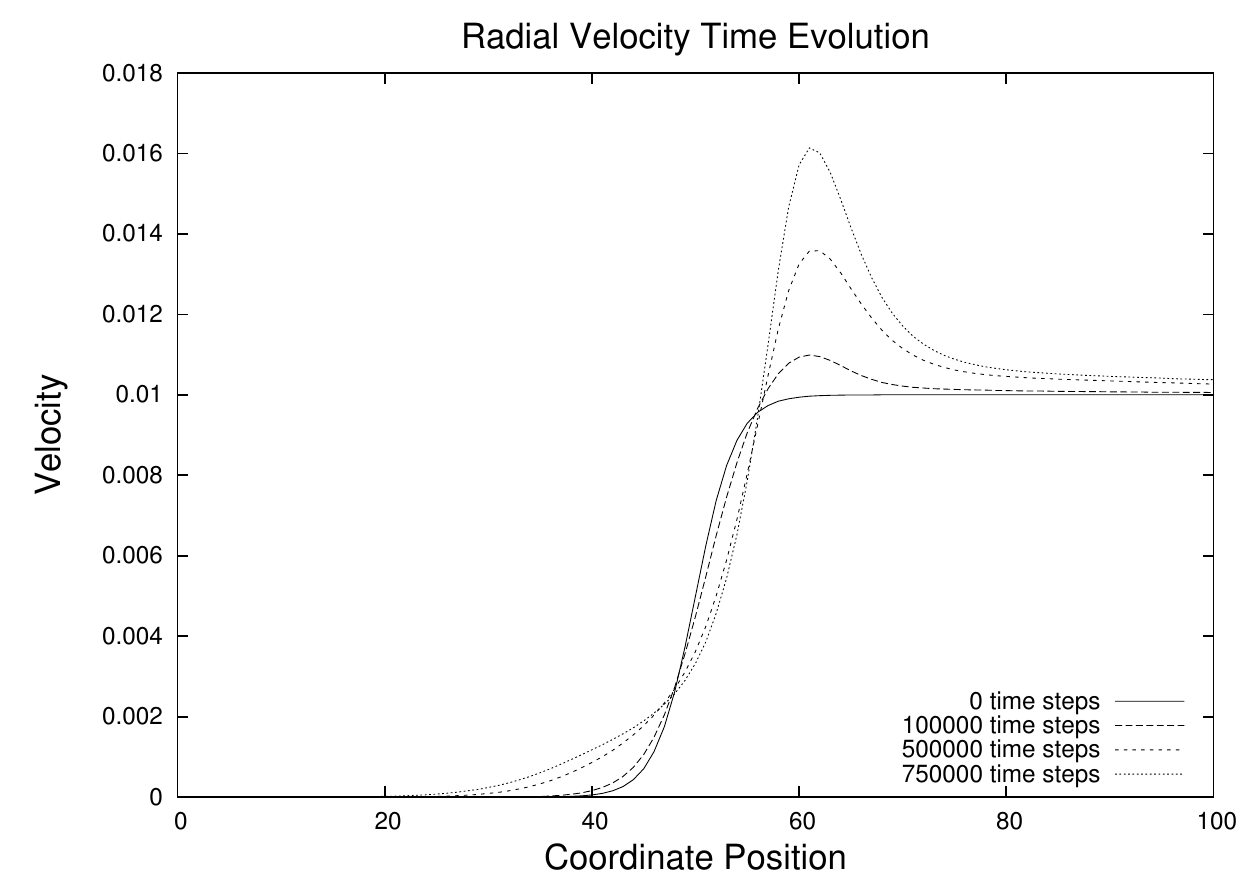}
        }
        \subfloat{
                \includegraphics[width=0.4\textwidth]{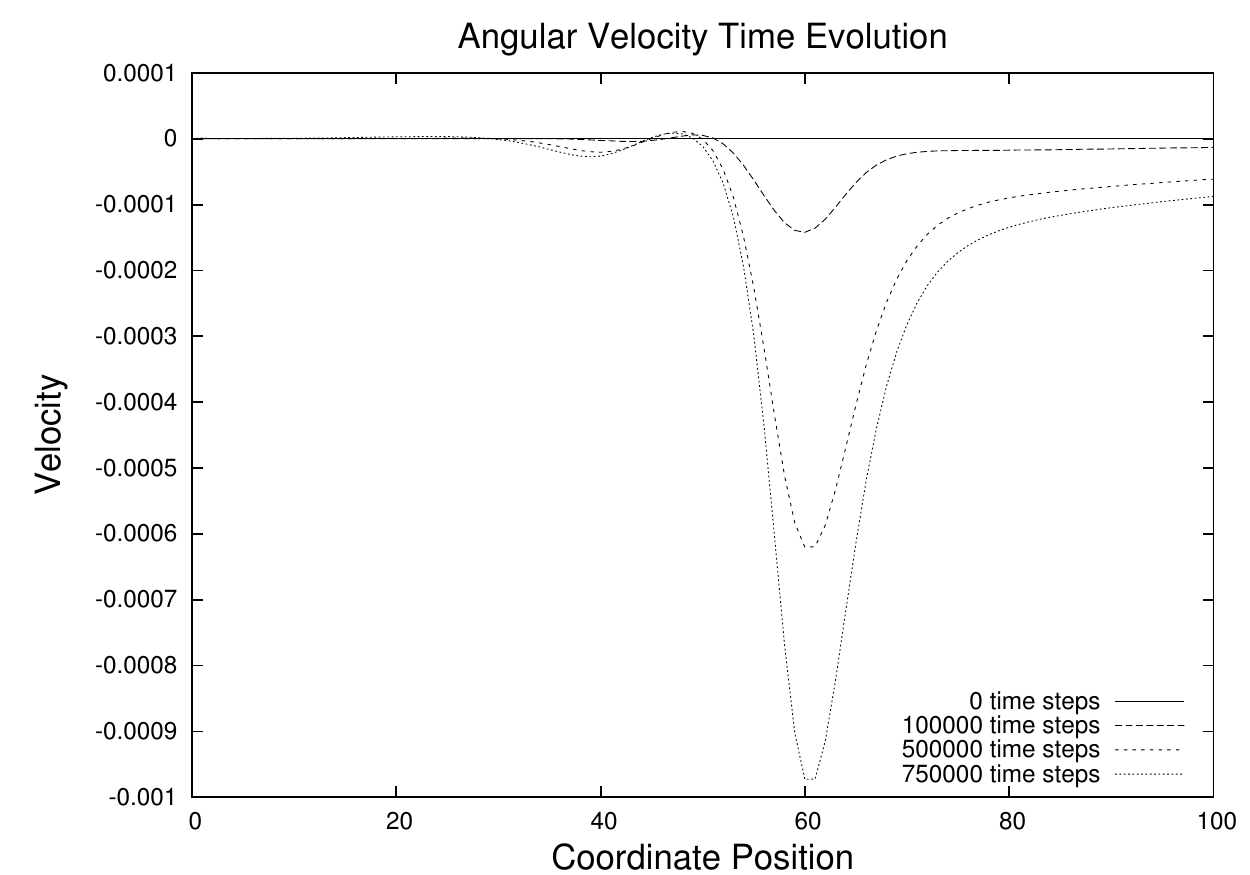}
        }
        \caption{Dynamics of the metric and velocity field at $\theta=\pi/2$ for a $2$-gaussian initial metric and a circularly symmetric velocity field.}\label{figure2gaussDynamics2} 
\end{figure}

\begin{figure}
    \centering
        \subfloat{
                \includegraphics[width=0.4\textwidth]{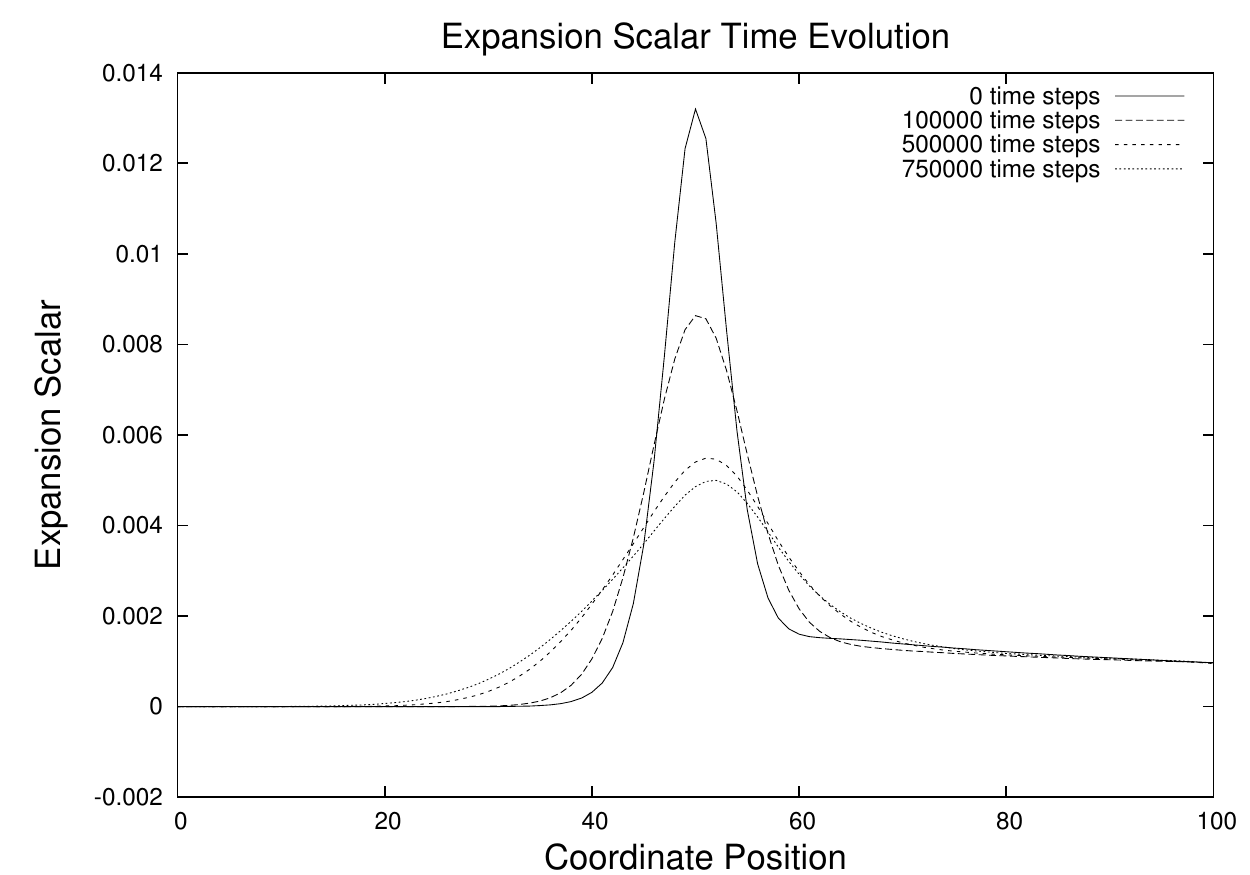}
        }
        \subfloat{
                \includegraphics[width=0.4\textwidth]{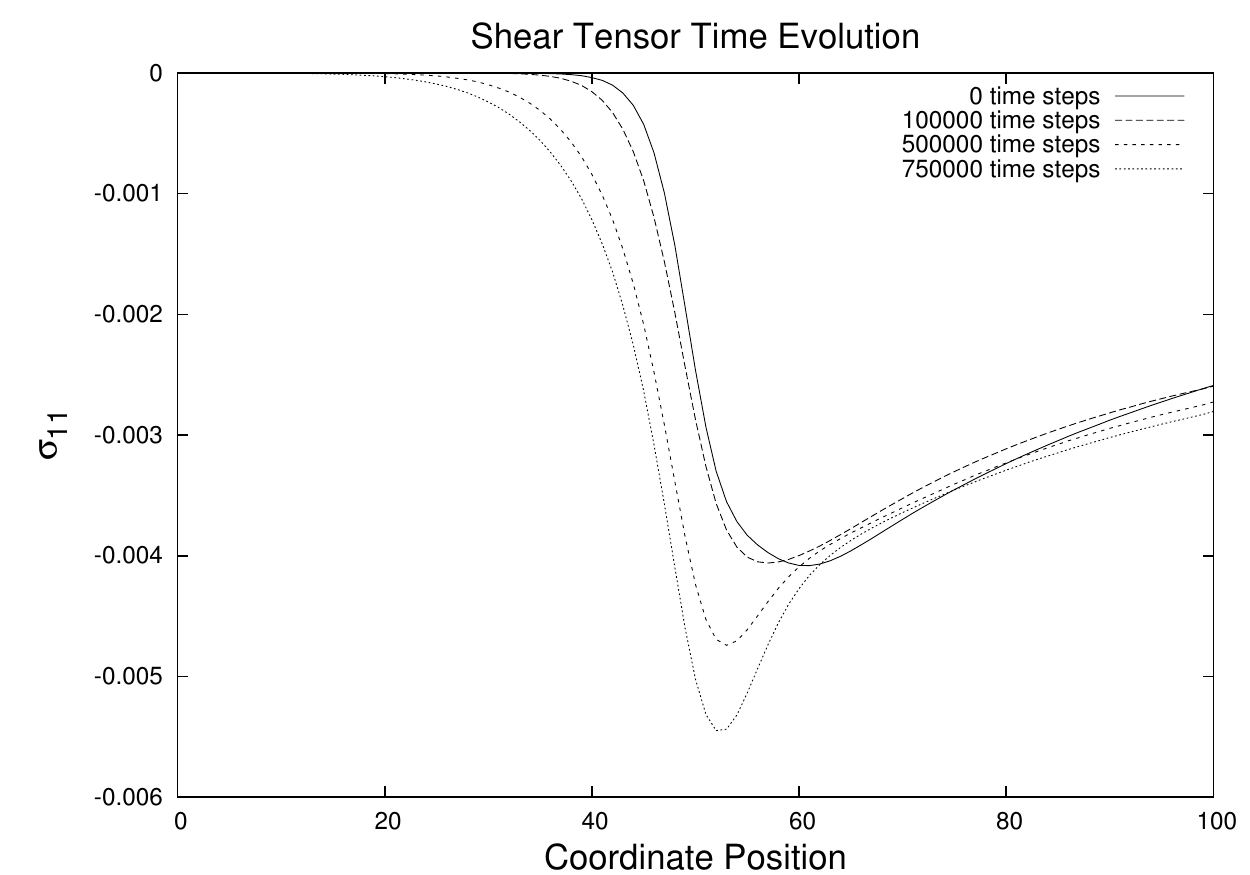}
        }

        \subfloat{
                \includegraphics[width=0.4\textwidth]{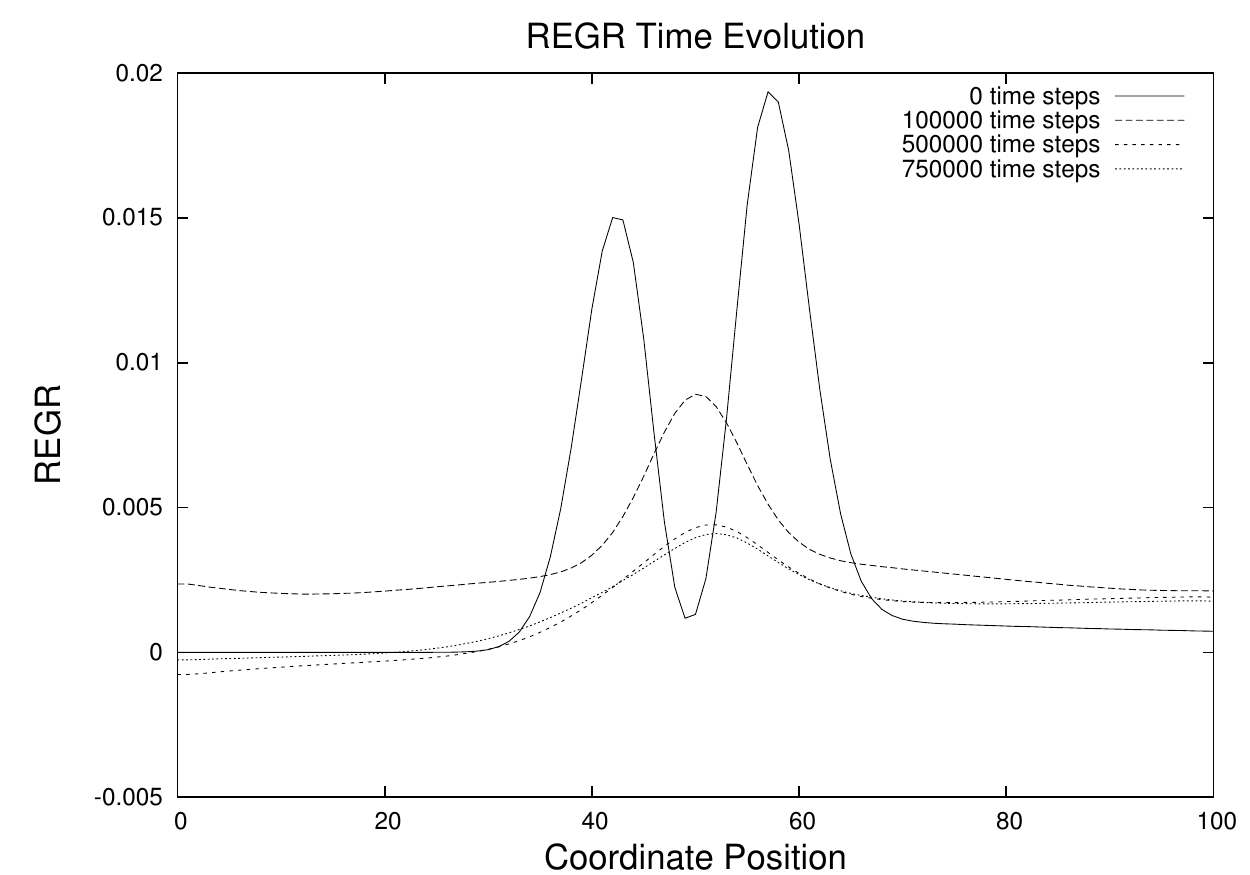}
        }
        \subfloat{
                \includegraphics[width=0.4\textwidth]{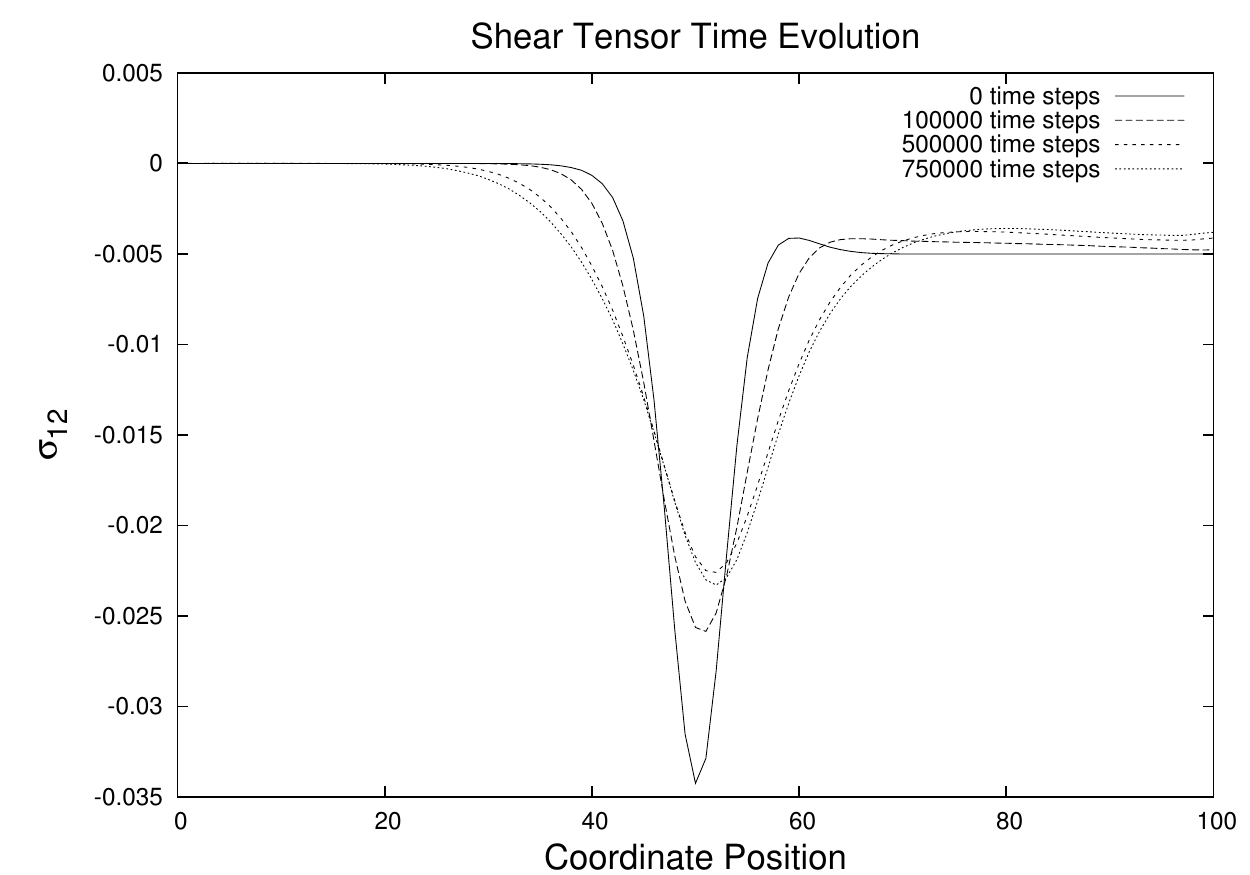}
        }

        \subfloat{
                \includegraphics[width=0.4\textwidth]{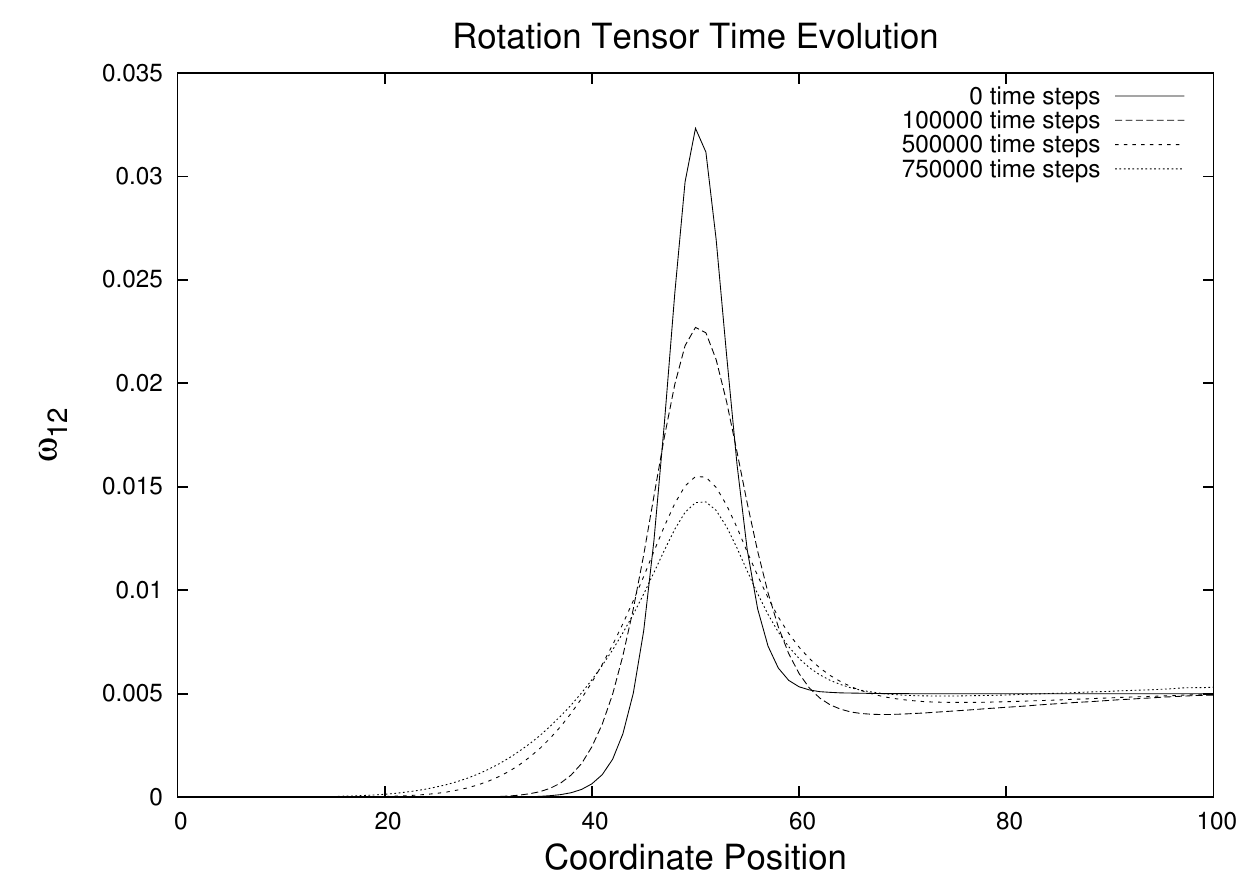}
        }
        \subfloat{
                \includegraphics[width=0.4\textwidth]{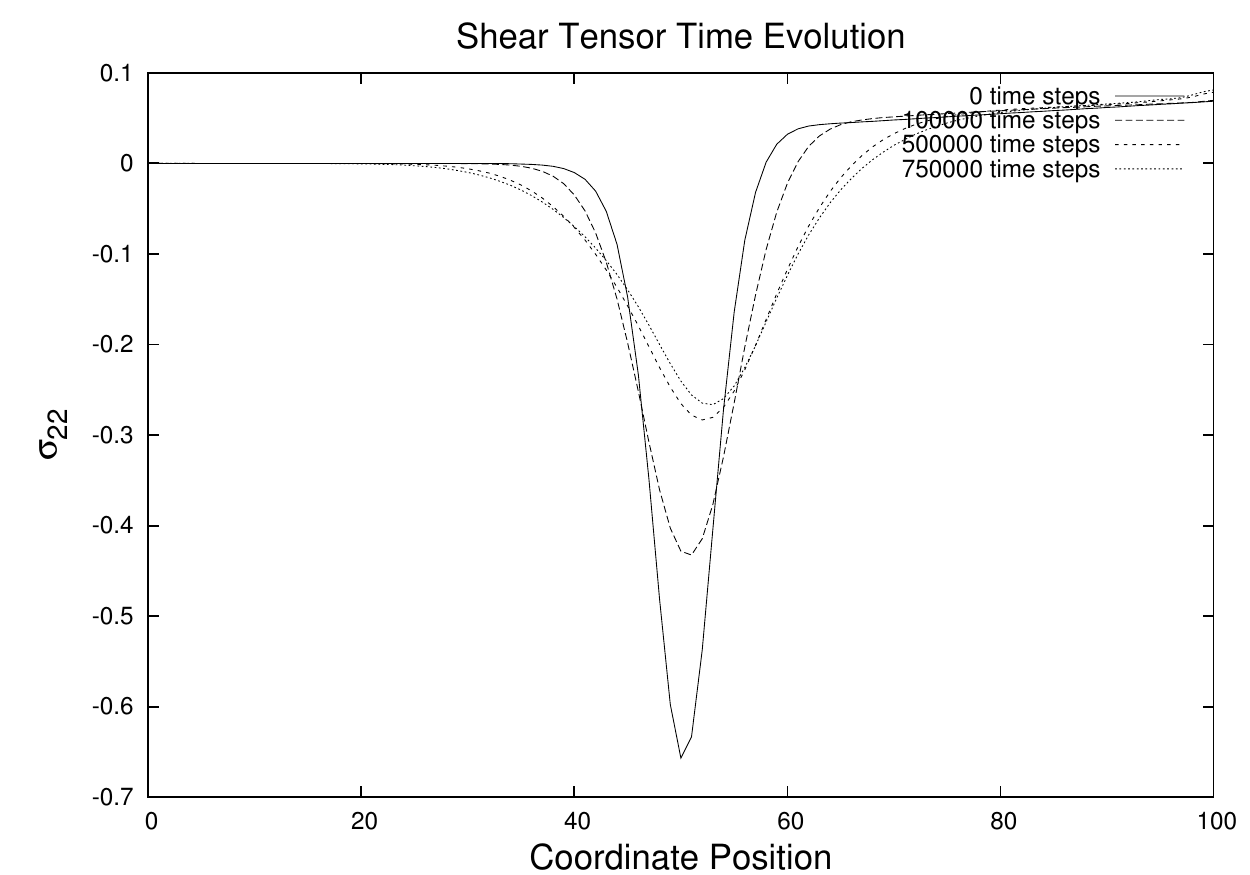}
        }
        \caption{Deformation tensors at $\theta=\pi/4$ for a $2$-gaussian initial metric and a circularly symmetric velocity field.}\label{figure2gaussDefs4} 
\end{figure}

\begin{figure}
    \centering
        \subfloat{
                \includegraphics[width=0.4\textwidth]{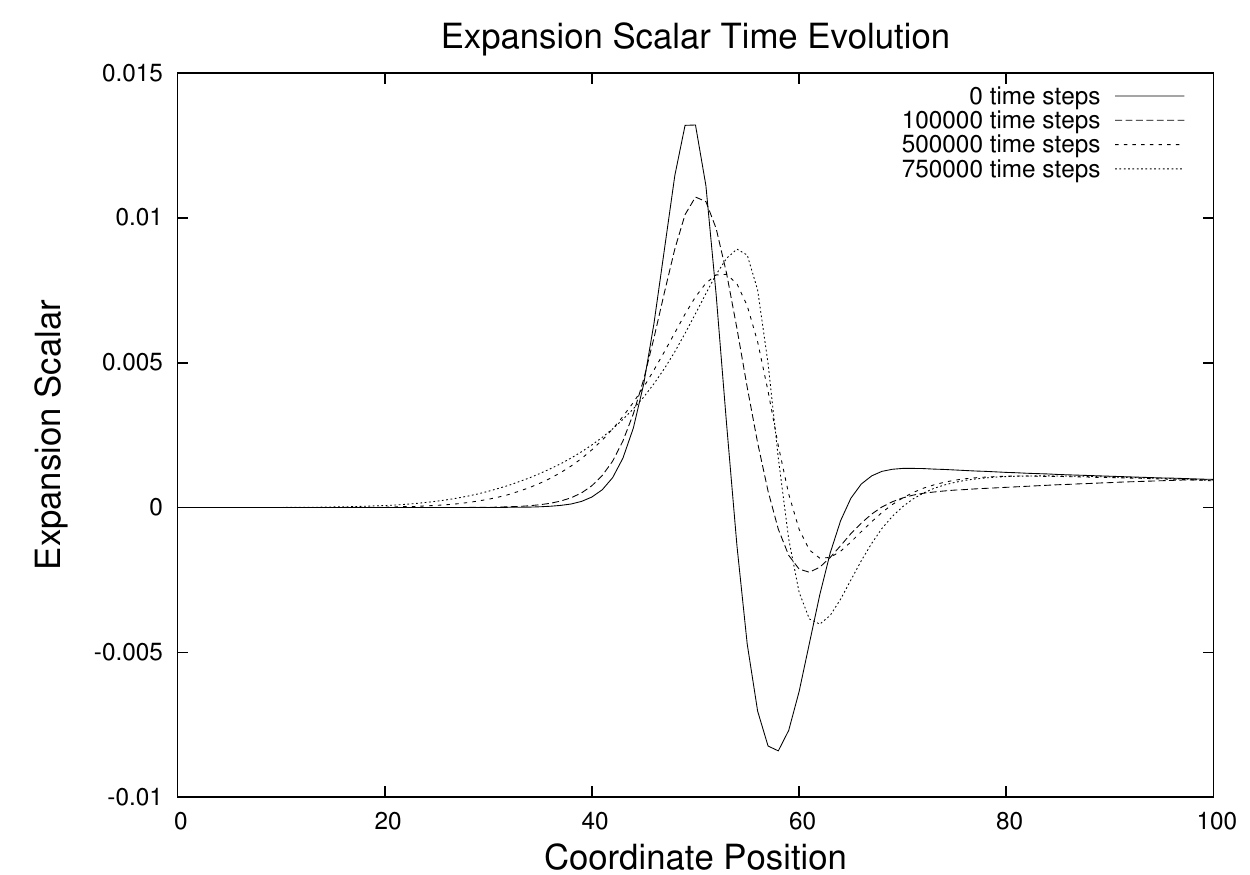}
        }
        \subfloat{
                \includegraphics[width=0.4\textwidth]{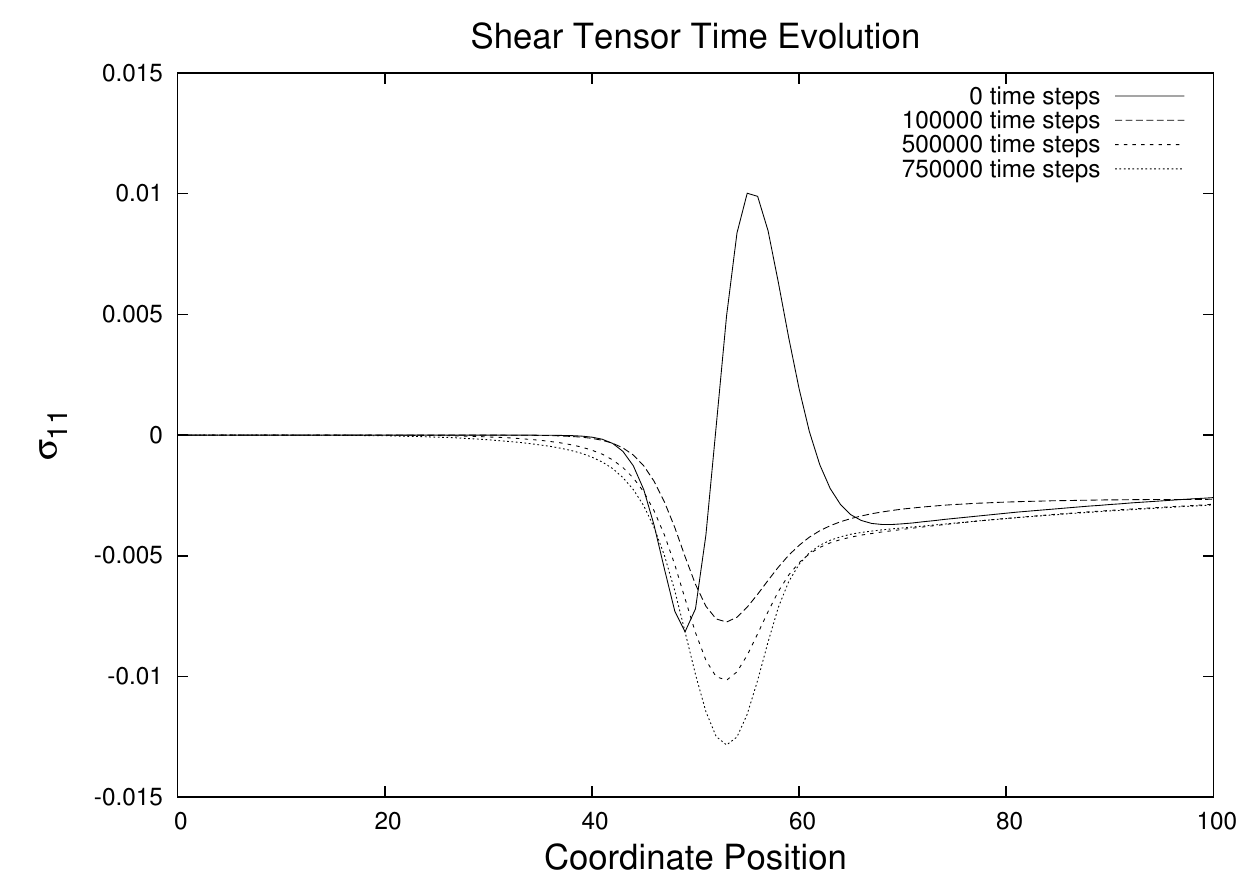}
        }

        \subfloat{
                \includegraphics[width=0.4\textwidth]{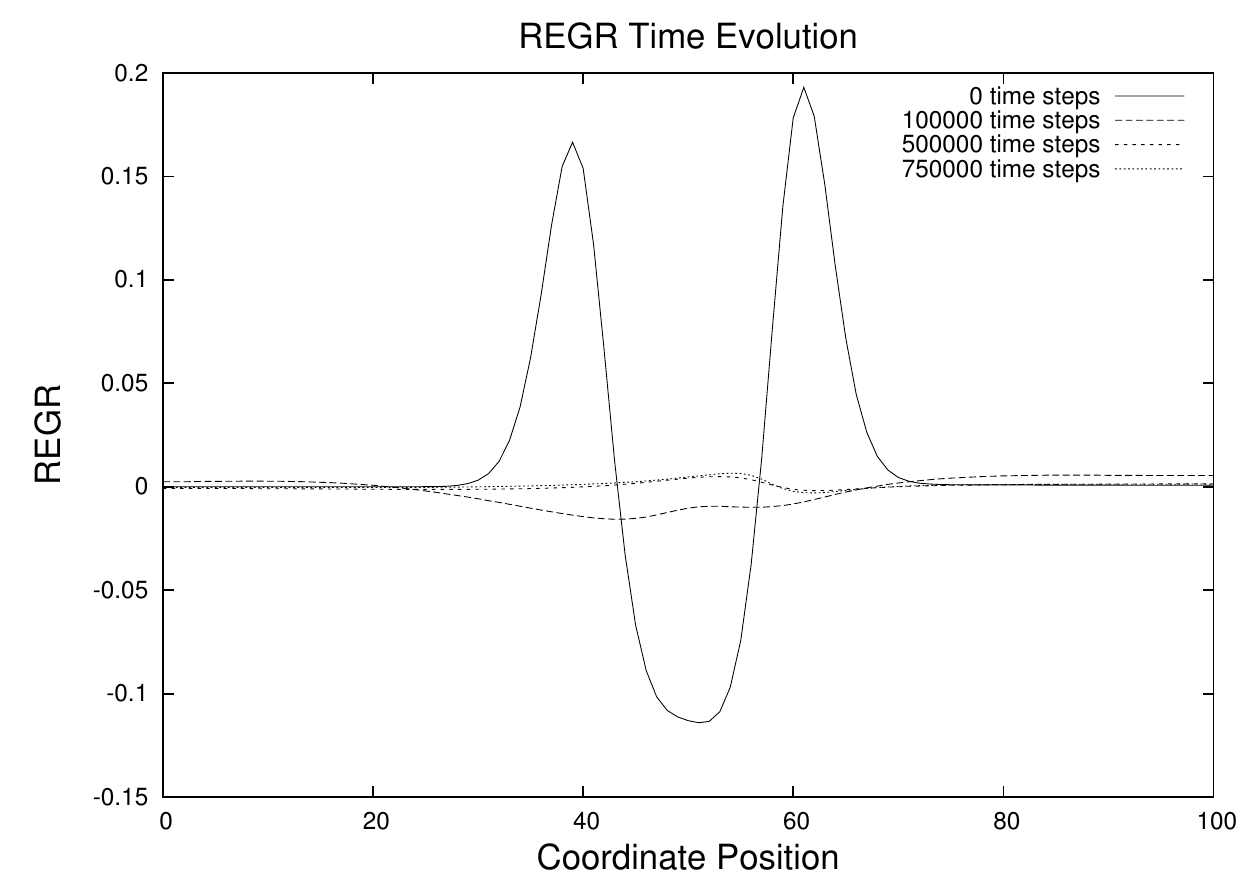}
        }
        \subfloat{
                \includegraphics[width=0.4\textwidth]{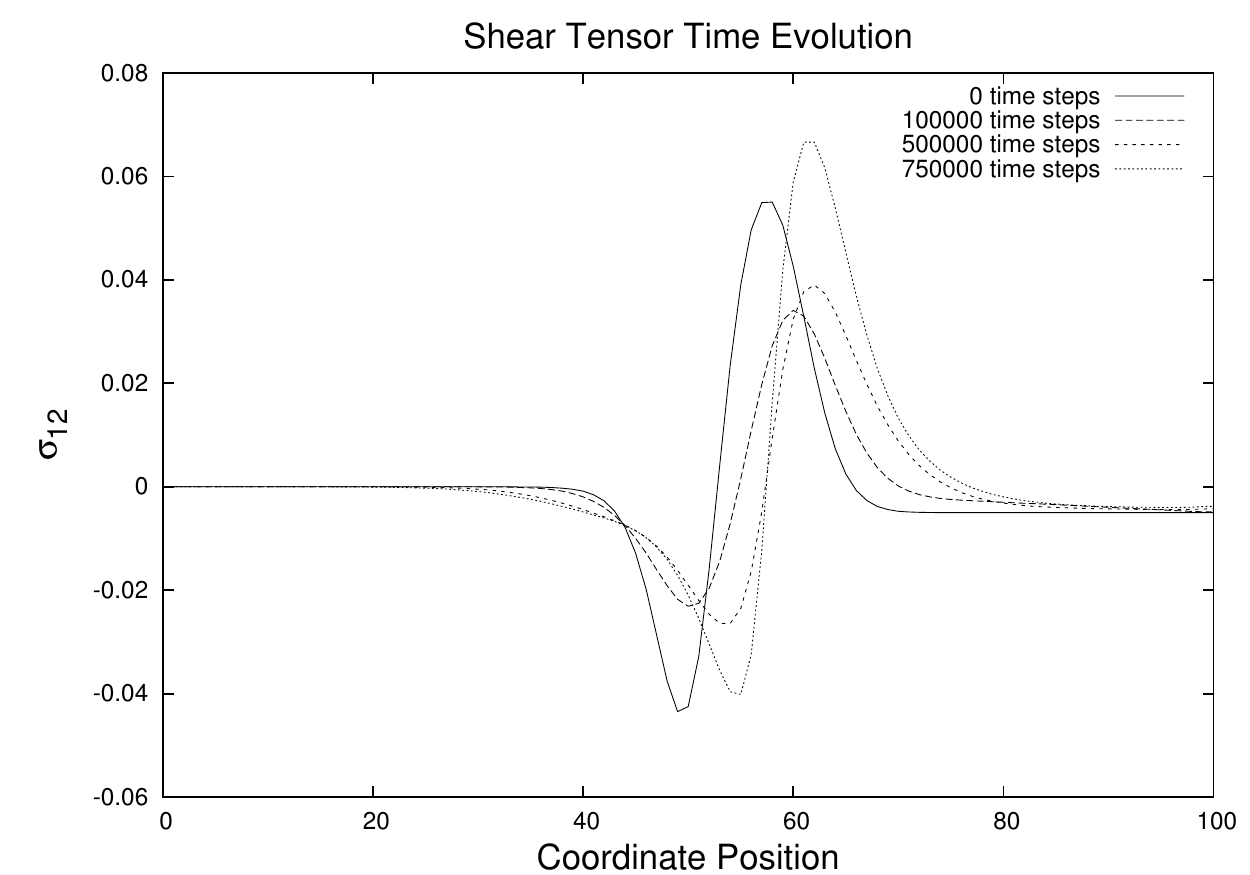}
        }

        \subfloat{
                \includegraphics[width=0.4\textwidth]{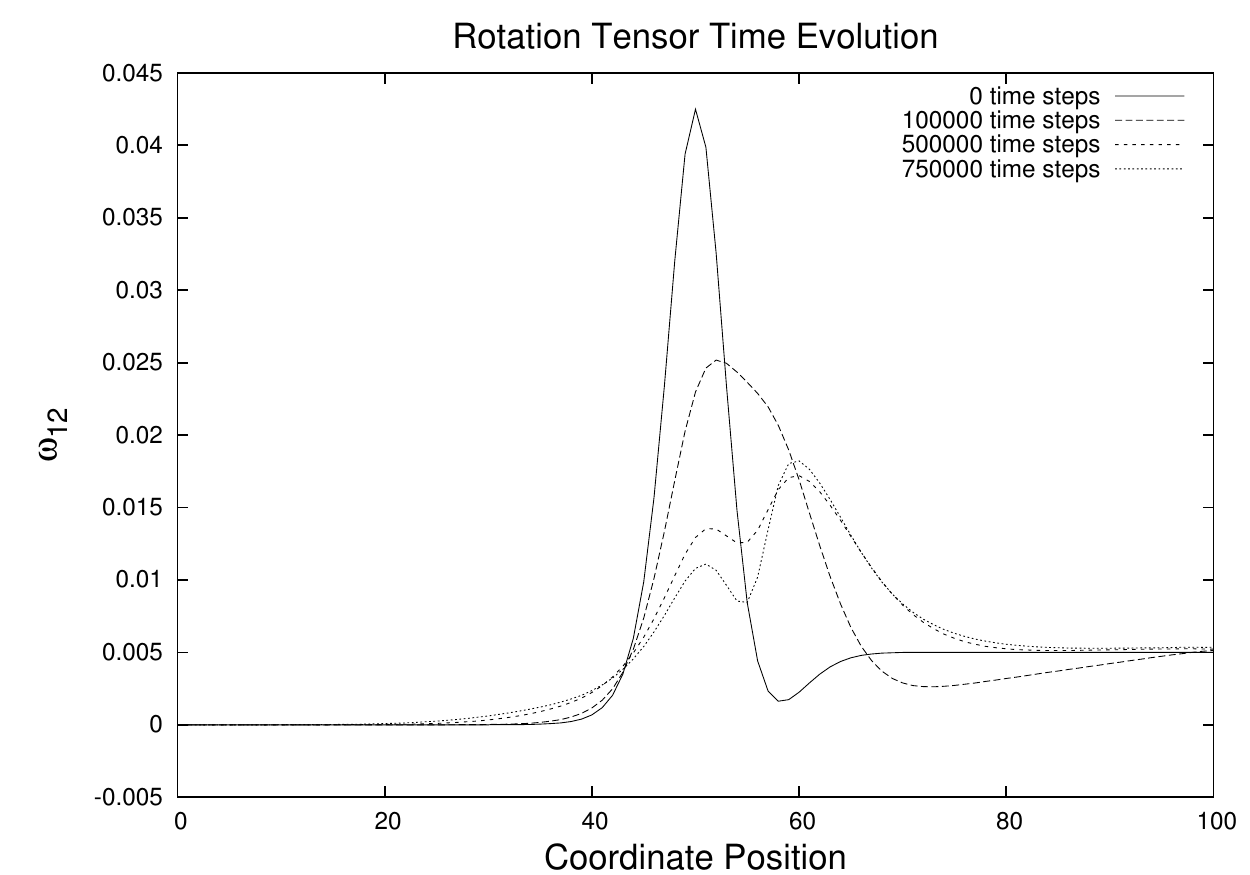}
        }
        \subfloat{
                \includegraphics[width=0.4\textwidth]{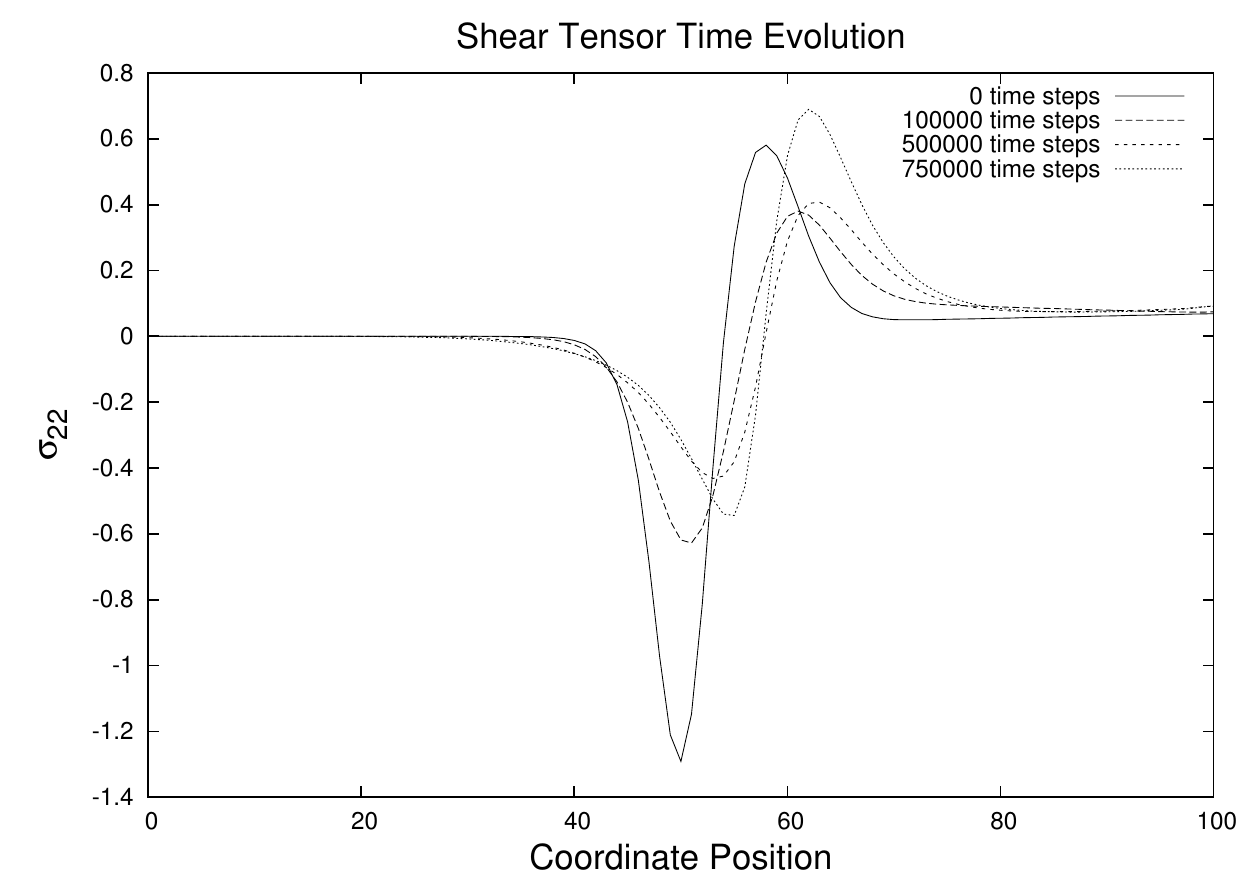}
        }
        \caption{Deformation tensors at $\theta=\pi/2$ for a $2$-gaussian initial metric and a circularly symmetric velocity field.}\label{figure2gaussDefs2} 
\end{figure}

\begin{figure}[t]
    \centering
		\subfloat{
            \includegraphics[width=0.5\textwidth]{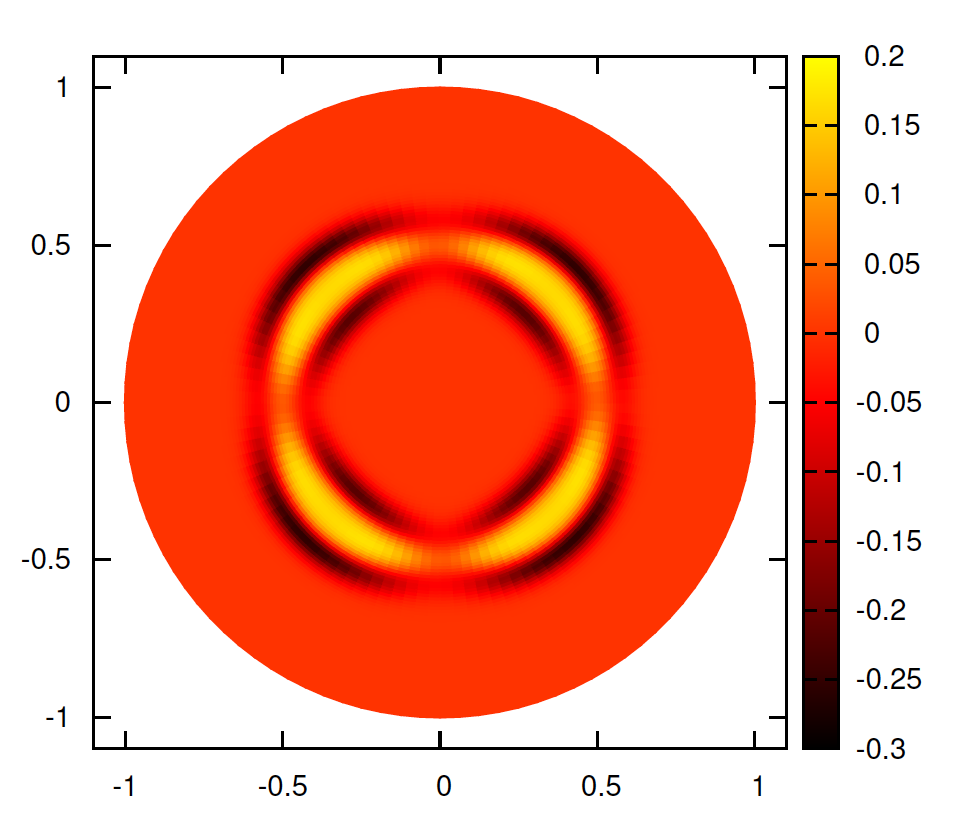}
        }

        \subfloat{                
            \includegraphics[width=\textwidth]{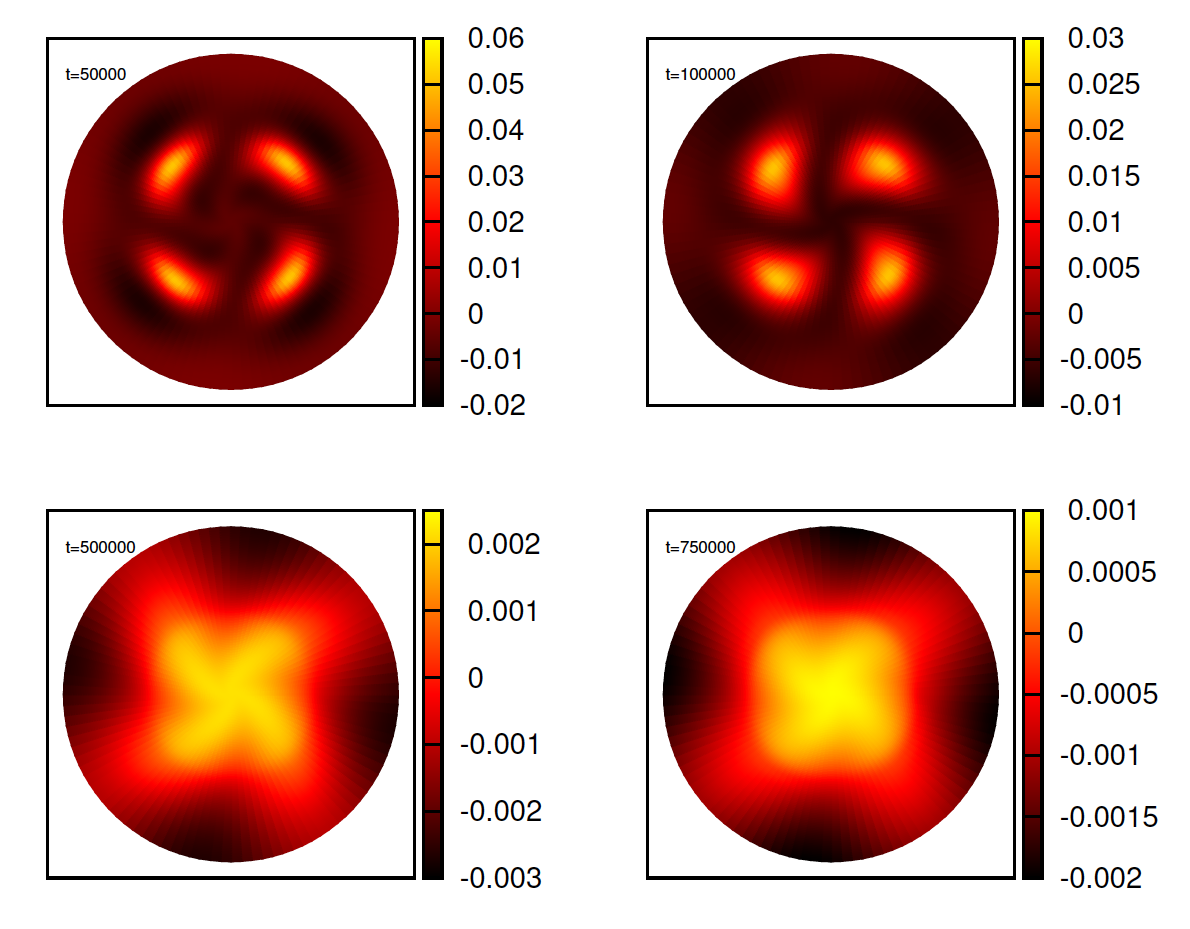}
        }
        \caption{Time evolution of the scalar curvature for a $4$-gaussian initial metric and a circularly symmetric velocity field.}\label{figure4gaussCurvature} 
\end{figure}

\begin{figure}
	\centering
	\includegraphics[width=0.9\textwidth]{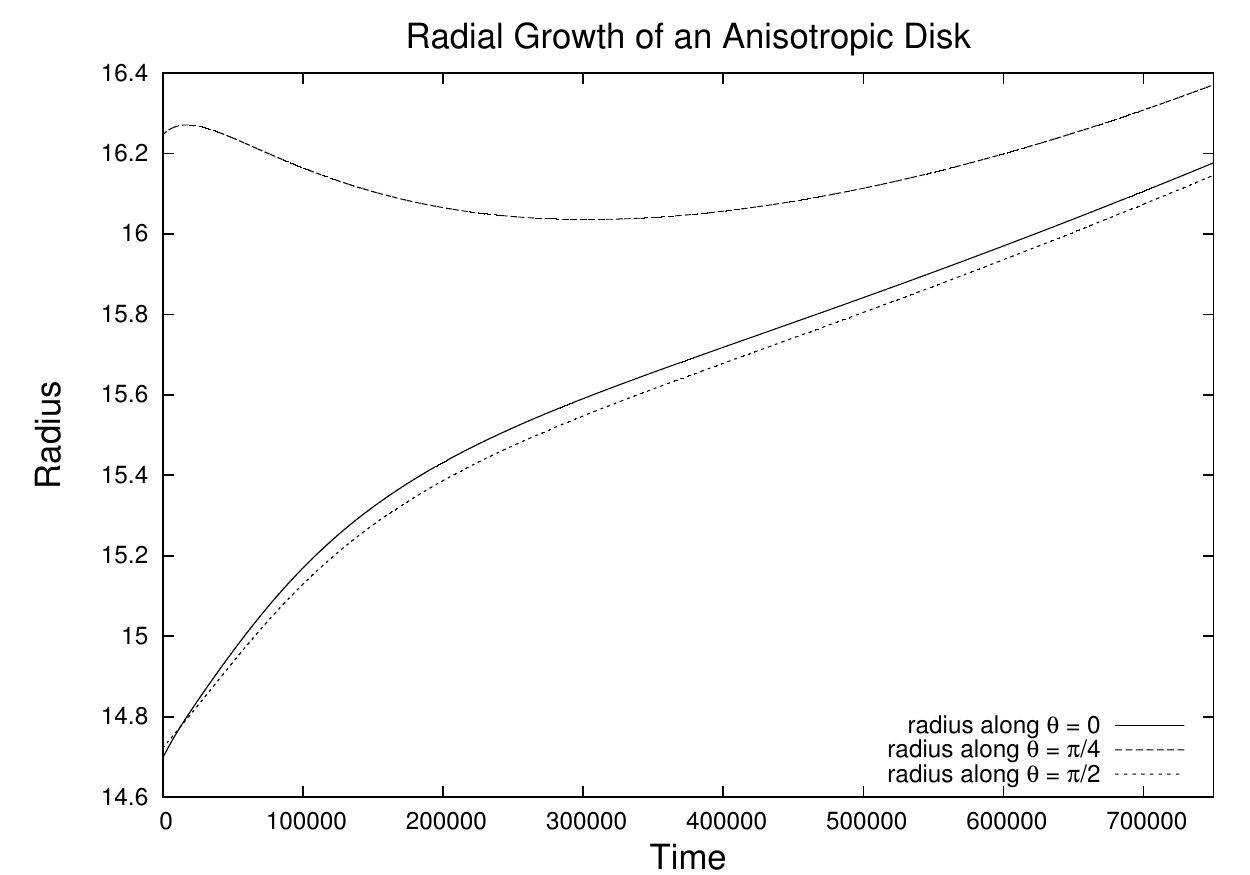}
	\caption{Radial growth of a $4$-gaussian initial metric and a circularly symmetric velocity field along different $\theta$ directions.}
	\label{figure4gaussRadius}
\end{figure}

\begin{figure}
    \centering
        \subfloat{
                \includegraphics[width=0.6\textwidth]{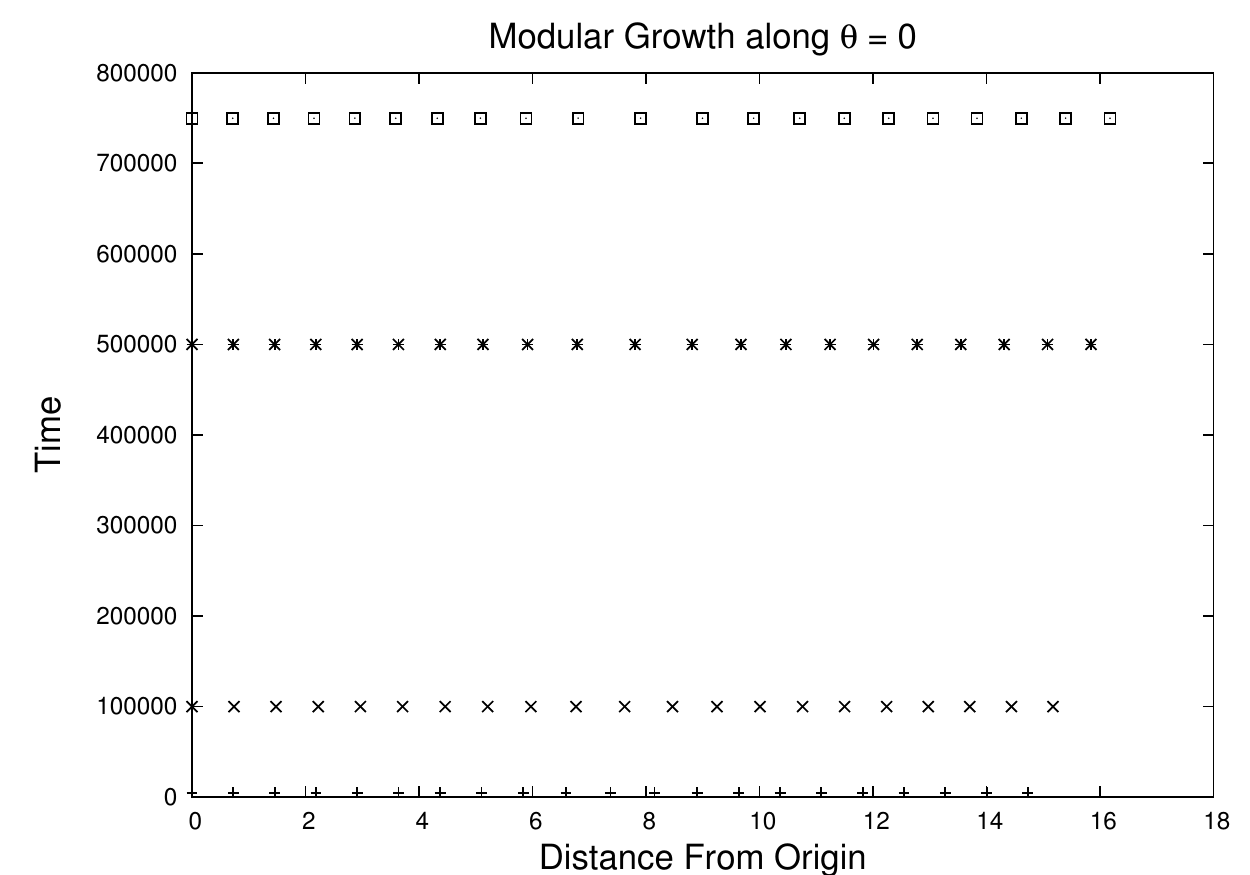}
        }

        \subfloat{
                \includegraphics[width=0.6\textwidth]{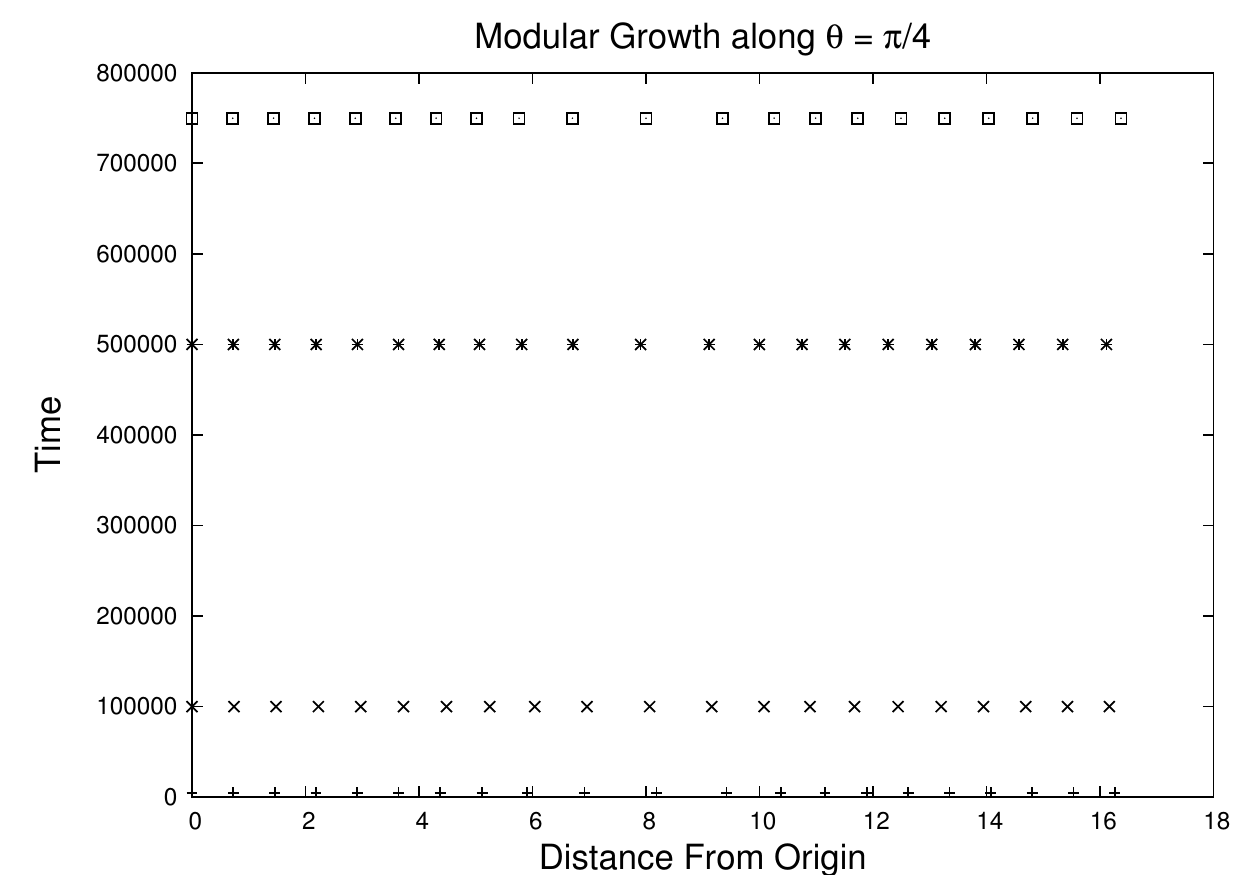}
        }

        \subfloat{
                \includegraphics[width=0.6\textwidth]{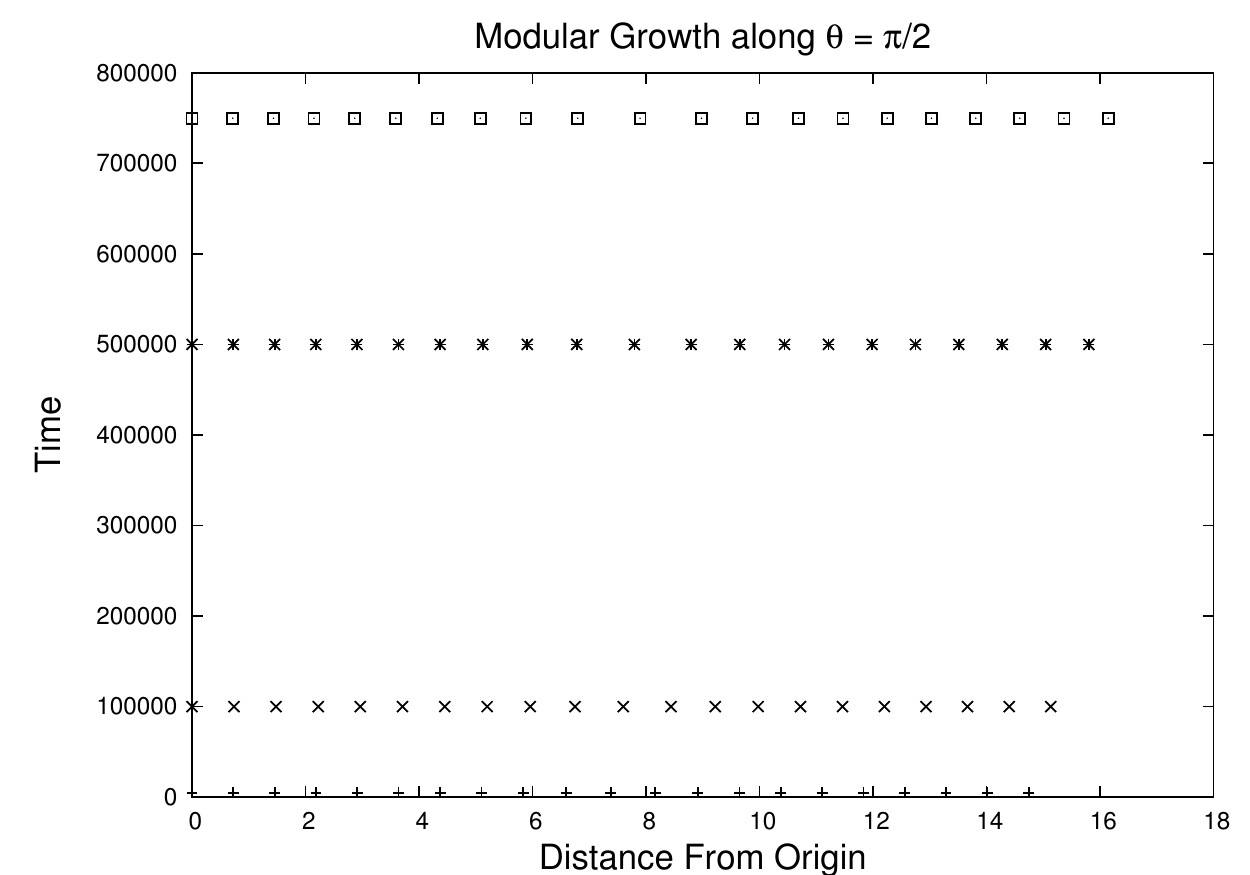}
        }
        \caption{Modular growth at different $\theta$ directions for a $4$-gaussian initial metric and a circularly symmetric velocity field.}\label{figure4gaussModular} 
\end{figure}

\begin{figure}
    \centering
        \subfloat{
                \includegraphics[width=0.4\textwidth]{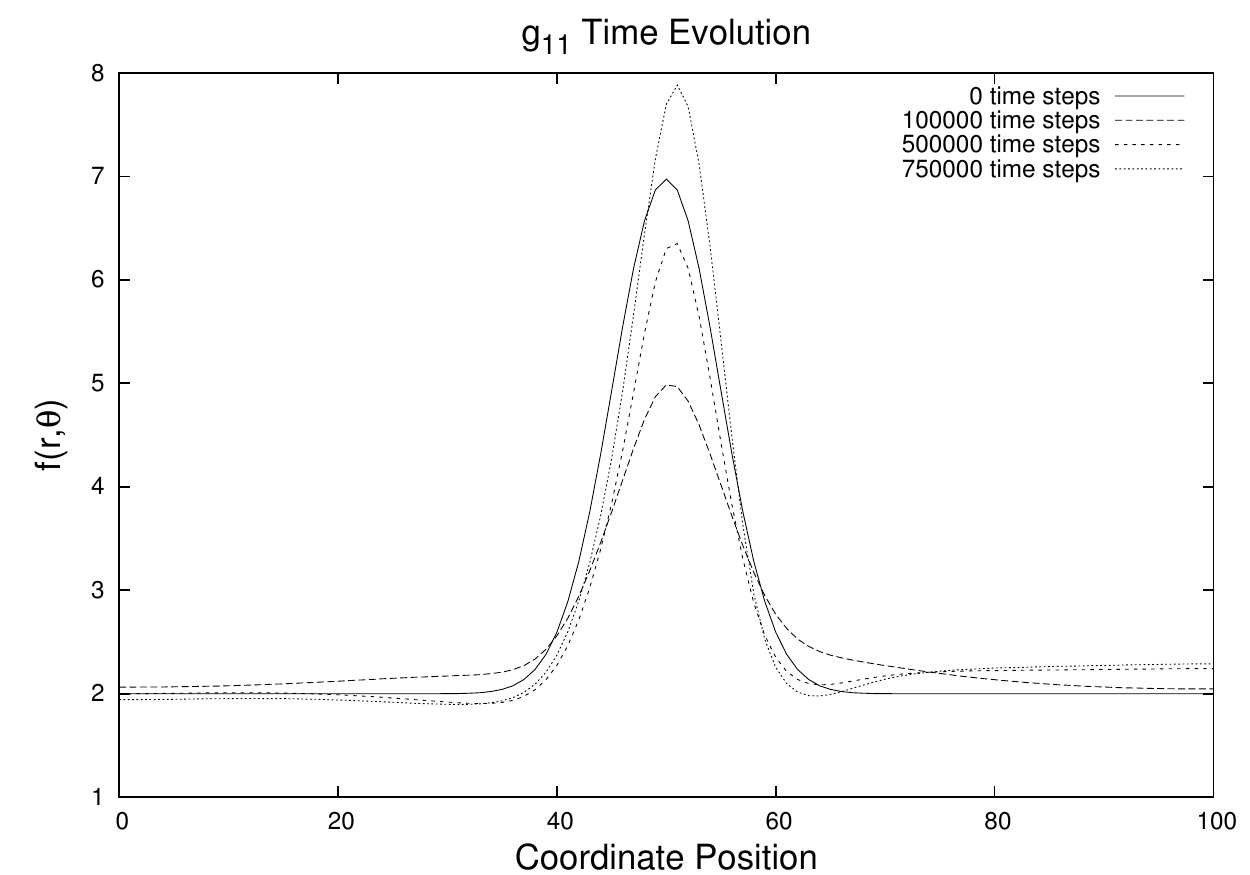}
        }
        \subfloat{
                \includegraphics[width=0.4\textwidth]{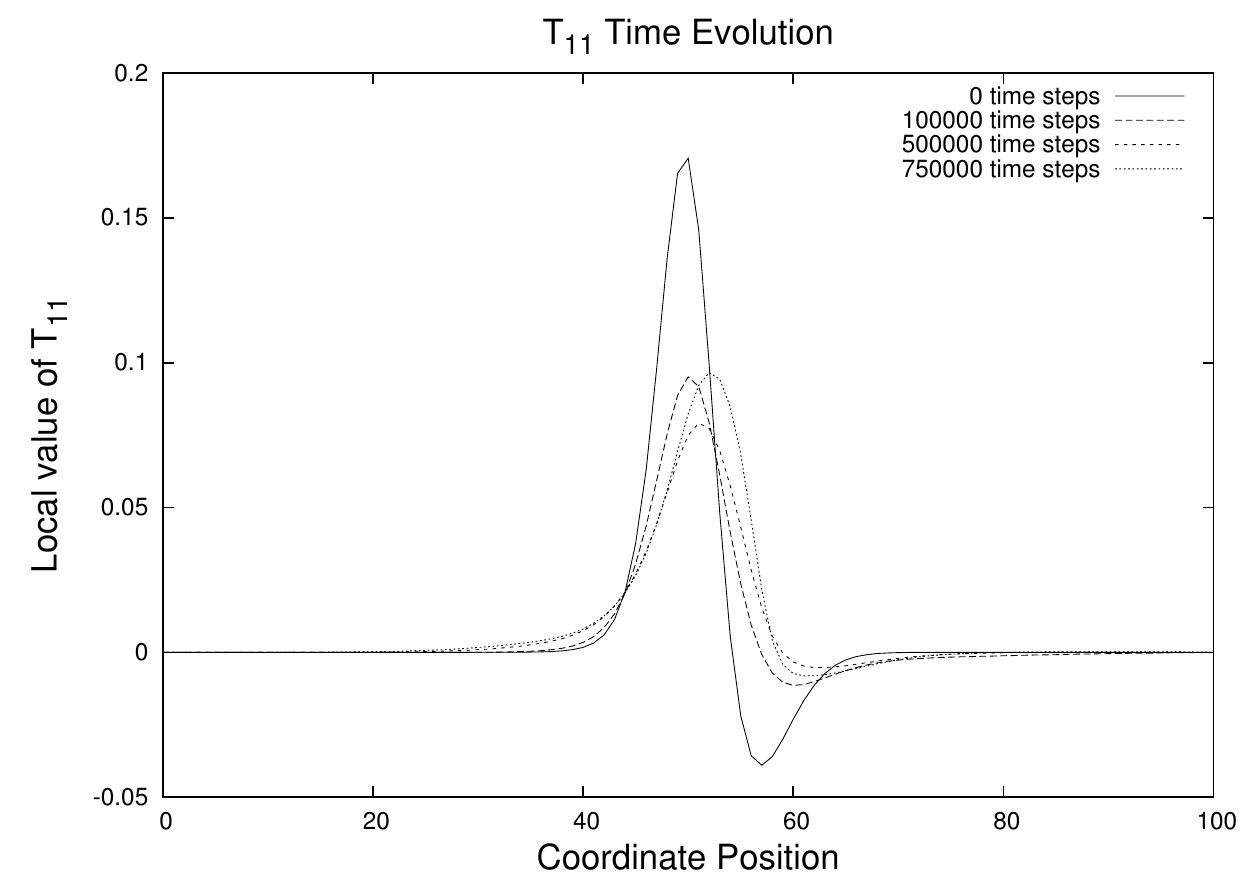}
        }

        \subfloat{
                \includegraphics[width=0.4\textwidth]{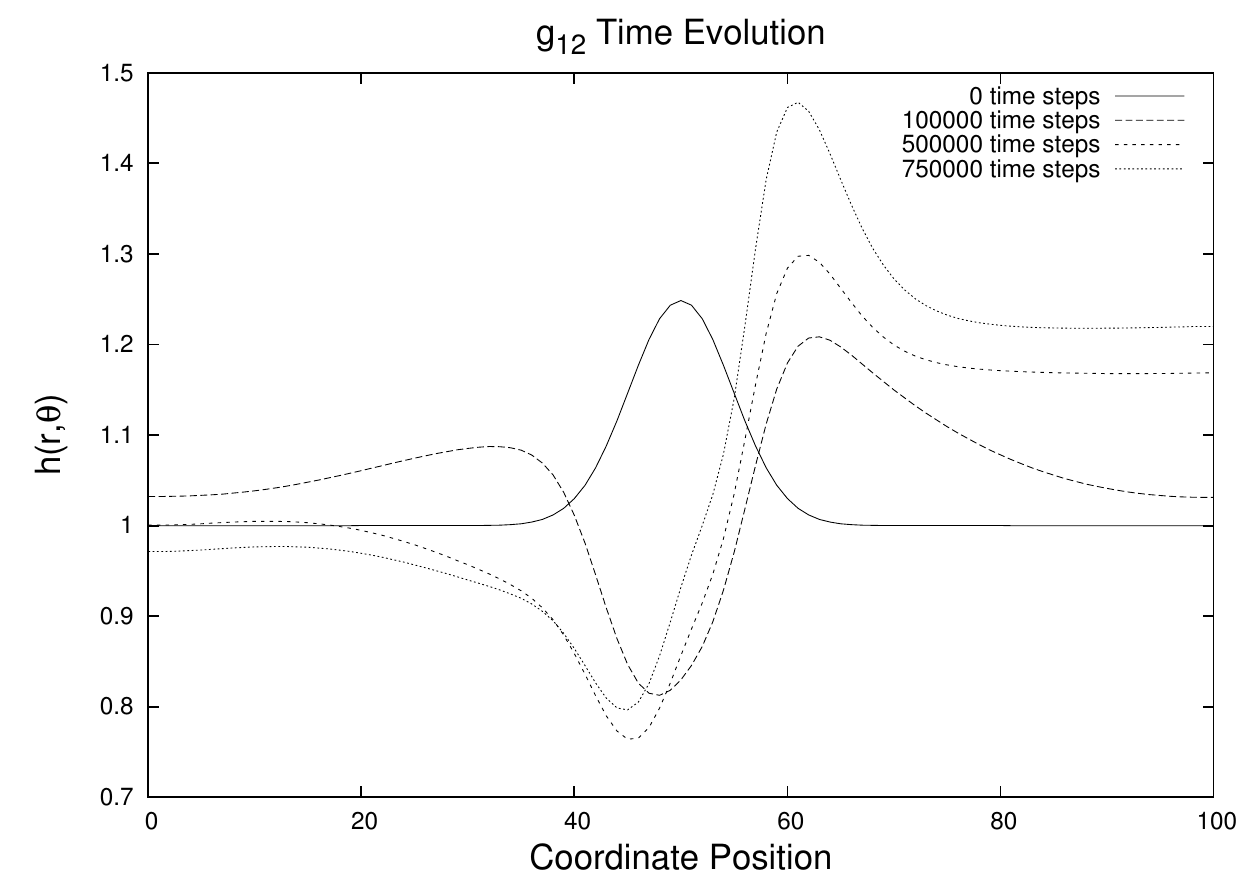}
        }
        \subfloat{
                \includegraphics[width=0.4\textwidth]{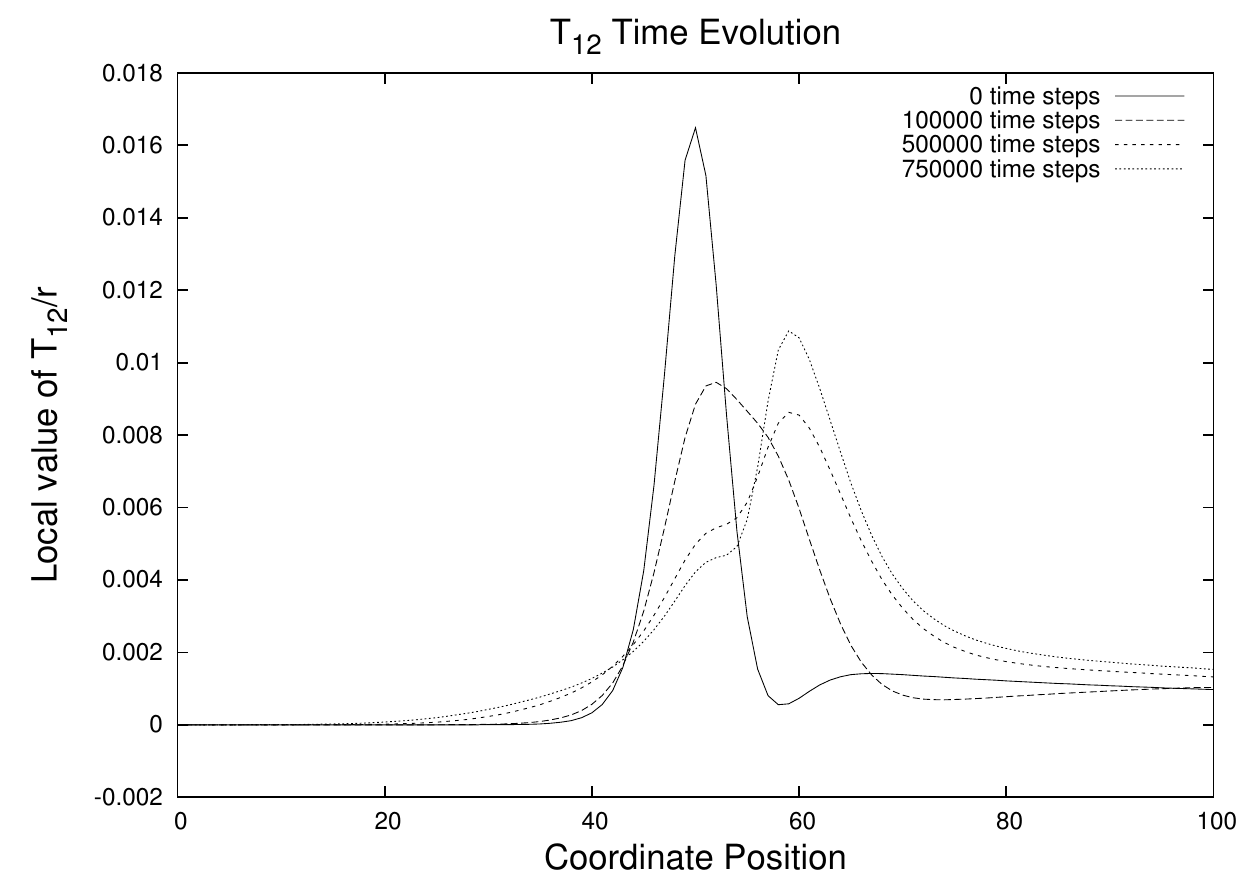}
        }

        \subfloat{
                \includegraphics[width=0.4\textwidth]{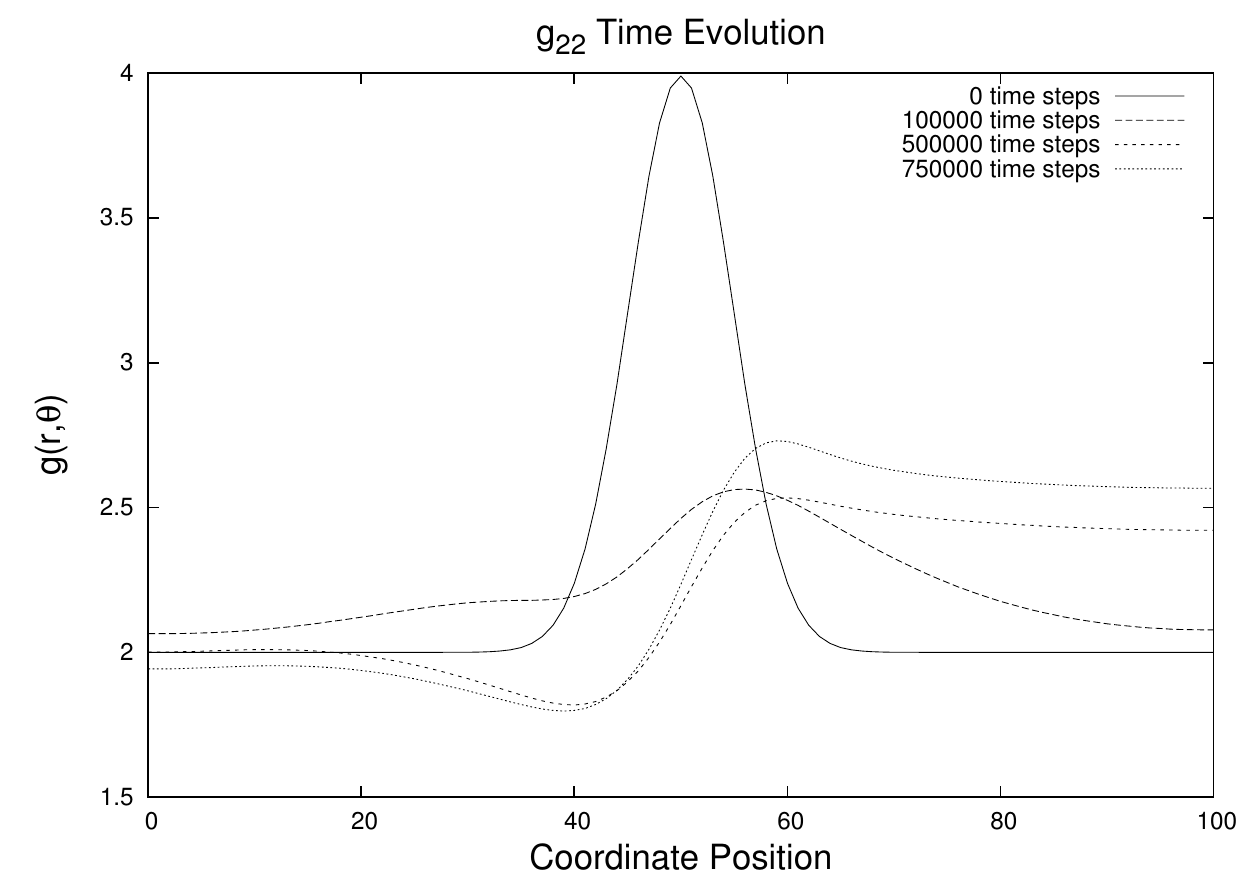}
        }
        \subfloat{
                \includegraphics[width=0.4\textwidth]{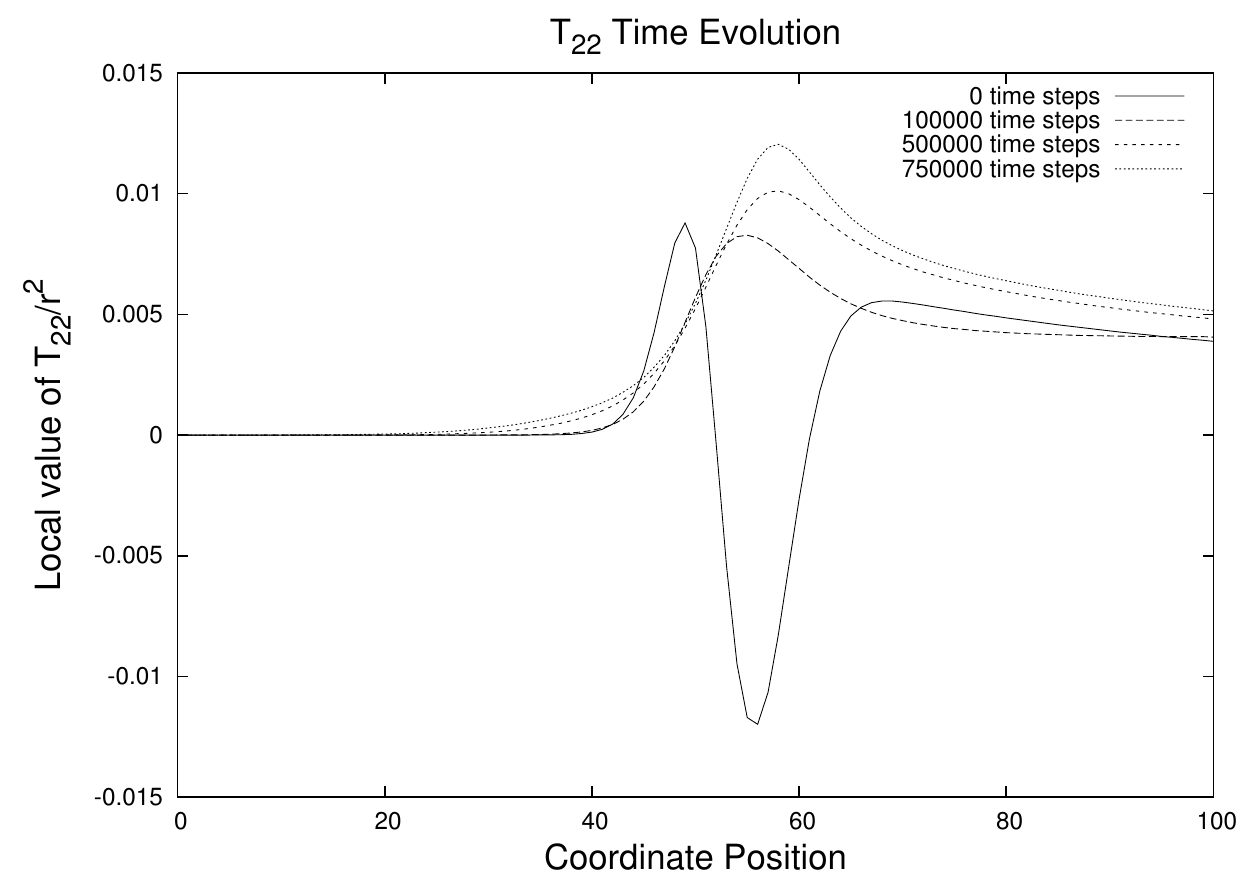}
        }
        
        \subfloat{
                \includegraphics[width=0.4\textwidth]{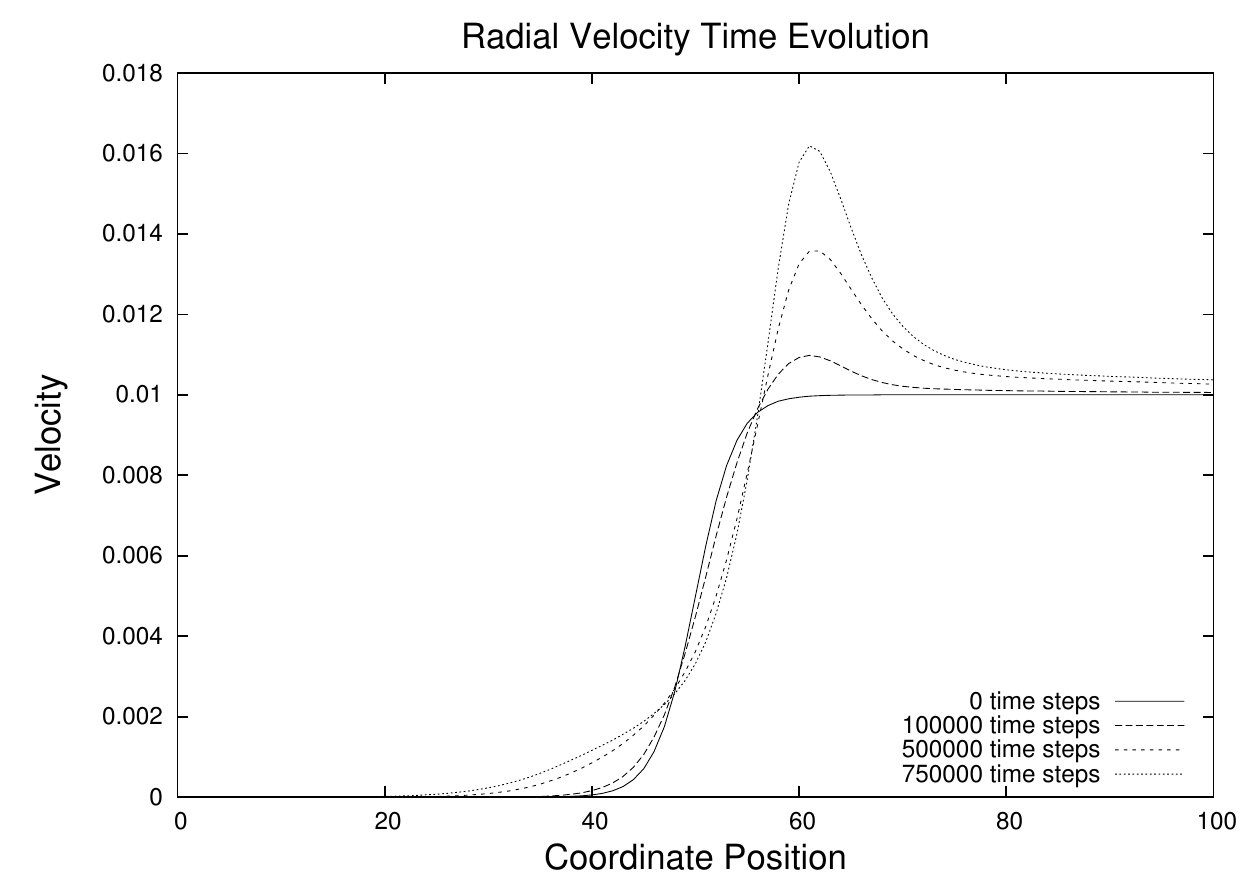}
        }
        \subfloat{
                \includegraphics[width=0.4\textwidth]{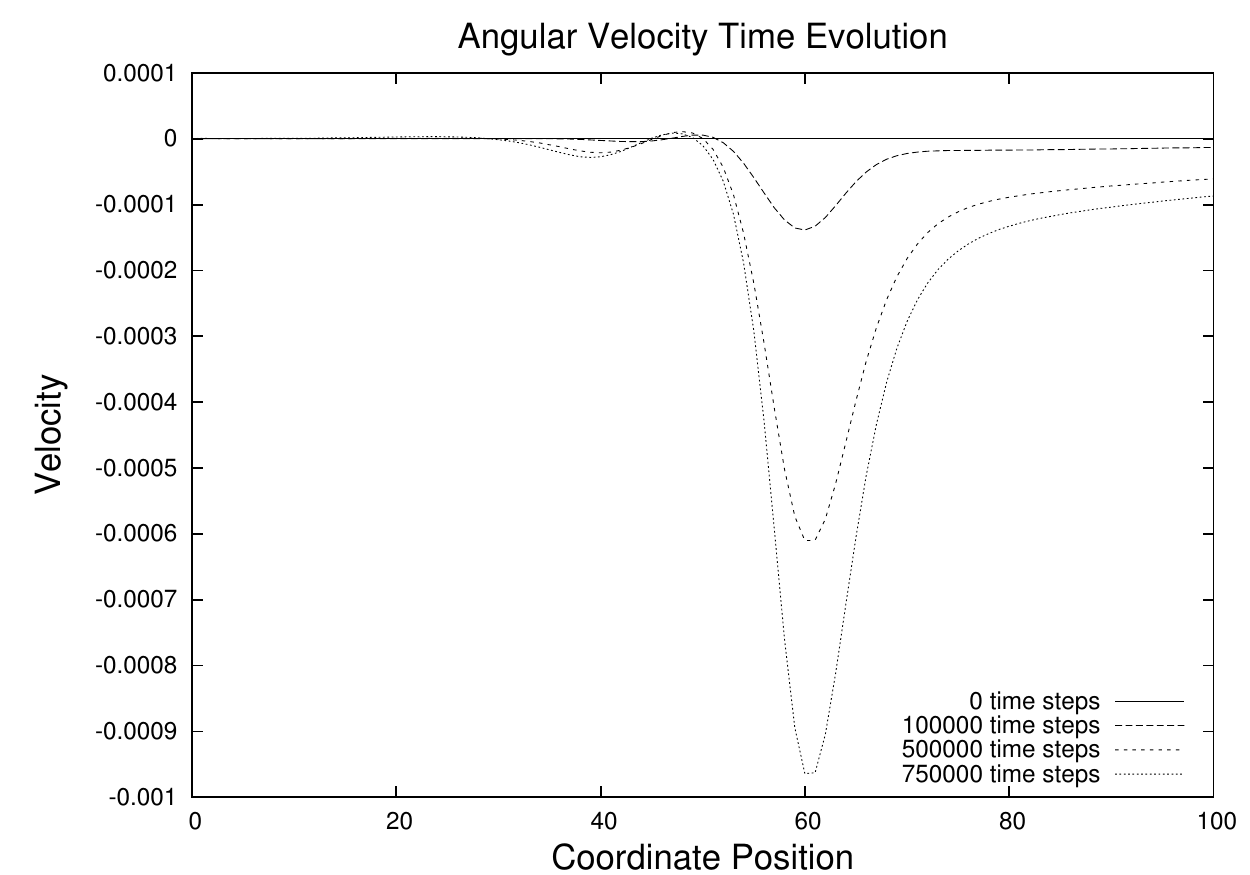}
        }
        \caption{Dynamics of the metric and velocity field at $\theta=\pi/4$ for a $4$-gaussian initial metric and a circularly symmetric velocity field.}\label{figure4gaussDynamics4} 
\end{figure}

\begin{figure}
    \centering
        \subfloat{
                \includegraphics[width=0.4\textwidth]{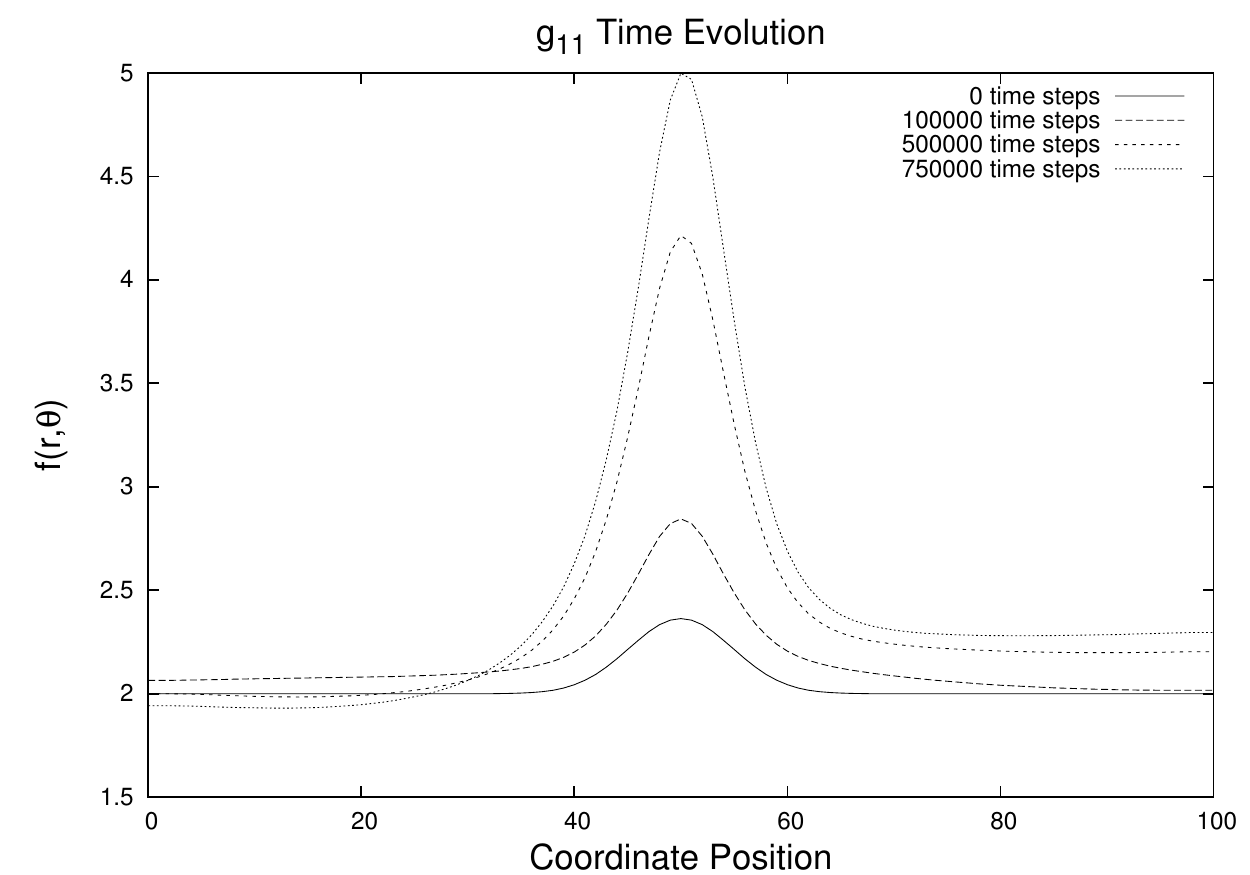}
        }
        \subfloat{
                \includegraphics[width=0.4\textwidth]{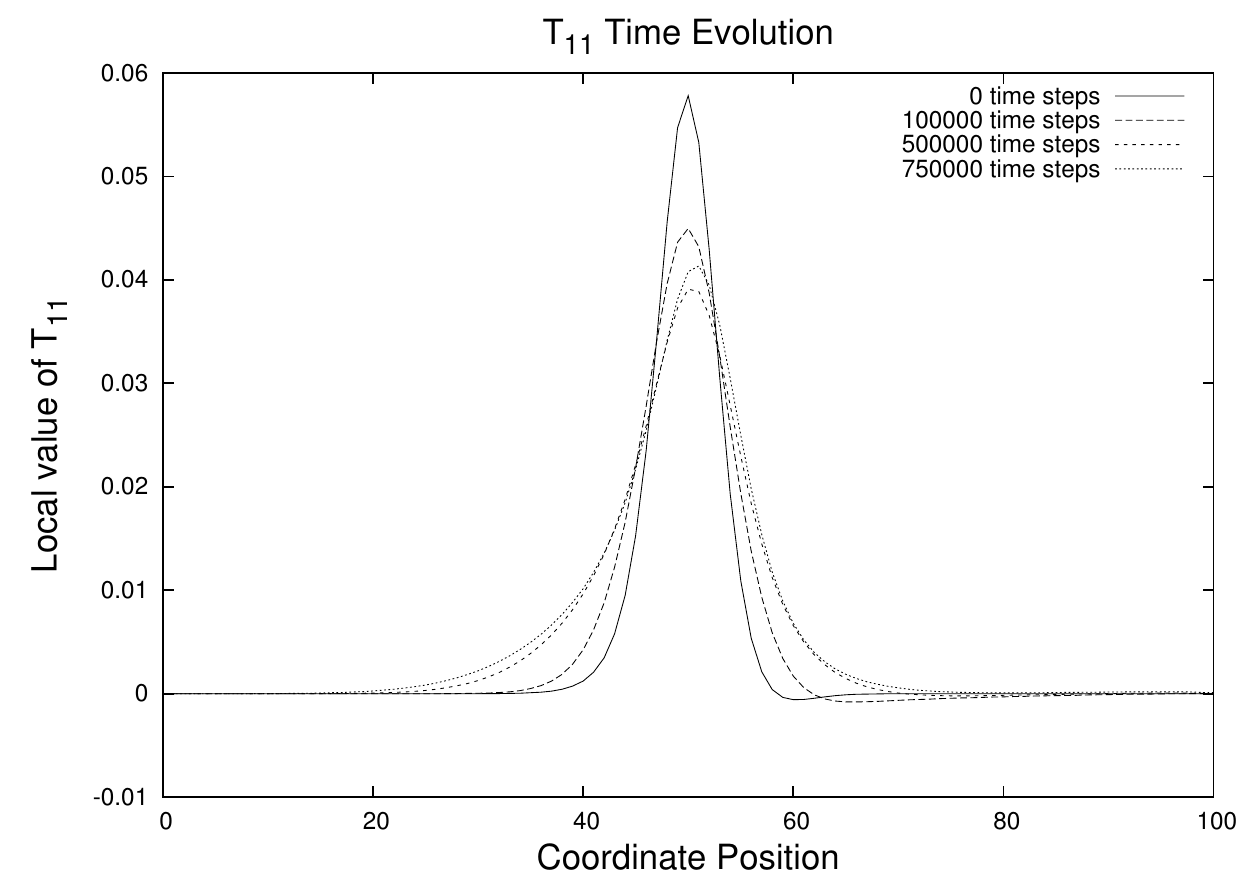}
        }

        \subfloat{
                \includegraphics[width=0.4\textwidth]{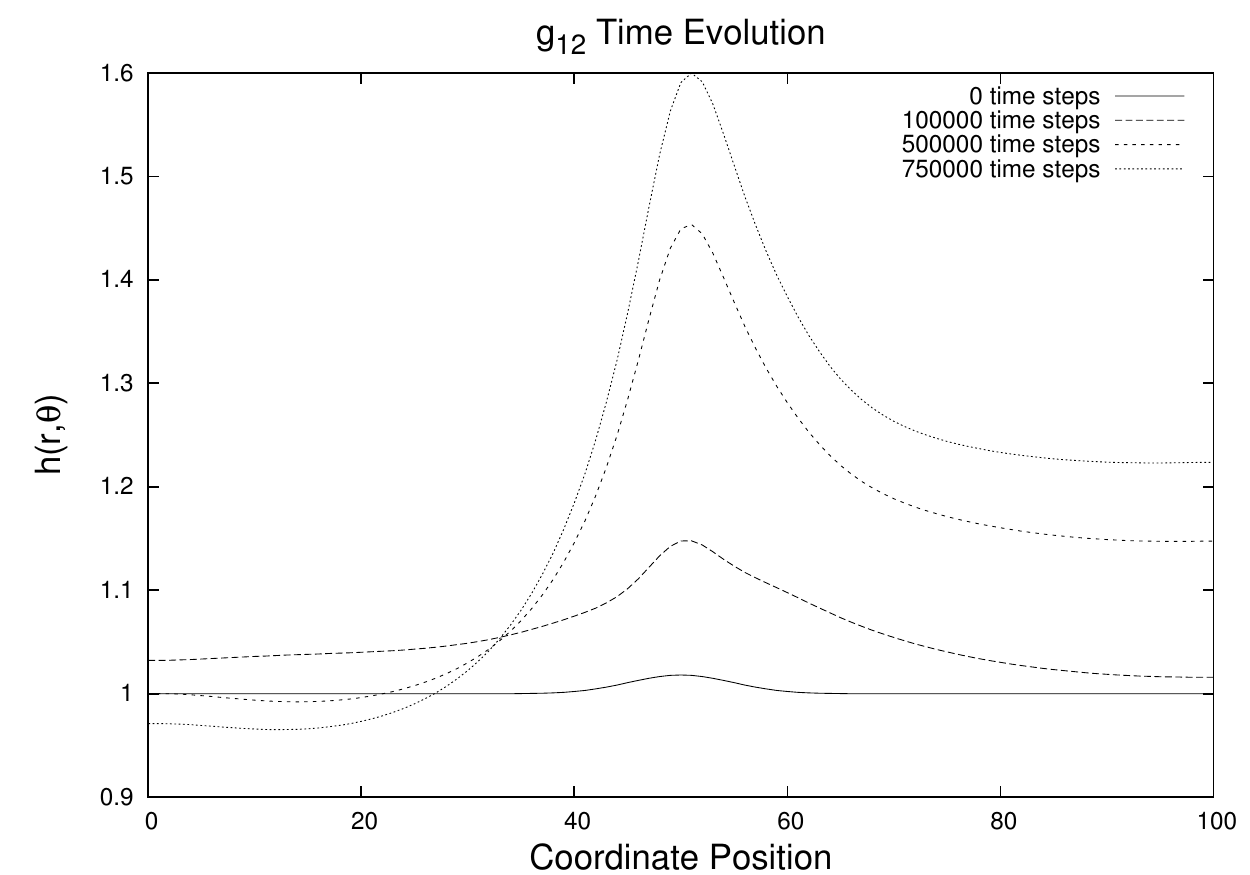}
        }
        \subfloat{
                \includegraphics[width=0.4\textwidth]{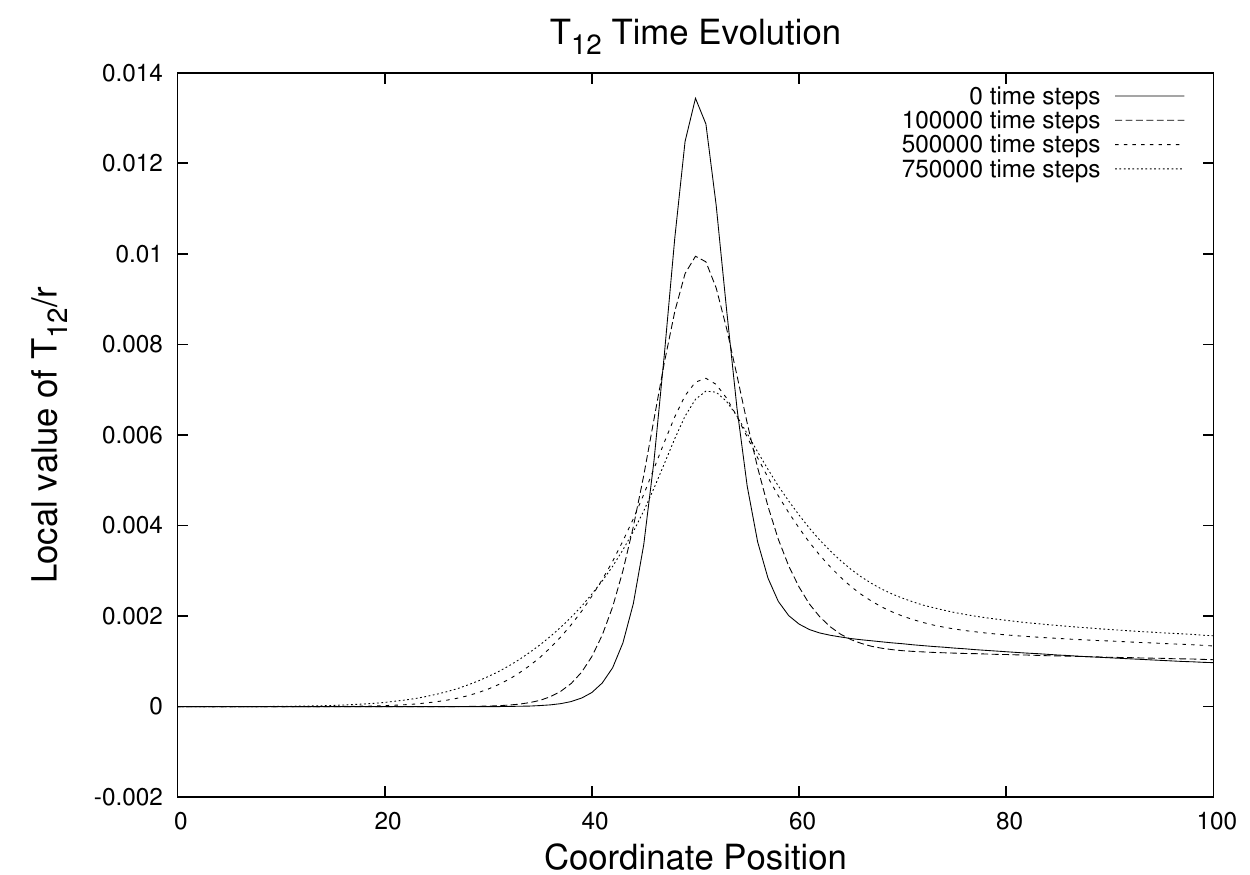}
        }

        \subfloat{
                \includegraphics[width=0.4\textwidth]{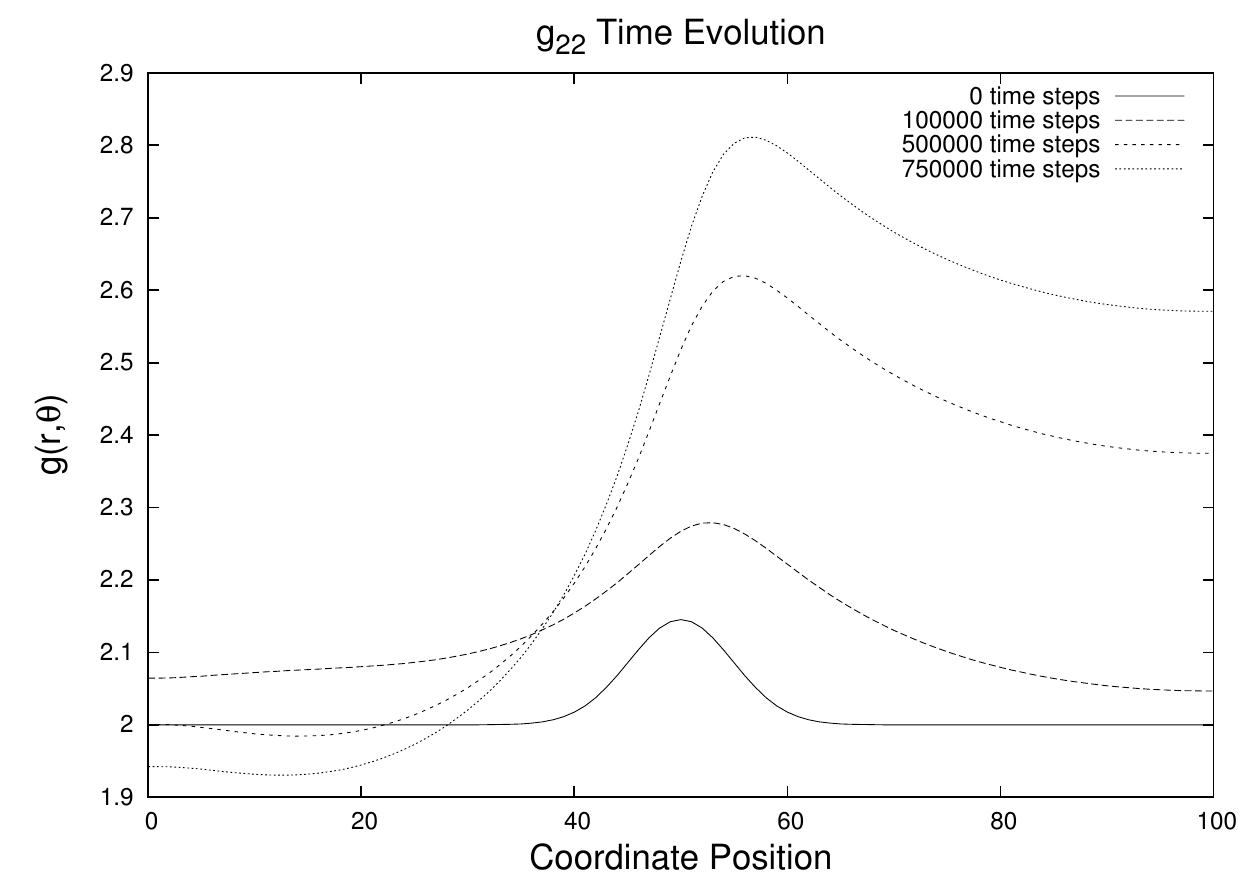}
        }
        \subfloat{
                \includegraphics[width=0.4\textwidth]{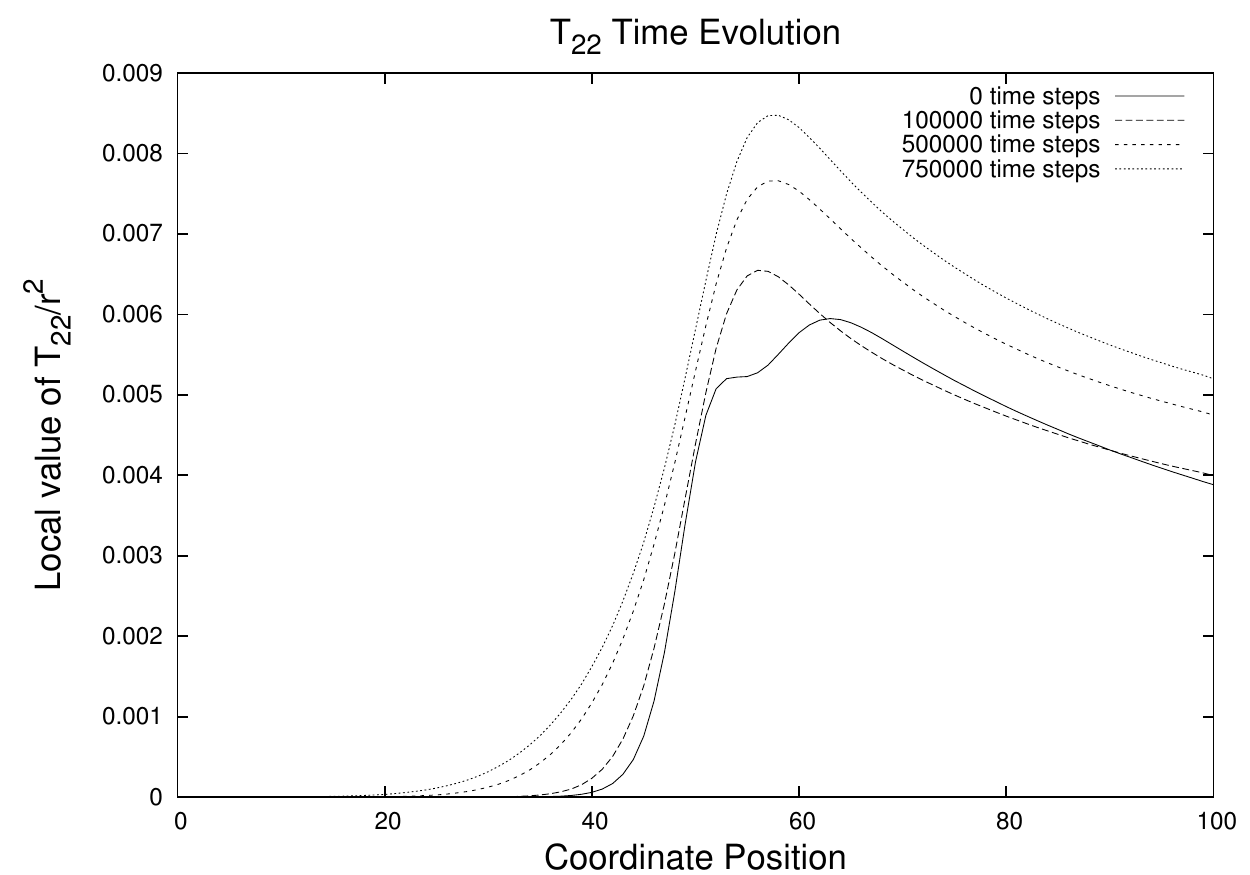}
        }
        
        \subfloat{
                \includegraphics[width=0.4\textwidth]{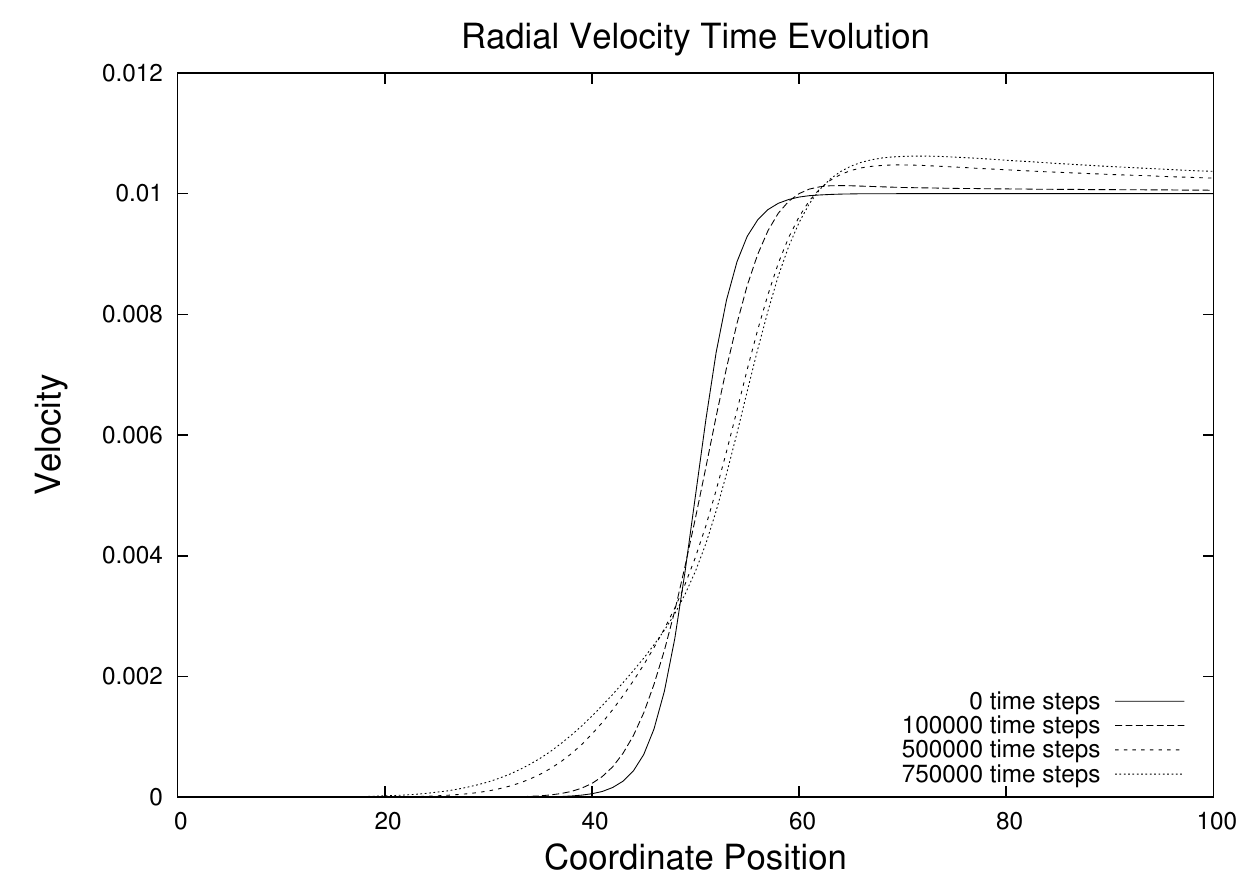}
        }
        \subfloat{
                \includegraphics[width=0.4\textwidth]{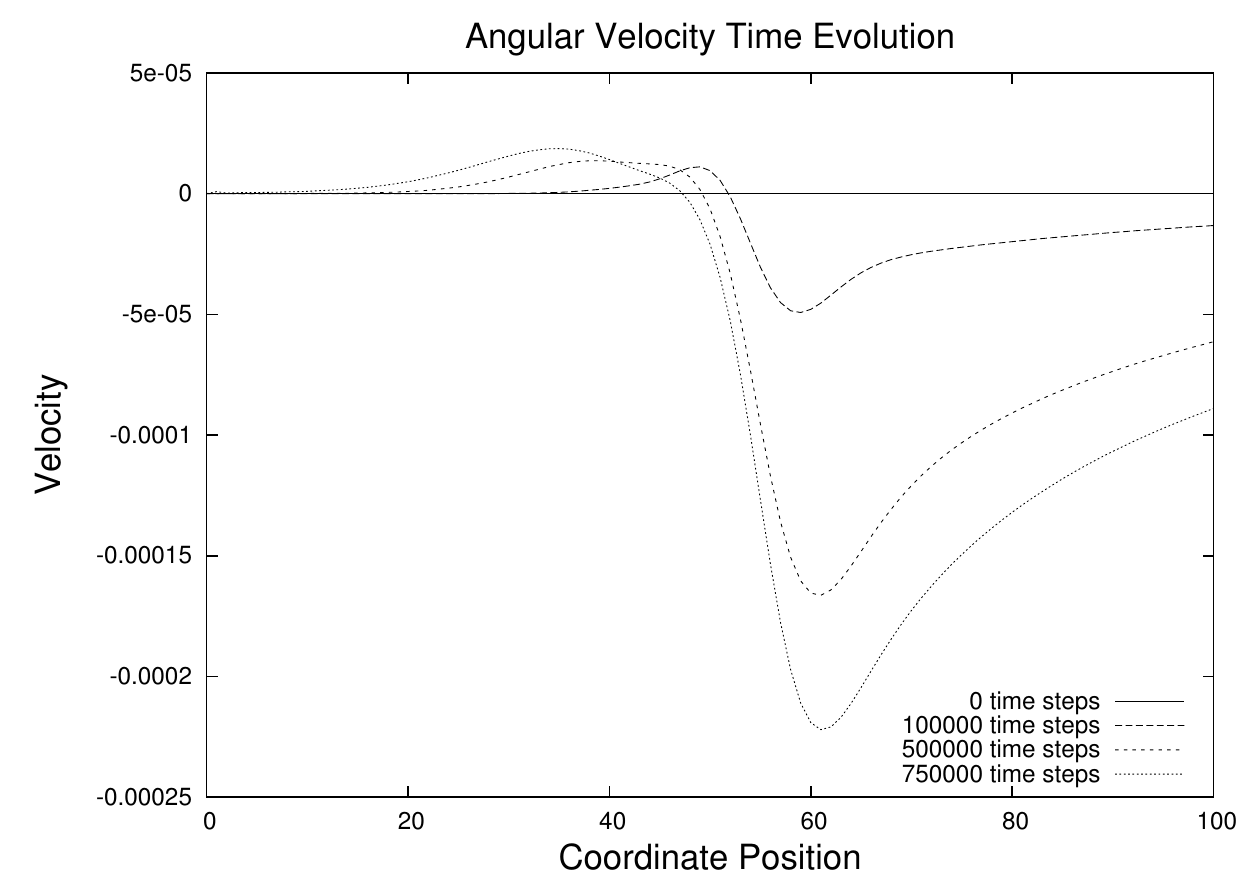}
        }
        \caption{Dynamics of the metric and velocity field at $\theta=\pi/2$ for a $4$-gaussian initial metric and a circularly symmetric velocity field.}\label{figure4gaussDynamics2} 
\end{figure}

\begin{figure}
    \centering
        \subfloat{
                \includegraphics[width=0.4\textwidth]{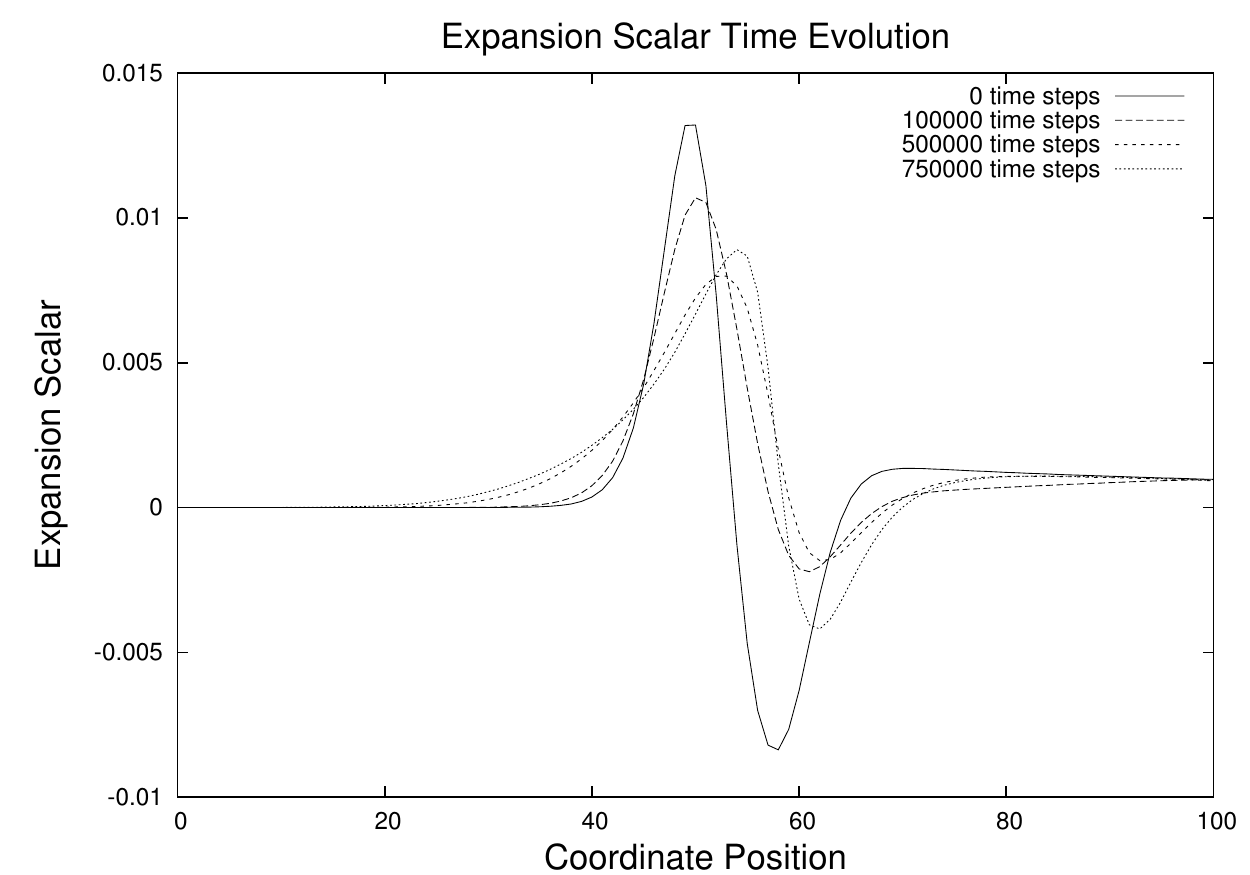}
        }
        \subfloat{
                \includegraphics[width=0.4\textwidth]{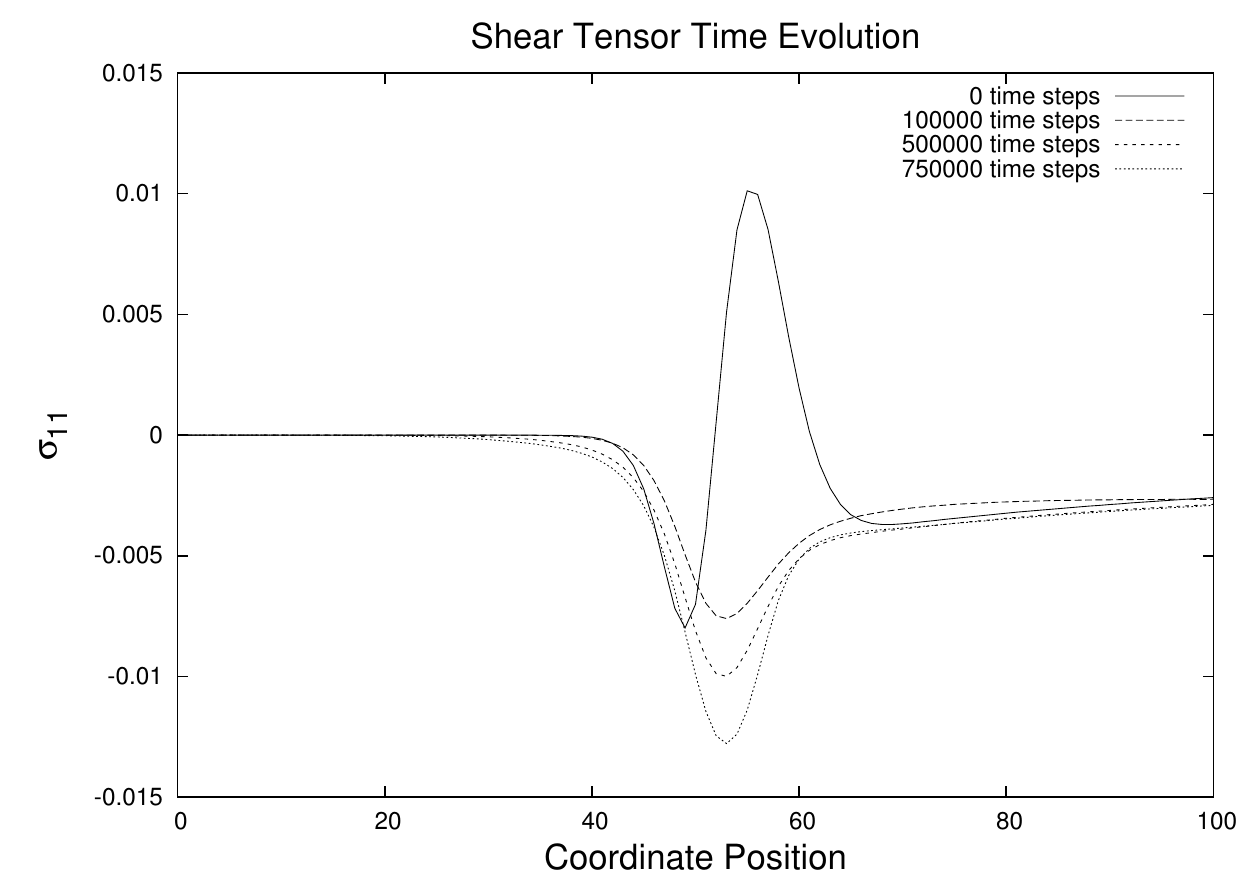}
        }

        \subfloat{
                \includegraphics[width=0.4\textwidth]{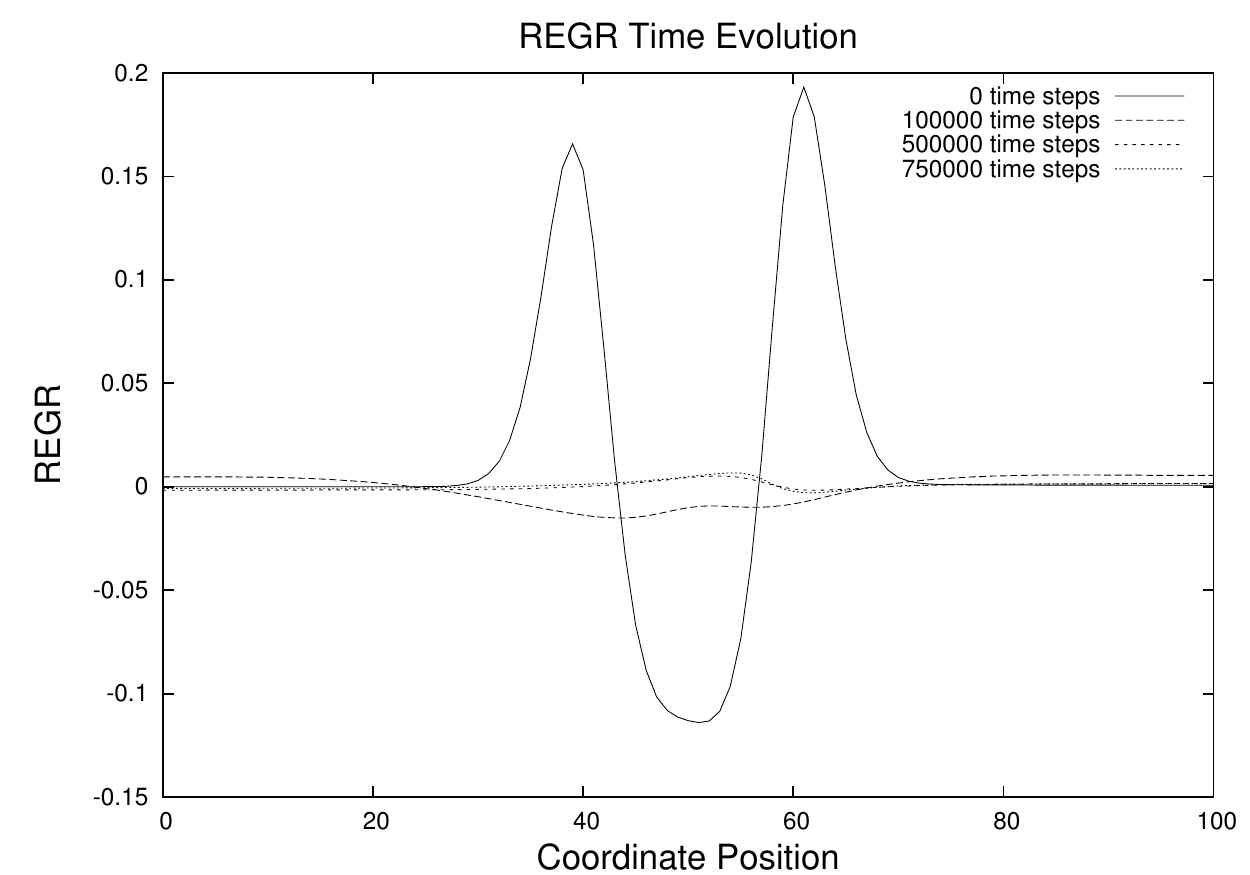}
        }
        \subfloat{
                \includegraphics[width=0.4\textwidth]{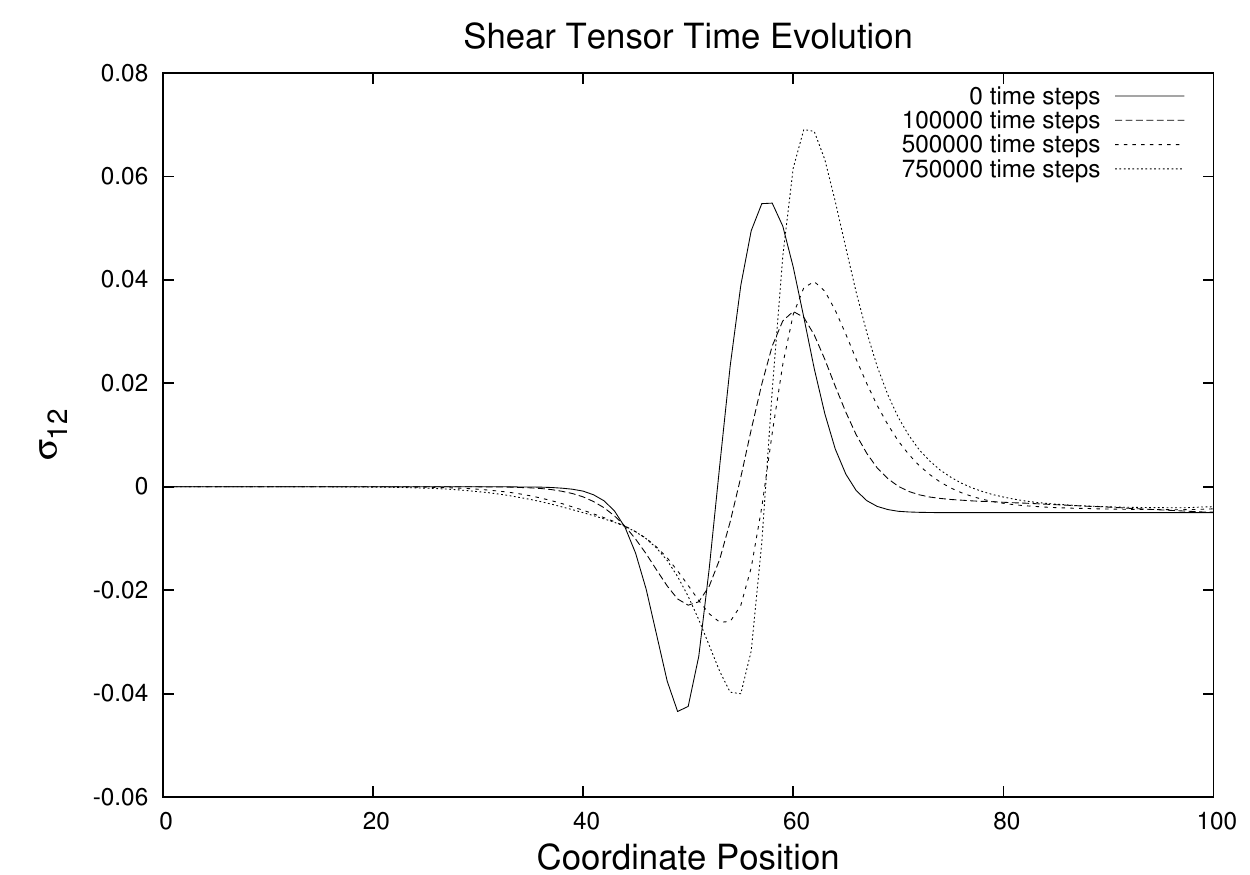}
        }

        \subfloat{
                \includegraphics[width=0.4\textwidth]{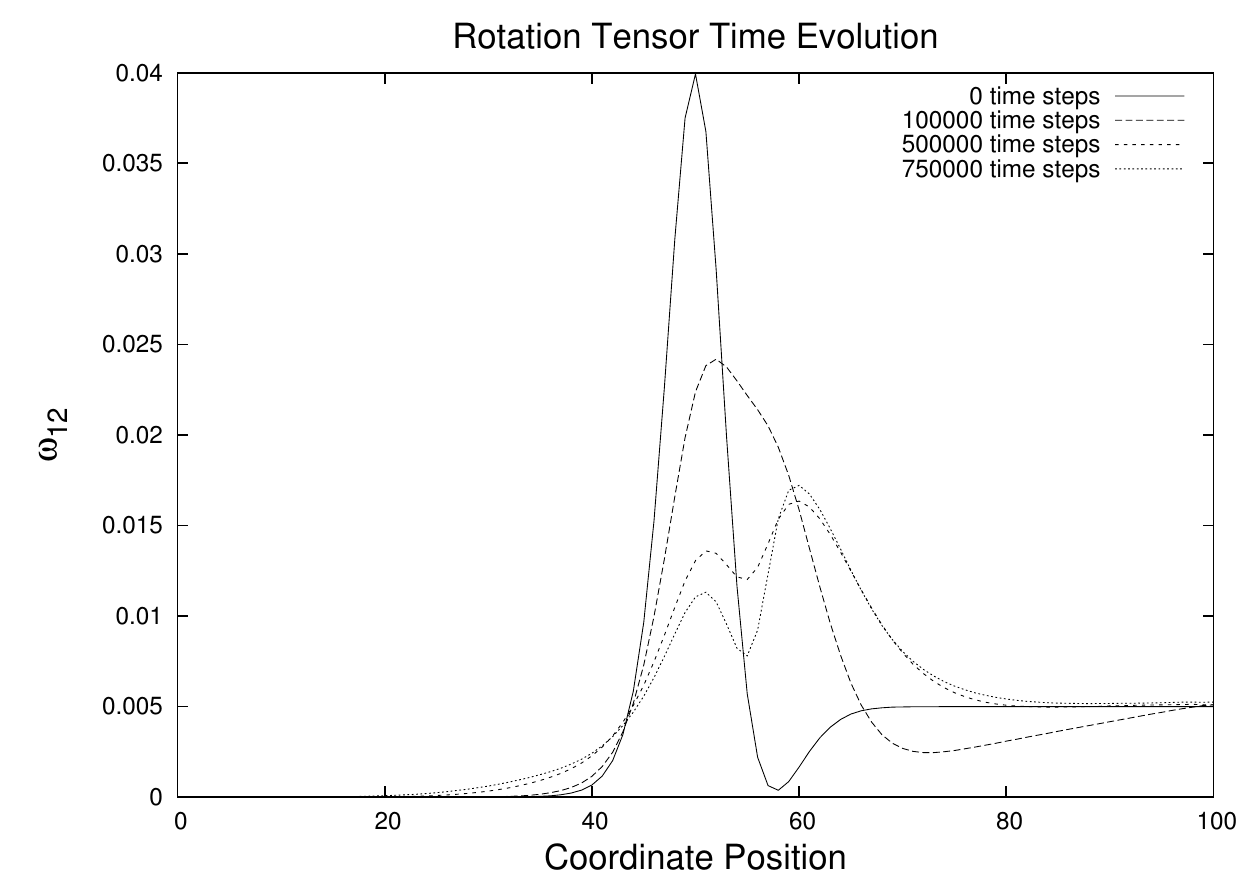}
        }
        \subfloat{
                \includegraphics[width=0.4\textwidth]{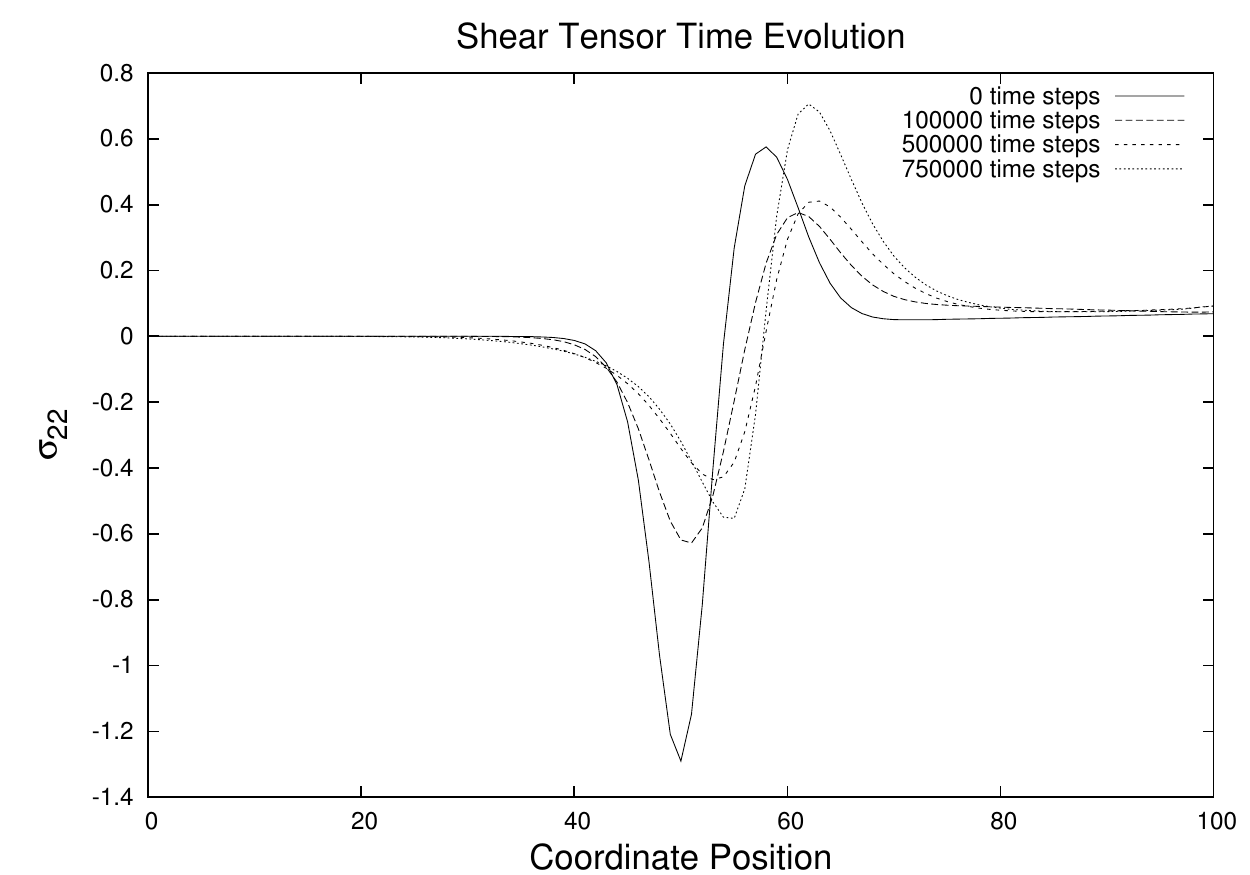}
        }
        \caption{Deformation tensors at $\theta=\pi/4$ for a $4$-gaussian initial metric and a circularly symmetric velocity field.}\label{figure4gaussDefs4} 
\end{figure}

\begin{figure}
    \centering
        \subfloat{
                \includegraphics[width=0.4\textwidth]{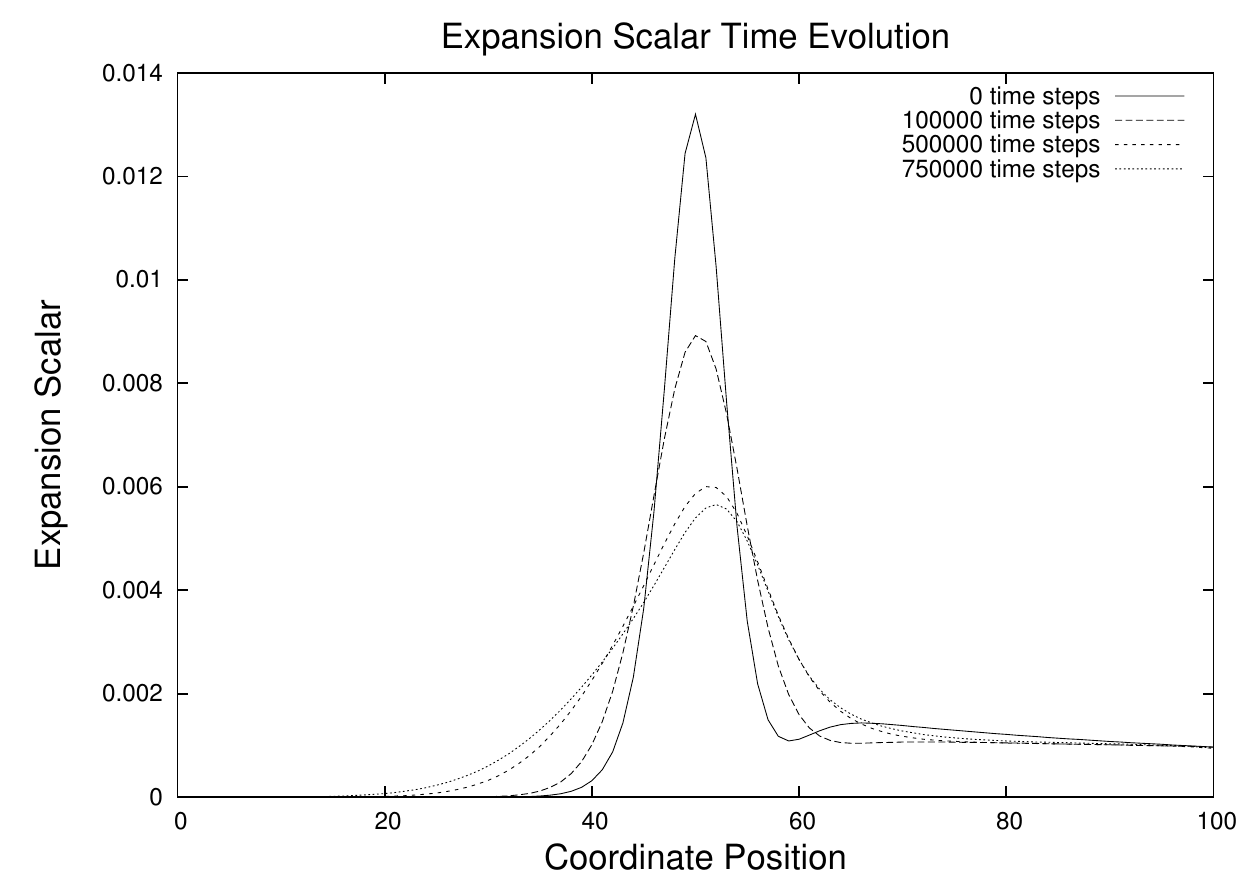}
        }
        \subfloat{
                \includegraphics[width=0.4\textwidth]{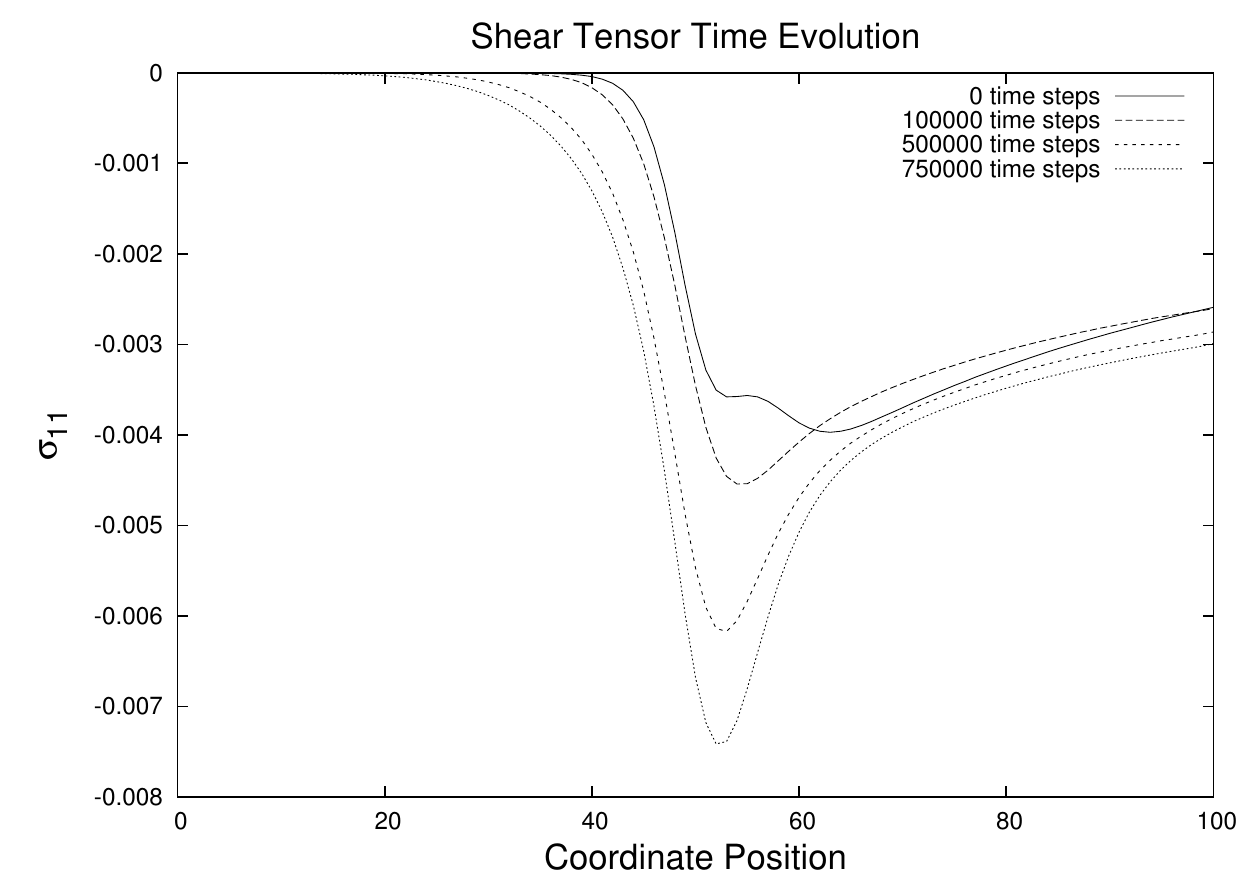}
        }

        \subfloat{
                \includegraphics[width=0.4\textwidth]{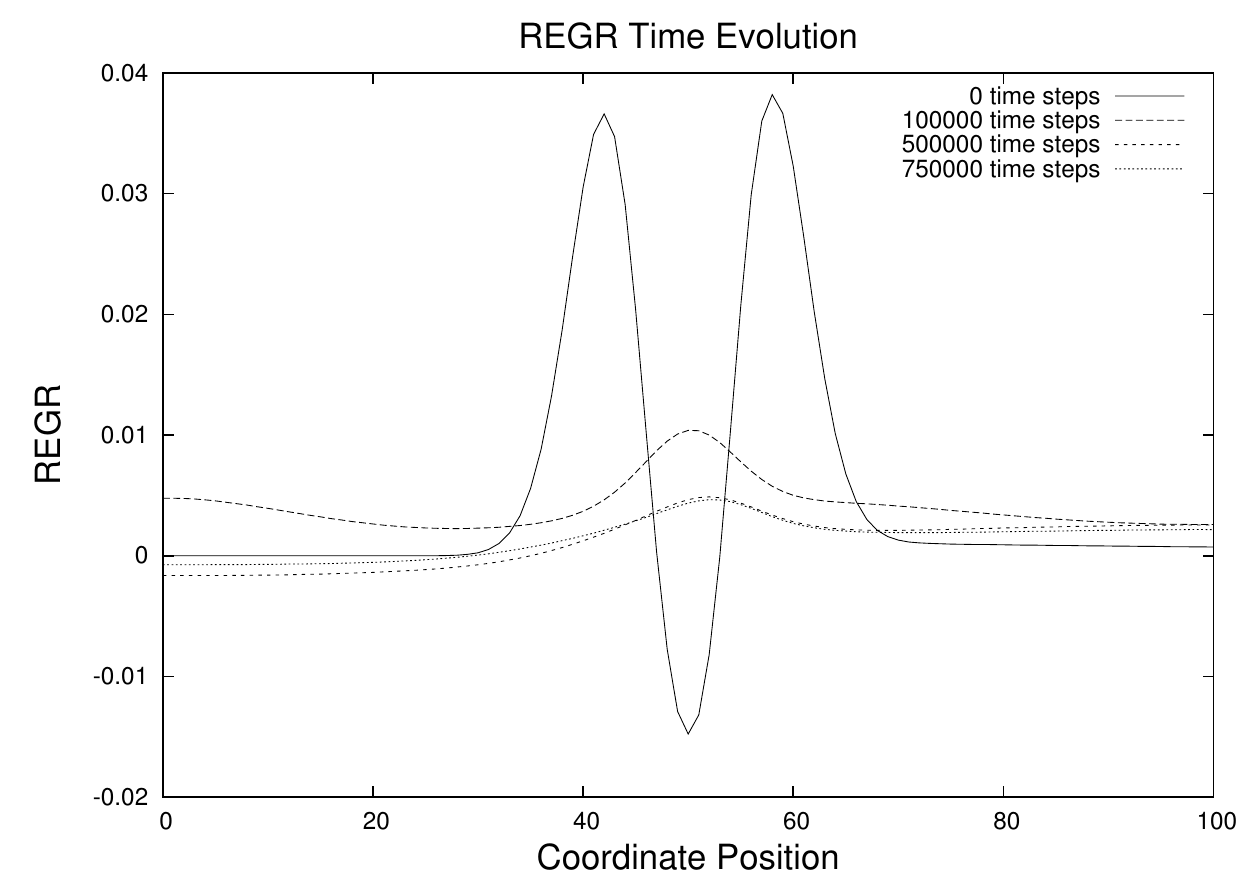}
        }
        \subfloat{
                \includegraphics[width=0.4\textwidth]{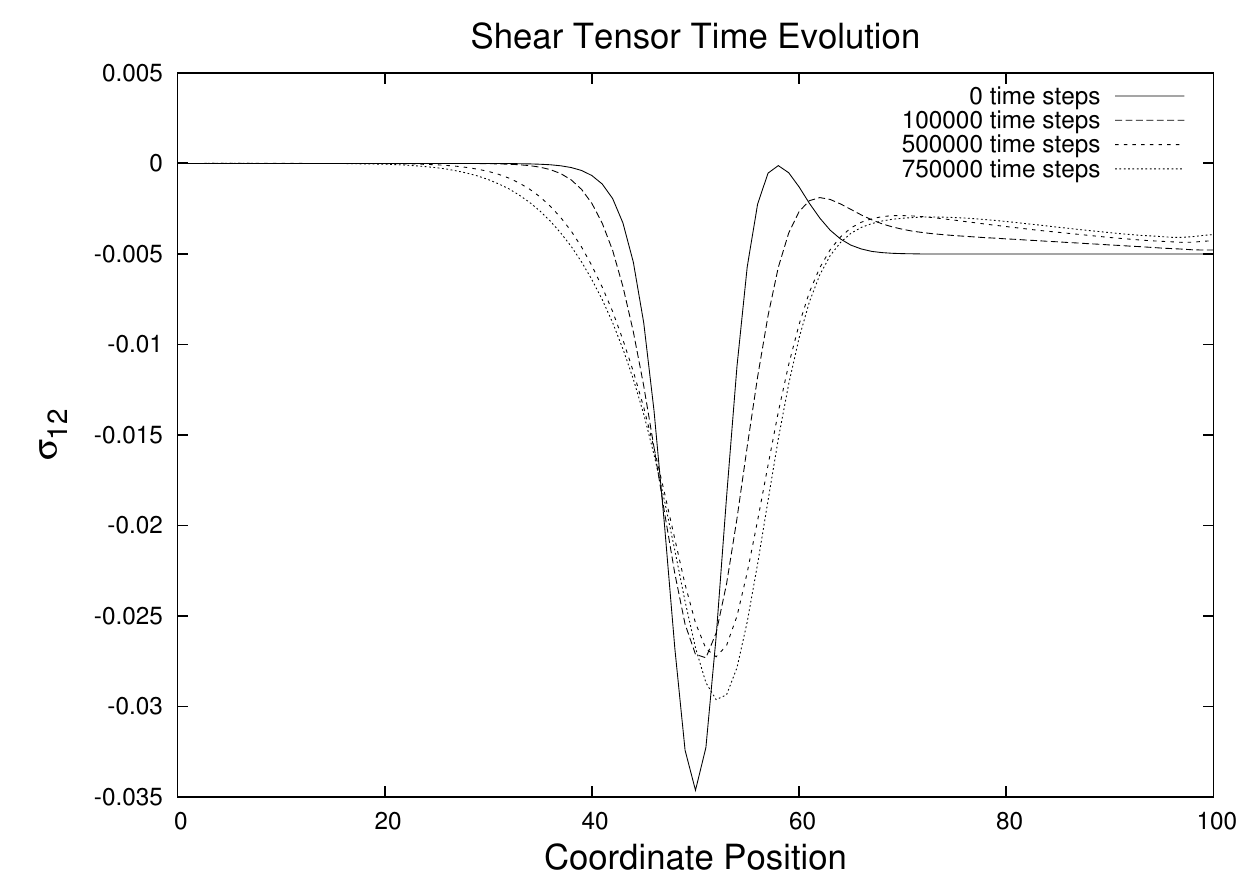}
        }

        \subfloat{
                \includegraphics[width=0.4\textwidth]{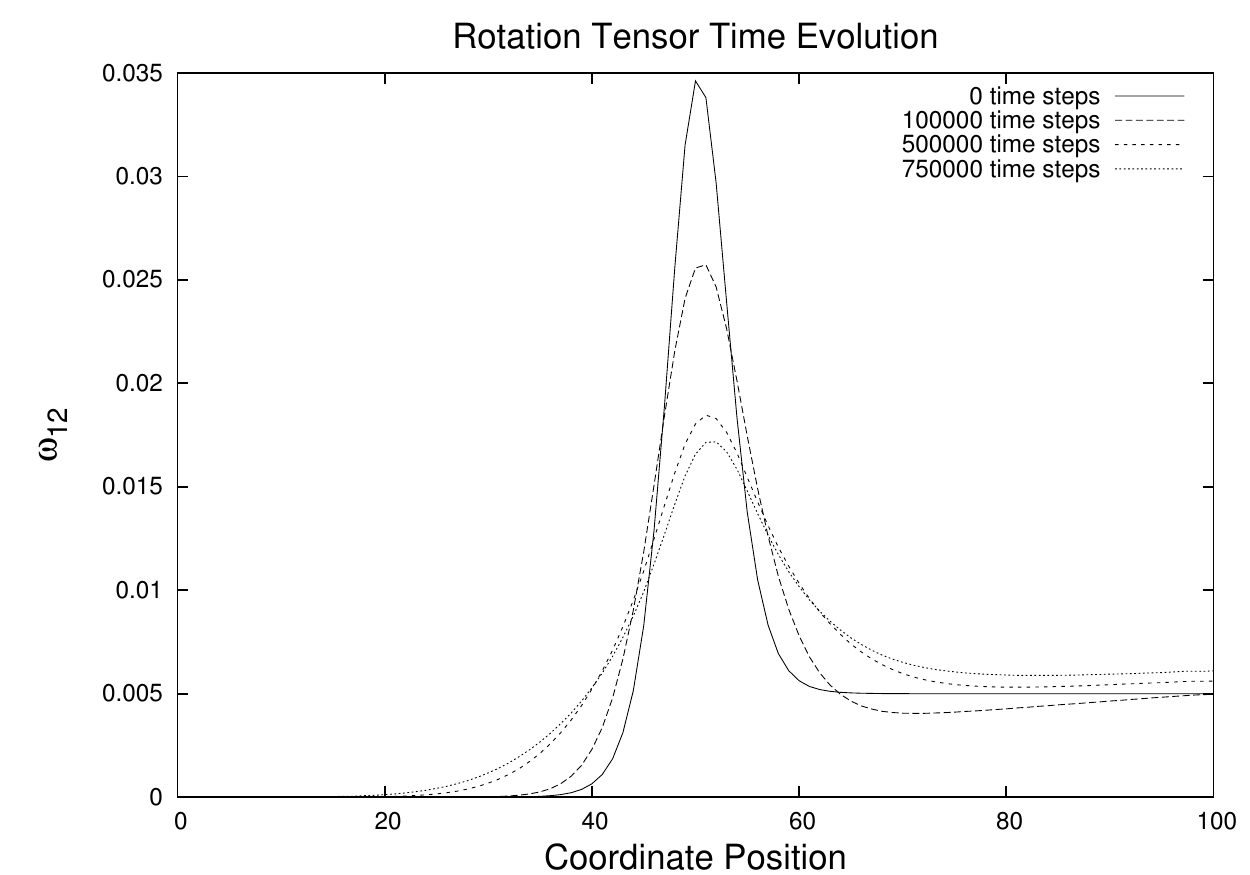}
        }
        \subfloat{
                \includegraphics[width=0.4\textwidth]{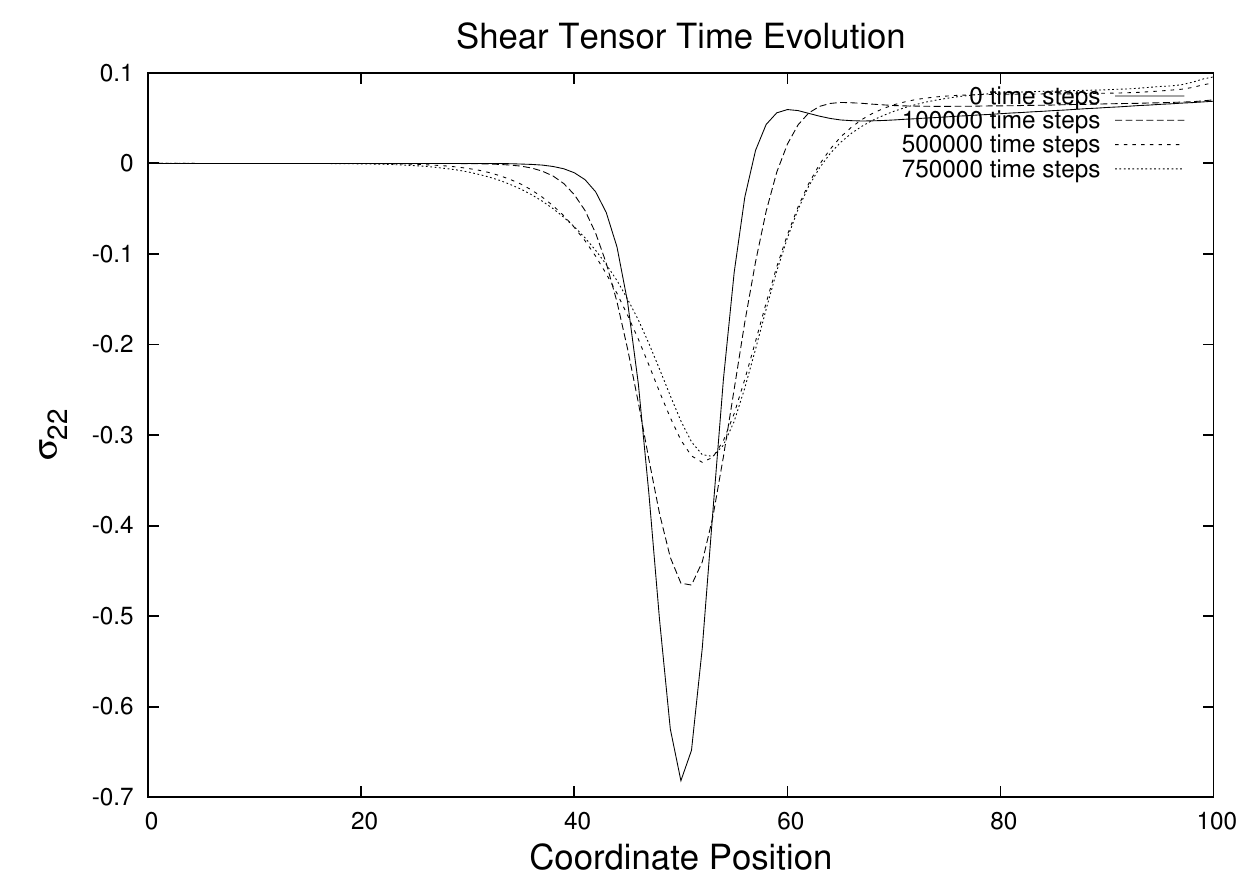}
        }
        \caption{Deformation tensors at $\theta=\pi/2$ for a $4$-gaussian initial metric and a circularly symmetric velocity field.}\label{figure4gaussDefs2} 
\end{figure}

Figures \ref{figure2gaussCurvature} to \ref{figure4gaussDefs2} show the results of these simulations.

When a circularly symmetric velocity field is coupled to the dynamics of an initially curved surface, the stagnation in
radial growth seen in Section \ref{sectionRF4gauss} is reversed. Even the shrinking solutions for the
$4$-Gaussian at $\theta=\pi/4$ are reversed dramatically (see Figure \ref{figureRF4radius}).
Additionally, the introduction of a circularly symmetric velocity field seems to smooth
out the curvature of the metric more than just the action of the Ricci flow by itself
(compare Figure \ref{figureRF4gaussCurvature} with Figure \ref{figure4gaussCurvature}). 
This again indicates that a monotonically increasing 
velocity field drives growth beyond what Ricci flow alone can accomplish.

Another important feature of the data is that the angular velocity $v^2$ starts having 
behaviour. This is not possible in isotropic simulations, and shows the influence
that the metric has on the velocity field.



\clearpage

\chapter{Make your own curved geometry!}
\label{DIYgeo}

\begin{figure}[t]
    \centering
            \includegraphics[width=1.0\textwidth]{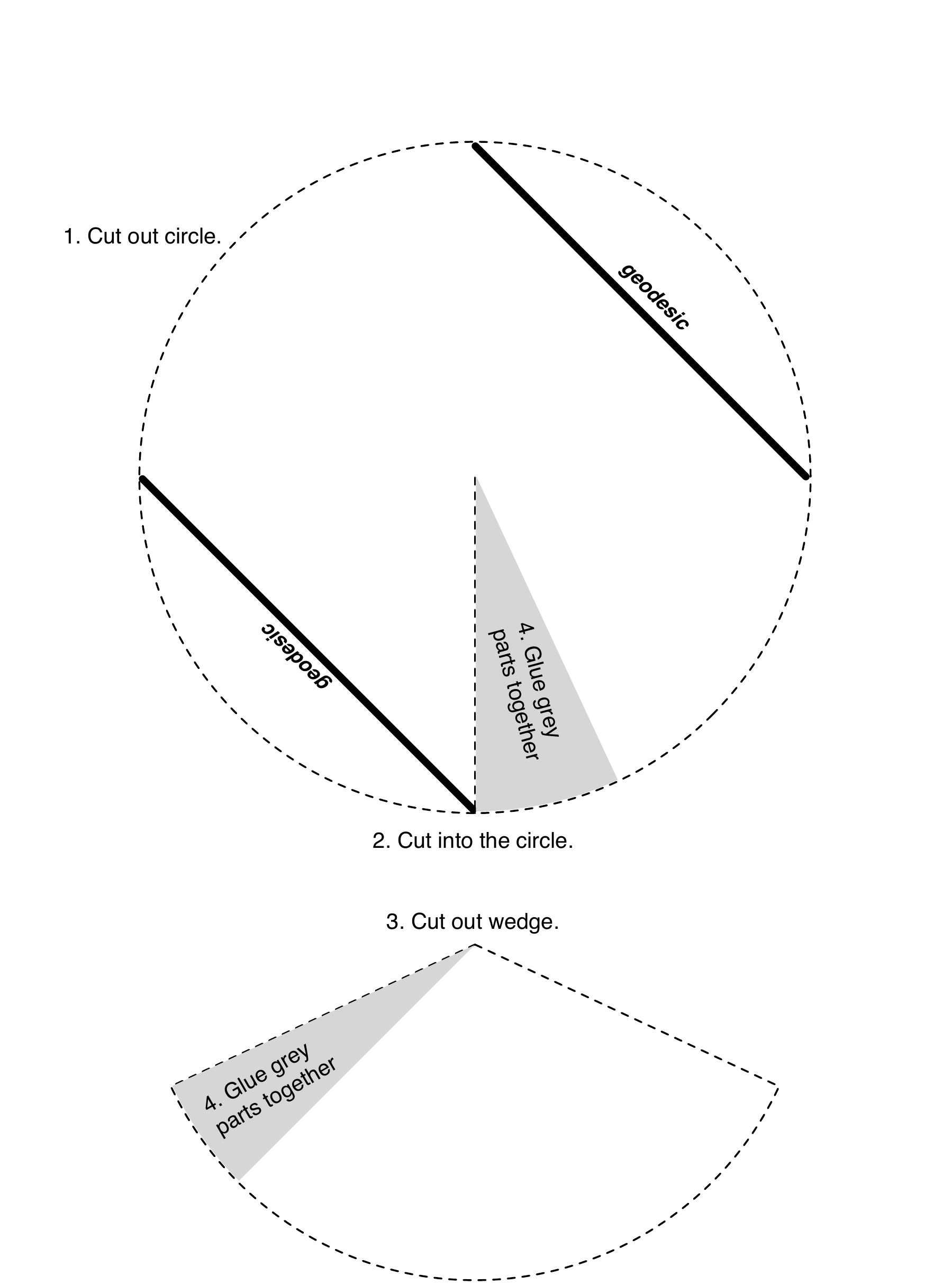}
\end{figure}

\bibliography{lit.bib}{}

\begin{thebibliography}{10}

\bibitem{turtleGeo}
Harold Abelson and Andrea DiSessa.
\newblock {\em Turtle geometry: The computer as a medium for exploring
  mathematics}.
\newblock MIT press, 1986.

\bibitem{exitFromProliferation}
Megan Andriankaja, Stijn Dhondt, Stefanie De~Bodt, Hannes Vanhaeren, Frederik
  Coppens, Liesbeth De~Milde, Per M{\"u}hlenbock, Aleksandra Skirycz, Nathalie
  Gonzalez, Gerrit~TS Beemster, et~al.
\newblock Exit from proliferation during leaf development in arabidopsis
  thaliana: A not-so-gradual process.
\newblock {\em Developmental Cell}, 22(1):64--78, 2012.

\bibitem{AudolyBoudaoud}
B~Audoly and A~Boudaoud.
\newblock {Self-similar structures near boundaries in strained systems}.
\newblock {\em {Physical Review Letters}}, {91}({8}), {Aug 22} {2003}.

\bibitem{avery1933structure}
George~S Avery~Jr.
\newblock Structure and development of the tobacco leaf.
\newblock {\em American Journal of Botany}, pages 565--592, 1933.

\bibitem{basu2012dblREGRpeak}
Paramita Basu and Anupam Pal.
\newblock A new tool for analysis of root growth in the spatio-temporal
  continuum.
\newblock {\em New Phytologist}, 195(1):264--274, 2012.

\bibitem{beemsterBaskin}
Gerrit~TS Beemster and Tobias~I Baskin.
\newblock Analysis of cell division and elongation underlying the developmental
  acceleration of root growth in arabidopsis thaliana.
\newblock {\em Plant Physiology}, 116(4):1515--1526, 1998.

\bibitem{bergmann2013topological}
Peter~G Bergmann and Venzo De~Sabbata.
\newblock {\em Topological properties and global structure of space-time}.
\newblock Springer, 2013.

\bibitem{braam2005}
Janet Braam.
\newblock In touch: plant responses to mechanical stimuli.
\newblock {\em New Phytologist}, 165(2):373--389, 2005.

\bibitem{brunet2013}
Thibaut Brunet, Adrien Bouclet, Padra Ahmadi, D{\'e}mosth{\`e}ne Mitrossilis,
  Benjamin Driquez, Anne-Christine Brunet, Laurent Henry, Fanny Serman,
  Ga{\"e}lle B{\'e}alle, Christine M{\'e}nager, et~al.
\newblock Evolutionary conservation of early mesoderm specification by
  mechanotransduction in bilateria.
\newblock {\em Nature Communications}, 4, 2013.

\bibitem{newRootDataCurvature}
Andr{\'e}s Chavarr{\'\i}a-Krauser, Kerstin~A Nagel, Klaus Palme, Ulrich Schurr,
  Achim Walter, and Hanno Scharr.
\newblock Spatio-temporal quantification of differential growth processes in
  root growth zones based on a novel combination of image sequence processing
  and refined concepts describing curvature production.
\newblock {\em New Phytologist}, 177(3):811--821, 2008.

\bibitem{geneticsOfGeometry}
E~Coen, AG~Rolland-Lagan, M~Matthews, JA~Bangham, and P~Prusinkiewicz.
\newblock {The genetics of geometry}.
\newblock {\em {Proceedings of the National Academy of Sciences}},
  {101}({14}):{4728--4735}, {Apr 6} {2004}.

\bibitem{buddedLeaves}
Etienne Couturier, Sylvain~Courrech Du~Pont, and St{\'e}phane Douady.
\newblock A global regulation inducing the shape of growing folded leaves.
\newblock {\em PloS One}, 4(11):e7968, 2009.

\bibitem{kinFlowsOnCurvedSurfaces}
Anirvan Dasgupta, Hemwati Nandan, and Sayan Kar.
\newblock Kinematics of flows on curved, deformable media.
\newblock {\em International Journal of Geometric Methods in Modern Physics},
  6(04):645--666, 2009.

\bibitem{Day2000organSizeControl}
SJ~Day and PA~Lawrence.
\newblock {Measuring dimensions: the regulation of size and shape}.
\newblock {\em {Development}}, {127}({14}):{2977--2987}, {Jul} {2000}.

\bibitem{instabilityOfGels}
Julien Dervaux and Martine~Ben Amar.
\newblock Mechanical instabilities of gels.
\newblock {\em Annu. Rev. Condens. Matter Phys.}, 3(1):311--332, 2012.

\bibitem{anisotropicDiskGrowth}
Julien Dervaux and Martine Ben~Amar.
\newblock {Morphogenesis of growing soft tissues}.
\newblock {\em {Physical Review Letters}}, {101}({6}), {Aug 8} {2008}.

\bibitem{dinverno}
Ray d'Inverno.
\newblock {\em Introducing Einstein's Relativity}.
\newblock Clarendon Press, Oxford, 1992.

\bibitem{softMatterBook}
Masao Doi.
\newblock {\em Soft Matter Physics}.
\newblock Oxford University Press, 2013.

\bibitem{cellSizesDonnelly}
Petra~M Donnelly, Dario Bonetta, Hirokazu Tsukaya, Ronald~E Dengler, and
  Nancy~G Dengler.
\newblock Cell cycling and cell enlargement in developing leaves of
  arabidopsis.
\newblock {\em Developmental Biology}, 215(2):407--419, 1999.

\bibitem{embeddingAndElasticity}
Efi Efrati, Eran Sharon, and Raz Kupferman.
\newblock The metric description of elasticity in residually stressed soft
  materials.
\newblock {\em Soft Matter}, 9(34):8187--8197, 2013.

\bibitem{SilkSciAm}
Ralph~O Erickson and Wendy~Kuhn Silk.
\newblock The kinematics of plant growth.
\newblock {\em Scientific American}, 242(5):134--151, 1980.

\bibitem{raven}
Ray~F. Evert and Susan~E. Eichhorn.
\newblock {\em Raven Biology of Plants}.
\newblock W.H. Freeman and Company, 2013.

\bibitem{green1996transductions}
Paul~B Green.
\newblock Transductions to generate plant form and pattern: an essay on cause
  and effect.
\newblock {\em Annals of botany}, 78(3):269--281, 1996.

\bibitem{Green1996Phyllotactic-pa}
PB~Green, CS~Steele, and SC~Rennich.
\newblock {Phyllotactic patterns: A biophysical mechanism for their origin}.
\newblock {\em {ANNALS OF BOTANY}}, {77}({5}):{515--527}, {MAY} {1996}.

\bibitem{rootsBook}
Peter~J Gregory.
\newblock {\em Plant roots: growth, activity and interactions with the soil}.
\newblock John Wiley \& Sons, 2008.

\bibitem{hamant2013widespread}
Olivier Hamant.
\newblock Widespread mechanosensing controls the structure behind the
  architecture in plants.
\newblock {\em Current opinion in plant biology}, 16(5):654--660, 2013.

\bibitem{harrison2012dynamic}
Lionel~G Harrison, Richard~J Adams, and David~M Holloway.
\newblock Dynamic regulation of growing domains for elongating and branching
  morphogenesis in plants.
\newblock {\em Biosystems}, 109(3):488--497, 2012.

\bibitem{hawkingBook}
Stephen Hawking.
\newblock {\em God created the integers: The mathematical breakthroughs that
  changed history}.
\newblock Running Press, 2007.

\bibitem{rootApexGTcoords}
Z~Hejnowicz and J~Karczewski.
\newblock Modeling of meristematic growth of root apices in a natural
  coordinate system.
\newblock {\em American Journal of Botany}, pages 309--315, 1993.

\bibitem{hejnowicz1984}
Z~Hejnowicz and John~A Romberger.
\newblock Growth tensor of plant organs.
\newblock {\em Journal of Theoretical Biology}, 110(1):93--114, 1984.

\bibitem{hejnowicz1995tissuesStrains}
Z~Hejnowicz and A~Sievers.
\newblock Tissue stresses in organs of herbaceous plants. i.
\newblock {\em Journal of Experimental Botany}, 46(289):1035--1035, 1995.

\bibitem{crochet}
David~W Henderson and Daina Taimina.
\newblock Crocheting the hyperbolic plane.
\newblock {\em The Mathematical Intelligencer}, 23(2):17--28, 2001.

\bibitem{hufnagelWingDisk}
Lars Hufnagel, Aurelio~A Teleman, Herv{\'e} Rouault, Stephen~M Cohen, and
  Boris~I Shraiman.
\newblock On the mechanism of wing size determination in fly development.
\newblock {\em Proceedings of the National Academy of Sciences},
  104(10):3835--3840, 2007.

\bibitem{monkGR}
Johannes Jaeger, David Irons, and Nick Monk.
\newblock Regulative feedback in pattern formation: towards a general
  relativistic theory of positional information.
\newblock {\em Development}, 135(19):3175--3183, 2008.

\bibitem{SIAMreview}
Gareth~Wyn Jones and S.~Jonathan Chapman.
\newblock {Modeling Growth in Biological Materials}.
\newblock {\em {SIAM Review}}, {54}({1}):{52--118}, {2012}.

\bibitem{Raychaudhuri}
Sayan Kar and Soumitra Sengupta.
\newblock The raychaudhuri equations: A brief review.
\newblock {\em Pramana}, 69(1):49--76, 2007.

\bibitem{kardar}
Mehran Kardar.
\newblock {\em Statistical physics of fields}.
\newblock Cambridge University Press, 2007.

\bibitem{KPZ}
Mehran Kardar, Giorgio Parisi, and Yi-Cheng Zhang.
\newblock Dynamic scaling of growing interfaces.
\newblock {\em Physical Review Letters}, 56(9):889, 1986.

\bibitem{GRNmodeling}
Guy Karlebach and Ron Shamir.
\newblock Modelling and analysis of gene regulatory networks.
\newblock {\em Nature Reviews Molecular Cell Biology}, 9(10):770--780, 2008.

\bibitem{genesAndGeoModel}
Richard Kennaway, Enrico Coen, Amelia Green, and Andrew Bangham.
\newblock Generation of diverse biological forms through combinatorial
  interactions between tissue polarity and growth.
\newblock {\em PLoS Computational Biology}, 7(6):e1002071, 2011.

\bibitem{thinSheetShapes}
Yael Klein, Efi Efrati, and Eran Sharon.
\newblock {Shaping of elastic sheets by prescription of non-Euclidean metrics}.
\newblock {\em {Science}}, {315}({5815}):{1116--1120}, {Feb 23} {2007}.

\bibitem{RDinBio}
Shigeru Kondo and Takashi Miura.
\newblock Reaction-diffusion model as a framework for understanding biological
  pattern formation.
\newblock {\em Science}, 329(5999):1616--1620, 2010.

\bibitem{kuchen2012generation}
Erika~E Kuchen, Samantha Fox, Pierre~Barbier de~Reuille, Richard Kennaway,
  Sandra Bensmihen, Jerome Avondo, Grant~M Calder, Paul Southam, Sarah
  Robinson, Andrew Bangham, et~al.
\newblock Generation of leaf shape through early patterns of growth and tissue
  polarity.
\newblock {\em Science}, 335(6072):1092--1096, 2012.

\bibitem{PDGinSAM}
Dorota Kwiatkowska.
\newblock Structural integration at the shoot apical meristem: models,
  measurements, and experiments.
\newblock {\em American Journal of Botany}, 91(9):1277--1293, 2004.

\bibitem{hamant2013microtubles}
Beno{\^\i}t Landrein and Olivier Hamant.
\newblock How mechanical stress controls microtubule behavior and morphogenesis
  in plants: history, experiments and revisited theories.
\newblock {\em The Plant Journal}, 75(2):324--338, 2013.

\bibitem{feelingGreen}
Gabriele~B Monshausen and Simon Gilroy.
\newblock Feeling green: mechanosensing in plants.
\newblock {\em Trends in cell biology}, 19(5):228--235, 2009.

\bibitem{rootGTnakielski}
Jerzy Nakielski.
\newblock The tensor-based model for growth and cell divisions of the root
  apex. i. the significance of principal directions.
\newblock {\em Planta}, 228(1):179--189, 2008.

\bibitem{genesAndCurvature}
Utpal Nath, Brian~CW Crawford, Rosemary Carpenter, and Enrico Coen.
\newblock Genetic control of surface curvature.
\newblock {\em Science}, 299(5611):1404--1407, 2003.

\bibitem{niklas1992plantMechBook}
Karl~J Niklas.
\newblock {\em Plant biomechanics: an engineering approach to plant form and
  function}.
\newblock University of Chicago press, 1992.

\bibitem{NumericsBook}
Richard~H Pletcher, John~C Tannehill, and Dale Anderson.
\newblock {\em Computational Fluid Mechanics and Heat Transfer}.
\newblock CRC Press, 2012.

\bibitem{PrusinkiewiczConstraints}
Przemyslaw Prusinkiewicz and Pierre~Barbier de~Reuille.
\newblock {Constraints of space in plant development}.
\newblock {\em {Journal of Experimental Botany}}, {61}({8}):{2117--2129}, {May}
  {2010}.

\bibitem{RHB}
Kenneth~Franklin Riley, Michael~Paul Hobson, and Stephen~John Bence.
\newblock {\em Mathematical methods for physics and engineering: a
  comprehensive guide}.
\newblock Cambridge University Press, 2006.

\bibitem{stressDecomp}
Edward~K Rodriguez, Anne Hoger, and Andrew~D McCulloch.
\newblock Stress-dependent finite growth in soft elastic tissues.
\newblock {\em Journal of Biomechanics}, 27(4):455--467, 1994.

\bibitem{genesAndShape}
Anne-Gaelle Rolland-Lagan, J~Andrew Bangham, and Enrico Coen.
\newblock Growth dynamics underlying petal shape and asymmetry.
\newblock {\em Nature}, 422(6928):161--163, 2003.

\bibitem{blackHoleEmbedding}
Joseph~D Romano and Richard~H Price.
\newblock Embedding initial data for black-hole collisions.
\newblock {\em Classical and Quantum Gravity}, 12(3):875, 1995.

\bibitem{quantumPhotosynthesis}
Elisabet Romero, Ramunas Augulis, Vladimir~I Novoderezhkin, Marco Ferretti, Jos
  Thieme, Donatas Zigmantas, and Rienk Van~Grondelle.
\newblock Quantum coherence in photosynthesis for efficient solar-energy
  conversion.
\newblock {\em Nature Physics}, 10(9):676--682, 2014.

\bibitem{3Ddress}
Jessica Rosenkrantz and Jesse Louis-Rosenberg.
\newblock Kinematics dress, 2014.

\bibitem{leafBulkModulus}
Takami Saito, Kouichi Soga, Takayuki Hoson, and Ichiro Terashima.
\newblock The bulk elastic modulus and the reversible properties of cell walls
  in developing quercus leaves.
\newblock {\em Plant and cell physiology}, 47(6):715--725, 2006.

\bibitem{acetabularia}
Kyle~A Serikawa and Dina~F Mandoli.
\newblock An analysis of morphogenesis of the reproductive whorl of
  acetabularia acetabulum.
\newblock {\em Planta}, 207(1):96--104, 1998.

\bibitem{leavesFlowersPlasticRuffling}
E~Sharon, M~Marder, and HL~Swinney.
\newblock {Leaves, flowers and garbage bags: Making waves}.
\newblock {\em {American Scientist}}, {92}({3}):{254--261}, {May-Jun} {2004}.

\bibitem{bucklingCascades}
E~Sharon, B~Roman, M~Marder, GS~Shin, and HL~Swinney.
\newblock {Mechanics: Buckling cascades in free sheets - Wavy leaves may not
  depend only on their genes to make their edges crinkle.}
\newblock {\em {Nature}}, {419}({6907}):{579}, {Oct 10} {2002}.

\bibitem{NEPmechanics}
Eran Sharon and Efi Efrati.
\newblock {The mechanics of non-Euclidean plates}.
\newblock {\em {Soft Matter}}, {6}({22}):{5693--5704}, {2010}.

\bibitem{wrinklingByGeometry}
Eran Sharon, Benoit Roman, and Harry~L. Swinney.
\newblock {Geometrically driven wrinkling observed in free plastic sheets and
  leaves}.
\newblock {\em {Physical Review E}}, {75}({4, 2}), {Apr} {2007}.

\bibitem{cornRootGrowth}
Robert~E Sharp, Wendy~Kuhn Silk, and Theodore~C Hsiao.
\newblock Growth of the maize primary root at low water potentials i. spatial
  distribution of expansive growth.
\newblock {\em Plant Physiology}, 87(1):50--57, 1988.

\bibitem{silk2014deposition}
Wendy~K Silk and Marie-B{\'e}atrice Bogeat-Triboulot.
\newblock Deposition rates in growing tissue: Implications for physiology,
  molecular biology, and response to environmental variation.
\newblock {\em Plant and Soil}, 374(1-2):1--17, 2014.

\bibitem{silk1979}
Wendy~Kuhn Silk and Ralph~O Erickson.
\newblock Kinematics of plant growth.
\newblock {\em Journal of Theoretical Biology}, 76(4):481--501, 1979.

\bibitem{contMechBook}
Anthony James~Merrill Spencer.
\newblock {\em Continuum mechanics}.
\newblock Courier Corporation, 2004.

\bibitem{godin2008}
Szymon Stoma, Mikael Lucas, J{\'e}r{\^o}me Chopard, Marianne Schaedel, Jan
  Traas, and Christophe Godin.
\newblock Flux-based transport enhancement as a plausible unifying mechanism
  for auxin transport in meristem development.
\newblock {\em PLoS Computational Biology}, 4(10):e1000207, 2008.

\bibitem{sugimoto2003big}
Keiko Sugimoto-Shirasu and Keith Roberts.
\newblock ``big it up'': endoreduplication and cell-size control in plants.
\newblock {\em Current Opinion in Plant Biology}, 6(6):544--553, 2003.

\bibitem{taiz}
Lincoln Taiz, Eduardo Zeiger, Ian~Max M{\o}ller, and Angus Murphy.
\newblock {\em Plant Physiology and Development}.
\newblock Sinauer Associates, Incorporated, 2015.

\bibitem{thompson1917}
D.W. Thompson.
\newblock {\em On Growth and Form}.
\newblock Cambridge University Press, 1917.

\bibitem{turingTest}
Nathan Tompkins, Ning Li, Camille Girabawe, Michael Heymann, G~Bard Ermentrout,
  Irving~R Epstein, and Seth Fraden.
\newblock Testing turing's theory of morphogenesis in chemical cells.
\newblock {\em Proceedings of the National Academy of Sciences},
  111(12):4397--4402, 2014.

\bibitem{toppingRF}
Peter Topping.
\newblock {\em Lectures on the Ricci flow}, volume 325.
\newblock Cambridge University Press, 2006.

\bibitem{turing1952}
Alan~Mathison Turing.
\newblock The chemical basis of morphogenesis.
\newblock {\em Philosophical Transactions of the Royal Society of London.
  Series B, Biological Sciences}, 237(641):37--72, 1952.

\bibitem{rootImaging}
Corine~M van~der Weele, Hai~S Jiang, Krishnan~K Palaniappan, Viktor~B Ivanov,
  Kannappan Palaniappan, and Tobias~I Baskin.
\newblock A new algorithm for computational image analysis of deformable motion
  at high spatial and temporal resolution applied to root growth. roughly
  uniform elongation in the meristem and also, after an abrupt acceleration, in
  the elongation zone.
\newblock {\em Plant Physiology}, 132(3):1138--1148, 2003.

\bibitem{rootImagingDoublePeakedGT}
A~Walter, H~Spies, S~Terjung, R~K{\"u}sters, N~Kirchge{\ss}ner, and U~Schurr.
\newblock Spatio-temporal dynamics of expansion growth in roots: automatic
  quantification of diurnal course and temperature response by digital image
  sequence processing.
\newblock {\em Journal of Experimental Botany}, 53(369):689--698, 2002.

\bibitem{inkjetLeaves}
Lisheng Wang, Simon~T Beyer, Quentin~CB Cronk, and Konrad Walus.
\newblock Delivering high-resolution landmarks using inkjet micropatterning for
  spatial monitoring of leaf expansion.
\newblock {\em Plant methods}, 7(1):1--10, 2011.

\end{thebibliography}
\bibliographystyle{plain}

\end{document}